\documentclass[a4paper,10pt]{article}
\usepackage{amsmath, amsthm, amssymb}
\usepackage{epsfig}
\usepackage{longtable}
\usepackage{graphicx}
\usepackage{amssymb}
\usepackage{epstopdf}
\usepackage{enumerate}
\usepackage{subfig}
\usepackage{setspace}
\usepackage{undertilde}
\usepackage[ot2enc]{inputenc}
\usepackage[russian,english]{babel}
\usepackage[top=3.75cm,bottom=3.75cm,right=3cm,left=3cm,bindingoffset=1cm,twoside=true]{geometry}
\newtheorem{theorem}{Theorem}[section]
\newtheorem{lemma}[theorem]{Lemma}
\newtheorem{corollary}[theorem]{Corollary}
\newtheorem{proposition}[theorem]{Proposition}
\newtheorem{conjecture}[theorem]{Conjecture}
\theoremstyle{definition}
\newtheorem{definition}[theorem]{Definition}

\theoremstyle{remark}
\newtheorem{remark}[theorem]{Remark}

\newcommand{\1}{\mathbf{1}}
\newcommand{\0}{\mathbf{0}}
\newcommand{\bZ}{\mathbf{Z}}
\newcommand{\Tr}{\mathrm{Tr}}
\newcommand{\Pf}{\mathrm{Pf}}
\newcommand{\sgn}{\mathrm{sgn}}
\newcommand{\vol}{\mathrm{vol}}
\newcommand{\erf}{\mathrm{erf}}
\newcommand{\erfc}{\mathrm{erfc}}
\newcommand{\qdet}{\mathrm{qdet}}
\newcommand{\bulk}{\mathrm{bulk}}
\newcommand{\edge}{\mathrm{edge}}
\newcommand{\fodd}{\mathbf{f}_{\mathrm{odd}}}
\newcommand{\odd}{\mathrm{odd}}
\newcommand{\Orb}{\mathrm{Orb}}
\newcommand\T{\rule[-1.4ex]{0.0pt}{4ex}}
\newcommand{\bX}{\mathbf{X}}
\newcommand{\bR}{\mathbf{R}}
\newcommand{\bA}{\mathbf{A}}
\newcommand{\bB}{\mathbf{B}}
\newcommand{\bC}{\mathbf{C}}
\newcommand{\bM}{\mathbf{M}}
\newcommand{\bP}{\mathbf{P}}
\newcommand{\bH}{\mathbf{H}}
\newcommand{\bT}{\mathbf{T}}
\newcommand{\bL}{\mathbf{L}}
\newcommand{\bQ}{\mathbf{Q}}
\newcommand{\bO}{\mathbf{O}}
\newcommand{\bD}{\mathbf{D}}
\newcommand{\bS}{\mathbf{S}}
\newcommand{\bE}{\mathbf{E}}
\newcommand{\bF}{\mathbf{F}}
\newcommand{\bG}{\mathbf{G}}

\newcommand{\bY}{\mathbf{Y}}
\newcommand{\bK}{\mathbf{K}}

\newcommand{\bt}{\mathbf{t}}
\newcommand{\bb}{\mathbf{b}}
\newcommand{\bc}{\mathbf{c}}
\newcommand{\be}{\mathbf{e}}
\newcommand{\bg}{\mathbf{g}}

\numberwithin{table}{subsection}

\allowdisplaybreaks[1]

\newcommand{\fullsub}[4]{
\begin{array}{ccc}
\vspace{-6pt}{#1}\hspace{-9pt}&{#2}&\hspace{-9pt}{#4}\\
&{}_{#3}&
\end{array}
}

\begin{document}


\section*{}
\pagenumbering{roman}
\thispagestyle{empty}

\begin{centering}
\vspace{60pt}A GEOMETRICAL TRIUMVIRATE OF REAL RANDOM MATRICES

\vspace{12pt} Anthony Mays

\vspace{280pt}Submitted in total fulfilment of the requirements of the degree of Doctor of Philosophy

\vspace{50pt}November 2011

\vspace{12pt}Department of Mathematics and Statistics

University of Melbourne

\vspace{80pt} {\small PRODUCED ON ARCHIVAL QUALITY PAPER}

\end{centering}

\newpage
\thispagestyle{empty}
${}$

\newpage

\section*{Abstract}
\addcontentsline{toc}{section}{Abstract}

The eigenvalue correlation functions for random matrix ensembles are fundamental descriptors of the statistical properties of these ensembles. In this work we present a five-step method for the calculation of these correlation functions, based upon the method of (skew-) orthogonal polynomials. This scheme systematises existing methods and also involves some new techniques. By way of illustration we apply the scheme to the well known case of the Gaussian orthogonal ensemble, before moving on to the real Ginibre ensemble. A generalising parameter is then introduced to interpolate between the GOE and the real Ginibre ensemble. These real matrices have orthogonal symmetry, which is known to lead to Pfaffian or quaternion determinant processes, yet Pfaffians and quaternion determinants are not defined for odd-sized matrices. We present two methods for the calculation of the correlation functions in this case: the first is an extension of the even method, and the second establishes the odd case as a limit of the even case.

Having demonstrated our methods by reclaiming known results, we move on to study an ensemble of matrices $\bY = \bA^{-1} \bB$, where $\bA$ and $\bB$ are each real Ginibre matrices. This ensemble is known as the \textit{real spherical ensemble}. By a convenient fractional linear transformation, we map the eigenvalues into the unit disk to obtain a rotationally invariant distribution of eigenvalues. The correlation functions are then calculated in terms of these new variables by means of finding the relevant skew-orthogonal polynomials. The expected number of real eigenvalues is computed, as is the probability of obtaining any number of real eigenvalues; the latter is compared to numerical simulation. The generating function for these probabilities is given by an explicit factorised polynomial, in which the zeroes are gamma functions.

We show that in the limit of large matrix dimension, the eigenvalues (after stereographic projection) are uniformly distributed on the sphere, a result which is part of a universality result called the \textit{spherical law}. By taking a different limit, we also show that the local behaviour of the eigenvalues matches that of the real Ginibre ensemble, which corresponds to the planar limit of the sphere.

Lastly, we examine the third ensemble in the triumvirate, the \textit{real truncated ensemble}, which is formed by truncating $L$ rows and columns from an $N\times N$ Haar distributed orthogonal matrix. By applying the five-step scheme and by averaging over characteristic polynomials we proceed to calculate correlation functions and probabilities analogously to the other ensembles considered in this work. The probabilities of obtaining real eigenvalues are again compared to numerical simulation. In the large $N$ limit (with small $L$) we find that the eigenvalues are uniformly distributed on the anti-sphere (after being suitably projected). This leads to a conjecture that, analogous to the circular law and the spherical law, there exists an \textit{anti-spherical law}. As we found for the spherical ensemble, we also find that in some limits the behaviour of this ensemble matches that of the real Ginibre ensemble.

\newpage

\section*{Declaration}
\addcontentsline{toc}{section}{Declaration}

This is to certify that:

\begin{description}
\item[(\textit{i})] {the thesis comprises only my original work towards the PhD except where indicated in the Preface,}
\item[(\textit{ii})]{due acknowledgement has been made in the text to all other material used,}
\item[(\textit{iii})]{the thesis is fewer than 100,000 words in length, exclusive of tables, maps, bibliographies and appendices.}
\end{description}

\vspace{80pt} Signed,

\vspace{60pt} \hspace{160pt} ANTHONY MAYS

\newpage

\section*{Preface}
\addcontentsline{toc}{section}{Preface}

The odd-dimensional method of Chapters \ref{sec:odd_from_even} and \ref{sec:Gin_oddfromeven} was joint work with Peter Forrester, and began as an offshoot of his Australian Research Council (ARC) project on the integrability aspects of random matrix theory. Our work was originally published in \cite{FM09}.

Most of the content of Chapter \ref{sec:SOE} was also work with Peter Forrester, which was developed collaboratively and published jointly in \cite{FM11}. I have since reworked the paper into the present format so that it coheres with the overall structure of the thesis.

The remainder of the thesis is largely an attempt to unify various existing methods in the field, and as such a number of prior results have been included. This will hopefully have the additional benefit of providing a useful self-contained resource for students and others. Any previous results are, of course, clearly identified as such and the original references are cited.

\newpage

\section*{Acknowledgements}
\addcontentsline{toc}{section}{Acknowledgements}

Firstly, thanks are due to the staff of the Department of Mathematics and Statistics, University of Melbourne for the use of their Research Support Scheme and for providing the day-to-day needs of Ph.D. study life. For financial income, I am grateful for the support of the Australia Postgraduate Award.

Thanks also to those other groups who supported me during my studies: Australian Mathematical Sciences Institute (AMSI) for their summer and winter schools; Erwin\\Schr\"{o}dinger Institute (ESI), Vienna; Mathematical Sciences Research Institute (MSRI), Berkeley; American Institute of Mathematics (AIM), Palo Alto; and the University of Oregon, Eugene. In relation to the latter, particular thanks are given to Chris Sinclair for arranging the logistics of my trip and for his hospitality and stimulating discussions.

Thanks to Jonith Fischmann for pointing out corrections to Chapter \ref{sec:SOEcharpolys}, Craig Hodgson for a short tutorial about the Stiefel manifold, Peter Paule for providing a Mathematica version of the Zeilberger algorithm, and James Garza for housing me in Southern California. An especially big thanks to Anita Ponsaing for putting up with all my clowning. She was always ready to listen and help out, but most importantly she's been a great friend with whom I've enjoyed travelling around the country and around the world.

As my secondary supervisor, Jan de Gier was not intimately involved in my research, however he was often able to assist me on more general questions, as well as help with conference attendance and discuss future career prospects. He was also the consummate BBQ host.

But the major portion of my gratitude is reserved for my primary supervisor, Peter Forrester, who, after taking me on as an honours student who just wanted to juggle, continued with me into a Ph.D. on random matrix theory. There were very few instances, if any, where he could not provide truly insightful comments that would clear up muddled understanding and point the way forward. His guidance paved the road of understanding to this remarkably interesting field.

\newpage

\tableofcontents

\newpage
\listoffigures
\listoftables
\newpage

\pagenumbering{arabic}
\section{Introduction}

One feels the need to start at the beginning, by defining what a random matrix is. At its broadest, the term is self-explanatory: we pick a matrix at random (using a specific distribution) from a set of matrices. This set is defined by some desired attributes of the matrices, such as Hermitian, orthogonal, non-singular or Gaussian distributed entries. To study a `typical' matrix from the set, one thinks of picking a matrix randomly from an \textit{ensemble} of all matrices having the particular attributes of interest. However, as we are cautioned in \cite{Edelman1993}, we should not confuse a `typical' random matrix with `any old' matrix; the matrices under study here have a very rich structure. A short and very readable general introductory review of the field is found in \cite{Diaconis2005}, while \cite{ForSnaVer2003, MezSna2005, AkeBaiDiF2011} contain reviews of a more technical nature and \cite{GuhMGWei1997, Diaconis2003, deift2007} focus on the applications of random matrices. Standard texts include \cite{Deift2000, mehta2004, BaiSilver2006, Deift2009, AndGuiZei2009, forrester?}.

The main focus of the field of random matrix theory is to analyse the eigenvalue distribution of the ensemble, although the behaviour of the eigenvectors may also be of interest. We remind the reader that the eigenvalues of a matrix $\bA$ are the set of $\lambda$ that satisfy the equation $\det (\bA-\lambda \1)=0$, where $\1$ is the identity matrix. This determinant is a polynomial in $\lambda$, called the \textit{characteristic polynomial} of the matrix $\bA$, and so the eigenvalues of a matrix are also the zeroes of its corresponding characteristic polynomial. From this observation, we expect that there should be a close correspondence between results concerning eigenvalue distributions in random matrix theory and those of the distributions of zeroes in the theory of random polynomials. Indeed, this turns out to be true, although the relationship extends far beyond the characteristic polynomial. We will not pursue random polynomial theory here; the interested reader should see \cite{HKPV2009} and references therein.

It may be expected \textit{a priori} that the eigenvalues of a random matrix are scattered uniformly at random over their support (exhibiting the `clumpy' patterns typical of such data), however this is far from true and they instead display strongly correlated behaviour. In this thesis we develop a method for calculating correlation functions for several ensembles of matrices with real elements. Many of the results are new, although, since it is our hope that this work will form a useful part of the reference literature for those working with real random matrices, we have attempted to provide a self-contained treatment, which explains its voluminous nature. Our original contributions include: a streamlined method for the calculation of the correlation kernel, for both even- and odd-sized matrices (unpublished) in Chapters \ref{sec:GOE_steps} and \ref{sec:GOE_odd}; an alternative method for calculating correlation functions for odd-sized matrices \cite{FM09} in Chapter \ref{sec:odd_from_even}; and the calculation of the correlation functions for the real spherical ensemble \cite{FM11} in Chapter \ref{sec:SOE}. The methods developed and presented here have also been applied in the papers \cite{ForSinc2010} and \cite{Forrester2010a}. We have also provided various reworkings and reinterpretations of known results, as well as calculations and simulations of the probability of obtaining some number of real eigenvalues.

The study of random matrices can be traced back to Hurwitz \cite{Hurwitz1898} (which is included in \cite{Hurwitz1933}) where he presented a parameterisation of the orthogonal group and then computed its volume form in terms of generalized Euler angles, which are a standard set of co-ordinates describing the rotation of one co-ordinate frame relative to another. In \cite{PozZycKus1998} the authors discuss Hurwitz's parameterisation and then use it as a practical way to generate random orthogonal matrices.

A significant milestone was passed in 1928 with the paper by Wishart \cite{Wishart1928}, where the purpose of his study was to analyse the estimated variance of an underlying population by taking $N$ samples from the population. If one writes the normalised, centred variables as a vector $\utilde{x}=[(x_j-\bar{x}) /\sqrt{N}]_{j=1,...,N}$, where $\bar{x}$ is the mean of the sample, then the variance is given by $\utilde{x}^T \cdot \utilde{x}$. However, when there are multiple variates $x_j^{(1)}, x_j^{(2)},...,x_j^{(M)}$, then one needs to consider all possible dot products $\utilde{x}^{(l)}{}^T \cdot \utilde{x}^{(m)}$, $l,m =1,...,M$ of the corresponding vectors. These dot products can be conveniently written in the form $\bX^T \bX$, where $\bX$ is an $N\times M$ matrix with these vectors forming the columns, a structure which has become known as a Wishart matrix. Wishart's contribution was to find the distribution of these variances for general $M$; he did so by adapting a geometrical technique that had previously been used by Fisher to establish the $M=2$ case in \cite{Fisher1915}.

\begin{figure}[htp]
\begin{center}
\subfloat[]{\label{fig:PoiDist} \includegraphics[scale=0.7]{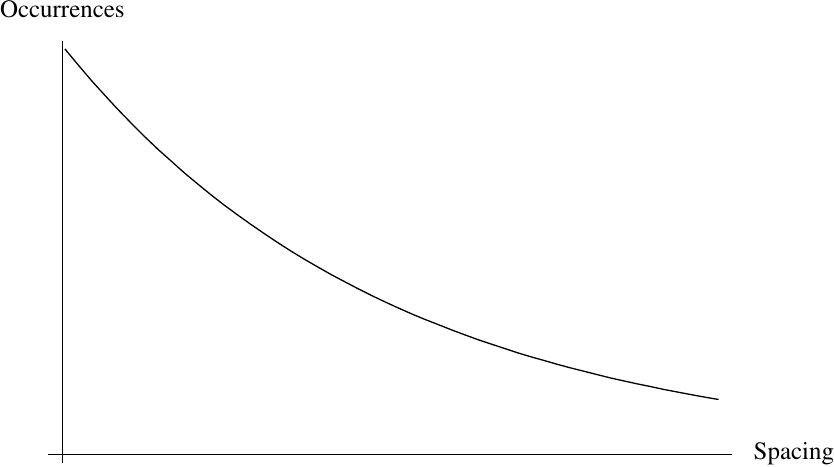}}
\qquad \qquad \subfloat[]{\label{fig:RMTDist} \includegraphics[scale=0.7]{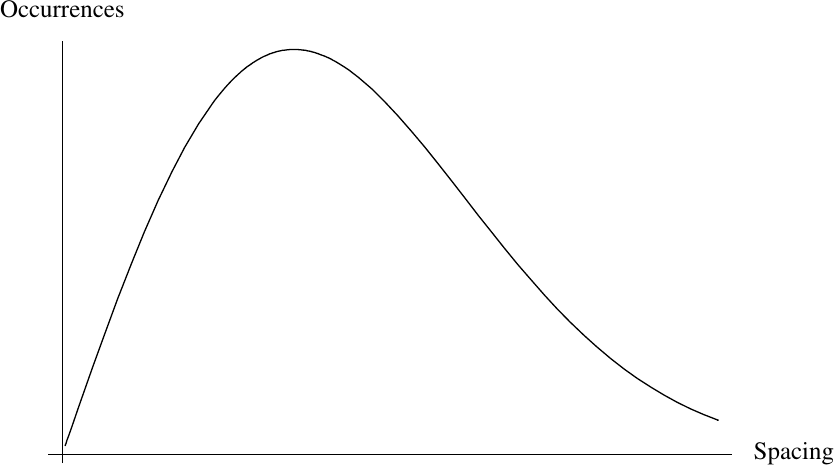}}
\caption[Typical spacing distributions for Poisson and random matrix processes.]{Spacing distributions for a) $e^{-x}$, representing a Poisson process, and b) $x {e^{-x^2}}$, representing a random matrix process.}
\end{center}
\end{figure}

One of the major technical achievements in the field came in 1939 with the (more or less) simultaneous calculation of various Jacobians, showing that they depend on a product of differences \cite{Fisher1939, Hsu1939, Roy1939, Girshick1939, Mood1951},
\begin{align}
\label{eqn:evalbeta} \prod_{1\leq j < k \leq N} |\lambda_k-\lambda_j|^{\beta}
\end{align}
(see \cite{Anderson2007} for a review of these calculations and a discussion of the timing of their publication). A Vandermonde factor in the eigenvalue distribution can be interpreted as repulsion between eigenvalues, where $\beta$ is the `strength' of the repulsion between them. This repulsion implies that the eigenvalues will tend to be more evenly spread over the support than if there were no interaction. In the latter case, where they are independent, then we have a Poisson process and one expects a spacing distribution like that in Figure \ref{fig:PoiDist}, in which case clumping of the points tends to occur. Numerical simulations on real symmetric matrices confirmed that the eigenvalues are inclined to repel \cite{Ros1958}, leading to a spacing distribution like Figure \ref{fig:RMTDist}, which turns out to be characteristic of determinantal processes, of which random matrix eigenvalue distributions is an example.

Interest in random matrices within the physics community began with Eugene Wigner in the 1950s. The problem being faced at the time was the analysis of the highly excited states of heavy nuclei. Modelling the problem as a set of interacting particles quickly leads to a set of unwieldy coupled equations. Rather, Wigner \cite{Wigner1951} suggested that a statistical approach might be more useful, and he conjectured that the distribution of the spacing between energy levels will be well approximated by the eigenvalue spacing distribution of large symmetric matrices (this statement became known as the \textit{Wigner surmise}) \cite{Wigner1957a, Wigner1957b}. This suggestion was based upon the physical reasoning that the nuclear energy levels corresponding to the same spin should repel, and that for small spacing the number of spacings should be approximately linearly dependent on the spacing distance (giving a graph something like Figure \ref{fig:RMTDist}); this expectation was also proposed by Landau and Smorodinsky \cite[Lecture 7]{LanSmo1955}. Experimental results such as those in \cite{PorTho1956, GurPev1956, BluPor1958} confirmed that this was true. In \cite{PorRos1960, RosPor1960} the authors comprehensively demonstrated that the repulsive nature of the energy levels could indeed be modeled by eigenvalues of symmetric matrices, and that the results matched Wigner's predictions. Further, they demonstrated that atomic spectra obey a similar repulsion, which was also confirmed in \cite{CaGe1983}, although the evidence tends to be less convincing than that of the nuclear levels. (For more evidence on nuclear energy levels and random matrix distributions, see \cite{BHP1983}.)

\begin{remark}
An excellent reference for historical information on nuclear and atomic spectra is \cite{Porter1965}, where many of these seminal papers are collected along with an introductory review of the theory by the editor.
\end{remark}

Random matrix eigenvalue distributions have also been compared to other quantum systems in a similar spirit; for instance, in \cite{Bohi1984} the spacing of eigenvalues in the quantum Sinai billiard are found to agree with those of a certain random matrix ensemble (the Gaussian orthogonal ensemble, to be introduced below). Another interpretation of these eigenvalues is as a Coulomb gas within a confining background potential; a viewpoint that Forrester employs in \cite{forrester?}, following Dyson \cite{dyson1962b}. Outside the arena of physics, a random matrix distribution has been favourably compared to the distribution of zeroes of the Riemann zeta function (see \cite{Diaconis2003, forrester?} for reviews and references), and in \cite{deift2007} Deift points to various social behaviours (boarding a plane, sorting playing cards, bus timetabling) that appear to obey Gaussian ensemble statistics.

\begin{remark}
Clearly, during the development of random matrix theory the comparison of theory with physical experiments and numerical simulations was a key factor in the progress of the field, and in this work we continue with this tradition. Since large numerical computations are relatively easy to perform on a desktop computer these days, we present plots of simulated spectra and numerical estimates of various probabilities, and compare them to the analytical results.
\end{remark}

In the series of papers \cite{dyson1962a, dyson1962b, dyson1962c} Dyson established that random matrix ensembles can naturally be classified into three classes corresponding to physical symmetries: time-reversal invariance with an even number of spins; time-reversal invariance with an odd number of spins; and systems without time-reversal invariance. He identified that each of these ensembles is connected to one of the classical groups studied by Weyl --- orthogonal, symplectic and unitary respectively --- by its invariance under conjugation by matrices from these groups. In \cite{Dyson1962TFW} Dyson deepens the argument, showing how this classification is isomorphic to that identified in Wigner's  similar work on time-inversion groups \cite{Wigner1959} (Wigner's original work was published, in German, in \cite{Wigner1932}, which was reprinted in \cite{Wightman1993}), and how these correspondences are fundamentally due to a theorem of Frobenius \cite[Section 11]{Dickson1914}, which states that there are exactly three associative division algebras over the real number field: the real numbers, the complex numbers, and the real quaternions (see Chapter \ref{sec:qdets_pfs} for a definition of a real quaternion). One finds that in each of these ensembles (and many since), the eigenvalue jpdf contains a Vandermonde product (\ref{eqn:evalbeta}). Dyson called this tripartite division the \textit{three-fold way} and found that they can be conveniently characterised by the parameter $\beta$ in (\ref{eqn:evalbeta}), with $\beta=1$ corresponding to the orthogonal ensemble, $\beta=2$ corresponding to the unitary ensemble and $\beta=4$ corresponding to the symplectic ensemble. In the case of matrices with Gaussian entries, this corresponds to real, complex and real quaternion matrices respectively, with the ensembles being called the Gaussian orthogonal ensemble (GOE), Gaussian unitary ensemble (GUE) and the Gaussian symplectic ensemble (GSE). In this work we will deal exclusively with ensembles of real matrices, that is, with $\beta=1$.

Dyson focused on what became known as the circular ensembles --- the circular orthogonal ensemble (COE), the circular unitary ensemble (CUE) and the circular symplectic ensemble (CSE) --- which consist of symmetric unitary, general unitary and self-dual unitary matrices respectively. These circular ensembles produced eigenvalues with compact support (the unit circle), which had the physical benefit of allowing a uniform probability distribution to be imposed. The matrices are drawn from the relevant invariant (or Haar) measure (see Chapter \ref{sec:step2} for more on this point). The unitary ($\beta=2$) ensemble turned out to be simplest, mathematically, to work with and in \cite{dyson1962c} a determinantal structure of its correlation functions was found. Dyson's seminal papers established the framework within which random matrix analysis was found to be naturally conducted; indeed in \cite{Mehta1967} Mehta applied the methods to the Gaussian ensembles and likewise found determinantal correlation functions.

The next step forward was contained in \cite{dyson1970}, where it was determined that the eigenvalue correlation functions for the $\beta=1$ and $\beta=4$ circular ensembles were given by quaternion determinants (see Chapter \ref{sec:qdets_pfs} for definitions) of matrices with $2\times 2$ matrix kernels (which reduced to determinants of $1\times 1$ kernels in the known $\beta=2$ case). This was shortly followed by the work in \cite{mehta1971} where Dyson's method was adapted to obtain similar results for the analogous Gaussian ensembles. The structural properties of these results has turned out to be another feature of random matrix studies; $\beta=2$ ensembles produce determinantal correlation functions, while $\beta=1$ and $\beta=4$ ensembles result in quaternion determinant or Pfaffian structures (where quaternion determinants and Pfaffians may, for the moment, be thought of as the square root of a determinant; see Chapter \ref{sec:Step3_GOE} for a more punctilious description). More recently however, Sinclair \cite{Sinclair2010} has shown that there is a Pfaffian structure for $\beta=2$, which does not appear to be a trivial rewriting of a determinant. In the same paper, that author goes on to establish that generalised Pfaffian structures (hyperpfaffians) occur for the Hermitian and circular ensembles for more general $\beta$ --- when $\beta=L^2$ ($L$ an integer) and $\beta=M^2+1$ ($M$ an odd integer). For a specific example of a $\beta=2$ Pfaffian correlation function see \cite{Kie2011}; also see \cite{ForSinc2010} for further applications of this idea.

Some years before the publication of these correlation functions, the Gaussian ensembles were generalised by Ginibre \cite{Gi65} by relaxing the Hermitivity constraint on the entries of the matrices. He defined three non-Hermitian ensembles of real, complex and real quaternion Gaussian entries. Although these ensembles do not obey the same invariance properties that the Gaussian ensembles do, they are often called the Ginibre orthogonal (GinOE), Ginibre unitary (GinUE) and Ginibre symplectic (GinSE) ensembles respectively, in analogue with the Gaussian ensembles. (In this work, we will refer to them as the real, complex and real quaternion Ginibre ensembles to keep in mind that invariance under the respective group is a key attribute of the Gaussian ensembles, and not of the Ginibre ensembles.)

While the eigenvalues of the (Hermitian) Gaussian ensembles, which all lay on the real line, had straightforward physical interpretation as energy levels, the spectra of the Ginibre matrices, which lay in the complex plane, did not have immediate physical motivation. However applications were forthcoming, and indeed it has been claimed (\cite{AK2007}) that non-Hermitian random matrices are now just as physically applicable as their Hermitian comrades. One of the first applications to be found for real non-Hermitian ensembles was in the work of May \cite{may1972}, where it was determined that the stability of a large biological web depended on all eigenvalues of a corresponding matrix having negative real part, and so analysis of the eigenvalue distribution was required.

\begin{figure}[htp]
\begin{center}
\subfloat[]{\label{fig:PoiDisk} \includegraphics[scale=0.4]{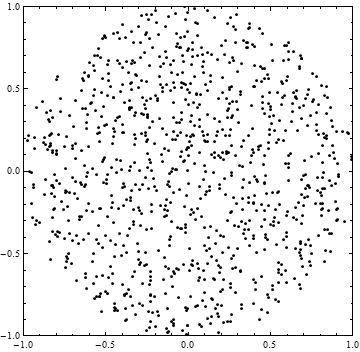}}
\qquad \qquad \qquad \subfloat[]{\label{fig:RMTDisk} \includegraphics[scale=0.4]{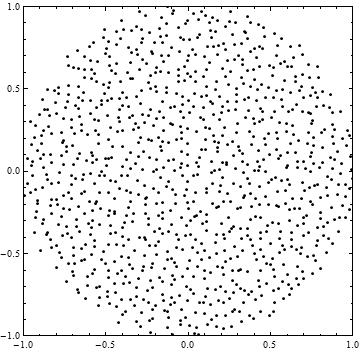}}
\caption[Simulation of a Poisson and random matrix point process in the disk.]{(a) 1000 points placed uniformly at random in the unit disk (a Poisson process), (b) Eigenvalues (scaled into the unit disk) of a $1000 \times 1000$ complex Ginibre matrix.}
\label{fig:PoiRMTDisk}
\end{center}
\end{figure}

As discussed above, the eigenvalues of Hermitian random matrices tend to repel, and it turns out that those of non-Hermitian matrices do so as well. In Figure \ref{fig:PoiRMTDisk} we compare a Poisson process in the unit disk (Figure \ref{fig:PoiDisk}) with an eigenvalue distribution over the same region (Figure \ref{fig:RMTDisk}); note the clumping of points in the former, and the more uniform distribution in the latter.
These eigenvalue distributions have been interpreted as a two-dimensional Coulomb gas \cite{forrester?}, or as describing a Voronoi tessellation of the plane \cite{LeCHo1990} that is more uniform than that given by a Poisson process \cite{HayQui2002}. This latter viewpoint can be applied to analyse situations where one expects, due to physical considerations, that there would be some repulsion between some entities such as trees in a forest, bird nesting sites or impurities in metals \cite{LeCHo1990}. Other uses of random non-Hermitian matrices have included synchronisation in random networks, statistical analysis of neurological activity, quantum chaos and polynuclear growth processes (see \cite{AK2007, KS2009, forrester?} for overviews and further references).

In his original paper Ginibre found that the eigenvalue jpdf for the (non-Hermitian) complex ensemble involved the product of differences (\ref{eqn:evalbeta}), and was structurally similar to that of the Gaussian ensembles. He went on to calculate the general $n$-point correlations for the complex case and found they were given by the determinant of a $1 \times 1$ kernel, again similar to the complex circular and Gaussian ensembles. In the case of the real quaternion matrices he was able to state the eigenvalue jpdf, but not to calculate the correlations as he lacked the quaternion determinant structure that Dyson would later introduce to the theory. The $1$-, $2$- and $3$-point functions for $\beta=4$ were identified by Mehta in \cite{Mehta1967} with the full correlations appearing many years later in the second edition of his book \cite{mehta1991}.

The real ensemble, however, proved much more difficult and is the subject of Chapter \ref{sec:GinOE} of the present work. First note from classical linear algebra or polynomial theory that a generic $N \times N$ real matrix has $0\leq k\leq N$ (with $k$ of the same parity as $N$) real eigenvalues and $(N-k)/2$ complex conjugate pairs of eigenvalues; so the eigenvalues come in two distinct species. This is a significant difference from the ensembles considered previously, where only one species was present: all eigenvalues are real for the Gaussian ensembles; they all lie on the unit circle for the circular ensembles; and the eigenvalues are general complex numbers for $\beta=2$ and strictly non-real complex for $\beta=4$ Ginibre ensembles. Ginibre was only able to establish the eigenvalue jpdf for the real ensemble in the restricted case that all eigenvalues were real, with the jpdf for general $k$ not appearing until \cite{LS91} and again independently in \cite{Ed97} where new methods of matrix decomposition were employed (see below). The correlation functions were yet longer in coming, needing a result from \cite{sinclair2006}, which established a quaternion determinant or Pfaffian form of the ensemble average, allowing Forrester and Nagao to calculate the real and complex correlation functions (for $N$ even) in \cite{FN07}. As for the GOE, the correlations had a quaternion determinant or Pfaffian structure with a $2\times 2$ kernel. These correlations were generalised to include real--complex cross-correlations in \cite{sommers2007} and independently in \cite{b&s2009}, again only for even dimensional matrices. The odd case was identified shortly afterwards by Sommers and Wieczorek \cite{sommers_and_w2008}, Forrester and the present author \cite{FM09} and Sinclair \cite{Sinc09} using three separate methods.

The difficulties that led to such a long delay in first the identification of the eigenvalue jpdf and then the full even and odd correlation functions for the real Ginibre ensemble were several. Classical results in linear algebra tell us that symmetric real matrices are orthogonally diagonalisable, that is, a symmetric matrix $\bS$ can be decomposed as
\begin{align}
\label{eqn:diagdecomp} \bS=\bO^{T}\bD\bO, 
\end{align}
where $\bD=\mathrm{diag} [\lambda_1,...,\lambda_N]$ is a diagonal matrix containing the eigenvalues and $\bO$ is a matrix whose columns are the corresponding orthonormal eigenvectors \cite{GoVl1996}. The integration over the orthogonal matrices (which gives us the volume of the orthogonal group $O(N)$) has a known evaluation, meaning that the dependence on the eigenvectors can be integrated out of the problem. However, asymmetric real matrices do not have this property; the diagonalising matrices are not orthogonal. Progress required the introduction of the Schur decomposition (see Chapter \ref{sec:Gejpdf}), where the diagonal structure (\ref{eqn:diagdecomp}) is forfeited, with $\bD$ being replaced by an upper triangular matrix with the eigenvalues of $\bS$ on the diagonal. The benefit is that the conjugating matrices are still orthogonal, and so they can be integrated over with known methods. This, of course, comes at the cost of requiring an extra $N(N-1)/2$ integrations over the upper triangular entries. The Schur decomposition method was employed in \cite{Ed97}, and a closely related form --- related via elementary row operations --- was used in \cite{LS91}, to obtain the joint distribution of the eigenvalues.

Yet even when the eigenvalue jpdf is established, there are more complications. In the case of the classical orthogonal ensembles, Dyson and Mehta were able to use an integration theorem (Proposition \ref{thm:integral_identities}) to calculate the correlation functions from the eigenvalue jpdf, however, as pointed out in \cite{AK2007}, this does not work for the real Ginibre ensemble. The distinction is that the eigenvalue jpdf for the real Ginibre ensemble pertains to one particular $(N,k)$ pair (recall that $k$ is the number of real eigenvalues), yet the system is only normalised for the sum over all $k$; Dyson's integration theorem does not seem applicable to this sum. In \cite{AK2007} the authors presented a Pfaffian integration formula to deal with this problem, but shortly thereafter the formulation of \cite{Sinc09} circumvented the problem entirely by presenting the ensemble average as a Pfaffian independent of $k$. Using this structure, and applying functional differentiation, the real--real and complex--complex correlations were established in \cite{FN08} via explicit calculation of the skew-orthogonal polynomials (see Chapters \ref{sec:skew_orthog_polys} and \ref{sec:Gsops}), which, as mentioned above, then led to the full correlations in \cite{sommers2007,b&s2009} in the restricted case that the system size is even. Yet, silver linings abound ---  this bipartite nature of the set of eigenvalues leads to a particularly interesting question about the real Ginibre ensemble that was raised in \cite{Ed97}: what is the probability of obtaining $k$ real eigenvalues from an $N \times N$ real matrix? We will investigate this for each of the non-Hermitian ensembles (Chapters \ref{sec:pnk}, \ref{sec:tGprobs}, \ref{sec:Ssops} and \ref{sec:TOEsops}).

In the case when the matrix dimension is odd, we must overcome more hurdles. Pfaffians are only defined for even-sized matrices and to adapt them to odd-size involves `bordering' by a new row and column or by removing a row and column from a computable even-sized system (see Chapters \ref{sec:GOE_odd} and \ref{sec:GinOE_odd}). These are not new ideas; they were used by de Bruijn in \cite{deB1955}, although the bordering procedure for Pfaffians can be traced back, at least, to Cayley in 1855 \cite{Cayley1855}, and the generation of an odd system from an even system has an even older pedigree, being found in Pfaff's original presentation of the theory in 1815 \cite{Pfaff1815} where he was motivated by reducing a set of ordinary differential equations in $2m$ variables to a set of equations in $2m-1$ variables (for a (somewhat) modern interpretation of these historical articles, see \cite{Muir1906,Muir1911}).

This even--odd asymmetry does not show itself in the $\beta=2$ or $\beta=4$ ensembles since the complex ensembles resulted in determinant structures which are insensitive to the parity of the matrix, while an $N \times N$ real quaternion matrix ensemble can be effectively viewed as a restricted class of $2N \times 2N$ complex matrices, and so they have an underlying even dimension. For the real Ginibre ensemble, we can explicitly identify the culprit. It turns out that since there is one real eigenvalue guaranteed to exist in an odd-sized real matrix (since the eigenvalues are real or complex-conjugate paired), this eigenvalue naturally forms the final row and column; it seems that the technical problems presented by the odd case are due to the fact that this preordained real eigenvalue exists at all. The problem arises when one attempts to apply the important method of integration over alternate variables, which was developed by de Bruijn and Mehta in \cite{deB1955, mehta1960, Mehta1967}, to obtain a Pfaffian expression for the partition function. This method pairs all the eigenvalues to allow one to overcome the asymmetry in the eigenvalue jpdf when $\beta=1$, but of course, in the case of $N$ odd, one eigenvalue must be unpaired and dealt with separately. This leads to other consequences, for example, given that the odd-sized matrix has at least one real eigenvalue, the probability of obtaining $k$ real eigenvalues is qualitatively different in the even and odd cases (for finite $N$) (see (\ref{eqn:GinOE_probsGF_pf}) and (\ref{eqn:GinOE_probs_odd})), although they are the same in the large $N$ limit.

The Ginibre ensembles can be generalised to the partially symmetric ensembles (see Chapter \ref{sec:tG}) by the incorporation of a parameter $-1<\tau<1$ (by convention). These ensembles interpolate between the symmetric/Hermitian/self-dual Gaussian ensembles ($\tau\to 1$) and ensembles of anti-symmetric/anti-Hermitian/anti-self-dual Gaussian matrices ($\tau\to -1$); $\tau=0$ corresponds to the Ginibre ensembles, with maximum asymmetry. With $\tau$ bounded away from $1$, then in the large $N$ limit the eigenvalue distributions and correlations are just scaled forms of those in the Ginibre ensembles. However, by carefully taking $\tau\to 1$ with increasing $N$ it is shown in \cite{FKS97} (where ensembles of complex matrices are discussed and this limit is called the \textit{weakly non-Hermitian} limit) that a new cross-over regime is obtained that interpolates between the apparently qualitatively different behaviours of the Ginibre and Gaussian ensembles. These partially-symmetric matrix ensembles have found application in the study of neural networks (see \cite{LS91} and references therein) and in quantum chaotic scattering \cite{FS03}. Another interesting review is contained in \cite{KS2009}. (For a review on related non-Hermitian ensembles as applied to quantum chromodynamics (QCD) see \cite{Ake2007}.)

Similar to Dyson's three-fold classification of the matrix ensembles --- having real, complex or real quaternion elements --- a new tripartite scheme has become apparent recently (see \cite{Krish2006,Krish2009} where it was introduced in the context of Gaussian analytic functions). From differential geometry we know that there are three distinct surfaces corresponding to constant Gaussian curvature $\kappa$: the plane ($\kappa=0$), the sphere ($\kappa >0$) and the anti- or pseudo-sphere ($\kappa <0$). For the Ginibre ensembles, one finds that in the limit of large matrix dimension the eigenvalues tend to uniform density on a (planar) disk (the so-called \textit{circular law}; see below), and we identify these ensembles with the plane. The sphere can be identified with the problem of generalised eigenvalues, that is, the set of $\lambda_j$ given by the solutions to the equation
\begin{align}
\label{eqn:genEvals} \det(\bB-\lambda\bA)=0,
\end{align}
where $\bA,\bB$ are some $N\times N$ matrices. Assuming that $\bA$ is invertible, these generalised eigenvalues are equivalent to the eigenvalues of the matrix $\bY=\bA^{-1}\bB$. In \cite{Krish2009} Krishnapur considers the case where $\bA$ and $\bB$ are complex Ginibre matrices. It turns out that these eigenvalues have uniform density on the sphere (under stereographic projection) and so ensembles of these matrices are appellated \textit{spherical ensembles}. Similar to the complex Ginibre ensemble, the complex spherical ensemble can be thought of as modelling a gas of charged particles, this time on a sphere; the works \cite{caillol81, FJM1992, forrester?} highlight the analogies. In \cite{FM09} Forrester and the present author analyse the analogous real ($\beta=1$) spherical ensemble, where the matrices $\bA, \bB$ are real Ginibre matrices. This is the subject of Chapter \ref{sec:SOE}.

The last in the geometrical triumvirate are the ensembles corresponding to the anti-sphere. By truncating a number $L$ of rows and columns from an $N \times N$ (complex) unitary matrix (that is, a matrix from the CUE) \.{Z}yczkowski and Sommers \cite{Z&S2000} form the complex \textit{truncated ensemble}. Various applications of these truncated unitary matrices have been found, such as quantum chaotic scattering and conductance (see \cite{FS03} and references therein and \cite{Forrester2006}) and the zeroes of Kac random polynomials \cite{Forrester2010a}. The analogous real ensemble, which was briefly discussed in \cite{Z&S2000}, is the truncation of real orthogonal matrices, the eigenvalue jpdf and correlation functions of which were contained in \cite{KSZ2010}. The analysis of these truncated ensembles is somewhat more intricate than those of Ginibre or spherical ensembles since the size of the truncation relative to the dimension of the unitary matrix leads to qualitatively different eigenvalue behaviour. For example, if the truncation is large then the eigenvalue statistics (under certain scaling) approach those of the real Ginibre ensemble, since the orthogonality constraint has small effect, but with a small truncation, then the orthogonality is strongly felt and the eigenvalues cluster near the unit circle. We find that the eigenvalues are uniformly distributed on the anti-sphere in the limit of large dimension, hence the correspondence with a surface having constant $\kappa <0$.

Since each of these surfaces has constant curvature, we can reasonably expect a uniform distribution of eigenvalues over some region of support. However, this should be contrasted with the work in \cite{FanTell2008} where they study the analogous problem on a particular surface (called \textit{Flamm's paraboloid}, which arises in general relativity) with non-constant curvature, resulting in a non-uniform density in the thermodynamic limit.

In the various analyses of the ensembles discussed above there are several techniques and approaches that are regularly employed. It is the purpose of this work to present a systematic approach that can be used for each of the real (corresponding to $\beta=1$) ensembles in the 12-part classification: the 3 symmetric/Hermitian/self-dual Gaussian ensembles and the 9 non-symmetric/non-Hermitian/non-self-dual ensembles corresponding to each of the surfaces of constant Gaussian curvature. Specifically, in Chapter \ref{sec:GOE_steps} we lay out a $5$-step scheme that is applicable to all the ensembles to be discussed, applying them to the GOE by way of illustration; the method can be broadly described as the \textit{(skew-) orthogonal polynomial method}, since knowledge of such polynomials allows explicit calculation of statistical quantities. Then in the following chapters we apply the $5$ steps to the real cases of the Ginibre ensemble (including the real partially symmetric ensemble as a generalisation), the spherical ensemble and the truncated ensemble. Only in the last of these (the real truncated ensemble) will we find that the scheme has some shortcomings, with another method, which has been applied in \cite{sommers_and_w2008,FM09,KSZ2010} (and is discussed in Chapters \ref{sec:SOEcharpolys} and \ref{sec:TOEkernelts}), seeming to be the more useful in that case.

It should be mentioned that this geometric classification is one of a number of classification schemes in the random matrix literature. The work by Altland and Zirnbauer (\cite{Zirn1996, AltZir1997}) classifies Hermitian random matrix ensembles by the requirement of symmetry under conjugation by various operators. A correspondence between this classification and the families of \textit{symmetric spaces} (which are also defined by their symmetries), as categorised by Cartan \cite{Car1926, Car1927}, is identified. This classification includes Dyson's `threefold way' and that of Verbaarschot \cite{Verb1994a}, where he identifies a `threefold way' for the \textit{chiral ensembles}. (We will briefly revisit chiral ensembles in Chapter \ref{sec:FW}.) In \cite{BerLeC2002} and \cite{Mag2008} the set of symmetries is broadened to include non-Hermitian matrix ensembles, which introduces a further twenty symmetric spaces (bringing the total to thirty classes). It turns out that the classification of the non-Hermitian cases is not as useful as that in the Hermitian cases --- for Hermitian ensembles, the form of the Jacobian for the change of variables to the eigenvalue jpdf is determined by its classification, however, this is not true for the non-Hermitian cases. We will not pursue these classifications any further here; the interested reader is referred to the original references as stated, or to \cite[Chapter 15.11]{forrester?}.

Lying behind all the results in the study of random matrices is the concept of \textit{universality}, which is analogous to the central limit theorem in classical probability theory. Universality refers to the observed phenomena that the statistics of high dimensional matrices tend to some unique behaviour, dependent only on some structural feature of the ensemble rather than on the particular distribution of the entries. In the case of the Gaussian ensembles, the statistics of Hermitian matrices with identically and independently distributed (iid) standard normal ($N[0,1]$) entries converge to those of Hermitian matrices with iid entries from any mean zero, unit variance distribution with the same value of $\beta$. An important result to come from the considerations of these random matrices is \textit{Wigner's semi-circle law} \cite{Wigner1957c} for the density of the eigenvalues (see Chapter \ref{sec:GOE_sums}). Although it did not provide good agreement with experiments on nuclear energy levels (as the spacing distribution did), since the energy levels certainly do not have a semi-circular distribution, it has proven to be ubiquitous in the study of random Hermitian matrices. A similar result in the case of the Ginibre matrices is the \textit{circular law} (Chapter \ref{sec:circlaw}) often attributed to Girko \cite{Girko1985a, Girko1985b}, which states that for large matrix size the eigenvalue densities of ensembles with iid entries drawn from a distribution with mean zero and unit variance converge to the uniform distribution on a disk of radius $\sqrt{N}$. We will also discuss an \textit{elliptical law} for the partially-symmetric ensembles (Chapter \ref{sec:tGkernelts}) and a \textit{spherical law} for the spherical ensembles (Chapter \ref{sec:SOElims}), which will lead us to the conjecture of an \textit{anti-spherical law} for the truncated ensembles by analogy (Chapter \ref{sec:uasc}).

We also find that, in various scaled limits, the real and complex members of the novempartite categorisation of ensembles (real, complex and real quaternion versions of Ginibre, spherical and truncated matrices) have identical behaviour. The real quaternion ($\beta=4$) cases of the spherical and truncated ensembles have yet to be explored, although they are expected to conform to the same behaviour. This is another interpretation of universality and we discuss it in Chapters \ref{sec:SOEsclims} and \ref{sec:uasc}.

Treating the concept of universality more generally we can find analogies of our results in the studies of random tensors \cite{Ma07}, random walks and random involutions \cite{BakFor2001}, the zeroes of random polynomials \cite{EK95, BakFor1997, Forrester2010a}, and, as discussed above, in seemingly unrelated physical applications: from nuclear energy levels, to Coulomb gases to car parking \cite{deift2006}. When ruminating on such contemplations, as with so many things in random matrix theory, we may invoke the spirit of Wigner; this time calling to mind his observation of the ``unreasonable effectiveness of mathematics" \cite{Wigner1960} (as did the authors of \cite{AK2007}). In the same way that the central limit theorem is the justification for the common appearance of the normal distribution in large ``real world" data sets, it seems that random matrix universality is pointing us to something fundamental (and yet fundamentally mystifying) in the relationship between the eigenvalues of a random matrix and the operation of our universe.

\newpage

\section{The Gaussian orthogonal ensemble}
\setcounter{figure}{0}
\label{sec:GOE_steps}

The eigenvalue pdf for a number of ensembles of random matrices with real entries is integrable. By this we mean that probabilistic quantities such as the generalised partition function and the correlation functions exhibit special structures leading to closed form expressions. Our concern with the detail of such calculation can be broken down into five steps:
\begin{enumerate}[I.]
\item{specification of the distribution of matrix elements;}
\item{changing variables to find the distribution of eigenvalues;}
\item{establishing a Pfaffian or quaternion determinant form of the generalised partition function;}
\item{finding the appropriate skew-orthogonal polynomials to simplify the Pfaffian or quaternion determinant;}
\item{calculating the correlations in terms of Pfaffians or quaternion determinants with explicit entries.}
\end{enumerate}
(For definitions of Pfaffians and quaternion determinants see Chapter \ref{sec:qdets_pfs}.) We will first illustrate the steps using the case of the Gaussian Orthogonal Ensemble (GOE), before applying them to the non-symmetric ensembles in the remaining chapters of this work.

\begin{remark}
There are, of course, methods that do not follow this structure, however, in this work we restrict our scope to these steps.
\end{remark}

\subsection{Step I: Joint probability density function of the matrix elements}

This first step, calculating the joint probability density function (jpdf), is essentially the statement of the problem, however, this does not mean that the distribution is necessarily obvious. Indeed, in the case of the spherical ensemble of Chapter \ref{sec:SOE} the element distribution is first specified as a product of the distribution of the component matrices; obtaining knowledge of the elements of the matrix product requires quite a deal of calculation. Further, as we shall see in Chapter \ref{sec:truncs}, for the case of anti-spherical ensemble the matrix distribution is non-analytic for large truncations. (Specifically, the normalisation in (\ref{eqn:TOEmpjdfnorm1}) does not exist when the number of truncated rows and columns is less than the number of rows and columns retained.)

\subsubsection{GOE element jpdf}

The formalism we use here follows that of \cite[Chapters 6 \& 7]{forrester?}. The GOE consists of real, symmetric matrices $\mathbf{X}$, containing Gaussian distributed elements. Specifically, the diagonal and strictly upper triangular elements individually have distributions
\begin{align}
\label{eqn:GOE_el_dist} \frac{1}{\sqrt{2\pi}}e^{-x^2_{jj}/2} \;\; \textrm{and}\;\; \frac{1}{\sqrt{\pi}}e^{-x^2_{jk}}
\end{align}
respectively. Elementary probability theory tells us that the probability of a set of multiple random, independent events is the product of these individual probability density functions. Thus, for our real, symmetric matrices with elements distributed as in (\ref{eqn:GOE_el_dist}), we have for the jpdf $P(\bX)$ of the entries of $\bX$
\begin{align}
\nonumber P(\mathbf{X})&=\prod_{j=1}^N \frac{1}{\sqrt{2\pi}}e^{-x^2_{jj}/2}\prod_{1\leq j< k \leq N} \frac{1}{\sqrt{\pi}}e^{-x^2_{jk}}=2^{-N/2}\pi^{-N(N+1)/4}\prod_{j,k=1}^Ne^{-x^2_{jk}/2}\\
\label{eqn:GOE_el_jpdf} &=2^{-N/2}\pi^{-N(N+1)/4}e^{-\sum_{j,k=1}^Nx^2_{jk}/2}=2^{-N/2}\pi^{-N(N+1)/4}e^{-(1/2)\mathrm{Tr}\mathbf{X}^2},
\end{align}
where Tr is the matrix trace. We note that $\int P(\bX) \prod_{1\leq j \leq k \leq N}dx_{jk}=1$, where each of the $N(N+1)/2$ integrals is over $(-\infty,\infty)$, and so $P(\bX)$ is, of course, a probability density function. The ensemble is called \textit{orthogonal} because $P(\bX) \prod_{1\leq j \leq k \leq N}dx_{jk}$ is unchanged by orthogonal conjugation; we will make this clear after Proposition \ref{prop:GOE_J}.

\begin{figure}[htp]
\begin{center}
\includegraphics[scale=0.7]{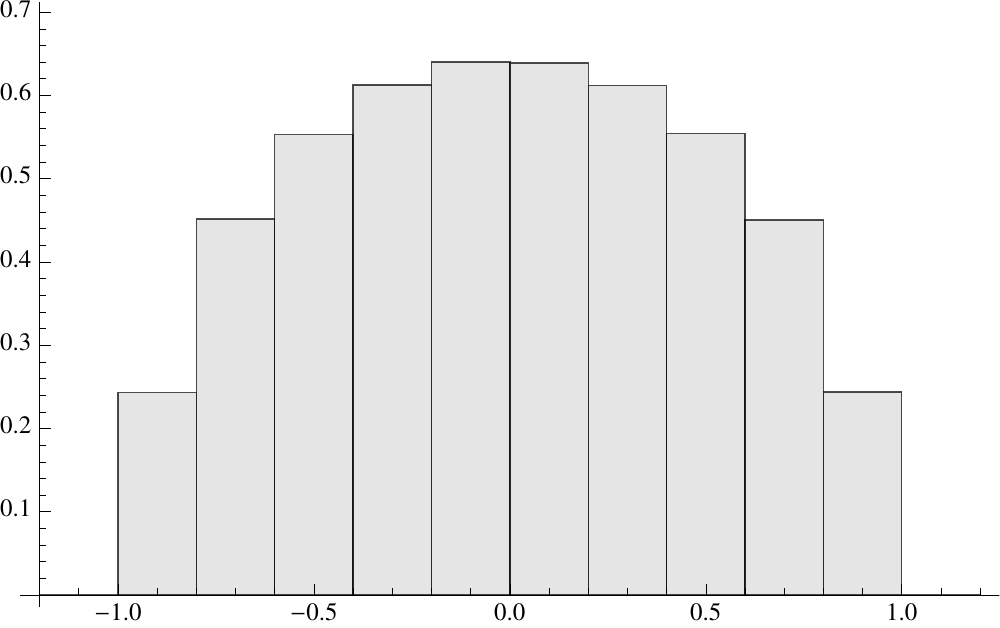}
\caption[Simulated GOE eigenvalue density.]{Eigenvalue density of $1000$ independent $500\times 500$ GOE matrices, normalised and scaled by $\sqrt{2N}$.}
\label{fig:GOE_eval_dens}
\end{center}
\end{figure}

We now have the elemental distribution but we are ultimately interested in the distribution of the eigenvalues of the matrices specified by (\ref{eqn:GOE_el_dist}). To gain some insight into the expected eigenvalue distribution we can simulate a sequence of random matrices and plot the resulting density. In Figure \ref{fig:GOE_eval_dens} we have plotted the eigenvalue density for $1000$ independent GOE matrices of size $500\times 500$.
\begin{remark}
Note that in Figure \ref{fig:GOE_eval_dens} we have normalised the density to have total mass one and scaled the $x$ axis so that the reader will be all the more awed when we compare the results to the known asymptotic limit in (\ref{eqn:wssl}).
\end{remark}


The candid pattern in the data suggests the integrable nature of the problem.

\subsection{Step II: Eigenvalue jpdf}
\label{sec:step2}

The goal here is to re-express the element jpdf in terms of the eigenvalues of the matrices that compose the ensemble. The idea is to separate the eigenvalues from the other independent variables --- relating to the eigenvectors --- in some fashion. Indeed, the key part of this step is to choose a convenient matrix decomposition that exposes the eigenvalues in such a way that the remaining degrees of freedom can be integrated over yielding a constant overall factor. This means that the problem is essentially one of changing variables and calculating the associated Jacobian.

\subsubsection{GOE eigenvalue jpdf}
\label{sec:GOE_eval_jpdf}

With $\vec{\lambda} =\{ \lambda_1,\dots,\lambda_N\}$ the eigenvalues of $\bX$, we see that we are looking to change variables and integrate (with somewhat loose notation) according to 
\begin{align}
\label{eqn:PX_QL1} \int P(\mathbf{X})\prod_{1\leq j \leq k \leq N}dx_{j,k}=Q(\vec{\lambda})\prod_{j=1}^Nd\lambda_j,
\end{align}
where $Q(\vec{\lambda})$ is the eigenvalue jpdf. We understand the integral in (\ref{eqn:PX_QL1}) to be over the $N(N-1)/2$ variables relating to the eigenvectors, leaving only the dependence on the eigenvalues $\lambda_j$. This will be made precise below.

\begin{remark}
Note that in this work we will consistently use $P$ to denote the probability distribution of a matrix (or the elements of the matrix), while $Q$ denotes the distribution of the eigenvalues of the matrix.
\end{remark}
\noindent To carry out the change of variables we use the decomposition (\ref{eqn:diagdecomp}) of real symmetric matrices to write
\begin{align}
\label{eqn:diag_decomp} \mathbf{X}=\mathbf{R LR}^T,
\end{align}
where $\mathbf{L}$ is diagonal, containing the $N$ eigenvalues of $\mathbf{X}$, and $\mathbf{R}$ is a real, orthogonal matrix whose columns are the corresponding normalised eigenvectors. The decomposition will be unique if \textit{i}) we order the eigenvalues in $\bL$, and \textit{ii}) we specify that the first row of $\bR$ is non-negative.

First, recall that the appropriate operation for products of differentials is the wedge product (although in (\ref{eqn:PX_QL1}) we used product notation since they are the same in this setting).
\begin{definition}
With $\bX=[x_{ij}]_{i,j=1,...,N}$, let $d\mathbf{X}$ be the matrix of differentials of the elements of $\mathbf{X}$,
\begin{align}
\nonumber d\mathbf{X}=\left[\begin{array}{cccc}
dx_{11}&dx_{12}&\cdot\cdot\cdot&dx_{1N}\\
dx_{21}&dx_{22}&\cdot\cdot\cdot&dx_{2N}\\
\vdots&\vdots&\ddots&\vdots\\
dx_{N1}&dx_{N2}&\cdot\cdot\cdot&dx_{NN}
\end{array}\right].
\end{align}
\end{definition}

\begin{definition}
Let $(d\mathbf{X})$ be the wedge product of the independent elements of $d\mathbf{X}$.
\end{definition}

In the case of $\mathbf{X}$ an $N \times N$ real, symmetric matrix we have
\begin{align}
\nonumber (d\mathbf{X})=\bigwedge_{1\leq j \leq k \leq N}dx_{jk},
\end{align}
and so we rewrite (\ref{eqn:PX_QL1}) as
\begin{align}
\label{eqn:PX_QL2} \int P(\bX)(d\bX)=Q(\vec{\lambda})(d\vec{\lambda}),
\end{align}
where, as in (\ref{eqn:PX_QL1}), the integral is understood to be over the variables relating to the eigenvectors.
\begin{remark}
Although products of differentials are understood to be wedge products, we will commonly use the notation of (\ref{eqn:PX_QL1}) when no confusion is likely.
\end{remark}
\noindent To proceed with the enterprise of calculating $Q(\vec{\lambda})$, first recall from multivariable calculus that in order to change variables from $\{ x_{j}\}_{j=1,...,N}$ to $\{ y_{j}\}_{j=1,...,N}$ we use the identity
\begin{align}
\label{eqn:COV} \bigwedge_{j=1}^Ndx_i=\left|J\right|\bigwedge_{j=1}^Ndy_j,
\end{align}
where $J$ is known as the \textit{Jacobian}, and is defined as
\begin{align}
\nonumber J:=\det \left[\frac{\partial x_{j}}{\partial y_{k}}\right]_{j,k=1,...,N}.
\end{align}
Comparing (\ref{eqn:PX_QL1}) with (\ref{eqn:COV}), and keeping in mind that the products of differentials in the former are in fact wedge products, we see that calculating the Jacobian is a key part of our program, yet it is not the whole program. In (\ref{eqn:COV}) there are an equal number of differentials on both sides of the equation, indeed the Jacobian would not even be defined if the number of differentials were not equal. Yet, on the right hand side (RHS) of (\ref{eqn:PX_QL1}) we see that there are $N$ independent differentials, while on the left hand side (LHS) there are $N(N+1)/2$, which, the incisive reader will note, is often considerably more than $N$. In fact, the change of variables equation we are actually interested in (ignoring constants for the moment) is
\begin{align}
\label{eqn:GOE_cov1} e^{-(1/2)\mathrm{Tr}\mathbf{X}^2}(d\bX)=|J|e^{-\sum_{j=1}^N\lambda_j^2/2}(d\vec{\lambda})(d\vec{p}),
\end{align}
where $\vec{p}=\{ p_1,...,p_{N(N-1)/2}\}$ are variables associated with the eigenvectors in the decomposition (\ref{eqn:diag_decomp}) and
\begin{align}
\label{def:GOE_J} J=\mathrm{det}\left[\begin{array}{cccc}
\frac{\partial x_{11}}{\partial\lambda_1}&\frac{\partial x_{12}}{\partial\lambda_1}&\cdot\cdot\cdot&\frac{\partial x_{NN}}{\partial\lambda_1}\\
\vdots&\vdots&\ddots&\vdots\\
\frac{\partial x_{11}}{\partial\lambda_N}&\frac{\partial x_{1N}}{\partial\lambda_N}&\cdot\cdot\cdot&\frac{\partial x_{NN}}{\partial\lambda_N}\\
\frac{\partial x_{11}}{\partial p_{1}}&\frac{\partial x_{12}}{\partial p_{1}}&\cdot\cdot\cdot&\frac{\partial x_{NN}}{\partial p_{1}}\\
\vdots&\vdots&\ddots&\vdots\\
\frac{\partial x_{11}}{\partial p_{N(N-1)/2}}&\frac{\partial x_{12}}{\partial p_{N(N-1)/2}}&\cdot\cdot\cdot&\frac{\partial x_{NN}}{\partial p_{N(N-1)/2}}
\end{array}\right].
\end{align}
Since the differentials $(d\vec{p})$ do not appear on the right hand side of (\ref{eqn:PX_QL1}), the variables $\{ p_1,...,p_{N(N-1)/2}\}$ are considered undesirables, and so we will integrate them out of the final expression, which will leave us with $Q(\vec{\lambda})$.

Before progressing, we establish a useful lemma, which we will repeatedly compel into service.
\begin{lemma} [\cite{muirhead1982} Theorem 2.1.6]
\label{lem:adma}
Let $\bZ:=[z_{j,k}]_{j,k=1,...,N}=\bA^T \bM \bA$ where\\$\bA:=[a_{j,k}]_{j,k=1,...,N}$ is a fixed, real, non-singular matrix and $\bM:=[m_{j,k}]_{j,k=1,...,N}$ is a real, symmetric matrix. Then
\begin{align}
\nonumber (d\bZ)&= \mathrm{det}(\bA)^{N+1}(d\bM)
\end{align}
\end{lemma}

\textit{Proof}: Firstly, noting that $\bA$ is fixed, we apply the product rule of differentiation to find that $d\bZ:=d(\bA^T\bM\bA)=\bA^Td\bM\bA$. Next we note that
\begin{align}
\label{eqn:p(a)} (d\bZ)=p(\bA)(d\bM),
\end{align}
where $p(\bA)$ is some polynomial in the $a_{j,k}$. This is clear from (\ref{eqn:COV}) since each $dz_{j,k}$ is a polynomial in the variables $dm_{j,k}$ with coefficients from $\bA$. Again using (\ref{eqn:p(a)}) we have that
\begin{align}
\nonumber ((\bA_1\bA_2)^Td\bM\bA_1\bA_2)=p(\bA_1\bA_2)(d\bM).
\end{align}
But we can also write
\begin{align}
\nonumber ((\bA_1\bA_2)^Td\bM\bA_1\bA_2)&=p(\bA_2)(\bA_1^Td\bM\bA_1)\\
\nonumber &=p(\bA_2)p(\bA_1)(d\bM)
\end{align}
and so we have the factorisation property
\begin{align}
\label{eqn:det_fact} p(\bA_1\bA_2)=p(\bA_1)p(\bA_2).
\end{align} 
From the working in \cite{MacD1943} we know that by considering the elementary rotation, stretching and shearing matrices, the only polynomial satisfying this property is
\begin{align}
\label{eqn:p=det} p(\bA)=(\det \bA)^m,
\end{align}
for some integer $m$. By taking $\bA=\mathrm{diag}[a,1,...,1]$ we find that $(d\bZ)=a^{N+1}(d\bM)$, and so $m=N+1$. Substituting this into (\ref{eqn:p=det}) and then (\ref{eqn:p=det}) into (\ref{eqn:p(a)}) gives the result.

\hfill $\Box$

We may now compute the Jacobian corresponding to the change of variables (\ref{eqn:PX_QL2}).

\begin{proposition}
\label{prop:GOE_J}
For $\mathbf{X}$ an $N \times N$ real, symmetric matrix we have
\begin{align}
\label{eqn:dX_jacob} (d\mathbf{X})=\prod_{1\leq j < k \leq N}|\lambda_k-\lambda_j|\bigwedge_{j=1}^N d\lambda_j \; (\mathbf{R}^Td\mathbf{R}),
\end{align}
where $\lambda_1,...,\lambda_N$ are the eigenvalues of $\bX$, and $\bR$ is real orthogonal.
\end{proposition}

\textit{Proof}: Applying the product rule of differentiation to (\ref{eqn:diag_decomp}) we have
\begin{align}
\nonumber d\bX=d\bR \bL \bR^T + \bR d\bL \bR^T + \bR \bL d\bR^T.
\end{align}
It will prove convenient to premultiply by $\bR^T$ and post multiply by $\bR$ giving
\begin{align}
\nonumber \bR^T d\bX \bR &=\bR^T d\bR \bL + \bL d\bR^T \bR+d\bL\\
\label{eqn:RTdXR} &=\bR^T d\bR \bL-\bL \bR^Td\bR+d\bL,
\end{align}
where we have used the fact that $\bR\bR^T=\1$ and the corollary $d\bR^T\bR=-\bR^Td\bR$. The convenience comes from the fact that now the first two terms in (\ref{eqn:RTdXR}) are products of the same matrices. Although it appears this convenience may come at the expense of complicating the LHS of (\ref{eqn:RTdXR}) we use Lemma \ref{lem:adma} to see that $(\bR^T d\bX \bR)=(\det \bR)^{N+1}(d\bX)$. Then when we recall that $\det \bR =\pm1$ and only the magnitude of the Jacobian is retained, we see that after taking wedge products, we have escaped penalty.

By explicit multiplication we have
\begin{eqnarray}
\nonumber &&\bR^T d\bR \bL -\bL \bR^Td\bR +d\bL = \\
\nonumber &&\left[\begin{array}{cccc}
d\lambda_1 & (\lambda_2-\lambda_1)\vec{r_1}^T\cdot d\vec{r}_2 &\cdot\cdot\cdot&(\lambda_N-\lambda_1)\vec{r_1}^T\cdot d\vec{r}_N\\
(\lambda_2-\lambda_1)\vec{r_1}^T\cdot d\vec{r}_2&d\lambda_2&\cdot\cdot\cdot&(\lambda_N-\lambda_2)\vec{r_2}^T\cdot d\vec{r}_N\\
\vdots&\vdots&\ddots&\vdots\\
(\lambda_N-\lambda_1)\vec{r_1}^T\cdot d\vec{r}_N&(\lambda_N-\lambda_2)\vec{r_2}^T\cdot d\vec{r}_N&\cdot\cdot\cdot&d\lambda_N
\end{array}\right],
\end{eqnarray}
where we have used the equalities $r_j^T\cdot dr_i = dr_i^T\cdot r_j = -r_i^T\cdot dr_j$ (the first equality follows since scalars are invariant under transposition, and the second is another consequence of $\bR \bR^T=\1$).

Taking wedge products of both sides of (\ref{eqn:RTdXR}) gives the result. Note that this tells us the natural choice for the variables $p_j$ in (\ref{def:GOE_J}) is such that each $dp_j$ is one of $\{ \vec{r}_i \cdot d\vec{r}_j\}_{i<j}$.

\hfill $\Box$

The key structural component of the eigenvalue jpdf is apparent from Proposition \ref{prop:GOE_J} --- the product of differences between eigenvalues. It is a ubiquitous occurrence in random matrix theory and is one of the unifying themes of the study.

In the previous section we mentioned that the reason for the appellation \textit{orthogonal} to describe the ensemble of real, symmetric Gaussian matrices is that the normalised quantity $P(\bX) (d\bX)$ is unchanged by orthogonal conjugation; we can now see why this is true. First note from (\ref{eqn:GOE_el_jpdf}) that $P(\bX)$ is invariant under any conjugation $\bX\to \bM^{-1}\bX\bM$ because of the cyclic property of the trace operator. Second, we examine the measure $(d\bX)$ from (\ref{eqn:dX_jacob}). The differential $n(n-1)/2$-form $(\bR^T d\bR)$ is invariant under the left operation $\bO\bR$, where $\bO$ is a fixed, orthogonal $N\times N$ matrix because
\begin{align}
\label{eqn:HMLinv} (\bR^T d\bR) \to \left( (\bO \bR)^Td(\bO\bR) \right)= \left(\bR^T \bO^T d(\bO)\bR +\bR^T \bO^T\bO  d(\bR)\right) = (\bR^T d\bR),
\end{align}
where the first term vanishes since $\bO$ is fixed. We also see that $(\bR^T d\bR)$ is invariant under right operation $\bR\bO^T$ using similar reasoning,
\begin{align}
\label{eqn:HMRinv} (\bR^T d\bR) \to \left(\bO \bR^T d(\bR) \bO^T\right) = (\bR^T d\bR),
\end{align}
where the equality follows from Lemma \ref{lem:adma} with $d\bM=\bR^T d\bR$, and $\bA=\bO^T$, since $\det \bO =1$. Note that (\ref{eqn:HMLinv}) and (\ref{eqn:HMRinv}) also imply that, for any operations $\bC\bR$ and $\bR\bC$, if $(\bR^Td\bR)$ is invariant then $\bC^T\bC=1=\det \bC$, or in other words, $(\bR^T d\bR)$ in only invariant under orthogonal transformation. So the GOE is well-defined by orthogonal invariance.

A measure with such invariance properties is called a left and/or right \textit{invariant measure}, or a (left/right) \textit{Haar measure}. (An excellent technical treatment of Haar measure is contained in \cite{nachbin1965}, while \cite{muirhead1982} is also very informative, specifically regarding the orthogonal group.)

\begin{definition} 
\label{def:Haar}
Let $G$ be a locally compact topological group and $H$ a Borel subgroup of $G$. If $\mu$ is a measure on $H$ and $\mu(h*H)=\mu(H)$ for all $h\in H$ then call $\mu$ a \textit{left invariant measure} or \textit{left Haar measure}.

Similarly, if $\mu(H)=\mu(H*h)$ then $\mu$ is called a \textit{right invariant measure} or \textit{right Haar measure}.
\end{definition}

It can be shown that left and right Haar measure is unique (up to a constant multiple), and in the case of the orthogonal group, the left and right Haar measures are the same (since $O(N)$ is compact). Combining this with the statements above we have that
\begin{align}
\label{eqn:Haar} (\bR^T d\bR)
\end{align}
is the unique Haar (or invariant) measure on $O(N)$, and we say the \textit{volume} of $O(N)$ is given by
\begin{align}
\label{def:volON}
\vol (O(N)):= \int_{O(N)} (\bR^T d\bR).
\end{align}
In order to complete the calculation of the eigenvalue jpdf we must calculate this volume, under the restriction that the first row of $\bR$ is positive. This amounts to integrating out the variables corresponding to the eigenvectors, which was the task implied by (\ref{eqn:PX_QL1}) and (\ref{eqn:PX_QL2}).
\begin{proposition}[\cite{muirhead1982} Corollary 2.1.16]
\label{prop:int_(RdR)}
Let $\bR$ be an $N \times N$ real, orthogonal matrix, with the first row of $\bR$ restricted to be positive, then we have
\begin{align}
\label{eqn:RTdR_integ} \int (\bR^T d\bR) = \frac{\pi^{N(N+1)/4}}{\prod_{j=1}^N\Gamma(j/2)}.
\end{align}
\end{proposition}

\textit{Proof}: Let $\bZ=[z_{j,k}]_{j,k=1,...,N}$ with the elements distributed as standard Gaussians
\begin{align}
\nonumber \frac{1}{\sqrt{2\pi}}e^{-z_{j,k}^2/2}.
\end{align}
(Note that $\bZ$ differs from $\bX$ as defined in (\ref{eqn:GOE_el_dist}) since $\bZ$ is not required to be symmetric.) The jpdf of the elements of $\bZ$ is therefore
\begin{align}
\nonumber P(\bZ)=(2\pi)^{-N^2/2}e^{-\sum_{j,k=1}^Nz_{j,k}^2/2}=(2\pi)^{-N^2/2}e^{-\mathrm{Tr}(\bZ^T\bZ)},
\end{align}
and, because the Gaussian distributions are normalised,
\begin{align}
\label{eqn:int_PZ=1} \int P(\bZ) (d\bZ)=1,
\end{align}
where the domain of integration is $\mathbb{R}^{N^2}$.

Now write $\bZ$ as $\bZ=\bH\bT$ where $\bH$ and $\bT$ are $N\times N$ matrices, $\bH$ is orthogonal and $\bT$ is upper triangular; this is known as a $\mathbf{QR}$ decomposition (here $\bQ=\bH$ and $\bT=\bR$) and can be accomplished by the Gram-Schmidt algorithm. To make the decomposition unique, the diagonal elements of $\bT$ are specified to be positive. With this decomposition we have
\begin{align}
\nonumber \Tr (\bZ^T\bZ)=\Tr(\bT^T\bT)=\sum_{1\leq j\leq k\leq N}t_{j,k}^2.
\end{align}
We also have
\begin{align}
\nonumber (d\bZ)=\prod_{j=1}^Nt_{j,j}^{N-j}(d\bT)(\bH^T d\bH),
\end{align}
which we can establish using the method of Proposition \ref{prop:GOE_J}. We see that (\ref{eqn:int_PZ=1}) becomes
\begin{align}
\label{eqn:hTdH1} \int \prod_{j=1}^Nt_{j,j}^{N-j}\prod_{1\leq j \leq k \leq N}e^{-t_{j,k}^2/2}dt_{j,k}\int (\bH^T d\bH)=(2\pi)^{N^2/2},
\end{align}
where the integrals over $t_{j,k}$ can be evaluated thusly
\begin{align}
\nonumber &\int \prod_{j=1}^Nt_{j,j}^{N-j}\prod_{1\leq j \leq k \leq N}e^{-t_{j,k}^2/2}dt_{j,k}\\
\nonumber &=\prod_{1\leq j < k \leq N}\int_{-\infty}^{\infty}e^{-t_{j,k}^2/2}dt_{j,k}\prod_{j=1}^N\int_{0}^{\infty}e^{-t_{j,j}^2/2}t^{N-j}_{j,j}dt_{j,j}\\
\nonumber &=\prod_{1\leq j < k \leq N}\sqrt{2\pi}\prod_{j=1}^N2^{j/2-1}\Gamma(j/2)\\
\label{eqn:hTdH2} &=2^{N^2/2-N}\pi^{N(N-1)/4}\prod_{j=1}^N\Gamma(j/2).
\end{align}
Substituting (\ref{eqn:hTdH2}) into (\ref{eqn:hTdH1}) we have
\begin{align}
\int (\bH^T d\bH)=\frac{2^N\pi^{N(N+1)/4}}{\prod_{j=1}^N\Gamma(j/2)}.
\end{align}

Since we specified the first row of $\bR$ to be positive, we divide through by $2^N$ (the number of possible signs in the first row) and we have the result.

\hfill $\Box$

\begin{remark}
Note that in the proof of Proposition \ref{prop:int_(RdR)} we found that
\begin{align}
\label{def:volO} \vol (O(N)) = \frac{2^N\pi^{N(N+1)/4}} {\prod_{j=1}^N\Gamma(j/2)}.
\end{align}
This will be of use in Chapter \ref{sec:truncs}.
\end{remark}

Combining Propositions \ref{prop:GOE_J} and \ref{prop:int_(RdR)} with (\ref{eqn:GOE_cov1}) we have the eigenvalue jpdf, which was first identified with the GOE (up to normalisation) in \cite{Wigner1957a}.

\begin{proposition}
\label{prop:GOE_eval_jpdf}
For $\bX$ an $N \times N$ real, symmetric matrix with iid Gaussian entries, the jpdf for the set $\vec{\lambda}=\{ \lambda_1,...,\lambda_N\}$, the eigenvalues of $\bX$, is
\begin{align}
\label{eqn:GOE_eval_jpdf} Q(\vec{\lambda})=2^{-3N/2}\prod_{j=1}^N\frac{e^{-\lambda_j^2/2}}{\Gamma(j/2+1)}\prod_{1\leq j < k \leq N}|\lambda_k-\lambda_j|.
\end{align}
\end{proposition}

\textit{Proof}: Substituting (\ref{eqn:dX_jacob}) into (\ref{eqn:PX_QL2}) using (\ref{eqn:RTdR_integ}) we almost have the result. The only extra concern is the constraints \textit{i} and \textit{ii} as discussed below (\ref{eqn:diag_decomp}). We have already accounted for the specification of the first row as positive at the end of the proof of Proposition \ref{prop:int_(RdR)}, while the relaxation of the ordering on the eigenvalues introduces a factor of $(N!)^{-1}$.

\hfill $\Box$

\begin{remark}
Note that we have implicitly ignored the matrices with repeated eigenvalues; we can do this since they form a set of measure zero inside the set of all GOE matrices. For the same reason we also ignore singular matrices and so we may take inverses with impunity.
\end{remark}

\subsection{Step III: Pfaffian form of generalised partition function}
\label{sec:Step3_GOE}

As mentioned after Proposition \ref{prop:GOE_J}, the main structural feature of (\ref{eqn:GOE_eval_jpdf}) is the product of differences (\ref{eqn:evalbeta}) with $\beta=1$, and this is one of the characteristic attributes of eigenvalue distributions where the entries of the matrix are real. (Recall from the Introduction that when the entries are complex then we find the same product raised to the power $\beta>1$.) A product of this form naturally leads to a determinantal expression via the identity
\begin{eqnarray}
\label{eqn:vandermonde}
\prod_{1\leq j<k\leq n}(x_k-x_j)=\mathrm{det}
\left[\begin{array}{ccccc}
1 & x_1 & x_1^2 & \cdot\cdot\cdot & x_1^{n-1}\\
1 & x_2 & x_2^2 & \cdot\cdot\cdot & x_2^{n-1}\\
\vdots & \vdots & \vdots & \ddots & \vdots\\
1 & x_n & x_n^2 & \cdot\cdot\cdot & x_n^{n-1}
\end{array}\right],
\end{eqnarray}
where (\ref{eqn:vandermonde}) is referred to as a \textit{Vandermonde determinant}. We can modify the identity to the form
\begin{eqnarray}
\label{eqn:vandermonde_polys}
\prod_{1\leq j<k\leq n}(x_k-x_j)=\mathrm{det}
\left[\begin{array}{ccccc}
p_0(x_1) & p_1(x_1) & p_2(x_1) & \cdot\cdot\cdot & p_{n-1}(x_1)\\
p_0(x_2) & p_1(x_2) & p_2(x_2) & \cdot\cdot\cdot & p_{n-1}(x_2)\\
\vdots & \vdots & \vdots & \ddots & \vdots\\
p_0(x_n) & p_1(x_n) & p_2(x_n) & \cdot\cdot\cdot & p_{n-1}(x_n)\\
\end{array}\right],
\end{eqnarray}
where $p_m(x)$ is a monic polynomial of degree $m$, by adding to each column appropriate multiples of the other columns. It will turn out that (\ref{eqn:vandermonde_polys}) is a more useful form for our desideratum.

This is all very gratifying, and will be crucial to the story that follows, however, in the case of GOE (and the other $\beta=1$ ensembles), we can look past the determinant and evince a deeper Pfaffian (or quaternion determinant) structure in a quantity called the \textit{generalised partition function}, from which we will calculate the correlation functions.
\begin{remark}
This Pfaffian structure also shows itself in the $\beta=4$ cases, although we shall not study them in this work. On the other hand, $\beta=2$ is traditionally analysed at the level of determinants and misses the Pfaffian substructure. While a determinant can always be rewritten trivially as a Pfaffian (see (\ref{eqn:chequer1}) and the surrounding discussion), in \cite{Sinclair2010} it is shown that there is a Pfaffian structure for $\beta=2$ which does not appear to be such a trivial rewriting, and in \cite{Kie2011} an explicit example of this has been found in the setting of chiral matrix ensembles (see also \cite{ForSinc2010}).
\end{remark}

\subsubsection{Quaternion determinants and Pfaffians}
\label{sec:qdets_pfs}

The \textit{quaternion determinant} was used by Dyson \cite{dyson1970} as a convenient notation for writing the eigenvalue correlation functions for the $\beta=1$ and $4$ cases of the circular ensemble. We will find that our correlations here similarly contain a quaternion structure and so we review some of the theory (\cite{mehta2004} contains a similar discussion). A good historical and technical overview is provided in \cite{dyson1972}.

A quaternion is analogous to a complex number, except it has four basis vectors instead of two. Typically they are written in the form $q=q_0+iq_1+jq_2+kq_3$, with the relations $i^2=j^2=k^2=ijk=-1$, and the $q_l$ are in general complex. Alternatively, quaternions can be represented as $2\times 2$ matrices $q=q_0\1+q_1\mathbf{e}_1+q_2\mathbf{e}_2+q_3\mathbf{e}_3$ using the Pauli spin matrices $\sigma_x,\sigma_y,\sigma_z$
\begin{align}
\nonumber &\1:=\left[\begin{array}{cc}
1 & 0\\
0 & 1
\end{array}\right], &&\mathbf{e}_1:=i\sigma_z=\left[\begin{array}{cc}
i & 0\\
0 & -i
\end{array}\right],\\
\nonumber &\mathbf{e}_2:=i\sigma_y=\left[\begin{array}{cc}
0 & 1\\
-1 & 0
\end{array}\right], &&\mathbf{e}_3:=i\sigma_x=\left[\begin{array}{cc}
0 & i\\
i & 0
\end{array}\right].
\end{align}
For $q_0=a+ib,q_1=c+id,q_2=e+if,q_3=g+ih$ we have
\begin{align}
\label{eqn:mat_quat} q=\left[ \begin{array}{cc}
w & x\\
y & z
\end{array}\right],
\end{align}
where $w=(a-d)+i(b+c),x=(e-h)+i(f+g),y=-(e+h)+i(g-f),z=(a+d)+i(b-c)$. The analogue of complex conjugation for quaternions we denote $\bar{q}=q_0-iq_1-jq_2-kq_3$, or in the matrix representation
\begin{align}
\nonumber \bar{q}=\left[ \begin{array}{cc}
z & -x\\
-y & w
\end{array}\right].
\end{align}
With the representation (\ref{eqn:mat_quat}) an $N \times N$ matrix with quaternion elements $\bQ=[q_{j,k}]$ can be viewed as a $2N \times 2N$ matrix with complex elements.

In the case that $q_0,q_1,q_2,q_3\in\mathbb{R}$ we say that $q$ is a \textit{real quaternion} and from (\ref{eqn:mat_quat}), with $\alpha=a+ic$ and $\beta=e+ig$, we have
\begin{align}
\label{def:real_quats} q=\left[ \begin{array}{cc}
\alpha & \beta\\
-\bar{\beta} & \bar{\alpha}
\end{array}\right],
\end{align}
with conjugate
\begin{align}
\nonumber \bar{q}=\left[ \begin{array}{cc}
\bar{\alpha} & -\beta\\
\bar{\beta} & \alpha
\end{array}\right].
\end{align}
A matrix $\bQ=[q_{j,k}]$, is said to be \textit{quaternion real} if all the quaternion elements $q_{j,k}$ are real quaternions.

We denote by $\bQ^D$ the matrix $[\bar{q}_{k,j}]$, and we call it the \textit{dual} of $\bQ$. If $\bQ=\bQ^D$ then $\bQ$ is said to be \textit{self-dual}.

\begin{definition} [Quaternion determinant]
Let $\bQ=[q_{j,k}]$ be an $N \times N$ self-dual matrix of $2\times 2$ real quaternions as in (\ref{def:real_quats}). The \textit{quaternion determinant} is defined by
\begin{equation}
\label{def:qdet} \mathrm{qdet}[\bQ]=\sum_{P\in S_N}(-1)^{N-l}\prod_1^l(q_{ab}q_{bc}\cdot\cdot\cdot q_{sa})^{(0)}.
\end{equation}
The superscript $(0)$ denotes the operation $\frac{1}{2}\mathrm{Tr}$ of the quantity in brackets. $P$ is any permutation of $(1,...,N)$ that consists of $l$ disjoint cycles of the form $(a\rightarrow b \rightarrow c \rightarrow \cdot\cdot\cdot \rightarrow s \rightarrow a)$.
\end{definition}

\begin{remark}
If the $q_{j,k}$ are scalar multiples of the identity, say $q_{j,k}=c_{j,k}\1_2$, then $\mathrm{qdet}[\bQ]=\mathrm{det}[\bC]$ where $\bC=[c_{j,k}]$.
\end{remark}

A structure that is closely related to the quaternion determinant is the Pfaffian.
\begin{definition} [Pfaffian]
\label{def:pfaff}
Let $\bX=[x_{ij}]_{i,j=1,...,2N}$, where $x_{ji}=-x_{ij}$, so that $\bX$ is an anti-symmetric matrix of even size. Then the \textit{Pfaffian} of $\bX$ is defined by
\begin{align}
\nonumber \mathrm{Pf} [\bX]&=\sum^*_{P(2l)>P(2l-1)}\varepsilon (P) x_{P(1),P(2)}x_{P(3),P(4)}\cdot\cdot\cdot x_{P(2N-1),P(2N)}\\
\label{def:Pf} &=\frac{1}{2^NN!}\sum_{P\in S_{2N}}\varepsilon (P) x_{P(1),P(2)}x_{P(3),P(4)}\cdot\cdot\cdot x_{P(2N-1),P(2N)},
\end{align}
where $S_{2N}$ is the group of permutations of $2N$ letters and $\varepsilon (P)$ is the sign of the permutation $P$. The * above the first sum indicates that the sum is over distinct terms only (that is, all permutations of the pairs of indices are regarded as identical).
\end{definition}

\begin{remark}
\label{rem:Pf_def}
In the second equality of (\ref{def:Pf}) the factors of $2$ are associated with the restriction $P(2l)>P(2l-1)$ while the factorial is associated with counting only distinct terms ($N!$ is the number of ways of arranging the $N$ pairs of indices $\{ P(2l-1),P(2l)\}$).
\end{remark}

\begin{remark}
At the risk of confusion, we shall use the terms \textit{skew-symmetric} and \textit{anti-symmetric} synonymously.
\end{remark}

In his 1815 publication, in an effort to solve certain classes of differential equations, Pfaff dealt with a structure that became what we know as Pfaffians. Determinants were not in common use at the time, and so they were not seen as being of a similar form. The treatment was formalised by Jacobi and recognised as an analogue of a determinant, indeed it was he who proved that skew-symmetric determinants of odd size are zero. As for nomenclature, Jacobi referred to Pfaff's Method (`\textit{Pfaffsche Methode}') in 1827, but when Cayley takes up the discussion in 1847 he refers to Jacobi (`\textit{les fonctions de M. Jacobi}'), before changing the eponym to Pfaff in a paper of 1854. The following relationship between a Pfaffian and a determinant  of a skew-symmetric matrix $\bA$,
\begin{align}
\label{eqn:pf_det} (\Pf \bA)^2 = \det \bA,
\end{align}
is also a classical result.

\begin{remark}
The historical information comes from Thomas Muir in \cite{Muir1906} and \cite{Muir1911}. He provides interesting contextualising commentary on many of the seminal papers in the theory of determinants.
\end{remark}

We can also trivially rewrite any determinant as a Pfaffian of a chequerboard matrix
\begin{align}
\label{eqn:chequer1} \det \bA =\Pf \; \tilde{\bA}_1,
\end{align}
where the $(2i-1)$-th row of $\tilde{\bA}_1$ is $[0, a_{i,1},0,a_{i,1}, ..., 0, a_{i,N}]$, with the remaining elements being determined by the required anti-symmetry. For example, with $N=3$,
\begin{align}
\label{eqn:N3chequer} \tilde{\bA}_1=\left[ \begin{array}{cccccc}
0 & a_{11} & 0 & a_{1,2} & 0 & a_{1,3}\\
-a_{11} & 0 & -a_{2,1} & 0 & -a_{3,1} & 0\\
0 & a_{2,1} & 0 & a_{2,2} & 0 & a_{2,3}\\
-a_{1,2} & 0 & -a_{2,2} & 0 & -a_{3,2} & 0\\
0 & a_{3,1} & 0 & a_{3,2} & 0 & a_{3,3}\\
-a_{1,3} & 0 & -a_{2,3} & 0 & -a_{3,3} & 0
\end{array}
\right].
\end{align}
A more compact description of this correspondence specifies the determinant matrix in terms of the chequerboard matrix
\begin{align}
\label{eqn:chequer} \Pf \tilde{\bA}_1=\det [\alpha_{2i-1,2j}]_{i,j=1,...,N},
\end{align}
where $\tilde{\bA}_1=[\alpha_{i,j}]_{i,j=1,...,2N}$. We will have use of (\ref{eqn:chequer}) in Chapter \ref{sec:pnk}.

Alternatively, we may rewrite the determinant as a Pfaffian with blocks of zeros on the diagonal
\begin{align}
\nonumber \det \bA = \Pf \left[\begin{array}{cc}
\0_{N\times N} & \tilde{\bA}_2\\
-\tilde{\bA}_2 & \0_{N\times N}
\end{array}\right],
\end{align}
where $\tilde{\bA}_2$ is given by elementary transformations of $\bA$ as so
\begin{align}
\nonumber \tilde{\bA}_2=\left[\begin{array}{c}
a_{2i-1,j}\\
-a_{2i,j}
\end{array} \right]_{i=1,...,N/2 \atop j=1,...,N}.
\end{align}
From these facts we note Pfaffian/quaternion determinantal processes are equivalent to determinantal processes, albeit with special structure.

Usefully, Pfaffians can be calculated using a form of Laplace expansion. To calculate a determinant, recall that we can expand along any row or column. For example, expand a matrix $A=[a_{ij}]_{i,j=1,...n}$ along the first row:
\begin{equation}
\nonumber \mathrm{det}[A]=a_{1,1}\mathrm{det}[A]^{1,1}-a_{1,2}\mathrm{det}[A]^{1,2} + \cdot\cdot\cdot (-1)^{n+1} a_{1,n}\mathrm{det}[A]^{1,n},
\end{equation}
where $\mathrm{det}[A]^{i,j}$ means the determinant of the matrix left over after deleting the $i$th row and $j$th column.

The analogous expansion for a Pfaffian involves deleting \textit{two} rows and \textit{two} columns each time. For example, expanding a skew-symmetric matrix $B=[b_{ij}]_{i,j=1,...n}$ ($n$ even) along the first row:
\begin{equation}
\nonumber \mathrm{Pf}[B]=b_{1,1}\mathrm{Pf}[B]^{1,1}-b_{1,2}\mathrm{Pf}[B]^{1,2} + \cdot\cdot\cdot (-1)^n b_{1,n}\mathrm{Pf}[B]^{1,n},
\end{equation}
where $\mathrm{Pf}[B^{i,j}]$ means the Pfaffian of the matrix left after deleting the $i$th  and $j$th rows and the $i$th and $j$th columns. Laplace expansion requires $n!$ calculations for a determinant, and $n!!=n\cdot (n-2)\cdot (n-4 )\cdot...$ in the case of a Pfaffian.

We can also identify quaternion determinant and Pfaffian analogues of a diagonal matrix. A determinant is most easily calculated if its matrix is diagonal, since then
\begin{align}
\nonumber \det\Big( \mathrm{diag} [a_1,...,a_N] \Big) =\prod_{j=1}^N a_j.
\end{align}
From (\ref{def:qdet}) we see that the analogous result for the quaternion determinant is
\begin{align}
\label{eqn:qdet_diag} \qdet\Big( \mathrm{diag}[a_1,a_1,...,a_{N/2}, a_{N/2}]\Big) = \prod_{j=1}^{N/2} a_j.
\end{align}

In the case of Pfaffians, however, clearly, diagonal matrices (with at least one non-zero element) are not skew-symmetric and so the Pfaffian of a diagonal matrix is undefined. However, we can define a suitably analogous matrix for a Pfaffian as
\begin{align}
\label{eqn:skew_diag_mat} \bA^{(D)}=\left[\begin{array}{cccc}
\bA_1 & \0 & \cdot\cdot\cdot & \0\\
\0 & \bA_2 & \cdot\cdot\cdot & \0\\
\vdots & \vdots & \ddots & \vdots\\
\0 & \0 & \cdot\cdot\cdot & \bA_{N/2}
\end{array}
\right],
\end{align}
where $\bA_j=\left[ \begin{array}{cc}
0 & a_j\\
-a_j & 0
\end{array}
\right]$ and $\0$ is the $2\times 2$ zero matrix. That is, the matrix has entries $\{a_1,...,a_{N/2} \}$ along the diagonal above the main diagonal, and $\{-a_1,...,-a_{N/2} \}$ on the diagonal just below the main diagonal, with zeros elsewhere. We call such a matrix \textit{skew-diagonal}. The analogy with the diagonal matrix of a quaternion determinant (\ref{eqn:qdet_diag}) comes from the fact that
\begin{align}
\label{eqn:pf_skew_diag_eval} \Pf \bA^{(D)}=\prod_{j=1}^{N/2}a_j.
\end{align}
Note that in (\ref{eqn:skew_diag_mat}) and (\ref{eqn:pf_skew_diag_eval}), we have implicitly assumed that $N$ is even. In the case that $N$ is odd there are additional technical details, which are dealt with in Chapter \ref{sec:GOE_odd}.

From the preceding we see that there is clearly a relationship between quaternion determinants and Pfaffians, and we would like to formalise this, but we first need the quaternion determinant analogue of (\ref{eqn:pf_det}) for $\bM$ a self-dual matrix \cite[Theorem 2]{dyson1970},
\begin{align}
\label{eqn:qdet2=det} (\qdet [\bM])^2=\det[\bM].
\end{align}
We also need to define
\begin{equation}
\label{def:Z2N} \bZ_{2N}:=\mathbf{1}_N\otimes \left[\begin{array}{cc}
0 & -1\\
1 & 0\\
\end{array}\right].
\end{equation}

\begin{proposition}
\label{prop:qdet=pf}
With $\bM$ a $2N \times 2N$ self-dual matrix and $\bZ_{2N}$ from (\ref{def:Z2N}) we have
\begin{align}
\nonumber \mathrm{Pf}[\bM\bZ^{-1}_{2N}]&=\mathrm{Pf}[\bZ^{-1}_{2N}\bM]=\mathrm{qdet} [\bM],\\
\label{eqn:qdet=pf} \mathrm{Pf}[\bM\bZ_{2N}]&=\mathrm{Pf}[\bZ_{2N}\bM]=(-1)^N\mathrm{qdet} [\bM].
\end{align}
\end{proposition}

\textit{Proof:} First we must be sure that the equations are well formed, that is, that if $\bM$ is a self-dual matrix then the result of operation by $\bZ_{2N}$ or $\bZ_{2N}^{-1}$ is anti-symmetric. Note that the tangible effect of right multiplication by $\bZ_{2N}^{-1}$ on any matrix $\bM$ is to interchange every pair of columns, and multiply the leftmost of each pair by $-1$. That of left multiplication is to interchange every pair of rows and multiply the bottom-most by $-1$. (Right/left multiplication by $\bZ_{2N}$ will also interchange each pair of columns/rows, but will multiply the \textit{rightmost} column/\textit{top-most} row of each pair by $-1$, since $\bZ_{2N}^{-1}=-\bZ_{2N}$.)

If the elements of $\bM$ are $\{ m_{j,k}\}_{j,k=1,...,2N}$ then self-duality implies $m_{2j-1,2k-1}=m_{2k,2j}$, $m_{2j-1,2k}=-m_{2k-1,2j}$, $m_{2j,2k-1}= -m_{2k,2j-1}$, $m_{2j,2k}= m_{2k-1,2j-1}$. If we now operate on this matrix with $\bZ_{2N}^{-1}$ on the right, the discussion above shows that the second index in each entry is switched from even to odd or vice-versa, with the extra condition that a change from even to odd picks up a negative sign. Applying this to the entries of $\bM$ we have $m_{2j-1,2k}=-m_{2k,2j-1},m_{2j-1,2k-1}=-m_{2k-1,2j-1}, m_{2j,2k}=-m_{2k,2j}, m_{2j,2k-1}= -m_{2k-1,2j}$, which the condition for an anti-symmetric matrix. A similar argument also works for the other operations.

From (\ref{eqn:pf_skew_diag_eval}) we see that $\Pf [\bZ_{2N}^{-1}]=1=(-1)^N\Pf [\bZ_{2N}]$, and so, by (\ref{eqn:pf_det}) and (\ref{eqn:qdet2=det}) we have that
\begin{align}
\nonumber (\Pf [\bM\bZ_{2N}^{-1}])^2=\det [\bM\bZ_{2N}^{-1}]=\det [\bZ_{2N}^{-1}\bM]=\det [\bM]=(\qdet [\bM])^{2}
\end{align}
and
\begin{align}
\nonumber (\Pf [\bM\bZ_{2N}])^2&=(-1)^{2N}\det [\bM\bZ_{2N}]=(-1)^{2N}\det [\bZ_{2N}\bM]=(-1)^{2N}\det [\bM]\\
&=((-1)^{N}\qdet [\bM])^{2}.
\end{align}
With $\bM$ the identity we establish the sign after taking the square root.

\hfill $\Box$

We can generalise the preceding result to any skew-diagonal matrix.

\begin{corollary}
\label{cor:qdet=pf}
For a skew-diagonal $2N\times 2N$ matrix $\bA$ as defined in (\ref{eqn:skew_diag_mat}) and a self-dual $2N\times 2N$ matrix $\bM$ we have
\begin{align}
\nonumber \Pf[\bM\bA]=\Pf[\bA\bM]=\Pf[\bA]\; \qdet[\bM]=\prod_{j=1}^{N/2}a_j\; \qdet[\bM].
\end{align}
\end{corollary}

\textit{Proof}: First note that $\bA=\mathbf{D}\bZ_{2N}^{-1}=\bZ_{2N}^{-1}\mathbf{D}$ where $\mathbf{D}=\mathrm{diag}[a_0,a_0,a_1,a_1,...,a_N,a_N]$. Then
\begin{align}
\nonumber \Pf[\bM\bA]=\Pf[\bA\bM]&=\Pf[\mathbf{D}\bZ_{2N}^{-1}\bM]\\
\nonumber &=\prod_{j=1}^{N/2}a_j \; \Pf[\bZ_{2N}^{-1}\bM]
\end{align}
where we have used (\ref{def:Pf}) for the last equality. The result now follows from Proposition \ref{prop:qdet=pf}.

\hfill $\Box$

With Proposition \ref {prop:qdet=pf} in mind, we see that quaternion determinants and Pfaffians are trivially related, a relation that will be exploited in this work. The only subtlety is that the Pfaffian matrix must be anti-symmetric and the quaternion matrix must be self-dual.

\subsubsection{Generalised partition function}

With $P(x_1,...,x_N)$ a probability density function, the average of the function $f(x_1,...,x_N)$ is
\begin{align}
\label{def:av} \langle f(x_1,...,x_N) \rangle_P:=\int_{\Omega}f(x_1,...,x_N)P(x_1,...,x_N)dx_1\cdot\cdot\cdot dx_N,
\end{align}
where $\Omega$ is the support of $P$. A special case is $f(x_1,...,x_N)=\prod_{j=1}^N u(x_j)$; choosing $u(x)=\chi_{x\in S}$, where $\chi_A=1$ if $A$ is true and $\chi_A=0$ otherwise, gives that (\ref{def:av}) equals the probability that all eigenvalues are in the set $S$. Our interest in $\langle \prod_{j=1}^N u(x_j) \rangle_P$ for general $u$ stems from its use in calculating correlation functions by applying functional differentiation (see (\ref{eqn:fnal_diff_correln}) below), although, it does have more general use. For instance, if $u(x)=1-\zeta \chi_{x_j\in J}$ then the probability that $n$ eigenvalues lie in the set $J$ is given by the $n$th derivative with respect to $\zeta$ (times some combinatorial factor). See \cite[Chapter 8]{forrester?} and \cite{tracy_and_widom1998} for more details.

\begin{definition}
\label{def:gen_part_fn}
Let $\mathcal{Q}(\mathbf{x})$ be the jpdf of the set $\mathbf{x}=\{x_1,...,x_N\}$ and define the generalised partition function of $\mathbf{x}$ as
\begin{align}
\label{def:single_gen_part_fn} Z_N[u]:=\Big\langle \prod_{j=1}^Nu(x_j) \Big\rangle_{\mathcal{Q}} = \int\prod_{j=1}^Nu(x_j)\; \mathcal{Q}(\mathbf{x})\; d\mathbf{x}.
\end{align}
In the case that $\mathbf{x}=\bigcup_{l=1}^m\mathbf{x}^{(l)}$, that is, $\mathbf{x}$ consists of multiple disjoint sets $\mathbf{x}^{(l)}=\{x_1^{(l)},...,x_{N_l}^{(l)}\}$, each containing elements of a different species, then define
\begin{align}
\label{def:multi_gen_part_fn} Z_N[u_1,...,u_m]:=\int\left(\prod_{j=1}^{N_1} u_1\left(x_j^{(1)}\right)\cdot\cdot\cdot\prod_{j=1}^{N_m} u_m\left(x_j^{(m)}\right)\right)\mathcal{Q}(\mathbf{x})\; d\mathbf{x}.
\end{align}
\end{definition}

The multiple disjoint sets of Definition \ref{def:gen_part_fn} correspond to the sets of eigenvalues in the ensemble. While (\ref{def:multi_gen_part_fn}) is unnecessarily general for a study of GOE (where there are only real eigenvalues), it will become relevant in the following chapters where we discuss matrices whose eigenvalues are either real, or non-real complex conjugate pairs.

It will turn out that the generalised partition function for GOE can be written in a convenient quaternion determinant or Pfaffian form, and then, in such a case, the correlation functions --- given as functional derivatives in (\ref{eqn:fnal_diff_correln}) below --- are a particularly terse quaternion determinant or Pfaffian expression.

\subsubsection{Pfaffian generalised partition function for GOE, $N$ even}
\label{sec:pf_gpf}

Before we proceed, note that Definition \ref{def:pfaff} only applies when the size of the matrix is even --- for now we will make this assumption. The case of $N$ odd will be dealt with in Chapter \ref{sec:GOE_odd}, where the particular problems presented by parity will be explored.

In the case of GOE we have only one species of eigenvalue (they are all real) and so we substitute (\ref{eqn:GOE_eval_jpdf}) into (\ref{def:single_gen_part_fn}) to find
\begin{align}
\nonumber Z_N[u]&=2^{-3N/2}\prod_{j=1}^N\frac{1}{\Gamma(j/2+1)} \int_{-\infty}^{\infty}d\lambda_1\cdot\cdot\cdot\int_{-\infty}^{\infty}d\lambda_N \\
\label{def:GOE_gpf1} &\times \prod_{j=1}^N u(\lambda_j)\: e^{-\lambda_j^2/2}\prod_{1\leq j < k \leq N}|\lambda_k-\lambda_j|.
\end{align}
Since $Q(\vec{\lambda})$ from Proposition \ref{prop:GOE_eval_jpdf} is a jpdf, we see that $Z_N[1]=1$.

Our task now is to express (\ref{def:GOE_gpf1}) in Pfaffian form. The method that will be used here and in the following chapters is known as the method of integration over alternate variables, which was introduced by de Bruijn \cite{deB1955} and applied to the present problem by Mehta \cite{mehta1960,Mehta1967}. The purpose of this method is to deal with the absolute value signs around the product of differences in the eigenvalue jpdf. The method is required since we note that (\ref{eqn:vandermonde_polys}) refers to a signed product of differences, and so we cannot apply (\ref{eqn:vandermonde_polys}) to (\ref{def:GOE_gpf1}) directly. However, if the eigenvalues have their ordering reinstated --- which is equivalent to a corresponding restriction to the domain of integration --- then the $|\cdot|$ can be removed and the identity applies. Integration over alternate variables is a technique to perform these integrals with ordered domain over a Vandermonde determinant. The method will be illustrated in the proof of the following proposition.

\begin{proposition}
\label{prop:GOE_gen_part_fn}
Let $Q(\vec{\lambda})$ be the eigenvalue jpdf for the GOE matrices, as given in (\ref{eqn:GOE_eval_jpdf}). For $N$ even the generalised partition function (as defined in (\ref{def:single_gen_part_fn})) for $Q(\vec{\lambda})$ is
\begin{align}
\label{eqn:GOE_GPF_even} Z_N[u]=\frac{N!}{2^N}\prod_{j=1}^N\frac{1}{\Gamma(j/2+1)}\Pf [\gamma_{jk}]_{j,k=1,...,N},
\end{align}
where
\begin{align}
\label{eqn:gammajk} \gamma_{jk}=\frac{1}{2}\int_{-\infty}^{\infty}dx\: e^{-x^2/2}\: u(x)\: p_{j-1}(x)\int_{-\infty}^{\infty}dy \: e^{-y^2/2}\: u(y)\: p_{k-1}(y)\: \sgn(y-x),
\end{align}
and $\{p_j(x)\}_{j,k=0,1,...,N-1}$ are arbitrary monic polynomials of degree $j$.
\end{proposition}

\textit{Proof}: We start by ordering the eigenvalues $-\infty < x_1 <\cdot\cdot\cdot < x_N < \infty$ (incurring a factor of $N!$) in (\ref{def:GOE_gpf1}) so that we can remove the $|\cdot|$ from the product of differences, putting it into Vandermonde form. With $A_N:= 2^{-3N/2}\prod_{j=1}^N\frac{1}{\Gamma(j/2+1)}$ this reordering gives
{\small
\begin{align}
\nonumber &Z_N[u]=A_NN!\int_{-\infty}^{x_2}dx_1 \int_{x_1}^{x_3}dx_2 \cdot\cdot\cdot \int_{x_{N-1}}^{\infty}dx_N\prod_{j=1}^Ne^{-x_j^2/2}\; u(x_j)\prod_{1\leq j < k \leq N}(x_k-x_j)\\
\nonumber &=A_NN!\int_{-\infty}^{x_2}dx_1\int_{x_1}^{x_3} dx_2\cdot\cdot\cdot \int_{x_{N-1}}^{\infty} dx_N \det \left[ e^{-x_j^2/2}u(x_j)p_{k-1}(x_j) \right]_{j,k=1,...,N}\\
\nonumber &=A_NN! \int_{-\infty}^{x_4}dx_2 \int_{x_2}^{x_6}dx_4 \cdot\cdot\cdot\int_{x_{N-2}}^{\infty}dx_N\\
\nonumber &\times \det \left[ \begin{array}{c}
\int_{-\infty}^{x_{2j}}e^{-x^2/2}u(x)p_{k-1}(x)dx\\
e^{-x_{2j}^2/2}u(x_{2j})p_{k-1}(x_{2j})
\end{array}\right]_{j=1,...,N/2 \atop k=1,...,N},
\end{align}
}where, for the second equality, we have used (\ref{eqn:vandermonde_polys}) and for the third we have made use of the observation that all dependence on $x_i$ occurs in row $i$ so the integrals can be applied individually to the relevant row of the determinant. The integrals over the odd numbered variables have been moved into the determinant and then, by adding the first row to the third row, and the first and third rows to the fifth row, and so on, all the integrals have lower terminal $-\infty$.

We see that the determinant is now symmetric in the variables $x_2,x_4,...,x_N$, and so we can remove the ordering $x_2 < x_4 < ... < x_N$ at the cost of dividing by $(N/2)!$. Expanding the determinant we find
\begin{align}
\nonumber Z_N[u]=A_N\frac{N!}{(N/2)!}\sum_{P\in S_N}\varepsilon(P)\prod_{l=1}^{N/2}\mu_{P(2l-1),P(2l)},
\end{align}
where
\begin{align}
\label{def:mu_GOE} \mu_{j,k}:=\int_{-\infty}^{\infty}dx\: e^{-x^2/2}\:u(x)\:p_{k-1}(x)\int_{-\infty}^xdy\: e^{-y^2/2}\:u(y)\:p_{j-1}(y),
\end{align}
and $\varepsilon(P)$ is the sign of the permutation $P$. By defining
\begin{align}
\nonumber \gamma_{j,k}:=\frac{1}{2}(\mu_{j,k}-\mu_{k,j}),
\end{align}
then we can restrict the sum to terms with $P(2l)>P(2l-1)$ and use the second equality in Definition \ref{def:pfaff} (recalling Remark \ref{rem:Pf_def}) to write
\begin{align}
\nonumber Z_N[u]=A_N2^{N/2} N!\sum_{P\in S_N \atop P(2l)>P(2l-1)}^*\varepsilon(P)\prod_{l=1}^{N/2}\gamma_{P(2l-1),P(2l)}.
\end{align}
Now using the first equality in Definition \ref{def:pfaff} we have the result.

\hfill $\Box$

\subsection{Step IV: Skew-orthogonal polynomials}
\label{sec:skew_orthog_polys}

As discussed above, if a Pfaffian is in skew-diagonal form (\ref{eqn:skew_diag_mat}), then it is easily calculated as the product of the upper diagonal entries. With the goal of achieving such a simplified form, we define an inner product $\langle p_j, p_k\rangle$ with a set of monic polynomials $p_0,p_1, \dots$ such that for $a_j \neq 0$
\begin{align}
\label{eqn:so_polys} \langle p_{2j},p_{2k}\rangle = \langle p_{2j+1},p_{2k+1}\rangle=0 &,&\langle p_{2j},p_{2k+1}\rangle=-\langle p_{2k+1},p_{2j}\rangle=\delta_{j,k}a_j.
\end{align}
Using these polynomials, the matrix $[\langle p_j, p_k\rangle]_{j,k=0,...,N-1}$ is in skew-diagonal form and its Pfaffian is given by (\ref{eqn:pf_skew_diag_eval}). We call the polynomials satisfying (\ref{eqn:so_polys}) \textit{skew-orthogonal polynomials}. If an appropriate inner-product can be defined such that the matrix in (\ref{eqn:GOE_GPF_even}) is of the form $[\langle p_j,p_k\rangle]_{j,k =0,...,N-1}$, and the corresponding skew-orthogonal polynomials can be found, then the calculation of $Z_N$ will be greatly simplified, and the correlation function may be computed. This, then, is the next task.

\subsubsection{Skew-orthogonal polynomials for GOE}
\label{sec:GOE_sops}

\begin{definition}
\label{def:GOE_soip}
Let $\langle p,q\rangle$ be the inner product defined by
\begin{align}
\label{eqn:GOE_soip} \langle p,q\rangle&:= \frac{1}{2}\int_{-\infty}^{\infty}dx\: e^{-x^2/2}\: p(x)\int_{-\infty}^{\infty}dy \: e^{-y^2/2}\: q(y)\: \sgn(y-x).
\end{align}
Also let $\{ R_j \}_{j=0,1,...}$ be a set of monic skew-orthogonal polynomials, satisfying the conditions (\ref{eqn:so_polys}) with respect to the inner product (\ref{eqn:GOE_soip}). (These are not unique as any replacement $p_{2m+1}(x) \mapsto p_{2m+1}(x) +c \: p_{2m}(x)$, where $c$ is some constant, leaves (\ref{eqn:so_polys}) unchanged by the linearity property of inner products.)
\end{definition}

\begin{remark}
There exist formulae for the skew-orthogonal polynomials, corresponding to various weight functions, in determinantal \cite{ForNagRai2006} and Pfaffian forms \cite{AkeKiePhi2010}, however the calculations implied by these methods may not be tractable (for instance, in the case of the truncated ensembles of Chapter \ref{sec:truncs}).
\end{remark}

We note that since the inner product (\ref{eqn:GOE_soip}) is just $\gamma_{jk}\big|_{u=1}$ of Proposition \ref{prop:GOE_gen_part_fn}, then the skew-orthogonal polynomials $R_0,R_1,...$ corresponding to this inner product will skew-diagonalise the matrix in (\ref{eqn:GOE_GPF_even}) with $u=1$. We will present these skew-orthogonal polynomials, and verify that they indeed satisfy (\ref{eqn:so_polys}) --- a derivation of these polynomials can be found in \cite{afnvm2000} and \cite[Chapter 6.4]{forrester?}, where use is made of facts pertaining to the $\beta=2$ and $\beta=4$ Gaussian ensembles. The skew-orthogonal polynomials for GOE turn out to be proportional to the \textit{Hermite polynomials}
\begin{align}
\nonumber H_n(x):=&\;(-1)^n e^{x^2}\frac{d^n}{dx^n}e^{-x^2}\\
\label{eqn:herm_polys} =&\;\sum_{m=0}^{\lfloor n/2\rfloor}(-1)^m2^{(n-m)}{n \choose 2m}\frac{(2m)!}{2^m m!}\; x^{(n-2m)},
\end{align}
where $n=0,1,...$ corresponds to the degree of the polynomial and $\lfloor x\rfloor$ is the floor function. Note that $H_n(x)$ is an even or odd function of $x$ depending on the parity of $n$. The Hermite polynomials have the remarkable recursive properties
\begin{align}
\label{eqn:recursive_herm} H_{n+1}(x)=2x\:H_{n}(x)-2n\:H_{n-1}(x)&,&&\frac{d}{dx}H_{n}(x)=2n\:H_{n-1}(x),
\end{align}
as well as satisfying the orthogonality condition
\begin{align}
\label{eqn:orthog_herm} \int_{-\infty}^{\infty}H_n(x)H_m(x)e^{-x^2}dx=\delta_{n,m}n! \:2^n\sqrt{\pi}.
\end{align}
Given that (\ref{eqn:GOE_soip}) also has a negative squared exponential weight, in light of (\ref{eqn:orthog_herm}) it is perhaps not surprising that Hermite polynomials appear as a result of skew-orthogonalising.

\begin{proposition}
\label{prop:GOE_soip}
Let $H_n(x)$ $(n=0,1,...)$ be the Hermite polynomials (\ref{eqn:herm_polys}). Skew-orthogonal polynomials corresponding to the inner product (\ref{eqn:GOE_soip}) are
\begin{align}
\nonumber R_{2j}(x)=\frac{H_{2j}(x)}{2^{2j}}&,&R_{2j+1}(x)&=\frac{H_{2j+1}(x)}{2^{2j+1}}-j\: \frac{H_{2j-1}(x)}{2^{2j-1}}\\
\label{eqn:sops} &&&=\frac{e^{x^2/2}}{2^{2j}}\frac{d}{dx}e^{-x^2/2}H_{2j}(x).
\end{align}
The normalisation is
\begin{align}
\label{eqn:GOE_norm} \langle R_{2j},R_{2j+1}\rangle = r_j=\frac{\Gamma (2j+1)}{2^{2j}}\sqrt{\pi}.
\end{align}
\end{proposition}

\textit{Proof}: The anti-symmetric condition is apparent from the presence of the sign function in (\ref{eqn:GOE_soip}). The conditions $\langle R_{2j},R_{2k}\rangle = \langle R_{2j+1},R_{2k+1}\rangle=0$ are easily checked: the inner integral (over $y$) will yield a function of opposite parity to the integrand of the outer integral, resulting in integration over an odd function from $-\infty$ to $\infty$, and so it is zero.

Now assume the polynomial degrees are of opposite parity. First, using the recursive properties (\ref{eqn:recursive_herm}) of Hermite polynomials we can establish the second equality of $R_{2j+1}(x)$ in (\ref{eqn:sops}). With this in hand we find that
\begin{align}
\nonumber \langle R_{2j},R_{2j+1}\rangle&=2^{-4j}\int_{-\infty}^{\infty}e^{-x^2}H_{2j}(x)H_{2k}(x)\:dx
\end{align}
and then from the orthogonality property (\ref{eqn:orthog_herm}) we have (\ref{eqn:GOE_norm}).

\hfill $\Box$

The immediate consequence of the polynomials in Proposition \ref{prop:GOE_soip} is that the Pfaffian in (\ref{eqn:GOE_GPF_even}) can be evaluated using (\ref{eqn:pf_skew_diag_eval}) as
\begin{align}
\label{eqn:pf=prodr} \Pf[\gamma_{j,k}]\Big|_{u=1}=\prod_{j=0}^{N/2-1}r_j.
\end{align}
Hence we can calculate $Z_N[1]$, which turns out to be $1$, as we knew it must be from the comment below (\ref{def:GOE_gpf1}). However, the important point is not that the generalised partition function has unit evaluation, but that the form it takes, using the skew-orthogonal polynomials, will be useful in the calculation of the correlation functions.
\begin{corollary}
\label{cor:ZN[1]}
With the generalised partition function $Z_N[u]$ from (\ref{def:GOE_gpf1}) and $r_0,...,r_{N/2-1}$ given by (\ref{eqn:GOE_norm}), we have
\begin{align}
\label{eqn:ZN[1]_eval} Z_N[1]=1=\frac{N!}{2^N}\prod_{j=1}^N\frac{1}{\Gamma(j/2+1)}\prod_{j=0}^{N/2-1}r_j.
\end{align}
\end{corollary}

\subsection{Step V: Correlation functions}
\label{sec:detPfcorrelns}

A statistic commonly of interest in random matrix systems is the eigenvalue density and higher order generalisations of the density, collectively called \textit{correlation functions}. A calculation of these correlation functions for various ensembles is a major aim of this work. We begin with the definition of correlation functions and then go on to discuss various tools and methods used in their calculation.

\begin{definition}
\label{def:integ_correlns}
For an ensemble of $N\times N$ matrices with eigenvalues $\lambda_1,...,\lambda_N$ in the set $\Omega$ and with eigenvalue jpdf $\mathcal{Q}(\lambda_1,...,\lambda_N)$ the \textit{$n$-point correlation function} of the positions $r_1,...,r_n$ is given by
\begin{align}
\nonumber \rho_{(n)}(r_1,...,r_n)&:=\frac{N(N-1)\cdot\cdot\cdot(N-n+1)}{Z_N[1]}\int_{\Omega}d\lambda_{n+1}\cdot\cdot\cdot\int_{\Omega}d\lambda_N \\
\label{eqn:integ_correlns} &\times  \mathcal{Q}(r_1,...,r_n,\lambda_{n+1},...,\lambda_N).
\end{align}
\end{definition}

The eigenvalue density is the $n=1$ case of (\ref{eqn:integ_correlns}). While the interpretation of the density (as the number of eigenvalues per unit volume) is clear, the higher order correlations are less perspicuous. A viewpoint in terms of conditional probabilities is that the ratio
\begin{align}
\nonumber \frac{\rho_{(n)}(r_1,...,r_n)}{\rho_{(n-1)}(r_1,...,r_{n-1})}
\end{align}
is equal to the eigenvalue density at $r_n$ given that there are eigenvalues at $r_1,...,r_{n-1}$.

One of the common ways to calculate the correlation functions is by using a recursion for integrals of quaternion determinants, known as the Dyson Integration Theorem \cite{dyson1970,mehta1976,AK2007}.

\begin{proposition} {\rm \textbf{Dyson Integration Theorem}}
\label{thm:integral_identities}
Let $f(x,y)$ be a function of real, complex or quaternion variables where
\begin{equation}
\nonumber \bar{f}(x,y)=f(y,x),
\end{equation}
with $\bar{f}$ being the function $f$, the complex conjugate of $f$ or the dual of $f$ depending on whether $x$ and $y$ are real, complex or quaternion respectively.

Also let 
\begin{eqnarray}
\label{eqn:proj_prop1} \int f(x,x) d\mu (x)&=&c,\\
\label{eqn:proj_prop2} \int f(x,y)f(y,z) d\mu(y)&=&f(x,z)+\lambda f(x,z) -f(x,z)\lambda,
\end{eqnarray}
for some suitable measure $d\mu$, a constant scalar $c$ and a constant quaternion $\lambda$.

Then for a matrix $F_{n\times n}=[f(x_i,x_j)]_{n\times n}$ we have
\begin{equation}
\nonumber \int \mathrm{qdet}[F_{n\times n}] d\mu(x_n) = (c-n+1)\hspace{3pt}\mathrm{qdet}[F_{(n-1) \times (n-1)}].
\end{equation}
\end{proposition}
\noindent (For a proof see Theorem 5.1.4 in \cite{mehta2004}.)

Examining Proposition \ref{thm:integral_identities} with (\ref{eqn:integ_correlns}) in mind, we see that it will be possible to calculate the correlation functions if the eigenvalue jpdf is in the form of a quaternion determinant (or Pfaffian). This is indeed possible (see \cite{dyson1970,mehta1976}), although this is not the approach we employ and we include it only for completeness. We do not use Dyson's theorem because for real non-symmetric ensembles (discussed in Chapters \ref{sec:GinOE}--\ref{sec:truncs}) we cannot satisfy (\ref{eqn:proj_prop1}) and (\ref{eqn:proj_prop2}), collectively called the projection property in \cite{AK2007}. In that paper the authors establish a generalised form of Dyson's theorem, which they call the \textit{Pfaffian integration theorem}, and use it to find the probability of obtaining some number of real eigenvalues in terms of zonal polynomials. Still, this does not suit our later purposes and we adopt instead a strategy first used in \cite{tracy_and_widom1998}, but with some significant modifications. This requires use of the general operator identity
\begin{align}
\label{eqn:1+AB} \det(\1+\mathbf{AB})=\det(\1+\mathbf{BA})
\end{align}
and the quaternion determinant analogue
\begin{align}
\label{eqn:qdet_1+AB} \qdet(\1+\mathbf{AB})=\qdet(\1+\mathbf{BA}),
\end{align}
(provided that the product $\mathbf{BA}$ is self-dual) where, for our purposes, $\1, \mathbf{A}$ and $\mathbf{B}$ are $N \times N$ matrices, and $\1$ is specifically the identity matrix. We will also employ  the \textit{Fredholm determinant} and its comrades the \textit{Fredholm quaternion determinant} and the \textit{Fredholm Pfaffian}.

\begin{definition}
\label{def:fred_D_QD_P}
Let $K$ be an integral operator with kernel $K(x,y)$ and $\lambda$ a complex parameter, then the \textit{Fredholm determinant} is defined by
\begin{align}
\nonumber \det[1+\lambda K]:=1+\sum_{s=1}^{\infty}\frac{\lambda^s}{s!}\int_{-\infty}^{\infty}dx_1\cdot\cdot\cdot\int_{-\infty}^{\infty}dx_s\det [K(x_j,x_k)]_{j,k=1,...,s}.
\end{align}
In the case that the matrix $[K(x_j,x_k)]_{j,k=1,...,s}$ is self-dual, we define the \textit{Fredholm quaternion determinant}
\begin{align}
\nonumber \qdet[1+\lambda K]:=1+\sum_{s=1}^{\infty}\frac{\lambda^s}{s!}\int_{-\infty}^{\infty}dx_1\cdot\cdot\cdot\int_{-\infty}^{\infty}dx_s\;\qdet [K(x_j,x_k)]_{j,k=1,...,s},
\end{align}
and, when $[K(x_j,x_k)]_{j,k=1,...,s}$ is anti-symmetric, the \textit{Fredholm Pfaffian}
\begin{align}
\nonumber \Pf[1+\lambda K]:=1+\sum_{s=1}^{\infty}\frac{\lambda^s}{s!}\int_{-\infty}^{\infty}dx_1\cdot\cdot\cdot\int_{-\infty}^{\infty}dx_s\;\Pf [K(x_j,x_k)]_{j,k=1,...,s}.
\end{align}
\end{definition}

\begin{remark}
We note that while the Fredholm determinant has been known for over a century, the Fredholm Pfaffian seems to have be been introduced in \cite{Rains2000}, and was then used in \cite{BK2007} to prove some variants of the Pfaffian integration theorem discussed above. The first mention of a Fredholm quaternion determinant in the literature appears to be in \cite{tracy_and_widom1998}; of course, by Corollary \ref{cor:qdet=pf}, it is a trivial rewriting of the Fredholm Pfaffian.
\end{remark}
\begin{remark}
Since these definitions involve infinite sums, there is the question of convergence. However, we sidestep this complication since we will only be using operators of finite rank, and thus, the sums are of finite length.
\end{remark}

Although we shall not tackle the $N$ odd case until Chapter \ref{sec:GOE_odd}, a key technical consideration in that case will be the square root of a Fredholm determinant. We would like the square root to be a Fredholm quaternion determinant or Pfaffian (depending on the attributes of $K$) in analogue with (\ref{eqn:pf_det}) and (\ref{eqn:qdet2=det}). However, we cannot invoke the `Freshman's dream' (that the power of a sum is the sum of the powers \cite{Hu90}) and so we require a more subtle approach. Instead, we first establish that Fredholm operators are limiting cases of some discretised form.

\begin{lemma}[\cite{WW63}, Chapter XI]
\label{lem:pf_qdet_kernel}
Let $\lambda$ be some complex variable and $\{x_1,...,x_m \}\in\mathbb{R}^m$, with fixed $\delta:=x_j-x_{j-1}$ for all $1\leq j\leq m$ (that is, $\delta$ is the constant distance between any two of the variables), considered as a discretisation of an interval $I$. Then, for some integral operator $K$ with kernel $K(x,y)$ supported on $I$ and
\begin{align}
\nonumber \tilde{K}_m(\delta)=\left[
\begin{array}{cccc}
\delta K(x_1,x_1) & \delta K(x_1,x_2) & \cdot\cdot\cdot & \delta K(x_1,x_m)\\
\delta K(x_2,x_1) & \delta K(x_2,x_2) & \cdot\cdot\cdot & \delta K(x_2,x_m)\\
\vdots & \vdots & \ddots & \vdots\\
\delta K(x_m,x_1) & \delta K(x_m,x_2) & \cdot\cdot\cdot & \delta K(x_m,x_m)
\end{array}
\right],
\end{align}
we have
\begin{align}
\label{eqn:freddet_lim} \lim_{\substack{\delta\rightarrow 0\\ m\rightarrow \infty}}\det [1-\lambda \tilde{K}_m(\delta)]=\det[1-\lambda K],
\end{align}
and, in the case that $\tilde{K}_m$ is self-dual,
\begin{align}
\nonumber \lim_{\substack{\delta\rightarrow 0\\ m\rightarrow \infty}}\qdet [1-\lambda \tilde{K}_m(\delta)]=\qdet[1-\lambda K],
\end{align}
or, in the case that $\tilde{K}_m$ is anti-symmetric,
\begin{align}
\nonumber \lim_{\substack{\delta\rightarrow 0\\ m\rightarrow \infty}}\Pf [1-\lambda \tilde{K}_m(\delta)]=\Pf [1-\lambda K].
\end{align}
\end{lemma}

\textit{Proof}: Expanding the LHS of (\ref{eqn:freddet_lim}) in powers of $\lambda$ we have
\begin{align}
\nonumber &\det\left[1-\lambda \tilde{K}_m(\delta)\right]\\
\nonumber &= 1-\lambda \sum_{p=1}^m \delta K(x_p,x_p)+\frac{\lambda^2}{2!}\sum_{p,q=1}^m\delta^2 \det\left[
\begin{array}{cc}
K(x_p,x_p) & K(x_p,x_q)\\
K(x_q,x_p) & K(x_q,x_q)
\end{array} \right]\\
\label{eqn:discrete_FH} &-\frac{\lambda^3}{3!}\sum_{p,q,r=1}^m\delta^3 \det\left[
\begin{array}{ccc}
K(x_p,x_p) & K(x_p,x_q) & K(x_p,x_r)\\
K(x_q,x_p) & K(x_q,x_q) & K(x_q,x_r)\\
K(x_r,x_p) & K(x_q,x_r) & K(x_r,x_r)
\end{array} \right]+\cdot\cdot\cdot
\end{align}
where the sum continues up to the $m$-th power of $\lambda$. Taking $m\to\infty$ and $\delta\to 0$ the sums become Riemann integrals and (\ref{eqn:discrete_FH}) becomes
\begin{align}
\nonumber &1-\lambda\int_I dx\; K(x,x)+\frac{\lambda^2}{2!}\int_I dx\int_I dy \det\left[
\begin{array}{cc}
K(x,x) & K(x,y)\\
K(y,x) & K(y,y)
\end{array} \right]+\cdot\cdot\cdot.
\end{align}
Recalling Definition \ref{def:fred_D_QD_P} establishes (\ref{eqn:freddet_lim}). Applying the same reasoning to the Fredholm quaternion determinant and Pfaffian we have the remaining results.

\hfill $\Box$

Observe that (\ref{eqn:freddet_lim}) allows us to intepret the LHS's of Definition \ref{def:fred_D_QD_P} as the product $\prod_{j=1}^{\infty} (1+\lambda \mu_j)$, where $\mu_j$ are the eigenvalues of $K$; for this infinite product to make sense we require some technical assumptions on $K$ (see \cite[Section XI.1]{WW63}).

Combining Lemma \ref{lem:pf_qdet_kernel} with (\ref{eqn:pf_det}) and (\ref{eqn:qdet2=det}) the desired square root relationships between the Fredholm operators now follow trivially; we quote them here as corollaries for ease of reference.

\begin{corollary}
\label{cor:sqrtFreds} Let $K$ be an integral operator with kernel $K(x,y)$ and with Fredholm determinant, Fredholm quaternion determinant and Fredholm Pfaffian as in Definition \ref{def:fred_D_QD_P}, then we have
\begin{align}
\label{eqn:det_qdet_kernel} \Big(\det [1+\lambda K]\Big)^{1/2}=\qdet [1+\lambda K]
\end{align}
in the case that $[K(x_j,x_k)]_{j,k=1,2,...}$ is self-dual, and
\begin{align}
\label{eqn:det_pf_kernel} \Big(\det [1+\lambda K]\Big)^{1/2}=\Pf [1+\lambda K]
\end{align}
in the case that $[K(x_j,x_k)]_{j,k=1,2,...}$ is anti-symmetric.
\end{corollary}

The utility of these Fredholm operators comes from an alternative form of the correlation functions: with $Z_N[a]$ from (\ref{def:single_gen_part_fn}) the $n$-point correlation function is 
\begin{align}
\label{eqn:fnal_diff_correln} \rho_{(n)}(r_{1},...,r_{n})=\frac{1}{Z_N[a]}\frac{\delta^n}{\delta a(r_1)\cdot\cdot\cdot \delta a(r_n)}Z_N[a]{\Bigg |}_{a=1}.
\end{align}
We can describe the equivalence between Definition \ref{def:integ_correlns} and (\ref{eqn:fnal_diff_correln}) in a heuristic fashion: while (\ref{eqn:integ_correlns}) relies on integrating over the density function to leave only the number of eigenvalues desired, (\ref{eqn:fnal_diff_correln}) starts by integrating over all eigenvalues (which is the partition function) and then ``undoes" a number of integrals equal to the number of eigenvalues one wishes to keep. This heuristic points to our intended use of the Fredholm operators (defined as sums of integrals); the functional differentiation will pick out only the particular term required.

The method that we develop here, particularly that in Proposition \ref{prop:pf_integ_op}, is inspired by the approaches of Forrester \cite[Chapter 5.2]{forrester?} and Tracy and Widom \cite{tracy_and_widom1998} (for the even case) where they use (\ref{eqn:1+AB}) to find the correlation functions for Hermitian ensembles, and the work of Borodin and Sinclair for both the even \cite{b&s2009} and odd \cite{Sinc09} cases. (In \cite{sommers2007} and \cite{sommers_and_w2008} the authors also use (\ref{eqn:fnal_diff_correln}) to obtain the eigenvalue correlations for $N$ even and odd (respectively), however the details are somewhat different to our techniques and we will not pursue their methods here.) In \cite{b&s2009} the authors use the Pfaffian identity
\begin{align}
\label{eqn:rains} \frac{\Pf(\bC^{-T}-\bA^T\bB\bA)}{\Pf(\bC^{-T})}=\frac{\Pf(\bB^{-T}-\bA\bC\bA^T)}{\Pf(\bB^{-T})},
\end{align}
which is due to Rains \cite{Rains2000}, where $\bB$ and $\bC$ are $2m\times 2m$ and $2n\times 2n$ anti-symmetric matrices respectively, and $\bA$ is any $2m\times 2n$ matrix. Proposition \ref{prop:pf_integ_op} unifies the approaches of Forrester, Tracy and Widom, with that of Borodin and Sinclair.

The advantage of our method is that all cases --- symmetric and asymmetric for both even and odd --- can be dealt with in the same framework, with only minor modifications and generalisations at each step. There are also hints that the method may also be applicable to ensembles with a higher number of distinct species of eigenvalue, such as the $*$-cosquare ensembles discussed in Chapter \ref{sec:FW}.

\subsubsection{GOE $n$-point correlations, $N$ even}
\label{sec:correlns_even_GOE}

\begin{definition}
\label{def:GOE_correln_kernel}
With $R_0(x),R_1(x),...$ the skew-orthogonal polynomials of Proposition \ref{prop:GOE_soip} and $r_j:=\langle R_{2j},R_{2j+1}\rangle$ the corresponding normalisations, let
\begin{align}
\label{def:GOEphi} \Phi_k(x):=\frac{1}{2}\int_{-\infty}^{\infty}dy\:R_k(y)\: e^{-y^2/2}\: \mathrm{sgn}(x-y).
\end{align}
Then let $\mathbf{f}(x,y)$ be the $2\times 2$ matrix
\begin{align}
\label{def:GOE_Qdcorrelnk} \mathbf{f}(x,y)=\left[\begin{array}{cc}
S(x,y) & \tilde{I}(x,y)\\
D(x,y) & S(y,x)
\end{array}\right],
\end{align}
where
\begin{align}
\nonumber S(x,y)&=\sum_{j=0}^{N/2-1}\frac{e^{-y^2/2}}{r_j}\Big( \Phi_{2j}(x)R_{2j+1}(y)-\Phi_{2j+1}(x)R_{2j}(y)\Big),\\
\nonumber D(x,y)&=\sum_{j=0}^{N/2-1}\frac{e^{-(x^2+y^2)/2}}{r_j}\Big( R_{2j}(x)R_{2j+1}(y)-R_{2j+1}(x)R_{2j}(y)\Big),\\
\nonumber \tilde{I}(x,y)&:=I(x,y)+\frac{1}{2}\mathrm{sgn}(y-x)\\
\nonumber &=\sum_{j=0}^{N/2-1}\frac{1}{r_j}\Big( \Phi_{2j+1}(x)\Phi_{2j}(y)-\Phi_{2j}(x)\Phi_{2j+1}(y)\Big)+\frac{1}{2}\mathrm{sgn}(y-x).
\end{align}
\end{definition}

Since the correlation functions will turn out to be quaternion determinants of matrices composed of blocks of (\ref{def:GOE_Qdcorrelnk}), $\mathbf{f}(x,y)$ is known as a \textit{correlation kernel}. From (\ref{eqn:qdet=pf}) we find the equivalent kernel for the correlations expressed as Pfaffians,
\begin{align}
\label{def:GOE_Pfcorrelnk}
\mathbf{f}(x,y)\bZ_2^{-1}=\left[\begin{array}{cc}
-\tilde{I}(x,y) & S(x,y)\\
-S(y,x) & D(x,y)
\end{array}\right].
\end{align}
The term \textit{Pfaffian kernel} is also used to refer to (\ref{def:GOE_Pfcorrelnk}).

Here we point out a simple relationship between the elements of $\mathbf{f}(x,y)$, which explains the choice of the appellations $D(x,y)$ and $I(x,y)$.

\begin{lemma}
\label{lem:GOE_D=S=I}
The elements of $\mathbf{f}(x,y)$ are thusly related
\begin{align}
\nonumber D(x,y)=\frac{\partial}{\partial x}S(x,y)&,& I(x,y)&=\frac{1}{2}\int_{-\infty}^{\infty}S(x,z)\sgn (z-y)dz\\
\nonumber &&&=-\int_x^y S(x,z)dz.
\end{align}
\end{lemma}

\textit{Proof}: The derivative for $D(x,y)$ can be done simply when one recalls the identity $\frac{\partial}{\partial x}\sgn(x-a)=2\delta(x-a)$. The first equality of $I(x,y)$ can be seen by inspection, and the second equality is verified by noting that the two sides agree along the line $y=x$, and the derivatives with respect to both $x$ and $y$ are equal (the differentiation can be accomplished using the method of differentiating under the integral sign). 

\hfill $\Box$

Using Definition \ref{def:GOE_correln_kernel}, we will rewrite the generalised partition function (\ref{eqn:GOE_GPF_even}) by applying the identity (\ref{eqn:qdet_1+AB}). First, recall that an \textit{integral operator} $T$ with kernel $K(x,y)$ supported on $I$ operates on a function $h$ thusly,
\begin{align}
\label{def:integop} T\:h[x]=\int_I K(x,y)h(y)dy,
\end{align}
with the convention that $y$ is always the variable of integration. The square brackets indicate that the resulting function (after the operation by the integral operator) is a function of $x$. We use the notation $a\otimes b$ to denote an operator with kernel $K(x,y)=a(x)b(y)$, that is, a kernel that factorises as separate functions of $x$ and $y$, in which case we have
\begin{align}
\label{eqn:integop_ab} a\otimes b\:h[x]=a(x)\int_I b(y)h(y)dy.
\end{align}

We now use (\ref{eqn:qdet_1+AB}) to convert the problem from a 2 dimensional function in variable-sized matrix to a variable dimensional function in a $2\times 2$ matrix, which is how we will present the correlation functions.

\begin{proposition}
\label{prop:pf_integ_op}
Let $\gamma_{jk}$ be as in (\ref{eqn:gammajk}) and $\mathbf{f}(x,y)$ be as in (\ref{def:GOE_Qdcorrelnk}). Then, by using the skew-orthogonal polynomials of Proposition \ref{prop:GOE_soip}, we have
\begin{align}
\label{prop:integ_op_even} \Pf [\gamma_{jk}]_{j,k=1,...,N}=\prod_{j=0}^{N/2-1}r_j\; \qdet [\1_2+ \mathbf{f}^T (\mathbf{u}-\1_2)],
\end{align}
where $\1_2$ is the $2\times 2$ identity matrix, and $\mathbf{f}^T(\mathbf{u}-\1_2)$ is the matrix integral operator with kernel $\mathbf{f}^T(x,y) \mathrm{diag}[u(y)-1,u(y)-1]$ (that is, we have a Fredholm quaternion determinant).
\end{proposition}

\textit{Proof}: In the definition of $\gamma_{jk}$ in (\ref{eqn:gammajk}) we let $u=\sigma+1$ and $\psi_j(x):=e^{-x^2/2}R_{j-1}(x)$, where $R_0,R_1,...$ are the skew-orthogonal polynomials (\ref{eqn:sops}), and denote by $\epsilon$ the integral operator with kernel $\mathrm{sgn}(x-y)/2$. We then have
{\small
\begin{align}
\nonumber \gamma_{jk}&=\gamma_{jk}\Big|_{u=1}-\int_{-\infty}^{\infty}\Big(\sigma(x)\psi_j(x)\epsilon \psi_k[x]-\sigma(x)\psi_k(x)\epsilon\psi_j[x]-\sigma(x)\psi_k(x)\epsilon(\sigma\psi_j)[x]\Big)dx\\
\nonumber &=\gamma^{(1)}_{jk}-\gamma^{(\sigma)}_{jk},
\end{align}
}with $\gamma_{jk}^{(1)}:= \gamma_{jk}\big|_{u=1}$ and $\gamma_{jk}^{(\sigma)}$ the remaining term. With the skew-orthogonal polynomials we see that $[\gamma^{(1)}_{jk}]$ is of the form (\ref{eqn:skew_diag_mat}), and so it can be written $\mathbf{D}\bZ^{-1}_N$ with $\bZ_N$ as in (\ref{def:Z2N}) and $\mathbf{D}=\mathrm{diag}[r_0,r_0,r_1,r_1,...,r_{N/2},r_{N/2}]$. We then have
\begin{align}
\nonumber \Pf[\gamma_{jk}]&=\Pf(\mathbf{D}\bZ^{-1}_N-[\gamma^{(\sigma)}_{jk}])\\
\nonumber &=\Pf\big{(}\mathbf{D}\bZ^{-1}_N(\1_N-\bZ_N \mathbf{D}^{-1} [\gamma^{(\sigma)}_{jk}]) \big{)}\\
\nonumber &=\prod_{j=0}^{N/2-1}r_j\; \qdet \big{(}\1_N- \bZ_N\mathbf{D}^{-1}[\gamma^{ (\sigma)}_{jk}] \big{)},
\end{align}
where, for the last equality, we have used Corollary \ref{cor:qdet=pf}. Defining
\begin{align}
\label{def:G_psi} G_{2j-1}(x):=\psi_{2j}(x),&& G_{2j}(x):=-\psi_{2j-1}(x),
\end{align}
and recalling the effect of operation by $\bZ_N$ (as discussed in the proof of Proposition \ref{prop:qdet=pf}) we find
\begin{align}
\nonumber &\Pf [\gamma_{jk}]_{j,k=1,...,N}=\prod_{j=0}^{N/2-1}r_j \; \qdet \Bigg[\delta_{j,k}+\\
\label{eqn_detgamma1} &\frac{1}{r_{\lfloor (j-1)/2 \rfloor}}\int_{-\infty}^{\infty}\Big(\sigma(x)G_j(x)\epsilon \psi_k[x]-\sigma(x)\psi_k(x)\epsilon G_j[x]-\sigma(x)\psi_k(x)\epsilon(\sigma G_j)[x]\Big)dx \Bigg]
\end{align}
with $\lfloor z \rfloor$ the floor function. Now let $\bA$ be the $N\times 2$ matrix-valued integral operator on $(-\infty, \infty)$ with kernel $\bZ^{-1}_N\mathbf{D}^{-1}\sigma(y) (\Omega\mathbf{E})^T$ where
\begin{align}
\label{eqn:omega_E} \Omega:=\left[
\begin{array}{cc}
-\epsilon \sigma & -1\\
1 & 0
\end{array}
\right]&&\mathrm{and} && \mathbf{E}:=\left[
\begin{array}{ccc}
\psi_1(y) & \cdot \cdot\cdot & \psi_N(y)\\
\epsilon\psi_1[y] & \cdot \cdot\cdot & \epsilon\psi_N[y]
\end{array}
\right].
\end{align}
(Care should be taken to note that the top-left element of $\Omega$ is an integral operator acting on the elements of $\mathbf{E}$.) Explicitly, the kernel of $\bA$ can be written
\begin{align}
\left[\begin{array}{cc}
\nonumber -\frac{\sigma(y)}{r_{\lfloor (j-1)/2 \rfloor}}\big(\epsilon G_j[y]+\epsilon(\sigma G_j)[y]\big)&\frac{\sigma(y)}{r_{\lfloor (j-1)/2 \rfloor}}G_j(y)
\end{array}\right]_{j=1,...,N},
\end{align}
which we include for clarity. Now with $\mathbf{B}=\mathbf{E}$ we have
\begin{align}
\label{eqn:1+ABdecomp} \1_N-\bZ_N\mathbf{D}^{-1}[\gamma^{(\sigma)}_{jk}]= \1_N+\bA\mathbf{B}
\end{align}
and we may make use of (\ref{eqn:qdet_1+AB}). With the definitions above, we see that $\1_2 + \bB\bA$ is the $2\times 2$ matrix integral operator
\begin{align}
\label{eqn:1+ba1} \left[\begin{array}{cc}
1-\sum_{j=1}^N\frac{1}{r_{\lfloor (j-1)/2 \rfloor}}\Big( \psi_j\otimes \sigma \epsilon G_j+\psi_j\otimes \sigma \epsilon (\sigma G_j)\Big) & \sum_{j=1}^N\frac{1}{r_{\lfloor (j-1)/2 \rfloor}}\psi_j\otimes \sigma G_j\\
-\sum_{j=1}^N\frac{1}{r_{\lfloor (j-1)/2 \rfloor}}\Big(\epsilon \psi_j\otimes \sigma\epsilon G_j+\epsilon \psi_j\otimes \sigma \epsilon(\sigma G_j) \Big) & 1+\sum_{j=1}^N\frac{1}{r_{\lfloor (j-1)/2 \rfloor}}\epsilon \psi_j\otimes \sigma G_j
\end{array}\right]
\end{align}
using the integral operator notation of (\ref{eqn:integop_ab}). To achieve the final result we should like to eliminate terms containing the factor $\epsilon\sigma$, since we will then be able to factor out $\sigma$ and make the impending functional differentiation straightforward. However, the appearance of the $\epsilon\sigma$ factor is only an apparent complication and will be dealt with by a judicious matrix factorisation. We have
\begin{align}
\label{eqn:ba} \bB\bA=\mathbf{E}\; \bZ^{-1}_N\mathbf{D}^{-1}\sigma(y) (\Omega\mathbf{E})^T =\mathbf{E}\; \bZ^{-1}_N\mathbf{D}^{-1}\sigma(y) \mathbf{E}^T\Omega^T,
\end{align}
and
\begin{align}
\label{eqn:OT_decomp} \Omega^T\left[
\begin{array}{cc}
1 & 0\\
-\epsilon\sigma & 1
\end{array}
\right]=\left[
\begin{array}{cc}
\epsilon\sigma & 1\\
-1 & 0
\end{array}
\right]\left[
\begin{array}{cc}
1 & 0\\
-\epsilon\sigma & 1
\end{array}
\right]=\bZ^{-1}_2
\end{align}
(where taking the transpose of $\Omega$ includes the interchange of the variables $x$ and $y$ in $\epsilon$). Replacing $\Omega^T$ using (\ref{eqn:OT_decomp}) we find that the kernel of (\ref{eqn:1+ba1}) factorises as
\begin{align}
\nonumber \1_2 + \bB\bA=\left(\left[
\begin{array}{cc}
1 & 0\\
-\epsilon\sigma & 1
\end{array}
\right]+\mathbf{E}\; \bZ^{-1}_N\mathbf{D}^{-1}\sigma(y) \mathbf{E}^T \bZ^{-1}_2\right) \left[
\begin{array}{cc}
1 & 0\\
\epsilon\sigma & 1
\end{array}
\right]
\end{align}
{\small \begin{align}
\label{eqn:1+ba2} =\left[\begin{array}{cc}
1-\sum_{j=1}^N\frac{1}{r_{\lfloor (j-1)/2 \rfloor}} \psi_j\otimes \sigma \epsilon G_j & \sum_{j=1}^N\frac{1}{r_{\lfloor (j-1)/2 \rfloor}}\psi_j\otimes \sigma G_j\\
-\epsilon\sigma-\sum_{j=1}^N\frac{1}{r_{\lfloor (j-1)/2 \rfloor}} \epsilon \psi_j\otimes \sigma\epsilon G_j & 1+\sum_{j=1}^N\frac{1}{r_{\lfloor (j-1)/2 \rfloor}}\epsilon \psi_j\otimes \sigma G_j
\end{array}\right] \left[
\begin{array}{cc}
1 & 0\\
\epsilon\sigma & 1
\end{array}
\right].
\end{align}
}The equality of (\ref{eqn:1+ba1}) and (\ref{eqn:1+ba2}) can also be checked directly by noting that $\psi_j\otimes\sigma\epsilon(\sigma G_j)=-\psi_j\otimes \sigma G_j(\epsilon \sigma)$ where, on the right hand side, the operator $\epsilon\sigma$ is understood to act before the larger integral operator; to wit 
\begin{align}
\label{eqn:eps_sig} \psi_j\otimes \sigma G_j(\epsilon \sigma) h[x]=\psi_j(x)\int_{-\infty}^{\infty}dy \sigma(y) G_j(y) \int_{-\infty}^{\infty}\sigma(z) h(z) \mathrm{sgn}(y-z)dz.
\end{align}
Similarly $\epsilon\psi_j\otimes\sigma\epsilon(\sigma G_j)=-\epsilon\psi_j\otimes\sigma G_j(\epsilon \sigma)$.

The right-most matrix in (\ref{eqn:1+ba2}) has quaternion determinant $1$ and, recalling that the transpose of a matrix integral operator involves both the transpose of the matrix itself and also transposition of the operator variables, we can identify the remaining matrix in (\ref{eqn:1+ba2}) with the right hand side of (\ref{prop:integ_op_even}). The only caveat is that we obtain the negative of $D$ and $\tilde{I}$ as defined in Definition \ref{def:GOE_correln_kernel}, however since they only appear as the product $D(\alpha,\beta) \tilde{I}(\alpha,\beta)$ in the expansion of the Fredholm quaternion determinant the result is unchanged.

\hfill $\Box$

With Proposition \ref{prop:pf_integ_op} in hand, we see from (\ref{eqn:GOE_GPF_even}) that
\begin{align}
\label{eqn:ZN_integ_op} Z_N[u]=\frac{N!}{2^N}\prod_{j=1}^N\frac{1}{\Gamma(j/2+1)} \prod_{j=0}^{N/2-1}r_j \: \qdet [\1_2+\mathbf{f}^T(\mathbf{u}-\1_2)],
\end{align}
and we can now establish the general $n$-point correlation functions. Note that (\ref{eqn:ZN_integ_op}) contains a Fredholm quaternion determinant, and so, by the discussion under (\ref{eqn:fnal_diff_correln}), we expect that we will apply functional differentiation to pick out the particular term corresponding to the desired correlation function.

In the following proposition we find a quaternion determinant expression for the correlation functions, and then apply Proposition \ref{prop:qdet=pf} in an `after-the-fact' manner to conveniently find the Pfaffian expression, which is how the correlations were written in \cite{b&s2009}. However, in the proof of Proposition \ref{prop:pf_integ_op} above, we can see a deeper structural connection between the two expressions. Recall that to obtain (\ref{eqn:1+ba2}) we used (\ref{eqn:OT_decomp}) to factorise $\Omega^T$ in (\ref{eqn:ba}). If we instead wrote
\begin{align}
\nonumber \1+\bB\bA=\Big((\Omega^{T})^{-1} + \mathbf{E}\; \bZ^{-1}_N\mathbf{D}^{-1}\sigma(y) \mathbf{E}^T \Big) \Omega^T,
\end{align}
then we are led directly to a Pfaffian form of (\ref{prop:integ_op_even}), and then to Pfaffian correlation functions. This is essentially how Rains' identity (\ref{eqn:rains}) comes into play.
\begin{remark}
In fact, (\ref{eqn:rains}) plays a more significant role in our proof of Proposition \ref{prop:pf_integ_op} than it may, at first, appear. The seemingly miraculous appearance of the matrices $\Omega$ and $\bE$ in (\ref{eqn:omega_E}) was inspired by Rains' identity. Recalling (\ref{eqn:pf_det}) and using some simple algebra, we see that (\ref{eqn:rains}) becomes
\begin{align}
\label{eqn:rains_mod} \det(\1-\bC^T\bA^T\bB\bA)^{1/2} =\det(\1-\bB^T\bA\bC\bA^T)^{1/2}.
\end{align}
In the case that $\bC$ is skew-diagonal, we have quaternion determinants, instead of square roots of determinants. Using (\ref{eqn:rains_mod}) one can conjecture the form of the required integral operators.
\end{remark}

\begin{proposition}
\label{prop:correlns_GOE_even}
With $\mathbf{f}(x,y)$ as in Definition \ref{def:GOE_correln_kernel}, the $n$-point eigenvalue correlations for the GOE, with $N$ even, are given by 
\begin{align}
\nonumber \rho_{(n)}(x_1,...,x_n)&=\qdet[\mathbf{f}(x_l,x_m)]_{l,m=1,...,n}\\
\label{eqn:correlns_GOE} &=\Pf[\mathbf{f}(x_l,x_m)\bZ_2^{-1}]_{l,m=1,...,n}.
\end{align}
\end{proposition}

\textit{Proof}: Recalling Definition \ref{def:fred_D_QD_P} we see that (\ref{eqn:ZN_integ_op}) becomes
{\small
\begin{align}
\nonumber &Z_N[u]=\frac{N!}{2^N}\prod_{j=1}^N\frac{1}{\Gamma(j/2+1)}\prod_{j=0}^{N/2-1}r_j\\
\nonumber &\times\bigg( 1+\sum_{k=1}^{\infty}\frac{1}{k!}\int_{-\infty}^{\infty}dx_1\: (u(x_1)-1) \cdot\cdot\cdot \int_{-\infty}^{\infty} dx_k \:(u(x_k)-1)\:\qdet[\mathbf{f}(x_l,x_m)]_{l,m=1,...,k} \bigg).
\end{align}
}Now to make use of (\ref{eqn:fnal_diff_correln}) we first note that only terms with $k\geq n$ survive the functional differentiation, and then any terms with $k>n$ will be killed off once $u=1$. So we are left with the $n!$ terms corresponding to $k=n$, giving
\begin{align}
\nonumber \rho_{(n)}(x_1,...,x_n)=\frac{\prod_{j=0}^{N/2-1}r_j}{\Pf [\gamma_{jk}]_{j,k=1,...,N}\Big|_{u=1}}\qdet [\mathbf{f}(x_l,x_m)]_{l,m=1,...,n}.
\end{align}
Recalling (\ref{eqn:pf=prodr}) we have the first equality in (\ref{eqn:correlns_GOE}), and by using (\ref{eqn:qdet=pf}) we have the second.

\hfill $\Box$

\subsubsection{Summation formulae for the kernel elements, $N$ even}
\label{sec:GOE_sums}

Here we will show that the sum $S(x,y)$ in Definition \ref{def:GOE_correln_kernel} can be performed explicitly. This will be of use in analysing the large $N$ limit of the density, and will give us the leading order behaviour that we see in Figure \ref{fig:GOE_eval_dens}. Further, since we have the inter-relationships of Lemma \ref{lem:GOE_D=S=I}, a closed form for $S(x,y)$ implies that all the correlation functions can be written in a closed form.

First we quote a classical result, known as the \textit{Christoffel--Darboux formula}.
\begin{proposition}
With a set of orthogonal polynomials $\{q_0(x),...,q_n(x) \}$ with highest degree coefficient $k_n$, and
\begin{align}
\nonumber (q_j, q_k):= \int_{-\infty}^{\infty} e^{-z^2} q_j(z) q_k(z) dz,
\end{align}
we have
\begin{align}
\label{eqn:CD} \sum_{j=0}^n \frac{q_j(x)q_j(y)} {(q_j, q_j)}=\frac{k_n}{(q_n, q_n) k_{n+1}} \frac{q_{n+1}(x)q_n(y) - q_n(x)q_{n+1}(y)}{x-y}.
\end{align}
\end{proposition}

(For proofs we refer the reader to \cite{Szego1959} and \cite{forrester?}.) By taking the limit $y\to\ x$ in (\ref{eqn:CD}) we also have the formula
\begin{align}
\label{eqn:CDdiff} \sum_{j=0}^n \frac{\big(q_j(x)\big)^2} {(q_j, q_j)}=\frac{k_n}{(q_n, q_n) k_{n+1}} \Big(q'_{n+1}(x)q_n(x) -q'_n(x)q_{n+1}(x)\Big),
\end{align}
where the apostrophe represents differentiation with respect to $x$. Note that in our case, we are using monic Hermite polynomials and so $k_j=1$ for all $j=0,...,n$.

Using (\ref{eqn:CD}), with the working given in \cite[Chapter 6.4.2]{forrester?}, we have
\begin{align}
\nonumber S(x,y)&=\frac{e^{-(x^2+y^2)/2}}{2^{N-1}\sqrt{\pi}\: \Gamma(N-1)}\frac{H_{N-1}(x)H_{N-2}(y)- H_{N-2}(x)H_{N-1}(y)}{x-y}\\
\label{eqn:GOEsumS} &+\frac{\: e^{-y^2/2}}{ 2\sqrt{\pi}\: \Gamma(N-1)}H_{N-1}(y)\Phi_{N-2}(x),
\end{align}
where the $H_j(x)$ are the Hermite polynomials (\ref{eqn:herm_polys}) and $\Phi_j(x)$ is from (\ref{def:GOEphi}). Since the density is defined as the one-point correlation we see from (\ref{eqn:correlns_GOE}) that
\begin{align}
\label{eqn:1pt_correln} \rho_{(1)}(x)=S(x,x).
\end{align}
By applying (\ref{eqn:CDdiff}) and (\ref{eqn:recursive_herm}) to (\ref{eqn:GOEsumS}) we have that
\begin{align}
\nonumber \rho_{(1)}(x)&=\frac{e^{-x^2}}{2^{N-2}\sqrt{\pi}\: \Gamma(N-1)}\Big((N-1)(H_{N-2}(x))^2- (N-2)H_{N-3}(x)H_{N-1}(x)\Big)\\
\label{eqn:GOEsumSdiff} &+\frac{\: e^{-x^2/2}}{ 2\sqrt{\pi}\: \Gamma(N-1)}H_{N-1}(x)\Phi_{N-2}(x).
\end{align}

\begin{figure}[htp]
\begin{center}
\includegraphics[scale=0.7]{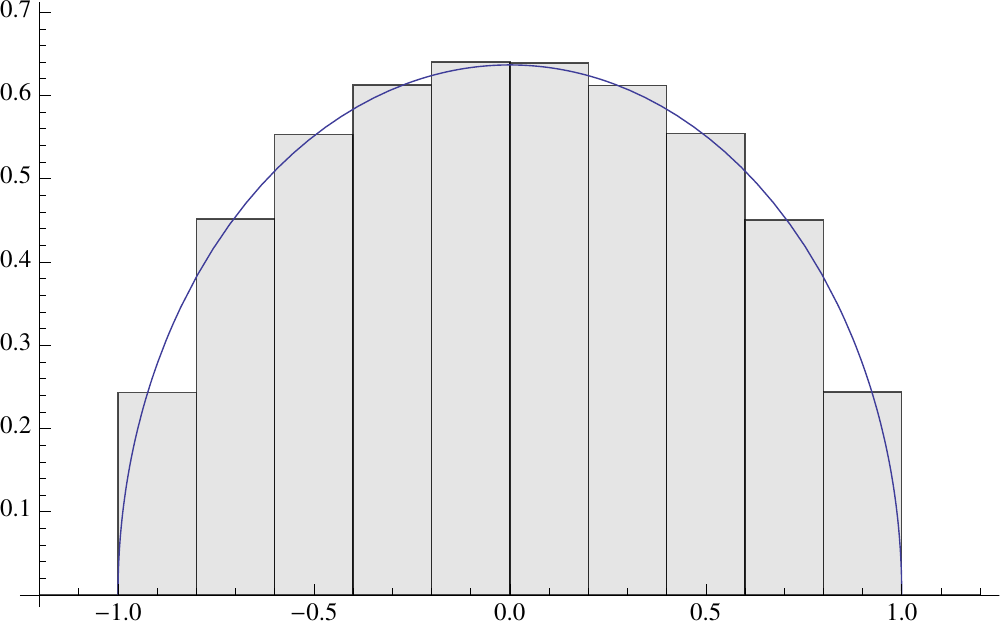}
\caption[Comparison of simulated GOE eigenvalue density with analytic prediction.]{Comparison of the simulated eigenvalue density of Figure \ref{fig:GOE_eval_dens} with the solid line given by the analytic result in (\ref{eqn:wssl}).}
\label{fig:GOEwWSSL}
\end{center}
\end{figure}

Using an electrostatic analogy, reasoning in \cite[Chapter 4.2]{mehta2004} and \cite[Chapter 1.4]{forrester?} concludes that the leading order behaviour of the eigenvalue density will be given by
\begin{align}
\label{eqn:wsslwN} \rho_{(1)}(x)=\frac{1}{\pi}\sqrt{2N-x^2},
\end{align}
which is a semi-circle of radius $\sqrt{2N}$. This suggests taking the normalised limit
\begin{align}
\label{eqn:wssl} \lim_{N\to \infty}\sqrt{\frac{2}{N}}\rho_{(1)}\left( \sqrt{2N}x \right)=\left\{ \begin{array}{ll}
\frac{2}{\pi}\sqrt{1-x^2}, & |x|<1,\\
0, & |x| \geq1.
\end{array}\right.
\end{align}
(One way to carry out this task is to use the so-called \textit{Plancherel-Rotach} asymptotic formula for Hermite polynomials. In \cite[Chapter 1.4.3]{forrester?}, the author obtains the semi-circular density by reframing the question as a Riemann-Hilbert problem.) In Figure \ref{fig:GOEwWSSL} we compare the simulated eigenvalue density of Figure \ref{fig:GOE_eval_dens} to (\ref{eqn:wssl}).
Using the knowledge that the density tends towards (\ref{eqn:wsslwN}), we scale the eigenvalues $\tilde{x}_j=\pi \rho x_j/\sqrt{N}$, where $\rho$ is the average bulk density, to obtain the bulk limiting form of the general correlation functions \cite{Gaudin1961} \cite[Chapter 7.8.1]{forrester?}
\begin{align}
\nonumber &\lim_{N\to\infty}\left( \frac{\pi \rho}{\sqrt{N}}\right)^n \rho_{(n)}\left( \tilde{x}_1,..., \tilde{x}_n\right)\\
\label{eqn:GOESbulk} &=\rho^{n} \qdet\left[\begin{array}{cc}
S^{\bulk}(x_j,x_k)&\tilde{I}^{\bulk}(x_j,x_k)\\
D^{\bulk}(x_j,x_k)&S^{\bulk}(x_k,x_j)
\end{array}
\right]_{j,k=1,...,n},
\end{align}
where
\begin{align}
\nonumber S^{\bulk}(x_j,x_k)&=\frac{\sin \pi \rho\: (x_j-x_k)}{\pi \rho\: (x_j-x_k)},\\
\nonumber \tilde{I}^{\bulk}(x_j,x_k)&=\frac{1}{\pi\rho}\int_{0}^{\pi \rho\: (x_j-x_k)}\frac{\sin t}{t}dt+\frac{1}{2\rho}\sgn(x_k-x_j),\\
\nonumber D^{\bulk}(x_j,x_k)&=\frac{\partial}{\partial x_j}\frac{\sin \pi \rho\: (x_j-x_k)}{\pi \rho\: (x_j-x_k)}.
\end{align}

The limiting density (\ref{eqn:wssl}) appears commonly in Hermitian and symmetric random matrix ensembles and is known as \textit{Wigner's semi-circle law}. It was first conjectured in the 1950s based upon numerical evidence before being shown analytically by Wigner in 1955 \cite{wigner1955} for a restricted class of matrices, and then found to apply to a broader range of matrices in \cite{wigner1958}. Various results since then, including \cite{Pastur1972, Joh1998, BaiSilver2006}, have established a quite general form of the law.

\begin{proposition}[Semi-circle law]
\label{prop:wssl}
Let $\bX=[x_{j,k}]_{j,k=1,...,N}$ be Hermitian, with iid entries (up to the required symmetry) $x_{j,k}$ drawn from any distribution of zero mean and variance $1$. Then the scaled eigenvalue density of $\bX$ tends to (\ref{eqn:wssl}) as $N\to\infty$.
\end{proposition}

\newpage

\section{The importance of being odd}
\setcounter{figure}{0}
\label{sec:GOE_odd}

As mentioned at the beginning of Chapter \ref{sec:qdets_pfs}, Definition \ref{def:pfaff} implies that the Pfaffian of a matrix $\bX$ is only defined if $\bX$ is of even dimension. In \cite{deB1955} de Bruijn discusses how the definition can be interpreted to include odd-sized matrices. We will stick with the convention that Definition \ref{def:pfaff} only applies to even-sized matrices, however his methods --- which involve bordering an odd-sized matrix with an extra row and column, or calculating the $N$ odd case by removing one variable to infinity in the $N+1$ case --- turn out to be similar to our development.

We see from Proposition \ref{prop:GOE_eval_jpdf} that the eigenvalue jpdf is insensitive to the parity of $N$, so the calculation up to that point does not need to be modified. But when Pfaffians are introduced in Step III (Chapter \ref{sec:Step3_GOE}) things go awry. It is at this point that we pick up the calculation, with $N$ now specified to be odd. We present two methods for calculating the correlation functions for odd-sized matrices: one in which an extra row and column are added to the generalised partition function (\ref{eqn:GOE_GPF_even}) (\cite{PanShu1991} and \cite{frahm_and_pichard1995} use similar constructions for the circular ensembles); while the other uses the known correlations for a $2N$-sized system and removes one eigenvalue off to infinity, leaving the correlations of a $(2N-1)$-sized system. Note that these techniques were not required for the original calculations of the GOE correlation functions for odd matrix size, since Proposition \ref{thm:integral_identities} was used. However, we develop these techniques here since Dyson's method is not applicable to the other ensembles that will be considered in the following chapters of the present work, and we hope to provide a unified treatment.

For the first approach (involving functional differentiation) the plan is to modify the method of integration over alternate variables in such a way as to generate an even-sized Pfaffian for $N$ odd, giving $Z_N[u]$. It is this modification to the alternate variable method that essentially distinguishes the treatment of $N$ odd from $N$ even in all of the $\beta=1$ ensembles. We will then rewrite $Z_N[u]$ as a Fredholm quaternion determinant and Fredholm Pfaffian and apply functional differentiation as in the even case to find the correlation functions. However, there is a further complication in establishing the Fredholm operators for $N$ odd: because of the structure of the odd generalised partition function, the calculation cannot be carried out strictly as a Pfaffian or quaternion determinant, since non-anti-symmetric and non-self-dual matrices are involved. Instead, we resort to a determinantal form and then use Corollary \ref{cor:sqrtFreds} after the fact.

\subsection{Pfaffian generalised partition function for GOE, $N$ odd}

Here we derive the $N$ odd analogue of Proposition \ref{prop:GOE_gen_part_fn} again using the method of integration over alternate variables. Recall from the proof of that proposition that the method relied on pairing up the rows (corresponding to pairing up the eigenvalues) in the Vandermonde determinant to symmetrise $Z_N[u]$. However, with $N$ odd, there is clearly a difficulty as there will be one unpaired row. Dealing with this row is something of a technical task, but it naturally leads to an even-sized Pfaffian, with an $N\times N$ block bordered by a row and column corresponding to the unpaired eigenvalue. This type of method was applied 

\begin{proposition}[\cite{Mehta1967}]
\label{prop:GOE_gen_part_fn_odd}
Let $Q(\vec{\lambda})$ be as in (\ref{eqn:GOE_eval_jpdf}). For $N$ odd the generalised partition function for $Q(\vec{\lambda})$ is
\begin{align}
\label{eqn:GOE_GPF_odd} Z_{N\:\odd}[u]=\frac{N!}{2^{N+1/2}}\prod_{j=1}^N\frac{1}{\Gamma(j/2+1)}\Pf \left[\begin{array}{cc}
[\gamma_{jk}] & [\nu_j]\\
\left[-\nu_k\right] & 0
\end{array}\right]_{j,k=1,...,N},
\end{align}
where $\gamma_{jk}$, $p_j(x)$ and $u$ are as in Proposition \ref{prop:GOE_gen_part_fn} and
\begin{align}
\label{def:GOE_nu} \nu_k:=\int_{-\infty}^{\infty}e^{-x^2/2}u(x)p_{k-1}(x)\:dx.
\end{align}
\end{proposition}

\textit{Proof:} As in the even case we order the eigenvalues in (\ref{def:GOE_gpf1}) according to $-\infty < x_1 <\cdot\cdot\cdot < x_N < \infty$, picking up a factor of $N!$, and let $A_N:= 2^{-3N/2}\prod_{j=1}^N\frac{1}{\Gamma(j/2+1)}$. We then have
{\small
\begin{align}
\nonumber Z_N[u]&=A_NN!\int_{-\infty}^{x_2}dx_1 \int_{x_1}^{x_3}dx_2 \cdot\cdot\cdot \int_{x_{N-1}}^{\infty}dx_N\prod_{j=1}^Ne^{-x_j^2/2}u(x_j)\prod_{1\leq j < k \leq N}(x_k-x_j)\\
\nonumber &=A_NN!\int_{-\infty}^{x_2} dx_1\int_{x_1}^{x_3} dx_2\cdot\cdot\cdot \int_{x_{N-1}}^{\infty}dx_N \det \left[ e^{-x_j^2/2}u(x_j)p_{k-1}(x_j) \right]_{j,k=1,...,N}\\
\nonumber &=A_NN!\int_{-\infty}^{x_4}dx_2\int_{x_2}^{x_6}dx_4 \cdot\cdot\cdot \int_{x_{N-3}}^{x_{N-1}} dx_{N-1}\\
\nonumber &\times \det \left[\begin{array}{c}
\left[ \begin{array}{c}
\int_{-\infty}^{x_{2j}}e^{-x^2/2}u(x)p_{k-1}(x)dx\\
e^{-x_{2j}^2/2}u(x_{2j})p_{k-1}(x_{2j})
\end{array}\right]\\
\int_{-\infty}^{\infty}e^{-x^2/2}u(x)p_{k-1}(x)dx
\end{array}\right]_{j=1,...,(N-1)/2 \atop k=1,...,N}\\
\nonumber &=A_N\frac{N!}{((N-1)/2)!}\int_{-\infty}^{\infty} dx_2\int_{-\infty}^{\infty} dx_4 \cdot\cdot\cdot \int_{-\infty}^{\infty}dx_{N-1}\\
\nonumber &\times \det \left[\begin{array}{c}
\left[ \begin{array}{c}
\int_{-\infty}^{x_{2j}}e^{-x^2/2}u(x)p_{k-1}(x)dx\\
e^{-x_{2j}^2/2}u(x_{2j})p_{k-1}(x_{2j})
\end{array}\right]\\
\int_{-\infty}^{\infty}e^{-x^2/2}u(x)p_{k-1}(x)dx
\end{array}\right]_{j=1,...,(N-1)/2 \atop k=1,...,N},
\end{align}
}where, for the third equality, we have added rows to make the integrals inside the determinant start at $-\infty$, and for the last equality we removed the ordering on the $(N-1)/2$ variables outside the determinant.

With $\mu_{jk}$ as in (\ref{def:mu_GOE}), expanding the determinant yields
\begin{align}
\nonumber Z_N[a]=A_N\frac{N!}{(N/2)!}\sum_{P\in S_N}\varepsilon(P)\nu_{P(N)}\prod_{l=1}^{(N-1)/2}\mu_{P(2l-1),P(2l)},
\end{align}
and restricting the sum to terms with $P(2l)>P(2l-1)$ we have
\begin{align}
\nonumber Z_N[a]=A_N2^{(N-1)/2} N!\sum_{P\in S_N \atop P(2l)>P(2l-1)}^*\varepsilon(P) \nu_{P(N)} \prod_{l=1}^{(N-1)/2} \gamma_{P(2l-1),P(2l)}.
\end{align}
Now, letting $\nu_{P(N),N+1}:=\nu_{P(N)}$ we use the first equality in Definition \ref{def:pfaff} and we have the result.

\hfill $\Box$

\subsection{Skew-orthogonal polynomials}
\label{sec:GOE_sops_odd}

Since (\ref{eqn:GOE_GPF_odd}) contains a Pfaffian, it will be simplest to calculate if we pick polynomials that skew-diagonalise the matrix as in the even case. However, we note that with the polynomials (\ref{eqn:sops}) $\nu_i\big|_{u=1} \neq 0$ for any $1\leq i \leq N$, instead we see that the matrix is of the form
\begin{align}
\label{eqn:skew_diag_mat_odd} \bA_o=\left[\begin{array}{ccc}
\bA & \0_{N-1} & \bb_{N-1}\\
\0_{N-1}^T & 0 & b_N\\
-\bb_{N-1}^T & -b_N & 0
\end{array}\right]
\end{align}
where $\bA$ is given by (\ref{eqn:skew_diag_mat}) and $\bb_{N-1}=[b_1\; b_2\;\cdot\cdot\cdot\; b_{N-1}]^T$. So this matrix differs from that of (\ref{eqn:skew_diag_mat}) in that it contains two extra rows and columns that border the $N-1\times N-1$ skew-diagonal matrix. While this matrix is not strictly skew-diagonal, it will serve our turn since by Laplace expansion
\begin{align}
\nonumber \Pf \bA_o = b_N\; \prod^{(N-1)/2}_{j=1}a_j.
\end{align}
and so we say the matrix $\bA_o$ is \textit{odd skew-diagonal}. However, a key difference between the even and odd skew-diagonal matrices is their inverses. An even skew-diagonal matrix can be written as $\mathbf{D}\bZ^{-1}_N$ where $\mathbf{D}$ is some diagonal matrix with every non-zero element repeated, and so its inverse is simply $\bZ_N\mathbf{D}^{-1}$ (a fact that was exploited in Proposition \ref{prop:pf_integ_op}). Yet an odd skew-diagonal matrix $\bA_o$ cannot be decomposed in such a fashion; the inverse is of the more complicated form
\begin{align}
\label{eqn:skew_inverse_odd} \bA_o^{-1}=\left[\begin{array}{ccc}
\bA^{-1} & \bc_{N-1} & \0_{N-1}\\
-\bc_{N-1}^T & 0 & -b_{N}^{-1}\\
\0_{N-1}^T & b_N^{-1} & 0
\end{array}\right]
\end{align}
where $\bc_{N-1}=[\frac{b_2}{a_1b_N}\;\frac{-b_1}{a_1b_N}\;\frac{b_4}{a_2b_N}\;\frac{-b_3}{a_2b_N}\;\cdot\cdot\cdot\; \frac{b_{N-1}}{a_{(N-1)/2}b_N}\; \frac{-b_{N-2}}{a_{(N-1)/2}b_N}]^T$, although we note that $\bA_o^{-1}$ is still anti-symmetric and the Pfaffian is
\begin{align}
\label{eqn:Pfinvo} \Pf \bA_o^{-1} = \left( b_N\; \prod^{(N-1)/2}_{j=1}a_j\right)^{-1},
\end{align}
as we should expect.

\subsection{GOE $n$-point correlations, $N$ odd}

\begin{definition}
\label{def:GOE_correln_kernel_odd}
With the definitions as used in Proposition \ref{def:GOE_correln_kernel} and $\nu_k$ as defined in (\ref{def:GOE_nu}), let $\fodd(x,y)$ be the $2\times 2$ matrix
\begin{align}
\label{eqn:GOE_correln_kernel} \fodd(x,y)=\left[\begin{array}{cc}
S_{\odd}(x,y) & \tilde{I}_{\odd}(x,y)\\
D_{\odd}(x,y) & S_{\odd}(y,x)
\end{array}\right],
\end{align}
where
\begin{align}
\nonumber S_{\odd}(x,y)&=\sum_{j=0}^{(N-1)/2-1}\frac{e^{-y^2/2}}{r_j}\Big( \hat{\Phi}_{2j}(x)\hat{R}_{2j+1}(y)-\hat{\Phi}_{2j+1}(x)\hat{R}_{2j}(y)\Big)\\
\nonumber &+\frac{e^{-y^2/2}}{\bar{\nu}_N}R_{N-1}(y),\\
\nonumber D_{\odd}(x,y)&=\sum_{j=0}^{(N-1)/2-1}\frac{e^{-(x^2+y^2)/2}}{r_j}\Big( \hat{R}_{2j}(x)\hat{R}_{2j+1}(y)-\hat{R}_{2j+1}(x)\hat{R}_{2j}(y)\Big),\\
\nonumber \tilde{I}_{\odd}(x,y)&:=I_{\odd}(x,y)+\frac{1}{2}\mathrm{sgn}(y-x)\\
\nonumber &=\sum_{j=0}^{(N-1)/2-1}\frac{1}{r_j}\Big( \hat{\Phi}_{2j+1}(x)\hat{\Phi}_{2j}(y)-\hat{\Phi}_{2j}(x)\hat{\Phi}_{2j+1}(y)\Big)+\frac{1}{2}\mathrm{sgn}(y-x)\\
\nonumber &+\frac{1}{\bar{\nu}_N}\Big( \Phi_{N-1}(x)-\Phi_{N-1}(y) \Big),
\end{align}
with $\bar{\nu}_{j}:=\nu_j\big|_{u=1}$ and
\begin{align}
\label{eqn:Rhat_GOE} \hat{R}_j(x)&:=R_j(x)-\frac{\bar{\nu}_{j+1}}{\bar{\nu}_N} R_{N-1}(x),\\
\nonumber \hat{\Phi}_j(x)&:=\frac{1}{2}\int_{-\infty}^{\infty}dy\: \hat{R}_j(y)\: e^{-y^2/2}\: \mathrm{sgn}(x-y).
\end{align}
\end{definition}
Again, the Pfaffian equivalent is
\begin{align}
\mathbf{f}_{\odd}(x,y)\bZ_2^{-1}=\left[\begin{array}{cc}
-\tilde{I}(x,y) & S(x,y)\\
-S(y,x) & D(x,y)
\end{array}\right].
\end{align}

To give away the ending, we will find that the correlations for the odd case are given by (\ref{eqn:correlns_GOE}) with $\mathbf{f}(x,y)$ replaced by $\fodd(x,y)$. The obvious way to obtain this result is to repeat the calculations of Chapter \ref{sec:correlns_even_GOE} using $Z_{N\:\odd}[u]$ of (\ref{eqn:GOE_GPF_odd}) instead of $Z_N[u]$ from (\ref{eqn:GOE_GPF_even}). The presentation of this method in the proof of Proposition \ref{prop:pf_integ_op} is such that, with a minor modification, we can proceed in the same fashion, highlighting the structural similarity between the even and odd cases. 

A perhaps more elegant approach is to use the known result for $N$ even and combine it with the physical intuition that a system containing an even number of interacting particles will tend to a system with one fewer particles if one of them is removed to infinity. This process works well for eigenvalues in an open set (such as here and in the Ginibre ensembles of Chapter \ref{sec:GinOE}), however if the eigenvalues are contained in a compact set (such as the spherical ensemble of Chapter \ref{sec:SOE}) then this method does not seem applicable.

\subsubsection{Functional differentiation method}

In the following proposition we will modify the proof of Proposition \ref{prop:pf_integ_op} to produce the odd analogue. The required modification is essentially the addition of an extra column to the matrix $\bE$ in (\ref{eqn:omega_E}). This approach is similar to that in \cite{Sinc09} where use was made of (\ref{eqn:rains}) while we use (\ref{eqn:1+AB}), although in that paper the matrix equivalent to $\bE$ (labelled $\bA$) is of much larger size: $(N+1)\times 2T$ where $T$ is some integer larger than $2N$.

\begin{proposition}
\label{prop:pf_integ_op_odd}
Let $\gamma_{jk}$ be as in (\ref{eqn:gammajk}), $\nu_k$ as in (\ref{def:GOE_nu}), with $\bar{\nu}_k:=\nu_k\big|_{u=1}$. Then, using the skew-orthogonal polynomials of Proposition \ref{prop:GOE_soip}, we have
\begin{align}
\label{prop:integ_op_odd} \Pf \left[\begin{array}{cc}
[\gamma_{jk}] & [\nu_j]\\
\left[-\nu_k\right] & 0
\end{array}\right]_{j,k=1,...,N}=\left(\bar{\nu}_N \prod_{j=0}^{(N-1)/2-1}r_j\right) \qdet [\1_2+\fodd^T(\mathbf{u}-\1_2)],
\end{align}
where $\fodd^T(\mathbf{u}-\1_2)$ is the matrix integral operator with kernel $\fodd^T(x,y)\; \mathrm{diag}[u(y)-1,u(y)-1]$.
\end{proposition}

\textit{Proof}: 
The proof for the odd case proceeds along the same lines as for the even case in Proposition \ref{prop:pf_integ_op}: we look for a pair of matrices $\bA$ and $\bB$ such that the left hand side of (\ref{prop:integ_op_odd}) can be expressed as $\1+\bA\bB$ and then apply (\ref{eqn:1+AB}).

First, for convenience, we define
\begin{align}
\nonumber \bC:=\left[\begin{array}{cc}
[\gamma_{jk}] & [\nu_j]\\
\left[-\nu_k\right] & 0
\end{array}\right]_{j,k=1,...,N}.
\end{align}
Then with $\psi_j:=e^{-x^2/2}R_{j-1}$ and $u=\sigma+1$ as in Proposition \ref{prop:pf_integ_op} we have
\begin{align}
\nonumber \bC=\bC^{(1)}-\bC^{(\sigma)},
\end{align}
where $\bC^{(1)}_{jk}:=\bC_{jk}\Big|_{u=1}$,
\begin{align}
\nonumber \bC^{(\sigma)}_{jk}:=\int_{-\infty}^{\infty}\Big(\sigma(x)\psi_j(x)\epsilon \psi_k[x]-\sigma(x)\psi_k(x) \epsilon\psi_j[x]-\sigma(x) \psi_k(x) \epsilon(\sigma \psi_j)[x] \Big)dx
\end{align}
for $1\leq j<k\leq N$ and
\begin{align}
\label{eqn:oddCnu} \bC^{(\sigma)}_{j,N+1}=-\int_{-\infty}^{\infty}e^{-x^2/2}\sigma(x)R_{j-1}(x)\; dx
\end{align}
for $1\leq j\leq N$, with the remaining elements being established by the anti-symmetry of $\bC$. Clearly
\begin{align}
\nonumber \Pf\; \bC&=\Pf\Big(\bC^{(1)}\big(\1_{N+1}-(\bC^{(1)})^{-1}\bC^{(\sigma)}\big)\Big).
\end{align}
From the discussion at the beginning of Chapter \ref{sec:GOE_sops_odd} we know that with the skew-orthogonal polynomials (\ref{eqn:sops}) $\bC^{(1)}$ is of the form (\ref{eqn:skew_diag_mat_odd}) and not of the form (\ref{eqn:skew_diag_mat}) and so we cannot apply Corollary \ref{cor:qdet=pf}. Instead we square both sides to obtain
\begin{align}
\det \bC = \left(\bar{\nu}_N \prod_{j=0}^{(N-1)/2-1}r_j\right)^2 \det \Big(\1_{N+1}-(\bC^{(1)})^{-1}\bC^{(\sigma)}\Big).
\end{align}
Recall $\Omega$ from (\ref{eqn:omega_E}) and extend the $2\times N$ matrix $\mathbf{E}$ with an extra column, defining
\begin{align}
\label{def:Eodd} \mathbf{E}_{\odd}:=\left[
\begin{array}{cccc}
\psi_1(y) & \cdot \cdot\cdot & \psi_N(y)&0\\
\epsilon\psi_1[y] & \cdot \cdot\cdot & \epsilon\psi_N[y]&-1
\end{array}
\right].
\end{align}
In analogy with the even case, we let $\bA$ be the $N+1\times 2$ integral operator on $(-\infty,\infty)$ with kernel $(\bC^{(1)})^{-1}\sigma(y)(\Omega \mathbf{E}_{\odd})^T$, although $(\bC^{(1)})^{-1}$ has the structure (\ref{eqn:skew_inverse_odd}). Carrying out the explicit computation of the kernel of $\bA$ we find the `hat' structure of (\ref{eqn:Rhat_GOE}) emerges naturally, so we define $\hat{\psi}$ and $\hat{G}$ to be the equivalent definitions used in Proposition \ref{prop:pf_integ_op} but replacing $R$ with $\hat{R}$.

The kernel of $\bA$ is then the $N+1\times 2$ matrix
\begin{align}
\left[\begin{array}{cc}
\nonumber [-\frac{\sigma(y)}{r_{\lfloor (j-1)/2 \rfloor}}\big(\epsilon \hat{G}_j[y]+\epsilon(\sigma \hat{G}_j)[y]\big)]&[\frac{\sigma (y)}{r_{\lfloor (j-1)/2 \rfloor}}\hat{G}_j(y)]\\
\sigma(y)\sum_{k=0}^{N-2}X_k\big(\epsilon G_k[y]+\epsilon(\sigma G_k)[y] \big)+\frac{\sigma(y)}{\bar{\nu}_N} & -\sigma(y)\sum_{k=0}^{N-2} X_k G_k(y)\\
\frac{\sigma(y)}{\bar{\nu}_N}\big(\epsilon \psi_N[y]+\epsilon(\sigma \psi_N)[y]\big)&-\frac{\sigma(y)}{\bar{\nu}_N}\psi_N(y)
\end{array}\right]_{j=1,...,N-1},
\end{align}
where $X_k:=\bar{\nu}_k/(r_{\lfloor(k-1)/2\rfloor}\bar{\nu}_N)$. With $\bB=\mathbf{E}_{\odd}$ we then have $\1_{N+1}-(\bC^{(1)})^{-1}\bC^{(\sigma)}=\1_{N+1}+\bA\bB$. Applying (\ref{eqn:1+AB}) then $\1_{N+1}+\bB\bA$ equals
\begin{align}
\label{eqn:1+ba1_odd}\left[\begin{array}{cc}
1+\kappa_{1,1} & \kappa_{1,2}\\
\kappa_{2,1} & 1+\kappa_{2,2}
\end{array}\right],
\end{align}
where
\begin{align}
\nonumber \kappa_{1,1}=&-\sum_{j=1}^{N-1}\frac{1}{r_{\lfloor (j-1)/2 \rfloor}}\Big( \hat{\psi}_j\otimes \sigma \epsilon \hat{G}_j+\hat{\psi}_j\otimes \sigma \epsilon (\sigma \hat{G}_j)\Big) + \frac{1}{\bar{\nu}_N}\psi_N\otimes \sigma,\\
\nonumber \kappa_{1,2}=&-\sum_{j=1}^{N-1}\frac{1}{r_{\lfloor (j-1)/2 \rfloor}}\Big(\epsilon \hat{\psi}_j\otimes \sigma\epsilon \hat{G}_j+\epsilon \hat{\psi}_j\otimes \sigma \epsilon(\sigma \hat{G}_j) \Big) +\frac{1}{\bar{\nu}_N}\epsilon\psi_N\otimes \sigma\\
\nonumber &-\frac{1}{\bar{\nu}_N}\left( 1\otimes\sigma\epsilon(\sigma\psi_N) +1\otimes\sigma\epsilon\psi_N\right),\\
\nonumber \kappa_{2,1}=&\sum_{j=1}^{N-1}\frac{1}{r_{\lfloor (j-1)/2 \rfloor}}\hat{\psi}_j\otimes \sigma \hat{G}_j,\\
\nonumber \kappa_{2,2}=&\sum_{j=1}^{N-1}\frac{1}{r_{\lfloor (j-1)/2 \rfloor}}\epsilon \hat{\psi}_j\otimes \sigma \hat{G}_j+\frac{1}{\bar{\nu}_N}1\otimes\sigma\psi_N.
\end{align}
With the decomposition of $\Omega^T$ given by (\ref{eqn:OT_decomp}) we factorise (\ref{eqn:1+ba1_odd}) as
{\small
\begin{align}
\nonumber \left[\begin{array}{c}
1-\sum_{j=1}^{N-1}\frac{1}{r_{\lfloor (j-1)/2 \rfloor}} \hat{\psi}_j\otimes \sigma \epsilon \hat{G}_j + \frac{1}{\bar{\nu}_N}\psi_N\otimes \sigma\\
-\epsilon\sigma-\sum_{j=1}^{N-1}\frac{1}{r_{\lfloor (j-1)/2 \rfloor}}\epsilon \hat{\psi}_j\otimes \sigma\epsilon \hat{G}_j +\frac{1}{\bar{\nu}_N}\left( \epsilon\psi_N\otimes\sigma -1\otimes\sigma\epsilon\psi_N\right)
\end{array}
\right.
\end{align}
\begin{align}
\left.
\label{eqn:1+ba2_odd} \begin{array}{c}
\sum_{j=1}^{N-1}\frac{1}{r_{\lfloor (j-1)/2 \rfloor}}\hat{\psi}_j\otimes \sigma \hat{G}_j\\
1+\sum_{j=1}^{N-1}\frac{1}{r_{\lfloor (j-1)/2 \rfloor}}\epsilon \hat{\psi}_j\otimes \sigma \hat{G}_j+\frac{1}{\bar{\nu}_N}1\otimes\sigma\psi_N
\end{array}\right]\left[
\begin{array}{cc}
1 & 0\\
\epsilon\sigma & 1
\end{array}
\right],
\end{align}
}which, as we saw in (\ref{eqn:1+ba1}) and (\ref{eqn:1+ba2}), eliminates the apparent complication of the factor $\epsilon\sigma$.

The matrix on the right in (\ref{eqn:1+ba2_odd}) has determinant $1$ and, recalling that $\fodd$ is invariant under the coincidental replacements $\tilde{I}(x,y)\rightarrow -\tilde{I}(x,y),D(x,y)\rightarrow -D(x,y)$, we have established the square of (\ref{prop:integ_op_odd}). Applying Corollary \ref{cor:sqrtFreds} then gives the result.

\hfill $\Box$

By comparing (\ref{eqn:1+ba1_odd}) and (\ref{eqn:1+ba2_odd}) to their counterparts (\ref{eqn:1+ba1}) and (\ref{eqn:1+ba2}) in the even case the similarity in the methods used is clear, which highlights the reason for this particular presentation. The key difference was that the matrix in the generalised partition function was skew-diagonalised in the even case, but not in the odd case.

\begin{corollary}
With $Z_{N\:\odd}[u]$ as in Proposition \ref{prop:GOE_gen_part_fn_odd} we have
\begin{align}
\nonumber Z_{N\:\odd}[u]=\frac{N!}{2^{N+1/2}}\prod_{j=1}^N\frac{1}{\Gamma(j/2+1)}\left(\bar{\nu}_N\prod_{j=0}^{(N-1)/2-1}r_j\right) \qdet [\1_2+\mathbf{f}_{\odd}^T(\mathbf{u}-\1_2)].
\end{align}
\end{corollary}

\textit{Proof}: Substitute (\ref{prop:integ_op_odd}) into (\ref{eqn:GOE_GPF_odd}).

\hfill $\Box$

Now, applying function differentiation as in Proposition \ref{prop:correlns_GOE_even}, we find the $N$ odd correlations.

\begin{proposition}
\label{prop:correlns_GOE_odd}
With $\mathbf{f}_{\mathrm{odd}}(x_j,x_k)$ from Definition \ref{def:GOE_correln_kernel_odd} the $n$th-order correlation function for GOE, with $N$ odd, is
\begin{align}
\nonumber \rho_{(n)}(x_1,...,x_n)&=\qdet [\mathbf{f}_{\mathrm{odd}}(x_l,x_m)]_{l,m=1,...,n}\\
\nonumber &=\Pf [\mathbf{f}_{\mathrm{odd}}(x_l,x_m)\bZ_2^{-1}]_{l,m=1,...,n}.
\end{align}
\end{proposition}

\subsubsection{Odd from even}
\label{sec:odd_from_even}

An alternative approach to the problem of deducing $N$ odd correlations is to take the known result in the even case and then somehow generate the odd case from that. To this end we can imagine that if one of the eigenvalues is removed to infinity, then we essentially have two independent systems: one of $N-1$ eigenvalues, and one with a single eigenvalue. The probability function is then the product of the individual probabilities. So the calculation of the odd case from that of the even with this `eigenvalue off to infinity' method will be a useful strategy if there exists an $f_N$ such that (\ref{eqn:GOE_eval_jpdf}) exhibits the factorisation
\begin{align}
\label{eqn:jpdfOEfactorisation}
\fullsub{Q(\vec{\lambda})}{\sim}{|\lambda_N|\rightarrow\infty}{f_N(\lambda_N)\; Q(\lambda_1,...,\lambda_{N-1}).}
\end{align}
Note that for finite $N$
\begin{align}
\nonumber Q(\vec{\lambda})&=2^{-3(N-1)/2}\prod_{j=1}^{N-1}\frac{e^{-\lambda_j^2/2}}{\Gamma(j/2+1)}\prod_{1\leq j < k \leq N-1}|\lambda_k-\lambda_j|\\
\nonumber &\times 2^{-3/2}\frac{e^{-\lambda_N^2/2}}{\Gamma (N/2+1)}\prod_{j=1}^{N-1}|\lambda_N-\lambda_j |.
\end{align}
So with
\begin{align}
\label{eqn:fN} f_N(x)=2^{-3/2}\frac{e^{-x^2/2}x^{N-1}}{\Gamma (N/2+1)}
\end{align}
(\ref{eqn:GOE_eval_jpdf}) satisfies (\ref{eqn:jpdfOEfactorisation}), and using (\ref{eqn:integ_correlns}) we then have
\begin{align}
\label{eqn:correlnOEfactorisation}
\fullsub{\rho_{(m)}^N(r_1,...,r_m)}{\sim}{|r_m|\rightarrow\infty}{Nf_N(r_m)\hspace{3pt}\rho_{(m-1)}^{N-1}(r_1,...,r_{m-1}),}
\end{align}
where the superscripts refer to the number of variables in the relevant distribution function. We now seek an interpretation of $N f_N(r_m)$.

\begin{lemma}
\label{lem:xm_to_infty}
Let
\begin{align}
\label{def:gen_jpdf} \mathcal{Q}(x_1,...,x_N):=\frac{1}{C_N}\prod_{j=1}^N e^{-V(x_j)}\prod_{1\leq j < k \leq N}|x_k-x_j|
\end{align}
be the joint probability density function of the variables $x_1,...,x_N$ and define
\begin{align}
\label{def:gen_fN} \mathcal{F}_N(x):=\frac{C_{N-1}}{C_N}x^{N-1}e^{-V(r)}.
\end{align}
Then, with $r:=x_1$, we have
\begin{align}
\nonumber \fullsub{\rho_{(1)}(r)}{\sim}{|r|\rightarrow\infty}{N\mathcal{F}_N(r).}
\end{align}
\end{lemma}

\textit{Proof}: Applying (\ref{eqn:integ_correlns}) to (\ref{def:gen_jpdf}) yields
{\small
\begin{align}
\nonumber \rho_1(r&)=N\frac{e^{-V(r)}}{C_N}\int_{-\infty}^{\infty}dx_2 \cdot\cdot\cdot \int_{-\infty}^{\infty} dx_N\prod_{j=2}^N |r-x_j|\; e^{-V(x_j)}\prod_{2\leq j < k \leq N}|x_k-x_j|\\
\nonumber &\mathop{\sim}\limits_{|r|\to\infty}N\frac{C_{N-1}}{C_N}r^{N-1}e^{-V(r)}\frac{1}{C_{N-1}}\int_{-\infty}^{\infty}dx_2\cdot\cdot\cdot\int_{-\infty}^{\infty}dx_N\\
\label{eqn:fN_proof} &\quad \times \prod_{j=2}^N e^{-V(x_j)}\prod_{2\leq j < k \leq N}|x_k-x_j|.
\end{align}
}Since $\mathcal{Q}(x_1,...,x_N)$ was defined as a probability density function in (\ref{def:gen_jpdf}) (that is, the integral is normalised to 1), the integral in (\ref{eqn:fN_proof}) equals $C_{N-1}$ and we have the result.

\hfill $\Box$

With (\ref{eqn:correlnOEfactorisation}) and Lemma \ref{lem:xm_to_infty} we have 
\begin{equation}
\label{eqn:finalOEfactorisation}
\fullsub{\rho_{(m)}^N(\lambda_1,...,\lambda_m)}{\sim}{|r_m|\rightarrow\infty}{\rho_{(1)}^N(r_m)\rho_{(m-1)}^{N-1}(r_1,...,r_{m-1}),}
\end{equation}
with the minor caveat that in the case that the eigenvalues are ordered (which they are in this case) one must be careful to remove only the largest eigenvalue off to infinity; however, this amounts to nothing more than a relabeling.

So we see that from knowledge of the $m$-point correlation with $N$ even, we can find the $(m-1)$-point correlation with $N$ odd, by factoring out the density of the largest eigenvalue and taking the limit. Our task now is to use (\ref{eqn:finalOEfactorisation}) to deduce Proposition \ref{prop:correlns_GOE_odd} from Proposition \ref{prop:correlns_GOE_even}; and for that we begin with humble row and column reduction.

Recalling (\ref{def:GOE_Pfcorrelnk}) we write out the Pfaffian in (\ref{eqn:correlns_GOE}), explicitly identifying the last two rows and columns, thusly
\begin{align}
\nonumber &\rho_{(m)}(x_1,...,x_m)=\\
\nonumber &\mathrm{Pf}\left[\begin{array}{cc}
\left[\begin{array}{cc}
-\tilde{I}(x_i,x_j) & S(x_i,x_j)\\
-S(x_j,x_i) & D(x_i,x_j)\\
\end{array}\right] & \left[\begin{array}{cc}
-\tilde{I}(x_i,x_m) & S(x_i,x_m)\\
-S(x_m,x_i) & D(x_i,x_m)\\
\end{array}\right]\\
&\\
\left[\begin{array}{cc}
-\tilde{I}(x_m,x_j) & S(x_m,x_j)\\
-S(x_j,x_m) & D(x_m,x_j)\\
\end{array}\right] & \left[\begin{array}{cc}
0 & S(x_m,x_m)\\
-S(x_m,x_m) & 0\\
\end{array}\right]\\
\end{array}\right]_{i,j=1,...,m-1},
\end{align}
for some fixed $m$. This matrix consists of four submatrices of sizes
\begin{itemize}
\item{Top left: $2(m-1)\times 2(m-1)$,}
\item{Top right: $2(m-1)\times 2$,}
\item{Bottom left: $2 \times 2(m-1)$,}
\item{Bottom right: $2 \times 2$.}
\end{itemize}
Applying elementary row and column operations yields
\begin{align}
\nonumber &\rho_{(m)}(x_1,...,x_m)=\\
\nonumber &S(x_m,x_m)\;\mathrm{Pf}\left[\begin{array}{cc}
\left[\begin{array}{cc}
\vspace{3pt}-\tilde{I}^*(x_i,x_j) & S^*(x_i,x_j)\\
-S^*(x_j,x_i) & D^*(x_i,x_j)\\
\end{array}\right] & \left[\begin{array}{cc}
\vspace{3pt}-\tilde{I}(x_i,x_m) & 0\\
-S(x_m,x_i) & 0\\
\end{array}\right]\\
&\\
\left[\begin{array}{cc}
\vspace{3pt}-\tilde{I}(x_m,x_j) & S(x_m,x_j)\\
0 & 0\\
\end{array}\right] & \left[\begin{array}{cc}
\vspace{3pt}0 & 1\\
-1 & 0\\
\end{array}\right]\\
\end{array}\right]_{i,j=1,...,m-1}\\
\label{eqn:Pf3}&=S(x_m,x_m)\;\mathrm{Pf}
\left[\begin{array}{cc}
\vspace{3pt}-\tilde{I}^*(x_i,x_j) & S^*(x_i,x_j)\\
-S^*(x_j,x_i) & D^*(x_i,x_j)\\
\end{array}\right]_{i,j=1,...,m-1},
\end{align}
where
\begin{align}
\nonumber D^*(x_i,x_j)&:=D(x_i,x_j)-\frac{D(x_i,x_m)S(x_m,x_j)}{S(x_m,x_m)}-\frac{S(x_m,x_i)D(x_m,x_j)}{S(x_m,x_m)},\\
\nonumber S^*(x_i,x_j)&:=S(x_i,x_j)-\frac{S(x_i,x_m)S(x_m,x_j)}{S(x_m,x_m)}-\frac{D(x_m,x_j)\tilde{I}(x_i,x_m)}{S(x_m,x_m)},\\
\nonumber \tilde{I}^*(x_i,x_j)&:=\tilde{I}(x_i,x_j)-\frac{S(x_i,x_m)\tilde{I}(x_m,x_j)}{S(x_m,x_m)}-\frac{S(x_j,x_m)\tilde{I}(x_i,x_m)}{S(x_m,x_m)}.
\end{align}
The second equality in (\ref{eqn:Pf3}) can be seen by using the Laplace expansion method for Pfaffians discussed in Chapter \ref{sec:qdets_pfs}. Recalling (\ref{eqn:1pt_correln}) we see that (\ref{eqn:Pf3}) factors out $\rho_{(1)}^N(x_m)$ as required by (\ref{eqn:finalOEfactorisation}). To reclaim Proposition \ref{prop:correlns_GOE_odd} we must then have
\begin{align}
\nonumber D^*(x_i,x_j)\Big|_{x_m\to\infty}&=D_{\odd}(x_i,x_j)\Big|_{N\to N-1},\\
\nonumber S^*(x_i,x_j)\Big|_{x_m\to\infty}&=S_{\odd}(x_i,x_j)\Big|_{N\to N-1},\\
\label{eqn:DSI*} \tilde{I}^*(x_i,x_j)\Big|_{x_m\to\infty}&=\tilde{I}_{\odd}(x_i,x_j)\Big|_{N\to N-1},
\end{align}
which is easily established when we note from Definition \ref{def:GOE_correln_kernel} that as $x_m\rightarrow \infty$
\begin{align}
\nonumber S(x_m,x_m)&\rightarrow \frac{e^{-x_m^2/2}}{r_{N/2-1}}R_{N-1}(x_m)\hspace{3pt}\frac{1}{2}\bar{\nu}_{N-1},\\
\nonumber D(x_i,x_m)&\rightarrow \frac{e^{-(x_i^2+x_m^2)/2}}{r_{N/2-1}}R_{N-2}(x_i)\hspace{3pt}R_{N-1}(x_m),\\
\nonumber S(x_i,x_m)&\rightarrow \frac{e^{-x^2/2}}{r_{N/2-1}}\Phi_{N-2}(x_i)\hspace{3pt}R_{N-1}(x_m),\\
\nonumber S(x_m,x_i)&\rightarrow \sum_{k=0}^{N/2-1}\frac{e^{-x_i^2/2}}{r_k}\Bigl[R_{2k+1}(x_i)\hspace{3pt} \frac{1}{2}\bar{\nu}_{2k+1} - R_{2k}(x_i)\hspace{3pt} \frac{1}{2}\bar{\nu}_{2k+2}\Bigr],\\
\label{SDI_lims} \tilde{I}(x_i,x_m)&\rightarrow \sum_{k=0}^{N/2-1}\frac{1}{r_k}\Bigl[\Phi_{2k+1}(x_i)\hspace{3pt} \frac{1}{2}\bar{\nu}_{2k+1} - \Phi_{2k}(x_i)\hspace{3pt} \frac{1}{2}\bar{\nu}_{2k+2}\Bigr]+\frac{1}{2}.
\end{align}

\begin{remark}
Note that it may appear to the reader that an error has been made: the limiting forms of $S(x_m,x_i)$ and $\tilde{I}(x_i,x_m)$ contain terms with factors of $\bar{\nu}_{N}$ whilst the odd forms of the kernel elements on the right hand side of (\ref{eqn:DSI*}) (with $N\to N-1$) have only $\bar{\nu}_{N-1}$ terms and lower. This seems to indicate that the sums in (\ref{SDI_lims}) should be restricted to $k=0,...,N/2-2$. However, this is only an apparent problem since the six terms corresponding to $k=N/2-1$ in the limiting forms of $D^*,S^*$ and $\tilde{I}^*$ are conjoined in a conspiracy of cancellation, resolving the problem.
\end{remark}

\newpage

\section{Real asymmetric ensemble}
\setcounter{figure}{0}
\label{sec:GinOE}

In this chapter we modify the Gaussian orthogonal ensemble of Chapter \ref{sec:GOE_steps} by relaxing the symmetry constraint on the elements of the matrices. As discussed in the introduction, the resulting ensemble was first formulated by Ginibre in 1965 \cite{Gi65}, where he also considered non-Hermitian complex and non-self-dual real quaternion matrices. As with the GOE, GUE and GSE these real, complex and real quaternion Ginibre ensembles correspond to $\beta=1,2$ and $4$ respectively. Recall that these ensembles do not obey the same invariance under orthogonal, unitary and symplectic groups, although they are sometimes denoted GinOE, GinUE and GinSE by analogy. In keeping with the theme of this work, we will only be looking at the case $\beta=1$ of real, asymmetric Gaussian matrices.

The effective difference between this ensemble and those considered earlier by Dyson and Mehta is that the eigenvalues are no longer constrained to a one dimensional support since, in general, a real matrix may have both real and complex conjugate paired eigenvalues. Given that the complex eigenvalues always come in conjugate pairs, we see that the number of real eigenvalues $k$ must be of the same parity as the size $N$ of the matrix. While these facts may be unsurprising when one knows a little linear algebra, it is remarkable since for us it means the real line is populated with eigenvalues despite having measure zero inside the support of the set of all eigenvalues. We will find in Chapter \ref{sec:Ginkernelts} that the expected number of real eigenvalues is proportional to $\sqrt{N}$ \cite{eks1994}.

The existence of both real and non-real complex eigenvalues is particular to the $\beta=1$ (real) case of Ginibre's ensembles, and is a significant complication. Indeed in his original paper, Ginibre was able to find the eigenvalue distribution for $\beta=4$ and both the eigenvalue distribution and the correlation functions for $\beta=2$. However, for $\beta=1$ he was only able to calculate the distribution in the restricted case that all the eigenvalues are real --- the full jpdf was not calculated until 1991 in \cite{LS91} and the full correlation functions not until 2008 \cite{FN07, sommers2007, sommers_and_w2008, b&s2009, FM09, Sinc09}. A major source of trouble is that the Dyson integration theorem (Proposition \ref{thm:integral_identities}) does not hold for real, asymmetric matrices, as pointed out in \cite{AK2007}. The effect of this bipartite set of eigenvalues is that the partition function is now a sum over the individual partition functions for each $k$.

Further, as for the GOE, the odd case again presents more difficulties, however it turns out that we can overcome them in exactly the same way: we find that there is naturally an extra row and column bordering the odd sized Pfaffian in the generalised partition function. Interestingly, this extra row and column have a more natural interpretation in the present setting: they correspond to the one real eigenvalue that is required to exist in an odd-sized real matrix (recalling from above that the eigenvalues are real or one of a complex conjugate pair).

\subsection{Element distribution}

According to the procedure outlined in Chapter \ref{sec:GOE_steps}, our first task is to specify the matrix element distribution. In the case of the real Ginibre ensemble this is particularly simple: for a matrix $\bX=[x_{j,k}]_{j,k=1,...,N}$ each element is independently drawn from a standard Gaussian distribution,
\begin{align}
\label{def:Gin_eldist} \frac{1}{\sqrt{2\pi}}e^{-x_{j,k}^2/2},
\end{align}
meaning that the element jpdf is
\begin{eqnarray}
\label{eqn:GinOE_eldist} P(\bX)=(2\pi)^{-N^2/2}\prod_{j,k=1}^{N}e^{-x^2_{j,k}/2}=(2\pi)^{-N^2/2}e^{-(\mathrm{Tr}\bX\bX^{T})/2}.
\end{eqnarray}

\begin{figure}[htp]
\begin{center}
\includegraphics[scale=0.5]{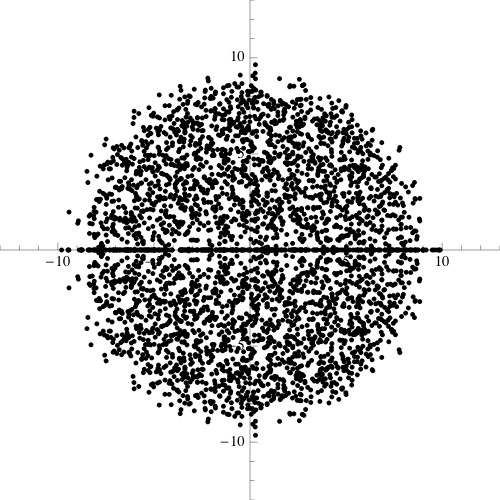}
\caption[Simulated real Ginibre eigenvalue plot.]{Plot of eigenvalues from 50 independent $75\times 75$ asymmetric real Gaussian matrices (3750 points in total). Note the density of eigenvalues along the real line and the reflective symmetry of the upper and lower half planes.}
\label{fig:GinOE_eval_plot}
\end{center}
\end{figure}

As mentioned above, the real Ginibre ensemble effectively contains two species of eigenvalues: real and non-real complex. To have a clear picture in our mind, we can produce a simulated eigenvalue plot; Figure \ref{fig:GinOE_eval_plot} clearly displays the finite probability of finding real eigenvalues, which were absent in the complex Ginibre ensemble of Figure \ref{fig:RMTDisk}.

\begin{figure}[htp]
\begin{center}
\subfloat[][$N=2$, $14206$ reals]{\includegraphics[scale=0.43]{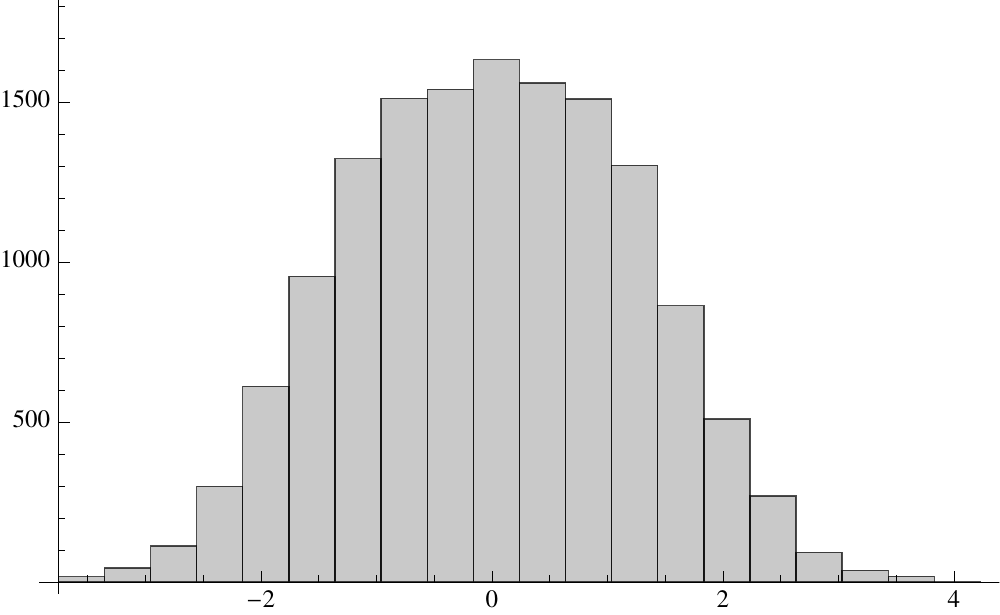}}
\;\subfloat[][$N=4$, $19490$ reals]{\includegraphics[scale=0.43]{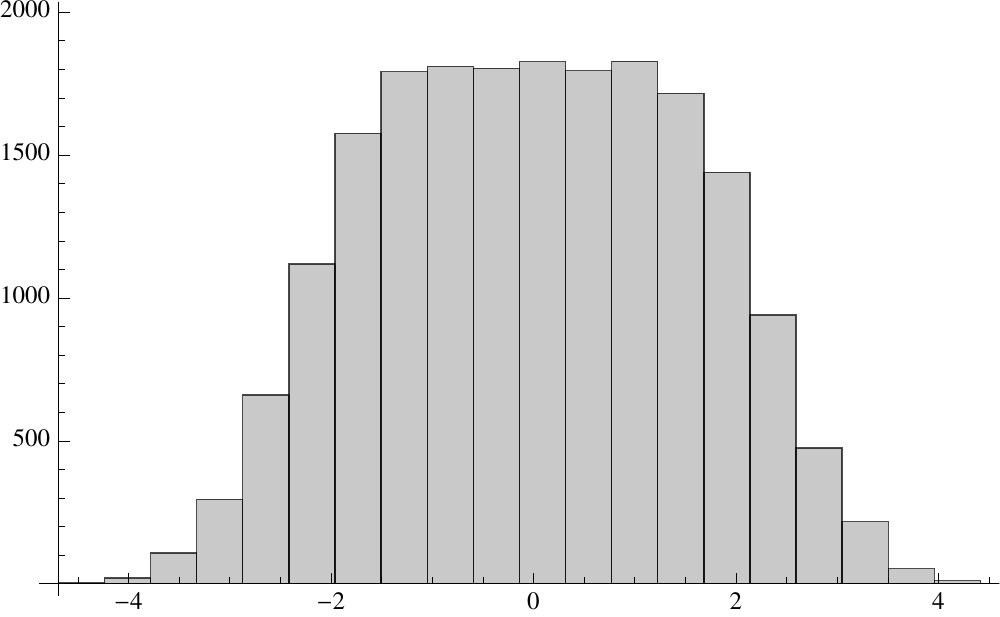}}
\;\subfloat[][$N=9$, $27960$ reals]{\includegraphics[scale=0.43]{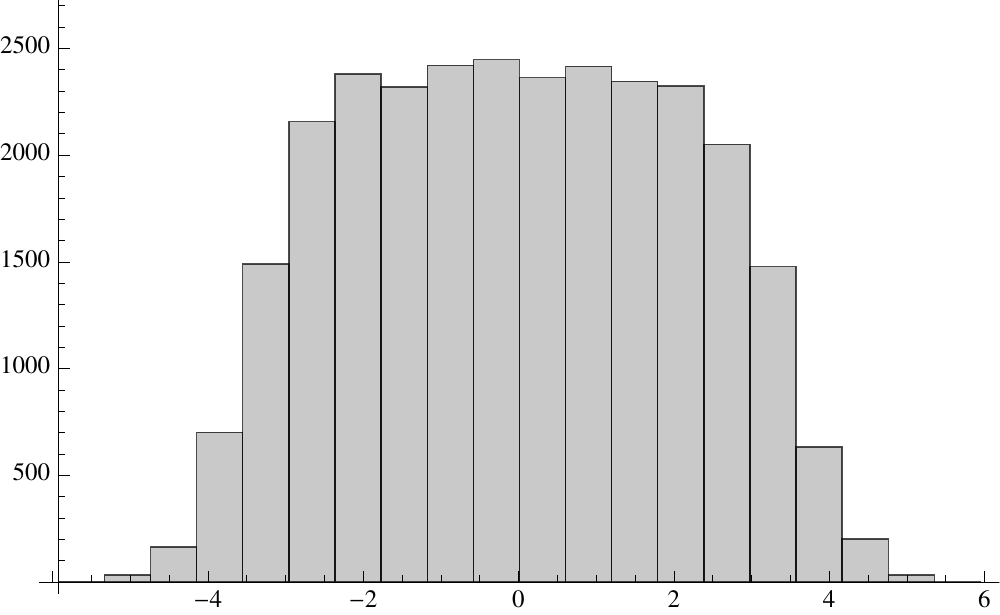}}

\subfloat[][$N=16$, $36104$ reals]{\includegraphics[scale=0.43]{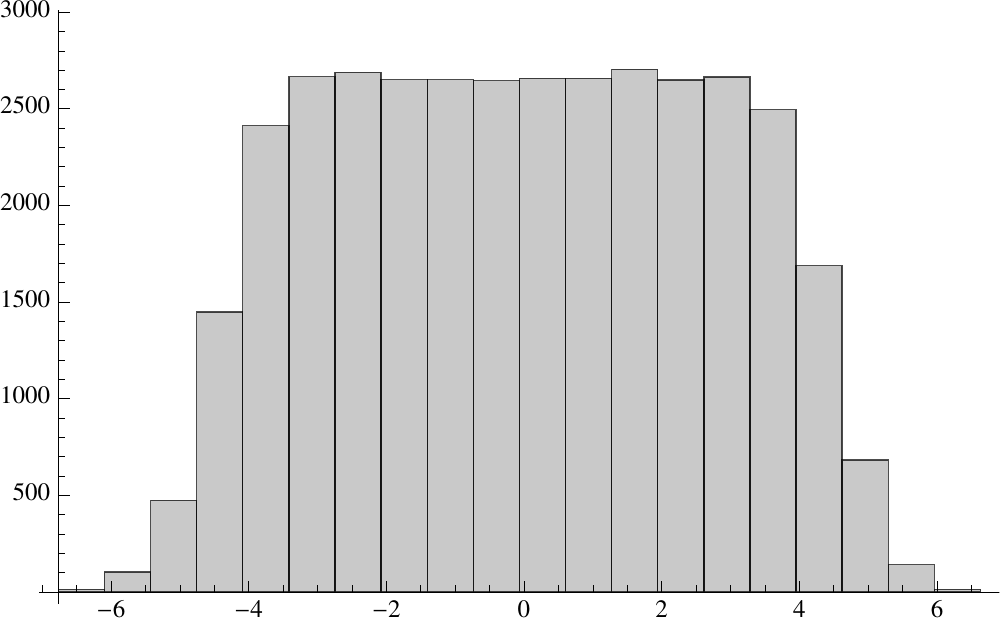}}
\;\subfloat[][$N=25$, $44300$ reals]{\includegraphics[scale=0.43]{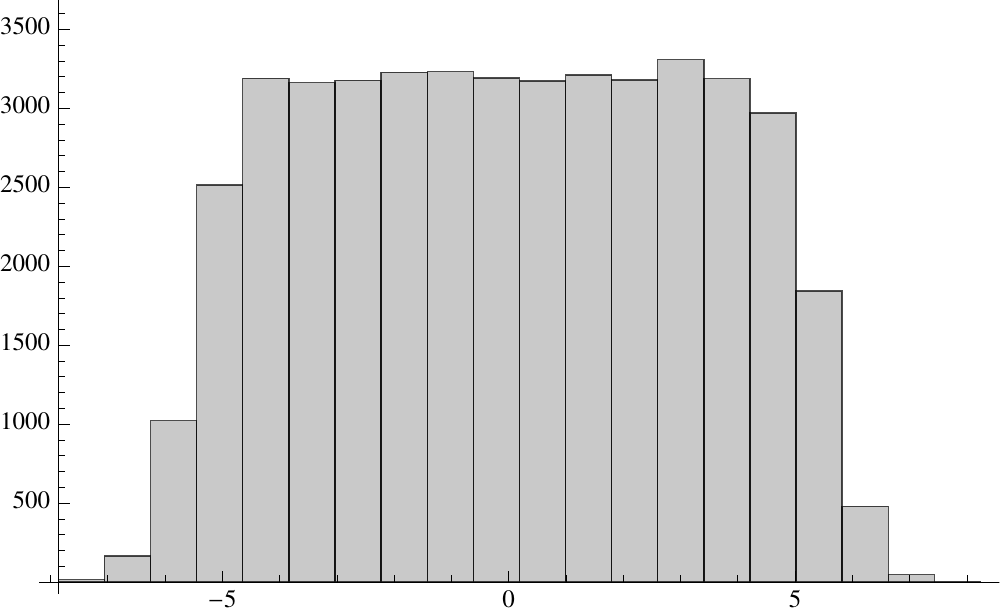}}
\;\subfloat[][$N=49$, $60774$ reals]{\includegraphics[scale=0.43]{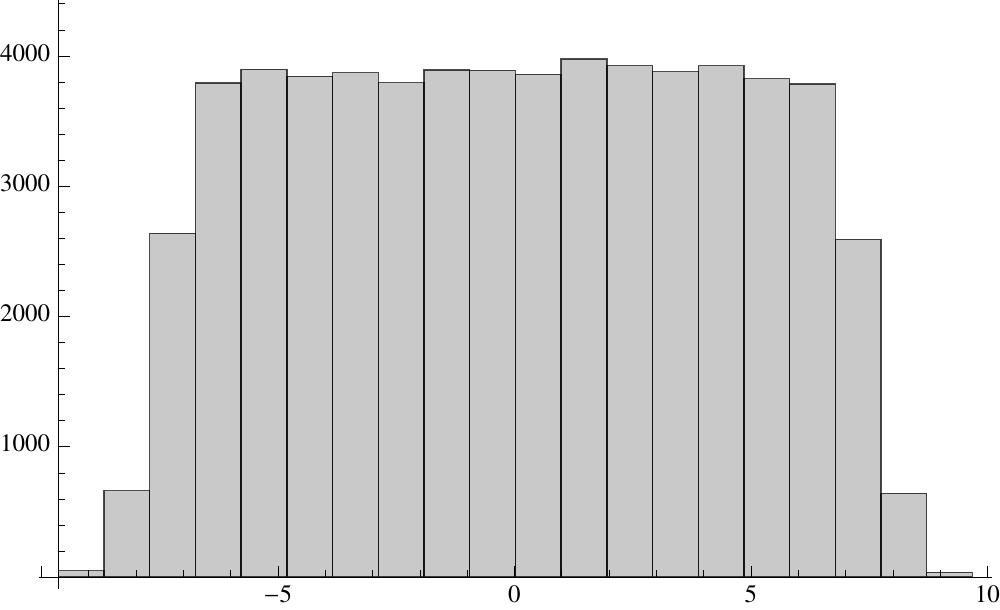}}

\subfloat[][$N=100$, $84418$ reals]{\includegraphics[scale=0.43]{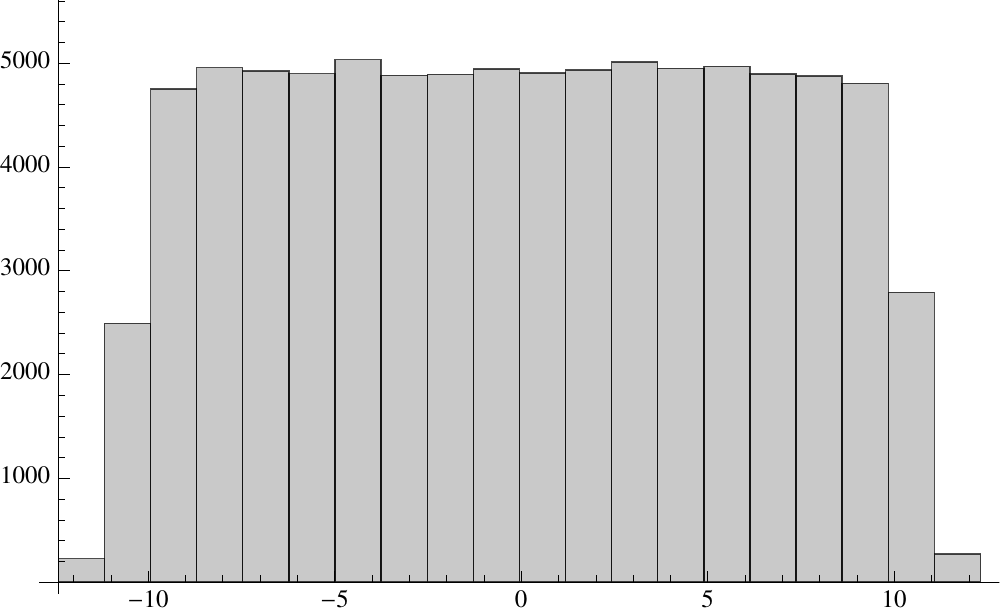}}
\caption[Simulated density of real eigenvalues in the real Ginibre ensemble.]{Real eigenvalue density for 10,000 instances of $N\times N$ real Ginibre matrices; the second value is the number of real eigenvalues in that simulation.}
\label{fig:GinOErdist}
\end{center}
\end{figure}

We can also generate plots analogous to Figure \ref{fig:GOE_eval_dens} for the GOE, of the density of these real eigenvalues for varying matrix dimension, see Figure \ref{fig:GinOErdist}.
Note that Figure \ref{fig:GinOErdist} also tells us the average number of real eigenvalues in these simulations for each matrix size; these compare favourably to the $\sqrt{N}$ behaviour of (\ref{eqn:bigNEN}), which is the expected number of these reals in the large $N$ limit \cite{eks1994}.

We will discuss the probability of finding real eigenvalues at length, so we make the following definition.
\begin{definition}
\label{def:probpnk}
Let $\bX$ be an $N\times N$ matrix. Define $p_{N,k}$ as the probability that there are $k$ real eigenvalues amongst the $N$ total eigenvalues of $\bX$.
\end{definition}

We can estimate the probability of finding some number of real eigenvalues in an $N\times N$ matrix from the real Ginibre ensemble by a Monte Carlo simulation, the results are contained in Table \ref{tab:GinOE_pnk}. Another feature to note is the apparently circular distribution of the eigenvalues in the complex plane. This is an illustration of the so-called \textit{Girko's Circular Law} which will be discussed in Chapter \ref{sec:circlaw}.

\begin{longtable}{|c|c||c|c||c|c|}
\hline
 & Simulated $p_{N,k}$ & & Simulated $p_{N,k}$ & & Simulated $p_{N,k}$\\
\hline
\hline
\endfirsthead

\hline
 & Simulated $p_{N,k}$ & & Simulated $p_{N,k}$ & & Simulated $p_{N,k}$\\
\hline
\hline
\endhead

$p_{2,2}$ & $0.70762$ & $p_{3,3}$ & $0.35349$&$p_{38,38}$&$0.00000$\\
$p_{2,0}$ & $0.29238$ & $p_{3,1}$ & $0.64651$&$p_{38,36}$&$0.00000$\\
\hline
$p_{4,4}$ & $0.12440$ & $p_{5,5}$ & $0.03116$&$p_{38,34}$ &$0.00000$\\
$p_{4,2}$ & $0.72249$ & $p_{5,3}$ & $0.51300$&$p_{38,32}$ &$0.00000$\\
$p_{4,0}$ & $0.15311$ & $p_{5,1}$ & $0.45584$&$p_{38,30}$ &$0.00000$\\
\cline{1-4}
$p_{6,6}$ & $0.00506$ & $p_{7,7}$ & $0.00073$&$p_{38,28}$ &$0.00000$\\
$p_{6,4}$ & $0.25231$ & $p_{7,5}$ & $0.08343$&$p_{38,26}$ &$0.00000$\\
$p_{6,2}$ & $0.64888$ & $p_{7,3}$ & $0.57908$&$p_{38,24}$ &$0.00000$\\
$p_{6,0}$ & $0.09375$ & $p_{7,1}$ & $0.33676$&$p_{38,22}$ &$0.00000$\\
\cline{1-4}
$p_{8,8}$ & $0.00006$ & $p_{9,9}$ & $0.00000$&$p_{38,20}$ &$0.00000$\\
$p_{8,6}$ & $0.02052$ & $p_{9,7}$ & $0.00359$&$p_{38,18}$ &$0.00000$\\
$p_{8,4}$ & $0.34469$ & $p_{9,5}$ & $0.14479$&$p_{38,16}$ &$0.00000$\\
$p_{8,2}$ & $0.57257$ & $p_{9,3}$ & $0.59056$&$p_{38,14}$ &$0.00001$\\
$p_{8,0}$ & $0.06216$ & $p_{9,1}$ & $0.26106$&$p_{38,12}$ &$0.00075$\\
\cline{1-4}
$p_{10,10}$ & $0.00000$ & $p_{11,11}$ & $0.00000$&$p_{38,10}$ &$0.01751$\\
$p_{10,8}$ & $0.00041$ & $p_{11,9}$ & $0.00001$&$p_{38,8}$ &$0.14862$\\
$p_{10,6}$ & $0.04477$ & $p_{11,7}$ & $0.01006$&$p_{38,6}$ &$0.4093$\\
$p_{10,4}$ & $0.41561$ & $p_{11,5}$ & $0.20768$&$p_{38,4}$ &$0.34735$\\
$p_{10,2}$ & $0.49503$ & $p_{11,3}$ & $0.58174$&$p_{38,2}$ &$0.07475$\\
$p_{10,0}$ & $0.04418$ & $p_{11,1}$ & $0.20051$&$p_{38,0}$ &$0.00171$\\
\hline
\caption[Simulated eigenvalue probabilities $p_{N,k}$ for the real Ginibre ensemble.]{Experimentally determined probabilities $p_{N,k}$ from simulations of 100,000 independent real Ginibre matrices.}
\label{tab:GinOE_pnk}
\end{longtable}

\subsection{Eigenvalue jpdf}
\label{sec:Gejpdf}

As in Chapter \ref{sec:GOE_eval_jpdf} one would like to diagonalise the matrix $\bX$ so that the $N^2-N$ degrees of freedom corresponding to the eigenvectors can be integrated over, leaving only the dependence on the eigenvalues. Since real symmetric matrices are diagonalised by orthogonal matrices, a fact we used in (\ref{eqn:diag_decomp}), integration over the elements of the diagonalising matrices (which amounts to computation of the volume of $O(N)$, the group of orthogonal $N\times N$ matrices) was a readily computable problem; this is the content of Proposition \ref{prop:int_(RdR)}. In \cite{Gi65} this was the approach of Ginibre as he attempted to calculate the eigenvalue distributions and correlation functions by diagonalising these non-Hermitian matrices, however he was unable to proceed very far in this direction for the $\beta=1$ (real asymmetric) matrices. The set of eigenvectors do not form an orthonormal basis, and so these matrices are not orthogonally diagonalisable as real symmetric matrices are. This means that the integral over the diagonalising matrices is not given by (\ref{eqn:RTdR_integ}), and in fact appears intractable.

Progress was made in 1991 \cite{LS91} and, independently, in \cite{Ed97}, where diagonal decomposition was abandoned in favour of upper triangular decomposition; in particular, Schur decomposition.
\begin{remark}
The decomposition used in \cite{LS91} was not that of Schur, however the triangular matrix they used was equivalent to that in (\ref{def:triangular_mat}) through elementary row and column operations.
\end{remark}
To express a matrix in this form, first note that if a general $N\times N$ real matrix $\bA$ has $k$ real eigenvalues then it has $(N-k)/2$ complex conjugate pairs of eigenvalues. The Schur decomposition of $\bA$ is then \cite{Schur1909}
\begin{align}
\label{eqn:GinOE_decomp} \bA=\bQ\bR\bQ^T,
\end{align}
where $\bQ$ is orthogonal and $\bR$ is the block upper triangular matrix
\begin{align}
\label{def:triangular_mat} \bR&= \left[\begin{array}{cccccc}
\lambda_1 & ... & R_{1,k} & R_{1,k+1} & ... & R_{1,m}\\
 & \ddots & \vdots & \vdots &  & \vdots \\
 &  & \lambda_k & R_{k,k+1} & ... & R_{k,m}\\
 &  &  & z_{k+1} & ... & R_{k+1,m}\\
 & 0 &  &  & \ddots & \vdots\\
 &  &  &  &  & z_m\\
\end{array}\right],&m=(N+k)/2,
\end{align}
where, on the diagonal, we have the real eigenvalues $\lambda_j$ and the $2\times2$ blocks
\begin{align}
\label{eqn:Schurz} z_j &=\left[\begin{array}{cc}
x_j & -c_j\\
b_j & x_j
\end{array}\right], \mathrm{where}\; b_j,c_j>0,
\end{align}corresponding to the complex eigenvalues $x_j \pm iy_j$, $y_j=\sqrt{b_jc_j}$. (This correspondence is clear since the eigenvalues of $z_j$ are $x_j\pm iy_j$.) Note that the dimension of $R_{i,j}$ depends on its position in $\bR$:
\begin{itemize}
\item{$1\times 1$ for $i,j\leq k$,}
\item{$1\times 2$ for $i\leq k, j>k$,}
\item{$2\times 2$ for $i,j>k$.}
\end{itemize}
(Proofs (in English) of the Schur decomposition are available in several places, for example see \cite[Theorem 7.1.3]{GoVl1996}.)

This decomposition is not yet unique for two reasons; first, we know that, in general, eigenvectors are unique only up to normalisation and direction. Since $\bQ$ corresponds to the eigenvectors and it is orthogonal, the normalisation constraint is already imposed. To fix the sign, we specify that the first row of $\bQ$ must be positive. Second, it is clear that the eigenvalues are at present arbitrarily ordered along the diagonal of $\bR$, so we choose the ordering
\begin{eqnarray}\label{9'}
\lambda_1 < \cdot\cdot\cdot < \lambda_k  \qquad {\rm and}
\qquad x_{k+1}<\cdot\cdot\cdot<x_m
\end{eqnarray}
and (\ref{eqn:GinOE_decomp}) is now a 1-1 correspondence.

\begin{remark}
\label{rem:Ginsing}
Note that, as we did for the GOE, we are ignore singular matrices and matrices with repeated eigenvalues since the set of these matrices has measure zero in the real Ginibre ensemble.
\end{remark}

The benefit of using the Schur decomposition in place of diagonalisation is that the integral over the conjugating matrices is given by (\ref{eqn:RTdR_integ}). Of course, the strictly upper triangular components of $\bR$ must now be integrated over (which will leave us with just the $1\times 1$ and $2\times 2$ diagonal blocks corresponding to the eigenvalues), but we shall see that the dependence on these variables factorises. Note that in Chapter \ref{sec:SOE} (for the real spherical ensemble), where we also use Schur decomposition, the situation is significantly more complicated as the integral over these upper triangular elements no longer factorises.

We will next calculate the Jacobian for the change of variables from the elements of the matrix $\bA$ to the eigenvalues of $\bA$ and find that as in Proposition \ref{prop:GOE_J} (the symmetric analogue) we will have a product of differences of eigenvalues. However, here the structure is slightly more complicated due to the existence of both real and non-real complex eigenvalues. In an attempt to arrest confusion, we first define the notation for this product.
\begin{definition}
Let $k$ be a positive integer. Then with $t_j=\lambda_j\in\mathbb{R}$ for $j\leq k$ and $t_j=w_j\in\mathbb{C}\backslash\mathbb{R}$ for $j>k$ define
\begin{align}
\label{def:GinOEpods} |\lambda(t_j)-\lambda(t_i)|:=\left\{\begin{array}{ll}
|\lambda_j-\lambda_i|, & \mathrm{for}\; k\geq j>i,\\
|w_j-\lambda_i||\bar{w}_j-\lambda_i|, & \mathrm{for}\; j>k\geq i,\\
|w_j-w_i||\bar{w}_j-w_i| &\\
\times |w_j-\bar{w}_i||\bar{w}_j-\bar{w}_i|, & \mathrm{for}\; j>i>k.
\end{array}\right.
\end{align}
\end{definition}
We can now state the asymmetric analogue of Proposition \ref{prop:GOE_J}.
\begin{proposition}[\cite{Ed97} Theorem 5.1]
\label{prop:GinJ}
Decomposing a real matrix $\bA$ according to the Schur decomposition (\ref{eqn:GinOE_decomp}) we have
\begin{align}
\nonumber (d\bA)&=2^{(N-k)/2}\prod_{j<p}|\lambda(R_{pp})-\lambda(R_{jj})| (d\tilde{\bR})(\bQ^Td\bQ)\\
\label{eqn:GinOE_J} &\times \prod_{j=1}^kd\lambda_j \prod_{l=k+1}^{(N+k)/2}|b_l-c_l |\;dx_ldb_ldc_l,
\end{align}
where $\tilde{\bR}$ is the strictly upper triangular part of (\ref{def:triangular_mat}) and $\bQ$ is an orthogonal matrix with the first row specified to be positive.
\end{proposition}

\textit{Proof}: In analogue with (\ref{eqn:RTdXR}) we begin with the decomposition (\ref{eqn:GinOE_decomp}) and apply the product rule of differentiation to find
\begin{align}
\nonumber \bQ^Td\bA\bQ=d\bR+\bQ^Td\bQ\bR-\bR\bQ^Td\bQ.
\end{align}
Now let $d\bO:=\bQ^Td\bQ$ and we see that the $ij$-th element of $\bQ^Td\bA\bQ$ is
\begin{align}
\nonumber dR_{ij}+dO_{ij}R_{jj}-R_{ii}dO_{ij}+\sum_{l<j}dO_{il}R_{lj}-\sum_{l>i}R_{il}dO_{lj},
\end{align}
or, specialising to particular cases,
\begin{align}
\nonumber &dO_{ij}R_{jj}-R_{ii}dO_{ij}+\sum_{l<j}dO_{il}R_{lj}-\sum_{l>i}R_{il}dO_{lj},&\mbox{ for } i>j,\\
\label{eqn:GinOE_wedges} &dR_{ii}+\sum_{l<i}dO_{il}R_{li}-\sum_{l>i}R_{il}dO_{li},&\mbox{ for } i=j,\\
\nonumber &dR_{ij}+dO_{ij}R_{jj}-R_{ii}dO_{ij}+\sum_{l<j}dO_{il}R_{lj}-\sum_{l>i}R_{il}dO_{lj},&\mbox{ for } i<j.
\end{align}
Noting that $dO_{ij}=-dO_{ji}$ (which is a consequence of the fact $\bQ\bQ^T=\1$) we find that taking the wedge product of the elements of $\bQ^Td\bA\bQ$ in the order $j=1,i=N,N-1,...,1$ then $j=2,i=N,N-1,...,1$ \textit{et cetera}, up to $j=N, i=N,N-1,...,1$ each of the differentials in the summations over $l$ in (\ref{eqn:GinOE_wedges}) have already been wedged and so, for the purposes of the wedge product, they can be ignored. Let $d\tilde{\bO}$ be the matrix $d\bO$ excluding the $2\times 2$ blocks $dO_{ii}$ along the diagonal, which have the form
\begin{align}
\nonumber \left[\begin{array}{cc}
0 & do_i\\
-do_i & 0
\end{array}\right],
\end{align}
for some $do_i$. Then the wedge product of the off-diagonal elements is
\begin{align}
\label{eqn:GinOEpods2} \prod_{j<p}|\lambda(R_{pp})-\lambda(R_{jj})|(d\tilde{\bO})(d\tilde{\bR}),
\end{align}
using the notation defined in (\ref{def:GinOEpods}). More explicitly, the $(d\tilde{\bR})$ factor in (\ref{eqn:GinOEpods2}) comes from the $dR_{ij}$ terms in (\ref{eqn:GinOE_wedges}) while the remaining factors are given by the terms $dO_{ij}R_{jj}-R_{ii}dO_{ij}$, each of which contributes
\begin{align}
\nonumber \begin{array}{rl}
|\lambda_j-\lambda_i| & \mathrm{for}\; k\geq j>i,\\
(x_j^2-\lambda_i^2)^2 +y_j^2 & \mathrm{for}\; j>k\geq i,\\
\left( (x_j-x_i)^2+(y_j-y_i)^2\right)\cdot \left( (x_j-x_i)^2+(y_j+y_i)^2\right)& \mathrm{for}\; j>i>k,
\end{array}
\end{align}
which we see is the same as (\ref{def:GinOEpods}) with $w_l=x_l+iy_l$. (We can also establish (\ref{eqn:GinOEpods2}) directly using \cite[Lemma 5.1]{Ed97}.) For $i=j\leq k$ we pick up just $d\lambda_i$, while for $i=j>k$, the middle row in (\ref{eqn:GinOE_wedges}) becomes
\begin{align}
\nonumber \left[\begin{array}{cc}
dx_i+(b_i-c_i)do_i & db_i\\
-dc_i& dx_i+(c_i-b_i)do_i
\end{array}\right]
\end{align}
and the wedge product of these elements is
\begin{align}
\nonumber 2^{(N-k)/2}\prod_{l=k+1}^{(N+k)/2}|b_l-c_l|do_ldx_ldb_ldc_l.
\end{align}
So then collecting the $do_i$ together with $(d\tilde{\bO})$ gives $(d\bO)$ and we have the result.

\hfill $\Box$

\begin{remark}
In the case that $k=N$ we note that (\ref{eqn:GinOE_J}) reduces (as expected) to (\ref{eqn:dX_jacob}), since $\tilde{\bR}$ is then the zero matrix.
\end{remark}

\begin{proposition}[\cite{LS91,Ed97}]
\label{prop:GinOE_eval_jpdf}
Let $\bA=[a_{ij}]$ be an $N\times N$ real matrix with standard Gaussian entries. Then if $\bA$ has sets of real and complex eigenvalues, $\Lambda=\{ \lambda_i \}_{i=1,...,k}$ and $W=\{ w_i,\bar{w}_i \}_{i=1,...,(N-k)/2}$ respectively, the eigenvalue jpdf is
\begin{equation}
\label{eqn:GinOEjpdf} Q_{N,k}(\Lambda,W) =C_{N,k} \prod_{i=1}^{k} e^{-\lambda_i^2/2} \prod_{j=1}^{(N-k)/2} e^{-(w_j^2+\bar{w}_j^2)/2}\:\mathrm{erfc}(\sqrt{2}| \mathrm{Im}(w_j)|)\: \big| \Delta (\Lambda\cup W)\big|,
\end{equation}
where $\Delta(\{x_i\}):=\prod_{i<j} (x_j-x_i)$ and
\begin{align}
\label{def:GinOECNK} C_{N,k}:=\frac{2^{-N(N-1)/4-k/2}}{k!((N-k)/2)!\prod_{l=1}^N\Gamma(l/2)}.
\end{align}
\end{proposition}

\textit{Proof}: We begin with the matrix element distribution (\ref{eqn:GinOE_eldist}). 
Let $\delta_j:=b_j-c_j$ then, using the Schur decomposition (\ref{eqn:GinOE_decomp}), we see that
\begin{align}
\label{eqn:GinOEexp_decomp} e^{-\mathrm{Tr}(\bA\bA^T)/2}=e^{-\sum_{i<j}r_{ij}^2/2} e^{-\sum_{j=1}^k\lambda_{j}^2/2} e^{-\sum_{j=1}^{(N-k)/2}x_j^2+y_j^2+\delta_j^2/2},
\end{align}
where $[r_{ij}]:=\tilde{\bR}$ are the strictly upper triangular elements of $\bR$. We can change variables from $b,c$ to $y,\delta$ with the equation
\begin{align}
\label{eqn:GinOE_covs} db_ldc_l=\frac{2y_l}{\sqrt{\delta_l^2+4y_l^2}}dy_l d\delta_l,
\end{align}
where $-\infty<\delta<\infty$ (and not $0<\delta<\infty$ as claimed in \cite[Lemma 5.2]{Ed97}). Taking the product of (\ref{eqn:GinOEexp_decomp}) and (\ref{eqn:GinOE_J}), using (\ref{eqn:GinOE_covs}), yields
\begin{align}
\nonumber &e^{-\mathrm{Tr}(\bA\bA^T)/2}(d\bA)=2^{(N-k)}e^{-\sum_{i<j}r_{ij}^2/2} e^{-\sum_{j=1}^k\lambda_{j}^2/2} e^{-\sum_{j=1}^{(N-k)/2}x_j^2+y_j^2+\delta_j^2/2}\\
\label{eqn:GinOEjpdf2} &\times\prod_{j<p}\big|\lambda(R_{pp})-\lambda(R_{jj})\big| (d\tilde{\bR})(\bQ^Td\bQ)\prod_{j=1}^kd\lambda_j \prod_{l=k+1}^{(N+k)/2}\frac{2\:|\delta_l |\; y_l}{\sqrt{\delta_l^2+4y_l^2}}dx_ldy_ld\delta_l.
\end{align}
Combining $\prod_{l=k+1}^{(N+k)/2}2y_l$ with the product of differences in (\ref{eqn:GinOEjpdf2}) gives $|\Delta (\Lambda \cup W)|$.

As we would like the end result expressed in terms of the eigenvalue variables $\lambda_l,x_l$ and $y_l$, we plan to integrate over the variables $\delta_l$. Since the variables $\delta_l$ and $y_l$ are coupled, this gives a function of $y_l$, which can be explicitly determined according to
\begin{align}
\nonumber \int_{\delta=-\infty}^{\delta=\infty}\frac{|\delta|\;e^{-\delta^2/2}}{\sqrt{\delta^2+4y^2}}\;d\delta&= 2\int_{\delta=0}^{\infty} \frac{\delta\: e^{-\delta^2/2}} {\sqrt{\delta^2+4y^2}}\;d\delta\\
\label{eqn:GinOE_erfc} &=\sqrt{2\pi}\;e^{2y^2}\mathrm{erfc}(\sqrt{2}\;|y|),
\end{align}
where the second equality in (\ref{eqn:GinOE_erfc}) is given in \cite[3.362.2]{GraRyz2000} after a change of variables (although in Edelman this evaluation included an erroneous factor of $2$, which cancelled the factor of $1/2$ introduced in relation to (\ref{eqn:GinOE_covs})). This result can be verified by checking that both sides agree at $y=0$ and that they have identical derivatives.

We need to integrate out the unwanted $N^2-N$ independent elements contained in $\bQ$ and $\tilde{\mathbf{R}}$: we have the integral over the orthogonal matrices from Proposition \ref{prop:int_(RdR)}, and the integrals over the $r_{ij}$ are simple Gaussians. Lastly, the factorials in the denominator of $C_{N,k}$ come from relaxing the ordering on the $k$ real eigenvalues and $(N-k)/2$ non-real complex conjugate pairs of eigenvalues.

\hfill $\Box$

\begin{remark}
\label{rem:Rtilde}
The integrals over $\tilde{\mathbf{R}}$ were quite straightforward in the proof of Proposition \ref{prop:GinOE_eval_jpdf} since each of them was a standard Gaussian. If this is not the case, however, then this calculation can be a significant technical hurdle. We will return to this point when we calculate the eigenvalue jpdf for the real, spherical ensemble in Chapter \ref{sec:Sevaldist}, where a more involved technique must be employed.
\end{remark}

In the following section we will find an expression for the probabilities of obtaining any number of real eigenvalues, however in the restricted case that $k=N$ (that is, we obtain all real eigenvalues) we can directly integrate the eigenvalue jpdf (\ref{eqn:GinOEjpdf}) by using the Selberg integral
\begin{align}
\label{def:Selb} S_N(\lambda_1,\lambda_2,\lambda):=\int_0^1 dt_1\cdot\cdot\cdot \int_0^1 dt_N \prod_{l=1}^N t_l^{\lambda_1}(1-t_l)^{\lambda_2}\prod_{1\leq j< l \leq N}|t_l-t_j|^{2\lambda},
\end{align}
which can be evaluated as the product of gamma functions \cite{Selb1944}
\begin{align}
\label{eqn:Selbint} S_N(\lambda_1,\lambda_2,\lambda)=\prod_{j=0}^{N-1}\frac{\Gamma(\lambda_1 + 1 +j\lambda) \Gamma(\lambda_2 +1 +j\lambda) \Gamma(1+\lambda(j+1))}{\Gamma(\lambda_1+\lambda_2+2 +\lambda(N+j-1)) \Gamma(1+\lambda)}.
\end{align}
With $k=N$ we have
\begin{align}
\label{eqn:pNN1} p_{N,N}=C_{N,N} \int_{-\infty}^{\infty} d\lambda_1 \cdot\cdot\cdot \int_{-\infty}^{\infty} d\lambda_N \prod_{i=1}^N e^{-\lambda_i^2/2} \prod_{j<l} |\lambda_l -\lambda_j|.
\end{align}
This multiple integral is known as a \textit{Mehta integral} \cite{mehta1991}, and by using (\ref{eqn:Selbint}), we find it has evaluation \cite[Chapter 4.7]{forrester?}
\begin{align}
\label{eqn:meEqn} 2^{3N/2}\prod_{j=1}^N \Gamma(j/2+1).
\end{align}
Substitution of (\ref{eqn:meEqn}) into (\ref{eqn:pNN1}) gives \cite[Corollary 7.1]{Ed97}
\begin{align}
\label{eqn:GinOEpNN} p_{N,N}=2^{-N(N-1)/4},
\end{align}
where we have used the Gamma function identity
\begin{align}
\nonumber \Gamma (n) \Gamma (n+1/2) = \frac{\sqrt{\pi} \: \Gamma (2n)} {2^{2n-1}}.
\end{align}

\subsection{Generalised partition function}
\label{sec:Ggpf}

\subsubsection{$N$ even}
\label{sec:Ggpfe}

With the eigenvalue distribution firmly in hand, we proceed to express the generalised partition function as a quaternion determinant or Pfaffian. As the anamnestic reader will recall, for the GOE we substituted the eigenvalue jpdf (\ref{eqn:GOE_eval_jpdf}) into (\ref{def:single_gen_part_fn}), the definition of $Z_N[u]$, and then applied the method of integration over alternate variables. However, as alluded to below Definition \ref{def:gen_part_fn}, the situation for real, asymmetric matrices is complicated by the existence of two species of eigenvalues, which means we must take (\ref{def:multi_gen_part_fn}), with $m=2$, as our definition of the generalised partition function. We then have
{\small
\begin{align}
\nonumber &Z_{k,(N-k)/2}[u,v]=C_{N,k}\int_{-\infty}^{\infty}d\lambda_1\cdot\cdot\cdot \int_{\infty}^{\infty}d\lambda_k \int_{\mathbb{R}^2_+}dw_1\cdot\cdot\cdot\int_{\mathbb{R}^2_+}dw_{(N-k)/2}\\
\label{eqn:GinOE_znk1} & \times\prod_{i=1}^ku(\lambda_i)\; e^{-\lambda_i^2/2}\prod_{j=1}^{(N-k)/2}v(w_j)\; e^{-(w_j^2+\bar{w}_j^2)/2}\mathrm{erfc}(\sqrt{2}|\mathrm{Im}(w_j)|) \; |\Delta (\Lambda\cup W)|.
\end{align}
}It will turn out that, as with the GOE in Chapter \ref{sec:pf_gpf}, the generalised partition function is dependent on the parity of $N$. Note that (\ref{eqn:GinOE_znk1}) is independent of the parity of $N$, but, as we did for the GOE, we proceed under the assumption that $N$ is even, postponing the odd case until Chapter \ref{sec:GinOE_genpartfn_odd}.

Since (\ref{eqn:GinOE_znk1}) is an integral over all the variables for a fixed number $k$ of real eigenvalues, we see that
\begin{align}
\label{eqn:Ginpnkzkn-k}
p_{N,k}=Z_{k,(N-k)/2}[1,1],
\end{align}
where $p_{N,k}$, from Definition \ref{def:probpnk}, is the probability of finding $k$ real eigenvalues from an $N\times N$ real Ginibre matrix. A generating function for these probabilities is then
\begin{align}
\label{def:GinOE_probsGF} Z_N(\zeta):=\sum_{k=0}^{N/2}\zeta^k Z_{2k,(N-2k)/2}[1,1].
\end{align}
But, of course, confining the correlations to a particular $k$ number of real eigenvalues is not in keeping with the realities of the problem, in which the number of real eigenvalues is not known \textit{a priori}. Consequently, we must introduce the summed-up generalised partition function
\begin{equation}
\label{eqn:summedup} Z_N[u,v]=\sum_{k=0}^N{}^{\sharp}\; Z_{k,(N-k)/2}[u,v],
\end{equation}
where $\sharp$ indicates that the sum is restricted to values of $k$ with the same parity as $N$ (in this case even). Our plan is to first find a quaternion determinant/Pfaffian form of (\ref{eqn:GinOE_znk1}), at which point the sum in (\ref{eqn:summedup}) is able to be performed quite simply.

Now we undertake the integration over alternate variables, with one further caveat: the ordering is no longer as simple as it was for GOE. However, the reader will see that this is just a small technical consideration and the procedure is not changed in any substantial way.

\begin{proposition}[\cite{sinclair2006, FN07}]
\label{prop:GinOE_gpf_even}
The generalised partition function $Z_{k,(N-k)/2}[u,v]$, for $k$ and $N$ even, can be written in the Pfaffian form
\begin{eqnarray}
\label{eqn:GinOE_gpf_even} Z_{k,(N-k)/2}[u,v]=\frac{2^{-N(N+1)/4}}{\prod_{l=1}^N\Gamma(l/2)}[\zeta^{k/2}]\Pf[\zeta\alpha_{j,l}+\beta_{j,l}],
\end{eqnarray}
where $[\zeta^n]$ means the coefficient of $\zeta^n$ and, with monic polynomials $\{ p_{i}(x)\}$ of degree i,
\begin{align}
\label{eqn:GinOE_alphabeta} \begin{split} \alpha_{j,l} & = \int_{-\infty}^{\infty}dx\; u(x)\; \int_{-\infty}^{\infty} dy\hspace{3pt}u(y)\; e^{-(x^2+y^2)/2} p_{j-1}(x)p_{l-1}(y)\hspace{3pt}\mathrm{sgn}(y-x),\\
\beta_{j,l} & = 2i\int_{\mathbb{R}_+^2}dw\;v(w)\; \mathrm{erfc}(\sqrt{2}|\mathrm{Im}(w)|)\; e^{-(w^2+\bar{w}^2)/2}\\
&\times \Bigl(p_{j-1}(w)p_{l-1}(\bar{w})-p_{l-1}(w)p_{j-1}(\bar{w})\Bigr).
\end{split}
\end{align}
\end{proposition}

\textit{Proof:} To remove the absolute value sign from the Vandermonde in (\ref{eqn:GinOE_znk1}) we start by ordering the real eigenvalues as $\lambda_1 < ... < \lambda_k$, picking up a factor of $k!$. Recall that $|\Delta (\Lambda\cup W)|= \prod_{1\leq j < l \leq N}|t_l- t_j|$, where $t_j\in \mathbb{R}$ for $j\leq k$, and $t_j\in \mathbb{C}\backslash \mathbb{R}$ for $j>k$. So with $l>k$, as long as $t_j\neq \bar{t}_l$, we have both $|t_l- t_j|$ and its complex conjugate in the Vandermonde product (recall the structure of (\ref{def:GinOEpods})) and so we can remove the absolute value. For the factors where $t_j= \bar{t}_l$ (of which there are $(N-k)/2$), then we can remove the absolute value as long as we multiply by $i$. Using (\ref{eqn:vandermonde_polys}) we have the Vandermonde determinant
\begin{align}
\label{eqn:vandermonde_GinOE} \Delta(\Lambda\cup W)=\mathrm{det}\left[ \begin{array}{c}
[p_{l-1}(\lambda_j)]_{j=1,...,k}\vspace{3pt}\\
\left[ \begin{array}{c}
p_{l-1}(w_j)\\
p_{l-1}(\bar{w}_j)\end{array} \right]_{j=1,...,(N-k)/2}
\end{array} \right]_{l=1,...,N},
\end{align}
which we substitute into (\ref{eqn:GinOE_znk1}) giving
{\small
\begin{align}
\nonumber &Z_{k,(N-k)/2}[u,v]=C_{N,k}i^{(N-k)/2}\frac{k!}{(k/2)!}\int_{\mathbb{R}^2_+}dw_1 \cdot\cdot\cdot \int_{\mathbb{R}^2_+} dw_{(N-k)/2}\\
\nonumber &\times\prod_{j=1}^{(N-k)/2} v(w_l)\; e^{-(w_j^2+\bar{w}_j^2)/2}\; \mathrm{erfc}(\sqrt{2}|\mathrm{Im}(w_j)|)\int_{-\infty}^{\infty}d\lambda_2 \int_{-\infty}^{\infty}d\lambda_4 \cdot\cdot\cdot \int_{\infty}^{\infty} d\lambda_k\\
\label{eqn:GinOE_ioav_even} &\times  \det\left[ \begin{array}{c}
\left[ \begin{array}{c}
\int_{-\infty}^{\lambda_{2j}}e^{-x^2/2}u(x)p_{l-1}(x)dx\\
e^{-\lambda_{2j}^2/2}u(\lambda_{2j})p_{l-1}(\lambda_{2j})
\end{array}\right]_{j=1,...,k}\\
\left[ \begin{array}{c}
p_{l-1}(w_j)\\
p_{l-1}(\bar{w}_j)\end{array} \right]_{j=1,...,(N-k)/2}
\end{array} \right]_{l=1,...,N},
\end{align}
}where, as in Proposition \ref{prop:GOE_gen_part_fn}, we have again added appropriate rows to make all the integrals inside the determinant begin at $-\infty$. The $(k/2)!$ in the denominator comes from relaxing the ordering on the real eigenvalues $\lambda_2,\lambda_4,...,\lambda_k$.

Expanding the determinant we have
\begin{align}
\nonumber Z_{k,(N-k)/2}[u,v]=C_{N,k}\frac{k!}{(k/2)!}\sum_{P\in S_N}\varepsilon(P) \prod_{l=1}^{k/2} a_{P(2l-1),P(2l)} \prod_{l=k/2+1}^{N/2} b_{P(2l-1),P(2l)},
\end{align}
where
\begin{align}
\label{def:GinOEgpf_ab}
\begin{split}
a_{j,l}&:=\int_{-\infty}^{\infty}dx\; u(x) e^{-x^2/2}p_{l-1}(x)\int_{-\infty}^{x}dy\; u(y) e^{-y^2/2}p_{j-1}(y),\\
b_{j,l}&:=i\int_{\mathbb{R}_+^2}dw\; v(w)\;\mathrm{erfc}(\sqrt{2}|\mathrm{Im}(w)|)\;e^{-(w^2+\bar{w}^2)/2}\; p_{j-1}(w)p_{l-1}(\bar{w}).
\end{split}
\end{align}
Let
\begin{align}
\label{def:alpha_beta}
\begin{split}
\alpha_{j,l}&:=a_{j,l}-a_{l,j},\\
\beta_{j,l}&:=2(b_{j,l}-b_{l,j}),
\end{split}
\end{align}
then we have the restriction $P(2l)>P(2l-1)$ and we write
\begin{align}
\nonumber &Z_{k,(N-k)/2}[u,v]=C_{N,k}\frac{k!}{(k/2)! 2^{(N-k)/2}}\\
\nonumber &\times \sum_{P\in S_N \atop P(2l)>P(2l-1)}\varepsilon(P)\prod_{l=1}^{k/2} \alpha_{P(2l-1),P(2l)}\prod_{l=k/2+1}^{N/2}\beta_{P(2l-1),P(2l)}\\
\nonumber &=C_{N,k}\; \frac{k!((N-k)/2)!}{2^{(N-k)/2}} \sum_{P\in S_N \atop P(2l)>P(2l-1)}^* \hspace{-12pt} \varepsilon(P) \prod_{l=1}^{k/2} \alpha_{P(2l-1),P(2l)} \prod_{l=k/2+1}^{N/2}\beta_{P(2l-1),P(2l)}\\
\nonumber &= C_{N,k}\; \frac{k!((N-k)/2)!}{2^{(N-k)/2}}\; [\zeta^{k/2}] \Pf[\zeta \alpha_{j,l}+\beta_{j,l}]_{j,l=1,...,N},
\end{align}
where we have used Definition \ref{def:pfaff}, recalling Remark \ref{rem:Pf_def}. Note that the $(k-N)/2$ factors of $2$ are to compensate for the factor of $2$ introduced in the definition of $\beta$. The result now follows immediately on substitution of $C_{N,k}$ from (\ref{def:GinOECNK}).

\hfill $\Box$

Performing the sum in (\ref{eqn:summedup}) gives us that the generalised partition function for general $N$ and $k$ (recalling that $k$ must be of the same parity as $N$) is
\begin{align}
\label{eqn:even_gpfsum} Z_N[u,v]=\frac{2^{-N(N+1)/4}}{\prod_{l=1}^N\Gamma(l/2)}\Pf[\alpha_{j,l}+\beta_{j,l}]_{j,l=1,...,N}.
\end{align}
We also find that the generating function for the probabilities (\ref{def:GinOE_probsGF}) becomes
\begin{align}
\label{eqn:GinOE_probsGF_pf} Z_N(\zeta)=\frac{2^{-N(N+1)/4}}{\prod_{l=1}^N\Gamma(l/2)}\Pf[\zeta\alpha_{j,l}+\beta_{j,l}]_{j,l=1,...,N}\Big|_{u=v=1}.
\end{align}
From (\ref{eqn:GinOE_probsGF_pf}), (\ref{eqn:Ginpnkzkn-k}) and (\ref{def:GinOE_probsGF}) we can see that the probabilities for the extremal values of $k$ are
\begin{align}
\label{eqn:pNNeven} p_{N,N}=\frac{2^{-N(N+1)/4}}{\prod_{l=1}^N\Gamma(l/2)}\Pf[\alpha_{j,l}\big|_{u=1}]_{j,l=1,...,N}
\end{align}
for all real eigenvalues, and
\begin{align}
\nonumber p_{N,0}=\frac{2^{-N(N+1)/4}}{\prod_{l=1}^N\Gamma(l/2)}\Pf[\beta_{j,l}\big|_{v=1}]_{j,l=1,...,N}
\end{align}
for all complex eigenvalues.

\subsubsection{$N$ odd}
\label{sec:GinOE_genpartfn_odd}

As discussed at the beginning of Chapter \ref{sec:Ggpfe}, the generalised partition function is parity dependent, in direct analogy with the GOE. Interestingly, the intricacies introduced by the odd case have a more natural interpretation in the real Ginibre ensemble. Recall that the application of integration over alternate variables in Proposition \ref{prop:GOE_gen_part_fn_odd} was complicated by the extra unpaired row in the Vandermonde determinant, which led to the Pfaffian of an odd-sized matrix with a border row and column corresponding to this extra eigenvalue. We will see the same structure for the real Ginibre matrices, however, this border now directly corresponds to the single real eigenvalue that is guaranteed to exist in an odd-sized real, asymmetric matrix. This is a consequence of the fact the eigenvalues of real, symmetric matrices are real or a complex conjugate pair.

Aside from some small technical considerations, the procedure is otherwise identical to the $N$ odd case of the GOE: we begin with the parity insensitive eigenvalue jpdf (\ref{eqn:GinOEjpdf}), and apply a modified form of integration over alternate variables to give us an even-sized Pfaffian.

\begin{proposition}
\label{prop:GinOE_gpf_odd}
Let $\alpha_{j,l}$ and $\beta_{j,l}$ be as in Proposition \ref{prop:GinOE_gpf_even} and
$\nu_j$ as in (\ref{def:GOE_nu}). Then, for $N,k$ odd, the generalised partition function for real, Ginibre matrices is
\begin{align}
\label{eqn:GinOE_gpf_odd} Z_{k,(N-k)/2}^{\odd}[u,v]=\frac{2^{-N(N+1)/4}}{\prod_{l=1}^N\Gamma(l/2)}[\zeta^{(k-1)/2}]\mathrm{Pf}\left[\begin{array}{cc}
[\zeta \alpha_{j,l}+\beta_{j,l}] & [\nu_j]\\
\left[-\nu_l\right]& 0\\
\end{array}\right]_{j,l=1,...,N}.
\end{align}
\end{proposition}

\textit{Proof}: First we remove the absolute value from the Vandermonde in (\ref{eqn:GinOEjpdf}) in the same way as in Proposition \ref{prop:GinOE_gpf_even}, multiplying by $k!\; i^{(N-k)/2}$. Then, using (\ref{eqn:vandermonde_GinOE}), the odd analogue of (\ref{eqn:GinOE_ioav_even}) is
{\small
\begin{align}
\nonumber &Z_{k,(N-k)/2}^{\odd}[u,v]=C_{N,k}\:i^{(N-k)/2}\frac{k!}{((k-1)/2)!}\int_{\mathbb{R}^2_+}dw_1\cdot\cdot\cdot \int_{\mathbb{R}^2_+}dw_{(N-k)/2}\\
\nonumber &\times \prod_{j=1}^{(N-k)/2} v(w_l)\; e^{-(w_j^2+\bar{w}_j^2)/2}\; \mathrm{erfc}(\sqrt{2}|\mathrm{Im}(w_j)|)\int_{-\infty}^{\infty}d\lambda_2\int_{-\infty}^{\infty}d\lambda_4\cdot\cdot \cdot\int_{-\infty}^{\infty}d\lambda_{k-1}\\
\nonumber &\times\hspace{3pt}\det \left[\begin{array}{l}
\left[\begin{array}{c}
\int_{-\infty}^{\lambda_{2j}}u(\lambda)e^{-\lambda^2/2}p_{l-1}(\lambda)d\lambda \\
u(\lambda_{2j})e^{-\lambda_{2j}^2/2}p_{l-1}(\lambda_{2j})\end{array}\right]_{j=1,...,(k-1)/2} \vspace{6pt}\\
\left[\begin{array}{c}
p_{l-1}(w_j)\\
p_{l-1}(\bar{w}_j)
\end{array}\right]_{j=1,...,(N-k)/2}\vspace{6pt}\\
\int_{-\infty}^{\infty}u(\lambda)e^{-\lambda^2/2}p_{l-1}(\lambda)d\lambda
\end{array}\right]_{l=1,...,N},
\end{align}
}where we have shifted the row corresponding to the $k$th eigenvalue to the bottom row. (Since this involves an even number of row transpositions the determinant is unchanged.) Expanding the determinant, with $a_{j,l}, b_{j,l}$ from (\ref{def:GinOEgpf_ab}) and $\alpha_{j,l},\beta_{j,l}$ from (\ref{def:alpha_beta}), we have
\begin{align}
\nonumber &Z_{k,(N-k)/2}^{\odd}[u,v]=C_{N,k}\frac{k!}{((k-1)/2)!}\\
\nonumber &\times \sum_{P\in S_N}\varepsilon(P)\; \nu_{P(N)}\prod_{l=1}^{(k-1)/2}a_{P(2l-1),P(2l)} \prod_{l=(k+1)/2}^{(N-1)/2}b_{P(2l-1),P(2l)}\\
\nonumber &=C_{N,k}\frac{k!((N-k)/2)!}{2^{(N-k)/2}}\\
\nonumber &\times \sum_{P\in S_N \atop P(2l)>P(2l-1)}^*\varepsilon(P)\;\nu_{P(N)}\prod_{l=1}^{(k-1)/2} \alpha_{P(2l-1),P(2l)} \prod_{l=(k+1)/2}^{(N-1)/2} \beta_{P(2l-1),P(2l)},
\end{align}
and, with $C_{N,k}$ from (\ref{def:GinOECNK}), we have the result.

\hfill $\Box$

As for $N$ even, $Z_{k,(N-k)/2}^{\odd}[u,v]$ is the generalised partition function for only one part of the relevant problem and we in fact need to sum over all possible $k$. The analogues of (\ref{eqn:summedup}) and (\ref{eqn:even_gpfsum}) are then
\begin{align}
\nonumber Z_N^{\odd}[u,v]&:=\sum_{k=1}^N{}^{\sharp}\; Z_{k,(N-k)/2}^{\odd}[u,v]\\
\label{eqn:odd_gpfsum} &=\frac{2^{-N(N+1)/4}}{\prod_{l=1}^N\Gamma(l/2)}\mathrm{Pf}\left[\begin{array}{cc}
[\alpha_{j,l}+\beta_{j,l}] & [\nu_j]\\
\left[-\nu_l\right]& 0\\
\end{array}\right]_{j,l=1,...,N},
\end{align}
and those of (\ref{def:GinOE_probsGF}) and (\ref{eqn:GinOE_probsGF_pf}) for the probabilities of finding $k$ (odd) real eigenvalues are
\begin{align}
\nonumber Z_N^{\odd}(\zeta)&:=\sum_{k=0}^{(N-1)/2}\zeta^k Z_{2k+1,(N-1-2k)/2}[1,1]\\
\nonumber &=\sum_{k=0}^{(N-1)/2} \zeta^k p_{2k+1,(N-1-2k)/2}\\
\label{eqn:GinOE_probs_odd} &=\left. \frac{2^{-N(N+1)/4}}{\prod_{l=1}^N\Gamma(l/2)}\mathrm{Pf}\left[\begin{array}{cc}
[\zeta \alpha_{j,l}+\beta_{j,l}] & [\nu_j]\\
\left[-\nu_l\right]& 0\\
\end{array}\right]_{j,l=1,...,N} \right|_{u=v=1}.
\end{align}
The probabilities for the extremal values of $k$ are also analogous.

\subsection{Skew-orthogonal polynomials for the real Ginibre ensemble}
\label{sec:Gsops}

As we saw in the case of the GOE, the Pfaffian in (\ref{eqn:even_gpfsum}) will be most easily calculated if we can find the appropriate polynomials that skew-diagonalise the matrix, or, for (\ref{eqn:odd_gpfsum}), make it odd skew-diagonal as in (\ref{eqn:skew_diag_mat_odd}). Recall that in Definition \ref{def:GOE_soip} we defined a skew-inner product based on the double integrals $\gamma_{j,l}$ from Proposition \ref{prop:GOE_gen_part_fn}, which were the entries of the ($N$ even) generalised partition function. Here we will make the analogous definition, this time using the $\alpha_{j,l}$ and $\beta_{j,l}$ of Proposition \ref{prop:GinOE_gpf_even}.

\begin{definition}
\label{def:Ginip1}
Let $\{ p_j\}_{j=1,2,...}$ be a set of monic polynomials of degree $N$. Define the inner product
\begin{align}
\nonumber (p_j,p_l)&:=\int_{-\infty}^{\infty}dx\int_{-\infty}^{\infty}dy\; e^{-(x^2+y^2)/2}p_{j}(x)p_{l}(y)\hspace{3pt}\mathrm{sgn}(y-x)\\
\nonumber &+2i\int_{\mathbb{R}_+^2}dw\; \mathrm{erfc}(\sqrt{2}|\mathrm{Im}(w)|)\; e^{-(w^2+\bar{w}^2)/2}\Bigl(p_{j}(w)p_{l}(\bar{w})-p_{l}(w)p_{j}(\bar{w})\Bigr)\\
\label{def:GinOE_sip} &=\alpha_{j+1,l+1}+\beta_{j+1,l+1}\big|_{u=v=1},
\end{align}
with $\alpha_{j,l}$ and $\beta_{j,l}$ as in (\ref{eqn:GinOE_alphabeta}).
\end{definition}

We would like to find monic polynomials that satisfy the skew-orthogonality properties
\begin{align}
\label{eqn:GinOE_soprops} (p_{2j},p_{2l}) = (p_{2j+1},p_{2l+1})=0 &,&(p_{2j},p_{2l+1})=-(p_{2l+1},p_{2j})=\delta_{j,l}r_j,
\end{align}
although note that the anti-symmetry property is obvious from the definition of the inner product (\ref{def:GinOE_sip}). The polynomials for the real Ginibre ensemble were first presented in \cite{FN07}.

\begin{proposition}
\label{prop:Ginsops}
The skew-orthogonal polynomials for the real Ginibre ensemble are
\begin{align}
\label{eqn:GinOE_sopolys} p_{2j(x)}=x^{2j}&,&p_{2j+1}(x)=x^{2j+1}-2j\; x^{2j-1},
\end{align}
with normalisation
\begin{align}
\label{eqn:GinOE_sopolys_norm} (p_{2j},p_{2j+1})=r_j=2\sqrt{2\pi}\;\Gamma (2j+1).
\end{align}
\end{proposition}

The direct verification that these polynomials are in fact skew-orthogonal with respect to the inner product is somewhat of a chore; we will only sketch some of the salient points. In the case $(p_{2j},p_{2l})$ or $(p_{2j+1},p_{2l+1})$ we see that the integrand in $\beta_{2j+1,2l+1}$ and $\beta_{2j+2,2l+2}$ is odd (since $\erfc$ is an odd function), and so $\beta_{2j+1,2l+1}=\beta_{2j+2,2l+2}=0$. Also, $\alpha_{2j+1,2l+1}=\alpha_{2j+2,2l+2}=0$ by the same reasoning as in Proposition \ref{prop:GOE_soip} (the inner integral produces an odd integrand for the outer integral). For the remaining properties, including the calculation of the normalisation, the reader is referred to \cite{FN08} for the details.
\begin{remark}
The details we have omitted from the verification of the skew-orthogonal polynomials involve finding recursions for the $\alpha$ and $\beta$ integrals of (\ref{def:GinOE_sip}). We will partially address this issue in the following, when we discuss the calculation of the probabilities $p_{N,k}$.
\end{remark}
\begin{remark}
The skew-orthogonal polynomials may also be found via an average over a characteristic polynomial (see \cite{AkeKiePhi2010}) or indirectly using knowledge of the average of the product of two characteristic polynomials; we will pursue this latter method further in Chapters \ref{sec:SOEcharpolys} and \ref{sec:TOEsops}.
\end{remark}

\subsubsection{Probability of $k$ real eigenvalues}
\label{sec:pnk}

Recall from (\ref{eqn:Ginpnkzkn-k}) that $p_{N,k}$, the probability of obtaining $k$ real eigenvalues from an $N\times N$ real, Gaussian matrix, is given by $Z_{k,(N-k)/2}[1,1]$. In order to calculate the probabilities using (\ref{eqn:GinOE_gpf_even}) and (\ref{eqn:GinOE_gpf_odd}) we substitute $\zeta\alpha_{j,l}+\beta_{j,l}= (\zeta-1)\alpha_{j,l}+\alpha_{j,l}+\beta_{j,l} =(\zeta-1)\alpha_{j,l}+ \delta_{j,l} r_{\lfloor (j-1)/2\rfloor}$. Recalling that $\alpha_{j,l}$ is anti-symmetric we can calculate the (non-zero) $\alpha_{j,l}\big|_{u=1}$ using the relation
\begin{align}
\label{eqn:arecurs} \alpha_{2j+1,2l+2}\big|_{u=1}=2l\: I_{l-1,j}-I_{l,j},
\end{align}
where $I_{j,l}$ satisfies the recursions
\begin{align}
\label{eqn:Irecurs}
\begin{split}
I_{j+1,l}=(2j+2)I_{j,l}-2\Gamma(j+l+3/2)\\
I_{j,l+1}=(2l+1)I_{j,l}+2\Gamma(j+l+3/2),
\end{split}
\end{align}
with $I_{0,0}=-2\sqrt{\pi}$ \cite{FN08}. Combining (\ref{eqn:Irecurs}) and (\ref{eqn:arecurs}) we have
\begin{align}
\label{eqn:Ginafinal} \alpha_{2j+1,2l+2}\big|_{u=1}=2\: \Gamma(j+l+1/2).
\end{align}

For $N$ odd we see from (\ref{eqn:GinOE_gpf_odd}) that we also need to calculate $\nu_j\big|_{u=1}$, however from its definition (\ref{def:GOE_nu}) we have $\nu_j\big|_{u=1}=0$ for $j$ even and for $j$ odd we integrate by parts to find
\begin{align}
\label{eqn:Gnu} \nu_j\big|_{u=1}=(j-2)!!\sqrt{2\pi}.
\end{align}
With the entries of the matrices so specified, we can then calculate the probabilities, although we still have an unwieldy $N\times N$ Pfaffian to deal with. This situation can be improved somewhat by noting that with the polynomials (\ref{eqn:GinOE_sopolys}) (or indeed with any set of alternating even and odd functions) the Pfaffian matrix takes on a chequer pattern like (\ref{eqn:N3chequer}), and so it can be written as an $N/2\times N/2$ determinant using (\ref{eqn:chequer}). Generally an order $p$ determinant can be computed in floating point arithmetic using $\mathrm{O}(p^3)$ operations, although the bit length of intermediate values can become exponentially long, and ill-conditioning can result if this is truncated  \cite{Stewart1973}. Alternatively, computer algebra can be used. The results of our calculations appear in Table \ref{tab:pnkxact_sim} of Appendix \ref{app:GinOE_kernel_elts} where they are compared to the results of the simulations in Table \ref{tab:GinOE_pnk}. Note that the exact results for $N=1,...,9$ appeared in \cite[Table 1]{Ed97}, while those for $N=12$ are listed in \cite[Table 2]{AK2007}.

The probability $p_{N,N}$ (that all eigenvalues are real) in (\ref{eqn:GinOEpNN}), which we calculated directly from the eigenvalue jpdf, can, of course, be obtained from (\ref{eqn:pNNeven}) or the odd equivalent from (\ref{eqn:GinOE_probs_odd}). The formula (\ref{eqn:GinOEpNN}) makes precise what we see experimentally in Table \ref{tab:GinOE_pnk}: the chance of finding all real eigenvalues rapidly decreases with $N$, yet for any finite $N$ we still have a non-zero probability that they are all real. Another interesting fact (which we can guess at from the table) established in the same work \cite[Corollary 7.2]{Ed97} is that all the probabilities are of the form $r+\sqrt{2}\:s$ where $r$ and $s$ are rational numbers.

We are now also in a position to quantify $E_N$, the expected number of real eigenvalues. We see from (\ref{def:GinOE_probsGF}) that (for $N$ even) this will be given by
\begin{align}
\label{eqn:Ginxnreals} E_N=2\frac{\partial}{\partial \zeta}Z_N(\zeta)\Big|_{\zeta=1}.
\end{align}
We will find, however, that we can calculate the expected value quite easily once we have the correlation functions and so we delay discussion of $E_N$ until Chapter \ref{sec:Ginkernelts}.

\begin{remark}
In (\ref{eqn:Ginxnreals}) we have required that $N$ be even, however this is just an artifact of the definition of the generating function $Z_N$. If the power of $\zeta$ in (\ref{def:GinOE_probsGF}) were $2k$ and in (\ref{eqn:GinOE_probs_odd}) it were $2k+1$ then we could use (\ref{eqn:Ginxnreals}) (without the factor of 2) for both even and odd cases. As it stands, for $N$ odd we have
\begin{align}
\nonumber E_N^{\odd}=2\frac{\partial}{\partial \zeta}Z^{\odd}_N(\zeta)+ \zeta^{-1}Z^{\odd}_N(\zeta) \Bigg|_{\zeta=1}.
\end{align}
\end{remark}

\subsection{Eigenvalue correlations for $N$ even}
\label{sec:Gincorrlnse}

As we have stressed, the difficulty we face in the case of the real Ginibre ensemble is the occurrence of two distinct species of eigenvalues. Consequently (recall the discussion below Proposition \ref{thm:integral_identities}) the real Ginibre ensemble does not satisfy (\ref{eqn:proj_prop1}) nor (\ref{eqn:proj_prop2}). If one attempts to apply that theorem it turns out that because the appropriate partition function is (\ref{eqn:summedup}) (a sum over the possible values of $k$) it is not possible to normalise the result of the integration on the left hand side, and so the right hand side does not eventuate. More details about this are contained in \cite{AK2007}, where the authors propose a way to integrate Pfaffians that avoids these complications. They then apply this method to calculate $p_{N,k}$ in terms of zonal polynomials. We will not pursue their method here since by using the generalised partition function with the relevant skew-orthogonal polynomials we are able to find a simple form for the computation of $p_{N,k}$, and further, we can push on to calculate the correlation functions with the same tools.

Recall that in Proposition \ref{prop:correlns_GOE_even} we found the correlation functions were given as a quaternion determinant with the $2\times 2$ correlation kernel $\mathbf{f}(x,y)$ from (\ref{def:GOE_Qdcorrelnk}). In that case (GOE matrices), all the eigenvalues are of a single species, and so the $2\times 2$ block represents the correlations between any pair of eigenvalues. In the present asymmetric case, the eigenvalues are now in two disjoint sets: real, and non-real complex. (We may use the term \textit{complex} to refer to these non-real complex eigenvalues if no confusion is likely.) So we will not be surprised to discover that a separate $2\times 2$ block is required for each pairing of eigenvalue species. Indeed, the correlation functions are built up from real-real, real-complex, complex-real and complex-complex $2\times 2$ blocks, each of which has the same structure as $\mathbf{f}(x,y)$.

\begin{definition}
\label{def:GinOE_kernel}
Let $p_0,p_1,...$ be the skew-orthogonal polynomials (\ref{eqn:GinOE_sopolys}) and $r_0,r_1,...$ the corresponding normalisations. With $N$ even define
\begin{align}
\nonumber S(\mu,\eta)&=2\sum_{j=0}^{\frac{N}{2}-1}\frac{1}{r_j}\Bigl[q_{2j}(\mu)\tau_{2j+1}(\eta)-q_{2j+1}(\mu)\tau_{2j}(\eta)\Bigr],\\
\nonumber D(\mu,\eta)&=2\sum_{j=0}^{\frac{N}{2}-1}\frac{1}{r_j}\Bigl[q_{2j}(\mu)q_{2j+1}(\eta)-q_{2j+1}(\mu)q_{2j}(\eta)\Bigr],\\
\nonumber \tilde{I}(\mu,\eta)&=2\sum_{j=0}^{\frac{N}{2}-1}\frac{1}{r_j}\Bigl[\tau_{2j}(\mu)\tau_{2j+1}(\eta)-\tau_{2j+1}(\mu)\tau_{2j}(\eta)\Bigr]+\epsilon(\mu,\eta)\\
\nonumber &=: I(\mu,\eta) +\epsilon(\mu,\eta),
\end{align}
where
\begin{align}
\nonumber q_j(\mu) &= e^{-\mu^2/2}\hspace{2pt}\sqrt{\mathrm{erfc}(\sqrt{2}|\mathrm{Im}(\mu)|)}\hspace{2pt}p_j(\mu),\\
\nonumber \tau_j(\mu) &= 
\left\{ 
\begin{array}{ll}
-\frac{1}{2}\int_{-\infty}^{\infty}\mathrm{sgn}(\mu-z)\hspace{3pt}q_j(z)\hspace{3pt}dz, & \mu\in \mathbb{R},\\
iq_j(\bar{\mu}),  & \mu\in \mathbb{R}_2^+,
\end{array}
\right.\\
\nonumber \epsilon(\mu,\eta) &= 
\left\{ 
\begin{array}{ll}
\frac{1}{2}\mathrm{sgn}(\mu-\eta),  & \mu,\eta\in \mathbb{R},\\
0,  & \mathrm{otherwise}.\\
\end{array}
\right.
\end{align}
And, in terms of these quantities, define
\begin{align}
\label{def:GinOE_correlnK} \bK(\mu,\eta)=\left[
\begin{array}{cc}
S(\mu,\eta) & - D(\mu,\eta)\\
\tilde{I}(\mu,\eta) & S(\eta,\mu)
\end{array}
\right].
\end{align}
\end{definition}
There are specific $2\times 2$ blocks corresponding to each of the four types of pairs of reals and complexes. The reader may find it helpful to look at the explicit forms of the kernel elements for each of these cases; they are written out in Appendix \ref{app:GinOE_kernel_elts_even}. Also note that we have assumed $N$ is even; we will require a modification to the kernel elements in the case $N$ is odd, which will be dealt with in course.

In the restricted case that the eigenvalues are all real or all complex, the kernel (\ref{def:GinOE_correlnK}) was identified in \cite{FN07}, with the cross-correlations being furnished in  \cite{sommers2007} and, using notation similar to ours, independently in \cite{b&s2009}, although the latter uses Pfaffians instead of quaternion determinants. By Proposition \ref{prop:qdet=pf} the matrices of Pfaffians and quaternion determinants are related by a factor of $\bZ_2^{-1}$, so the Pfaffian kernel is
\begin{align}
\label{def:GinPfK} \left[
\begin{array}{cc}
S(\mu,\eta) & - D(\mu,\eta)\\
\tilde{I}(\mu,\eta) & S(\eta,\mu)
\end{array}
\right] \bZ_2^{-1}=\left[
\begin{array}{cc}
D(\mu,\eta) & S(\mu,\eta)\\
-S(\eta,\mu) & \tilde{I}(\mu,\eta)
\end{array}
\right],
\end{align}
which is identical to that in \cite{b&s2009}.

As in Lemma \ref{lem:GOE_D=S=I} for the GOE we have some relationships between the kernel elements, although here there are more options to consider since there are four pairs of real and complex eigenvalues. These relationships are easily verified (particularly when the explicit forms in the appendix are kept in mind) and so no proof is given.
\begin{lemma}
\label{lem:Gin_s=d=i}
With the functions $S,D$ and $\tilde{I}$ as given in Definition \ref{def:GinOE_kernel}, and using the convention that $x,y\in \mathbb{R}$ and $w,z\in \mathbb{C}\backslash \mathbb{R}$ we have
\begin{align}
\label{eqn:Gin_s=d=i}
\begin{split}\tilde{I}_{r,r}(x,y)&=-\int_{x}^{y}S_{r,r}(t,y)dt+\frac{1}{2}\mathrm{sgn}(x-y),\\
\tilde{I}_{c,r}(w,x)&=-\tilde{I}_{r,c}(x,w)=iS_{c,r}(\bar{w},x),\\
\tilde{I}_{c,c}(w,x)&=iS_{c,c}(\bar{w},z),\\
D_{r,r}(x,y)&=-\frac{\partial}{\partial y}S_{r,r}(x,y),\\
D_{r,c}(x,w)&=-D_{c,r}(w,x)=-iS_{r,c}(x,\bar{w}),\\
D_{c,r}(w,x)&=-D_{r,c}(x,w)=-\frac{\partial}{\partial x}S_{c,r}(w,x),\\
D_{c,c}(w,z)&=-iS_{c,c}(w,\bar{z}),
\end{split}
\end{align}
where the subscripts $r$ and $c$ are to clarify the domain.
\end{lemma} 

\begin{remark}
Note the `missing' relation --- from the apparent symmetries it is expected that $\tilde{I}_{r,c}(x,w)$ could be calculated as some integral of $S_{r,c}(x,w)$. This can be done for the even case, however it cannot be done in the odd case because of the extra term that appears in $\tilde{I}_{r,c}(x,w)$, which is dependent on the complex variable (see Appendix \ref{app:GinOE_kernel_elts_odd}). Of course, we are still able to obtain $\tilde{I}_{r,c}(x,w)$ by its anti-symmetry and so the missing relation does not affect the formulation.
\end{remark}

In this section we will present two methods for calculating the correlation functions; both based on functional differentiation and both beginning with $Z_N[u,v]$ from (\ref{eqn:even_gpfsum}). The first method produces a Fredholm quaternion determinant with a $4\times 4$ kernel, instead of a $2\times 2$ kernel as in (\ref{eqn:ZN_integ_op}), where each $2\times 2$ block relates to a different pairing of reals and complexes. This method highlights the separate treatments required for the different pairings, however, it needs more general functional differentiation and Fredholm operators. The second method, which is more in keeping with the literature on the topic \cite{sommers2007,b&s2009,Sinc09}, uses a perhaps less natural approach where a generalised variable is used to stand for both real and complex variables as required, although it results in a $2\times 2$ kernel and allows the use of the existing functional differentiation and Fredholm operators from the GOE case. Of course, both methods result in the same correlation functions with kernel given by (\ref{def:GinOE_correlnK}), and each is easily generalised to the odd case using the same method as that in Proposition \ref{prop:pf_integ_op_odd}.

\begin{remark}
Although we say that the $2\times 2$ kernel method is more in keeping with the literature on the topic, a variant of the $4\times 4$ kernel method can be found in \cite[Chapter 6.7]{forrester?}.
\end{remark}

\subsubsection{Two component $4\times 4$ kernel method}

We would like to apply functional differentiation again (as in the case of the GOE in Chapter \ref{sec:correlns_even_GOE}) to find the correlation functions $\rho_{(n_1,n_2)}(x_{1},...,x_{n_1},w_{1},...,w_{n_2})$ with $n_1$ real and $n_2$ non-real complex eigenvalues. However, it seems (\ref{eqn:fnal_diff_correln}) is inadequate for our needs since it contains only one species of eigenvalue; we require instead the formula
{\small
\begin{align}
\nonumber &\rho_{(n_1,n_2)}(x_{1},...,x_{n_1},w_{1},...,w_{n_2}):=\\
\label{eqn:GinOEfnal_diff_correln} &\frac{1}{Z_N[u,v]}\frac{\delta^{n_1+n_2}}{\delta u(x_1)\cdot\cdot\cdot \delta u(x_{n_1})\delta v(w_1)\cdot\cdot\cdot \delta v(w_{n_2})}Z_N[u,v]{\Big |}_{u=v=1}.
\end{align}
}This formula augurs well, however, we will also require a generalised form of the Fredholm operators to make use of it.
\begin{definition}
\label{def:GinOE_Freds}
Let $\lambda$ be a constant parameter and $K_G$ an integral operator with kernel
\begin{align}
\nonumber \left[\begin{array}{cc}
\kappa_{11}(x,y) & \kappa_{12}(x,z)\\
\kappa_{21}(w,y) & \kappa_{22}(w,z),
\end{array}\right],
\end{align}
having two species of variable, $\{x,y\}$ and $\{ w,z\}$, and
\begin{align}
\nonumber K(x_i,x_j,w_m,w_n)_{s,t}:=\left[\begin{array}{cc}
\kappa_{11}(x_i,x_j) & \kappa_{12}(x_i,w_n)\\
\kappa_{21}(w_m,x_j) &\kappa_{22}(w_m,w_n)
\end{array} \right]_{i,j=1,...,s \atop m,n=1,...,t}.
\end{align}
Then
\begin{align}
\nonumber \det[1+\lambda K_G]:=&\sum_{s=0}^{\infty}\sum_{t=0}^{\infty}\frac{\lambda^{s+t}}{s!t!}\int dx_1\cdot\cdot\cdot\int dx_s \int dw_1\cdot\cdot\cdot \int dw_t\\
\nonumber &\times \det K(x_i,x_j,w_m,w_n)_{s,t},
\end{align}
\begin{align}
\nonumber \qdet[1+\lambda K_G]:=&\sum_{s=0}^{\infty}\sum_{t=0}^{\infty}\frac{\lambda^{s+t}}{s!t!}\int dx_1\cdot\cdot\cdot\int dx_s \int dw_1\cdot\cdot\cdot \int dw_t\\
\nonumber &\times \qdet \:K(x_i,x_j,w_m,w_n)_{s,t},
\end{align}
when $K(x_i,x_j,w_m,w_n)_{s,t}$ is self-dual, and
\begin{align}
\nonumber \Pf[1+\lambda K_G]:=&\sum_{s=0}^{\infty}\sum_{t=0}^{\infty}\frac{\lambda^{s+t}}{s!t!}\int dx_1\cdot\cdot\cdot\int dx_s \int dw_1\cdot\cdot\cdot \int dw_t\\
\nonumber &\times \Pf \: K(x_i,x_j,w_m,w_n)_{s,t},
\end{align}
when $K(x_i,x_j,w_m,w_n)_{s,t}$ is anti-symmetric. The first term $(s=t=0)$ in each sum is taken to be $1$.
\end{definition}
By following the same method of proof as in Lemma \ref{lem:pf_qdet_kernel}, we can establish its analogue in this case, which then gives us the analogue of Corollary \ref{cor:sqrtFreds}.
\begin{corollary}
With the definitions of Definition \ref{def:GinOE_Freds} we have
\begin{align}
\nonumber \left(\det[1+\lambda K_G]) \right)^{1/2}=\qdet[1+\lambda K_G]
\end{align}
in the case that $K(x_i,x_j,w_m,w_n)_{s,t}$ is self-dual, and
\begin{align}
\nonumber \left(\det[1+\lambda K_G]) \right)^{1/2}=\Pf[1+\lambda K_G]
\end{align}
when $K(x_i,x_j,w_m,w_n)_{s,t}$ is anti-symmetric.
\end{corollary}

We can see that with (\ref{eqn:GinOEfnal_diff_correln}) applied to the Fredholm operators in Definition \ref{def:GinOE_Freds}, we will be able to pick out the term in the expansion of the integral operator corresponding to any number of real and complex eigenvalues. The only difficulty that remains then is to express the generalised partition function as a Fredholm quaternion determinant, in analogue with (\ref{eqn:ZN_integ_op}).

We will use the integral operator definitions of (\ref{def:integop}) and (\ref{eqn:integop_ab}), where it will be recalled that, by convention, $y$ is the variable of integration. Here we use the convention that we integrate the real variable $y$ or the complex variable $z$. (We never have both $y$ and $z$ in the same expression so the convention can be applied consistently.)

\begin{proposition}
\label{prop:4x4_fred}
Let $x,y\in\mathbb{R}$ and $w,z\in\mathbb{C}\backslash\mathbb{R}$. Then, with $\alpha_{ij}$ and $\beta_{ij}$ as in Proposition \ref{prop:GinOE_gpf_even}, we have
\begin{align}
\label{eqn:4x4_fred} \Pf[\alpha_{ij} +\beta_{ij}]_{i,j=1,...,N}=\prod_{j=0}^{N/2-1}r_j\;\qdet [\1_4+ \bK_G(\mathbf{t}-\1_4)],
\end{align}
where $\bK_G(\mathbf{t}-\1_4)$ is the $4\times 4$ matrix integral operator with kernel
\begin{align}
\nonumber \left[\begin{array}{ll}
\bK(x,y) & \bK(x,z)\\
\bK(w,y) & \bK(w,z)
\end{array}\right] \left[\begin{array}{cccc}
u(y)-1 & 0 & 0 & 0\\
0 & u(y)-1 & 0 & 0\\
0 & 0 & v(z)-1 & 0\\
0 & 0 & 0 & v(z)-1\\
\end{array}\right],
\end{align}
with $\bK(\mu,\eta)$ as in (\ref{def:GinOE_correlnK}).
\end{proposition}

\textit{Proof}: Let
\begin{align}
\nonumber u=&\; \sigma+1,\\
\nonumber v=&\; \eta+1,\\
\nonumber \psi_j(x):=&\; e^{-x^2/2}p_{j-1}(x)\\
\label{def:GFredqdetPrf} \phi_j(z):= &\; \sqrt{\mathrm{erfc}(\sqrt{2}|\mathrm{Im}(z)|)}\: e^{-z^2/2}p_{j-1}(z).
\end{align}
We still require the integral operator $\epsilon$ as used in Proposition \ref{prop:pf_integ_op} for the real integrals, but we also need to expand it to include the complex case,
\begin{align}
\label{def:rc_eps} \epsilon f [\eta]=\left\{ \begin{array}{cl}
\frac{1}{2}\int_{\mathbb{R}}\psi(y)\:\sgn(\eta-y)dy, & \eta \in \mathbb{R},\\
-i\; \phi(\bar{\eta}), & \eta \in \mathbb{C}\backslash \mathbb{R}.
\end{array}\right.
\end{align}
So then
\begin{align}
\nonumber \alpha_{j,l}+\beta_{j,l}&=(\alpha_{j,l}+\beta_{j,l})\Big|_{u=v=1}\\
\nonumber &-2\int_{-\infty}^{\infty}\sigma(x)\big(\psi_j(x)\epsilon \psi_k[x]-\psi_k(x)\epsilon\psi_j[x]-\psi_k(x)\epsilon(\sigma\psi_j)[x]\big)dx\\
\nonumber &-2\int_{\mathbb{R}_+^2} \eta(z) \big(\phi_j(z)\epsilon\phi_l[z]-\epsilon\phi_j[z]\phi_l(z)\big)\; dz\\
\label{def:abtau} &=(\alpha_{j,l}+\beta_{j,l})^{(1)}-(\alpha_{j,l}+\beta_{j,l})^{(\tau)},
\end{align}
where $(\alpha_{j,l}+\beta_{j,l})^{(1)}:=(\alpha_{j,l}+\beta_{j,l})\Big|_{u=v=1}$ and $(\alpha_{j,l}+\beta_{j,l})^{(\tau)}$ is the two remaining terms. By using the skew-orthogonal polynomials we can decompose $\left[(\alpha_{j,l}+\beta_{j,l})^{(1)}\right]_{j,l=1,...,N}$ as in the proof of Corollary \ref{cor:qdet=pf} with $\bD\bZ_N^{-1}$, using $\bZ_N$ from (\ref{def:Z2N}) and\\ $\mathbf{D}=\mathrm{diag}[r_0,r_0,r_1,r_1,...,r_{N/2},r_{N/2}]$.

Recall $G_j(x)$ from (\ref{def:G_psi}) and let
\begin{align}
\nonumber H_{2j-1}(z):=\phi_{2j}(z),&& H_{2j}(z):=-\phi_{2j-1}(z),
\end{align}
then
\begin{align}
\nonumber &\Pf[\alpha_{j,l}+\beta_{j,l}]_{j,l=1,...,N}=\prod_{j=0}^{N/2-1}r_j\; \qdet\Big[\delta_{j,l}\\
\nonumber &+\frac{2}{r_{\lfloor (j-1)/2 \rfloor}}\int_{-\infty}^{\infty}\sigma(x)\big(G_j(x)\epsilon \psi_k[x]-\psi_k(x)\epsilon G_j[x]-\psi_k(x)\epsilon(\sigma G_j)[x]\big)dx\\
\nonumber &+\frac{2}{r_{\lfloor (j-1)/2 \rfloor}}\int_{\mathbb{R}_+^2} \eta(z)\big(H_j(z)\epsilon\phi_l[z]-\epsilon H_j[z]\phi_l(z)\big)\; dz\Big].
\end{align}
Now if we define
\begin{align}
\label{def:GinOE_om_F} \Omega_G:=\left[
\begin{array}{cccc}
-\epsilon \sigma & -1 & 0 & 0\\
1 & 0 & 0 & 0\\
0 & 0 & 0 & -1\\
0 & 0 & 1 & 0
\end{array}
\right]&&\mathrm{and}&& \mathbf{F}:=\left[
\begin{array}{ccc}
\psi_1(y) & \cdot \cdot\cdot & \psi_N(y)\\
\epsilon\psi_1[y] & \cdot \cdot\cdot & \epsilon\psi_N[y]\\
\phi_1(z) & \cdot\cdot\cdot & \phi_N(z)\\
\epsilon\phi_1[z] & \cdot\cdot\cdot & \epsilon\phi_N[z]
\end{array}
\right]
\end{align}
then we can proceed in much the same way as for Proposition \ref{prop:pf_integ_op}, with $\Omega_G$ and $\bF$ replacing $\Omega$ and $\bE$ respectively. To wit, let $\bA$ be the integral operator with kernel\\
$2\bZ_N^{-1}\bD^{-1}(\tau(\sigma,\eta) \Omega_G \bF)^T$, where $\tau(\sigma,\eta)=\mathrm{diag}[\sigma(y),\sigma(y),\eta(z),\eta(z)]$, and $\bB=\bF$. Then
\begin{align}
\nonumber \1_N-\bZ_N\bD^{-1}[(\alpha_{j,l}+\beta_{j,l})^{(\tau)}]=\1_N + \bA\bB
\end{align}
(cf. (\ref{eqn:1+ABdecomp})) and we apply (\ref{eqn:qdet_1+AB}). Since the kernel of $\bA$ is of the form
\begin{align}
\left[\begin{array}{ll}
\nonumber -\frac{2\sigma(y)}{r_{\lfloor (j-1)/2 \rfloor}}\big(\epsilon G_j[y]+\epsilon(\sigma G_j)[y]\big)&\frac{2\sigma(y)}{r_{\lfloor (j-1)/2 \rfloor}}G_j(y)
\end{array}\right.\\
\nonumber \left. \begin{array}{rr}
-\frac{2\eta(z)}{r_{\lfloor (j-1)/2 \rfloor}}\epsilon H_j[z]& \frac{2\eta(z)}{r_{\lfloor (j-1)/2 \rfloor}} H_j(z)
\end{array}\right]_{j=1,...,N}
\end{align}
(which is $N\times 4$) and $\bB$ is $4\times N$, after using (\ref{eqn:qdet_1+AB}) we are left with a $4\times 4$ matrix-valued integral operator, with kernel
\begin{align}
\nonumber \1_4+\bB\bA= \left[\begin{array}{cc}
\tilde{\kappa}_{r,r} & \tilde{\kappa}_{r,c}\\
\tilde{\kappa}_{c,r} & \tilde{\kappa}_{c,c}
\end{array}\right],
\end{align}
where
\begin{align}
\nonumber \tilde{\kappa}_{r,r}&=\left[\begin{array}{l}
1-\sum_{j=1}^N\frac{2}{r_{\lfloor (j-1)/2 \rfloor}}\Big(\psi_j\otimes \sigma \epsilon G_j + \psi_j \otimes \sigma\epsilon(\sigma G_j) \Big)\\
-\sum_{j=1}^N\frac{2}{r_{\lfloor (j-1)/2 \rfloor}}\Big(\epsilon\psi_j\otimes \sigma \epsilon G_j + \epsilon\psi_j \otimes \sigma\epsilon(\sigma G_j) \Big)
\end{array}\right.\\
\nonumber &\qquad\qquad\qquad\qquad\left.\begin{array}{r}
\sum_{j=1}^N\frac{2}{r_{\lfloor (j-1)/2 \rfloor}}\psi_j\otimes \sigma G_j\\
1+\sum_{j=1}^N\frac{2}{r_{\lfloor (j-1)/2 \rfloor}}\epsilon \psi_j\otimes \sigma G_j
\end{array}\right],\\
\nonumber \tilde{\kappa}_{r,c}&=\left[\begin{array}{cc}
-\sum_{j=1}^N\frac{2}{r_{\lfloor (j-1)/2 \rfloor}} \psi_j\otimes \eta \epsilon H_j & \sum_{j=1}^N\frac{2}{r_{\lfloor (j-1)/2 \rfloor}}\psi_j\otimes \eta H_j\\
-\sum_{j=1}^N\frac{2}{r_{\lfloor (j-1)/2 \rfloor}} \epsilon \psi_j\otimes \eta\epsilon H_j & \sum_{j=1}^N\frac{2}{r_{\lfloor (j-1)/2 \rfloor}}\epsilon \psi_j\otimes \eta H_j
\end{array}\right],\\
\nonumber \tilde{\kappa}_{c,r}&=\left[\begin{array}{cc}
-\sum_{j=1}^N\frac{2}{r_{\lfloor (j-1)/2 \rfloor}}\phi_j\otimes \sigma \epsilon G_j & \sum_{j=1}^N\frac{2}{r_{\lfloor (j-1)/2 \rfloor}}\phi_j\otimes \sigma G_j\\
-\sum_{j=1}^N\frac{2}{r_{\lfloor (j-1)/2 \rfloor}}\epsilon \phi_j\otimes \sigma\epsilon G_j & \sum_{j=1}^N\frac{2}{r_{\lfloor (j-1)/2 \rfloor}}\epsilon \phi_j\otimes \sigma G_j
\end{array}\right],\\
\nonumber \tilde{\kappa}_{c,c}&=\left[\begin{array}{cc}
1-\sum_{j=1}^N\frac{2}{r_{\lfloor (j-1)/2 \rfloor}}\phi_j\otimes \eta \epsilon H_j & \sum_{j=1}^N\frac{2}{r_{\lfloor (j-1)/2 \rfloor}}\phi_j\otimes \eta H_j\\
-\sum_{j=1}^N\frac{2}{r_{\lfloor (j-1)/2 \rfloor}}\epsilon \phi_j\otimes \eta\epsilon H_j & 1+\sum_{j=1}^N\frac{2}{r_{\lfloor (j-1)/2 \rfloor}}\epsilon \phi_j\otimes \eta H_j
\end{array}\right].
\end{align}
Noting that
\begin{align}
\label{eqn:omg_decomp} \Omega_G^T\left[\begin{array}{cccc}
1 & 0 & 0 & 0\\
-\epsilon \sigma & 1 & 0 & 0\\
0 & 0 & 1 & 0\\
0 & 0 & 0 & 1
\end{array}
\right]=\bZ_4^{-1},
\end{align}
we can rewrite $\bB\bA$ as
\begin{align}
\nonumber \bB\bA&=2\bE\bZ_N^{-1}\bD(\tau(\sigma,\eta)\Omega_G\bF)^T
=2\bE\bZ_N^{-1}\bD\bF^T\Omega_G^T\tau(\sigma,\eta)\\
\label{eqn:GinBAfactrd} &=2\bE\bZ_N^{-1}\bD\bF^T\bZ_4^{-1}\tau(\sigma,\eta)\left[\begin{array}{cccc}
1 & 0 & 0 & 0\\
\epsilon \sigma & 1 & 0 & 0\\
0 & 0 & 1 & 0\\
0 & 0 & 0 & 1
\end{array}
\right],
\end{align}
where we are keeping with the convention of (\ref{eqn:eps_sig}), that the operator $\epsilon\sigma$ from $\Omega_G$ acts before the larger operator. We can now remove terms with factors $\epsilon\sigma$ by factorising $\1_4+\bB\bA$ thusly
\begin{align}
\label{eqn:1+ABGinOE4} \1_4+\bB\bA= \left[\begin{array}{cc}
\kappa_{r,r} & \kappa_{r,c}\\
\kappa_{c,r} & \kappa_{c,c}
\end{array}\right]\;\left[\begin{array}{cccc}
1 & 0 & 0 & 0\\
\epsilon \sigma & 1 & 0 & 0\\
0 & 0 & 1 & 0\\
0 & 0 & 0 & 1
\end{array}
\right],
\end{align}
where
\begin{align}
\nonumber \kappa_{r,r}&=\left[\begin{array}{cc}
1-\sum_{j=1}^N\frac{2}{r_{\lfloor (j-1)/2 \rfloor}}\psi_j\otimes \sigma \epsilon G_j & \sum_{j=1}^N\frac{2}{r_{\lfloor (j-1)/2 \rfloor}}\psi_j\otimes \sigma G_j\\
-\epsilon\sigma-\sum_{j=1}^N\frac{2}{r_{\lfloor (j-1)/2 \rfloor}}\epsilon \psi_j\otimes \sigma\epsilon G_j & 1+\sum_{j=1}^N\frac{2}{r_{\lfloor (j-1)/2 \rfloor}}\epsilon \psi_j\otimes \sigma G_j
\end{array}\right],\\
\nonumber \kappa_{r,c}&=\left[\begin{array}{cc}
-\sum_{j=1}^N\frac{2}{r_{\lfloor (j-1)/2 \rfloor}} \psi_j\otimes \eta \epsilon H_j & \sum_{j=1}^N\frac{2}{r_{\lfloor (j-1)/2 \rfloor}}\psi_j\otimes \eta H_j\\
-\sum_{j=1}^N\frac{2}{r_{\lfloor (j-1)/2 \rfloor}} \epsilon \psi_j\otimes \eta\epsilon H_j & \sum_{j=1}^N\frac{2}{r_{\lfloor (j-1)/2 \rfloor}}\epsilon \psi_j\otimes \eta H_j
\end{array}\right],\\
\nonumber \kappa_{c,r}&=\left[\begin{array}{cc}
-\sum_{j=1}^N\frac{2}{r_{\lfloor (j-1)/2 \rfloor}}\phi_j\otimes \sigma \epsilon G_j & \sum_{j=1}^N\frac{2}{r_{\lfloor (j-1)/2 \rfloor}}\phi_j\otimes \sigma G_j\\
-\sum_{j=1}^N\frac{2}{r_{\lfloor (j-1)/2 \rfloor}}\epsilon \phi_j\otimes \sigma\epsilon G_j & \sum_{j=1}^N\frac{2}{r_{\lfloor (j-1)/2 \rfloor}}\epsilon \phi_j\otimes \sigma G_j
\end{array}\right],\\
\nonumber \kappa_{c,c}&=\left[\begin{array}{cc}
1-\sum_{j=1}^N\frac{2}{r_{\lfloor (j-1)/2 \rfloor}}\phi_j\otimes \eta \epsilon H_j & \sum_{j=1}^N\frac{2}{r_{\lfloor (j-1)/2 \rfloor}}\phi_j\otimes \eta H_j\\
-\sum_{j=1}^N\frac{2}{r_{\lfloor (j-1)/2 \rfloor}}\epsilon \phi_j\otimes \eta\epsilon H_j & 1+\sum_{j=1}^N\frac{2}{r_{\lfloor (j-1)/2 \rfloor}}\epsilon \phi_j\otimes \eta H_j
\end{array}\right].
\end{align}
The matrix on the right of (\ref{eqn:1+ABGinOE4}) has unit quaternion determinant and so we have the result (up to the sign of the $D$ and $\tilde{I}$ entries, which, as discussed in Proposition \ref{prop:pf_integ_op}, leaves the result unchanged).

\hfill $\Box$

Substitution of (\ref{eqn:4x4_fred}) into (\ref{eqn:even_gpfsum}) gives us the desired form of the generalised partition function,
\begin{align}
\label{eqn:4x4ZN} Z_N[u,v]=\frac{2^{-N(N+1)/4}}{\prod_{l=1}^N\Gamma(l/2)}\prod_{j=0}^{N/2-1}r_j\;\qdet [\1_4+ \bK_G(\mathbf{t}-\1_4)],
\end{align} 
recalling that this Fredholm quaternion determinant involves a double sum, as defined in Definition \ref{def:GinOE_Freds}. So now we apply (\ref{eqn:GinOEfnal_diff_correln}), which will pick out the term in (\ref{eqn:4x4ZN}) corresponding to $n_1$ real eigenvalues and $n_2$ complex conjugate pairs of eigenvalues, that is, the sought correlation function. (The proof is readily adapted from that of Proposition \ref{prop:correlns_GOE_even}.)

\begin{proposition}[\cite{FN07, sommers2007, b&s2009}]
\label{prop:GinOE_evencorrelns}
Let $N$ be an even integer. Then, with $\bK(\mu,\eta)$ from (\ref{def:GinOE_correlnK}), the real Ginibre eigenvalue correlation function for $n_1$ real and $n_2$ non-real, complex conjugate pairs is
\begin{align}
\nonumber &\rho_{(n_1,n_2)}(x_1,...,x_{n_1},w_1,...,w_{n_2})=\qdet\left[\begin{array}{cc}
\bK(x_i,x_j) & \bK(x_i,w_m)\\
\bK(w_l,x_j) & \bK(w_l,w_m)
\end{array}\right]_{i,j=1,...,n_1 \atop l,m=1,...,n_2}\\
\nonumber &=\Pf\left(\left[\begin{array}{cc}
\bK(x_i,x_j) & \bK(x_i,w_m)\\
\bK(w_l,x_j) & \bK(w_l,w_m)
\end{array}\right]\bZ_{2(n_1+n_2)}^{-1}\right)_{i,j=1,...,n_1 \atop l,m=1,...,n_2}, \quad x_i\in \mathbb{R},  w_i \in \mathbb{R}_2^+.
\end{align}
\end{proposition}

\subsubsection{One component $2\times 2$ kernel method}

Here we find the correlation functions by using the observation that if we can treat both real and complex eigenvalues in the same manner, then we will be able to apply the method of Proposition \ref{prop:pf_integ_op} more or less directly. Conceptually, we integrate along the real line and then over upper complex plane, all with one (hopefully) convenient notation. This technique can be seen as a limit of that used in \cite{b&s2009}. A suggestion of this approach can be found in (\ref{def:rc_eps}), where we have defined a single operator that depends on the reality of the variable. The approach here then is to define all functions and operators such that they act appropriately on real or complex variables. Of course, we must find the same correlation functions as by any other method, so the end result will again be Proposition \ref{prop:GinOE_evencorrelns}. We include this method for two reasons: first, the idea of treating the real and complex eigenvalues together is commonly employed in the literature on the real Ginibre ensemble (and, as mentioned above, has already appeared briefly in (\ref{def:rc_eps})); and second, it highlights that our approach here is broadly applicable.

To this end let the uppercase variables $X,Y$ stand for the variables $x,y$ real or $w,z$ complex as required. Also let
\begin{align}
\nonumber f(X)=\left\{ \begin{array}{ll}
f_1(x),&x\in\mathbb{R},\\
f_2(w),&w\in\mathbb{C}\backslash \mathbb{R}.
\end{array}\right.
\end{align}
The key to this method is the measure: if $x$ is real and $w$ non-real complex, then we define the measure $\mu$ for the uppercase variables $X,Y$ as
\begin{align}
\label{def:GinOE_measure} \int f(X)\; d\mu(X)= \int_{-\infty}^{\infty}f_1(x)\; dx+\int_{\mathbb{R}_{+}^2}f_2(w)\;dw.
\end{align}
With these modifications, we can now treat the real and complex variables together.

\begin{proposition}
\label{prop:GinOE_fred_op}
Let $\alpha_{j,l}$ and $\beta_{j,l}$ be as in (\ref{eqn:GinOE_alphabeta}), and let $X,Y$ be real or non-real complex variables as required. Then, with the skew-orthogonal polynomials (\ref{eqn:GinOE_sopolys}) and corresponding normalisations (\ref{eqn:GinOE_sopolys_norm}), we have
\begin{align}
\label{eqn:2x2qdet} \Pf[\alpha_{j,l}+\beta_{j,l}]_{j,l=1,...,N}=\prod_{j=0}^{N/2-1}r_j\; \qdet [\1_2+\mathbf{K}(\mathbf{t}-\1_2)],
\end{align}
where $\bK(\bt-\1_2)$ is an integral operator with kernel $\bK(X,Y)\mathrm{diag}[t(Y)-1,t(Y)-1]$, with $\bK(x,y)$ as in (\ref{def:GinOE_correlnK}) and
\begin{align}
\nonumber t(Y)=\left\{ \begin{array}{cl}
u(Y), & Y \in \mathbb{R},\\
v(Y), & Y \in \mathbb{C}\backslash \mathbb{R}.
\end{array}\right.
\end{align}
\end{proposition}

\textit{Proof}: With $u, v, \psi_j(x)$ and $\phi_j(x)$ from (\ref{def:GFredqdetPrf}) let,
\begin{align}
\nonumber \varphi(X)=\left\{ \begin{array}{ll}
\psi(X), & X \in \mathbb{R},\\
\phi(X), & X \in \mathbb{C}\backslash \mathbb{R},
\end{array}\right. && \theta(X)=\left\{ \begin{array}{ll}
\sigma(X), & X \in \mathbb{R},\\
\eta(X), & X \in \mathbb{C}\backslash \mathbb{R},
\end{array}\right.
\end{align}
and recall the integral operator $\epsilon$ of (\ref{def:rc_eps}). Now, with the measure defined in (\ref{def:GinOE_measure}) and $(\alpha_{j,l}+\beta_{j,l})^{(1)}$ from (\ref{def:abtau}), we have
{\small
\begin{align}
\nonumber \alpha_{j,l}+\beta_{j,l}&=(\alpha_{j,l}+\beta_{j,l})\Big|_{u=v=1}\\
\nonumber &-2\int \theta(Y)\big(\varphi_j(Y)\epsilon \varphi_k[Y]-\varphi_k(Y)\epsilon\varphi_j[Y]-\varphi_k(Y)\epsilon(\theta \varphi_j)[Y]\big)d\mu(Y)\\
\nonumber &=:(\alpha_{j,l}+\beta_{j,l})^{(1)}-(\alpha_{j,l}+\beta_{j,l})^{(\theta)},
\end{align}
}where if $\epsilon$ appears to the left of $\theta$ in a term that corresponds to a complex number then it is understood to be zero.

With the matrix re-written in this form, where both the real and complex cases are treated simultaneously, we can now apply the same method of proof as in Proposition \ref{prop:pf_integ_op}, with $\varphi$ replacing $\psi$. Explicitly, using the skew-orthogonal polynomials we decompose $\left[(\alpha_{j,l}+\beta_{j,l})^{(1)}\right]_{j,l=1,...,N}$ as $\bD\bZ_N^{-1}$ recalling $\bZ_N$ from (\ref{def:Z2N}) and with\\$\mathbf{D}=\mathrm{diag}[r_0,r_0,r_1,r_1,...,r_{N/2},r_{N/2}]$. Letting
\begin{align}
\nonumber G_{2j-1}(t):=\varphi_{2j}(t)&,&G_{2j}(t):=-\varphi_{2j-1}(t),
\end{align}
we have
{\small
\begin{align}
\nonumber &\Pf[\alpha_{j,l}+\beta_{j,l}]_{j,l=1,...,N}=\prod_{j=0}^{N/2-1}r_j\; \qdet\Bigg[\delta_{j,l}\\
\nonumber &+\frac{2}{r_{\lfloor (j-1)/2 \rfloor}}\int_{-\infty}^{\infty}\theta(Y)\Big(G_j(Y)\epsilon \varphi_k[Y]-\varphi_k(Y)\epsilon G_j[Y]-\varphi_k(Y)\epsilon(\theta G_j)[Y]\Big)d\mu(Y)\Bigg].
\end{align}
}With $\Omega_{\theta}$ as in (\ref{eqn:omega_E}), but with $\theta$ replacing $\sigma$, and
\begin{align}
\bE_{\varphi}:=\left[
\begin{array}{ccc}
\varphi_1(Y) & \cdot \cdot\cdot & \varphi_N(Y)\\
\epsilon\varphi_1[Y] & \cdot \cdot\cdot & \epsilon\varphi_N[Y]
\end{array}
\right],
\end{align}
let $\bA$ be the $N\times 2$ matrix-valued integral operator on $\mathbb{C}$ with kernel $2\bZ_N^{-1}\bD^{-1}\theta(Y)(\Omega_{\theta} \bE_{\varphi})^T$ and $\bB=\bE_{\varphi}$. Then
\begin{align}
\nonumber \1_N-\bZ_N\bD^{-1}[\alpha_{j,l}+\beta_{j,l}]^{(\theta)}=\1_N+\bA\bB
\end{align}
and we may apply (\ref{eqn:qdet_1+AB}) to give
{\small
\begin{align}
\nonumber &\1_2+\bB\bA= \left[\begin{array}{l}
1-\sum_{j=1}^N\frac{2}{r_{\lfloor (j-1)/2 \rfloor}}\Big( \varphi_j\otimes \theta \epsilon G_j+\varphi_j\otimes \theta \epsilon (\theta G_j)\Big)\\
-\sum_{j=1}^N\frac{2}{r_{\lfloor (j-1)/2 \rfloor}}\Big(\epsilon \varphi_j\otimes \theta \epsilon G_j+\epsilon \varphi_j\otimes \theta \epsilon(\theta G_j) \Big)
\end{array}\right.\\
\nonumber &\qquad\qquad\qquad\qquad\qquad\qquad\qquad\qquad\left.\begin{array}{r}
\sum_{j=1}^N\frac{2}{r_{\lfloor (j-1)/2 \rfloor}}\varphi_j\otimes \theta G_j\\
1+\sum_{j=1}^N\frac{2}{r_{\lfloor (j-1)/2 \rfloor}}\epsilon \varphi_j\otimes \theta G_j
\end{array}\right]\\
\label{1+BA_GinOE} &=\left[\begin{array}{cc}
1-\sum_{j=1}^N\frac{2}{r_{\lfloor (j-1)/2 \rfloor}} \varphi_j\otimes \theta \epsilon G_j & \sum_{j=1}^N\frac{2}{r_{\lfloor (j-1)/2 \rfloor}}\varphi_j\otimes \theta G_j\\
-\epsilon \theta -\sum_{j=1}^N\frac{2}{r_{\lfloor (j-1)/2 \rfloor}} \epsilon \varphi_j\otimes \theta\epsilon G_j & 1+\sum_{j=1}^N\frac{2}{r_{\lfloor (j-1)/2 \rfloor}}\epsilon \varphi_j\otimes \theta G_j
\end{array}\right] \left[
\begin{array}{cc}
1 & 0\\
\epsilon \theta & 1
\end{array}
\right],
\end{align}
}where we have used (\ref{eqn:OT_decomp}) to obtain the second equality (with $\Omega$ replaced by $\Omega_{\theta}$). The matrix on the right hand side of (\ref{1+BA_GinOE}) has unit quaternion determinant and so we have the result (up to the sign of the $D$ and $\tilde{I}$ entries).
 
\hfill $\Box$

If we relabel $Z_N[u,v]$ in (\ref{eqn:even_gpfsum}) as $Z_N^{(\varphi)}[t]$, with $u,v$ appropriately replaced by $t$, then substitute in (\ref{eqn:2x2qdet}), we have
\begin{align}
\label{eqn:Z_NGinOE_mod} Z_N^{(\varphi)}[t]=\frac{2^{-N(N+1)/4}}{\prod_{l=1}^N\Gamma(l/2)}\prod_{j=0}^{N/2-1}r_j\; \qdet [\1_2+\mathbf{K}(\mathbf{t}-\1_2)],
\end{align}
which is analogous (\ref{eqn:ZN_integ_op}). We must also rewrite (\ref{eqn:fnal_diff_correln}) as
{\small
\begin{align}
\label{eqn:GinOE_fnaldiff_mod} \rho_{(m)}(X_1,...,X_m)= &\frac{1}{Z_N^{(\varphi)}[t]}\frac{\delta^{m}}{\delta t(X_1)\cdot\cdot\cdot \delta t(X_{m})}Z_N^{(\varphi)}[t]{\Big |}_{t=1},
\end{align}
}where $m=n_1+n_2$. Substitution of (\ref{eqn:Z_NGinOE_mod}) into (\ref{eqn:GinOE_fnaldiff_mod}), and then replacing $X_i$ with $x_j$ real or $w_j$ non-real complex as obliged then gives Proposition \ref{prop:GinOE_evencorrelns}.

\begin{remark}
It is clear that both methods can be generalised to a higher number of eigenvalue domains (here we have only $\mathbb{R}$ and $\mathbb{R}_{2}^{+}$). To use the $4\times 4$ method, an extra 2 rows and/or columns are added to the matrices $\bE$ and $\Omega$ in (\ref{eqn:omega_E}) for each new domain. For $l$ domains you would finish up with a $2l\times 2l$ kernel, with which to use with a suitably expanded Fredholm operator. While for the $2\times 2$ method, we may appropriately redefine the functions and measure to include all the cases. Of the two methods, the first seems the more transparent, where each row and column can be identified with a particular domain, however the second highlights the universal nature of the problem.
\end{remark}

\subsection{Eigenvalue correlations for $N$ odd}
\label{sec:GinOE_odd}

Roughly speaking, the method of integration over alternate variables and the evenness of Pfaffians are technically why $N$ odd requires a separate treatment, but the conceptual hurdle comes from the requirement that there must be at least one real eigenvalue in a real, odd-sized matrix. This also highlights one reason why the odd $\beta=2$ and $\beta=4$ Ginibre ensembles have not presented significantly more difficulties than their even counterparts --- in these cases the sets of matrices with real eigenvalues have measure zero in the eigenvalue support.

The odd case was first successfully dealt with in \cite{sommers_and_w2008} by invoking artificial Grassmannians, shortly followed by \cite{FM09} where the authors demonstrated that the $N$ odd correlations can be obtained as a limiting case of the correlations for $N$ even, as demonstrated in Chapter \ref{sec:odd_from_even}. Lastly, in \cite{Sinc09} it was shown how one obtains the $N$ odd correlations by modifying the approach taken in \cite{b&s2009}. As with the GOE we will first use a modified form of this latter approach (using Fredholm operators and applying functional differentiation) in Chapter \ref{sec:Gin_odd_fdiff}, and then look at obtaining the odd case as a limit of the even case in Chapter \ref{sec:Gin_oddfromeven}. Although functional differentiation is also employed in \cite{sommers_and_w2008}, the particulars of their method are sufficiently outside the scope of this work that we shall not investigate them here.

\begin{definition}
\label{def:GinOE_kernel_odd}
Let $p_0,p_1,...$ be the skew-orthogonal polynomials (\ref{eqn:GinOE_sopolys}) and $r_0,r_1,...$ the corresponding normalisations, and let $\nu_j$ be as in (\ref{def:GOE_nu}) and $\bar{\nu}_j =\nu_j\big|_{u=1}$ as it was in Proposition \ref{prop:pf_integ_op_odd}. Define
\begin{align}
\nonumber S^{\odd}(\mu,\eta)&=2\sum_{j=0}^{\frac{N-1}{2}-1}\frac{1}{r_j}\Bigl[\hat{q}_{2j}(\mu)\hat{\tau}_{2j+1}(\eta)-\hat{q}_{2j+1}(\mu)\hat{\tau}_{2j}(\eta)\Bigr]+\kappa(\mu,\eta),\\
\nonumber D^{\odd}(\mu,\eta)&=2\sum_{j=0}^{\frac{N-1}{2}-1}\frac{1}{r_j}\Bigl[\hat{q}_{2j}(\mu)\hat{q}_{2j+1}(\eta)-\hat{q}_{2j+1}(\mu)\hat{q}_{2j}(\eta)\Bigr],\\
\nonumber \tilde{I}^{\odd}(\mu,\eta)&=2\sum_{j=0}^{\frac{N-1}{2}-1}\frac{1}{r_j}\Bigl[\hat{\tau}_{2j}(\mu)\hat{\tau}_{2j+1}(\eta)-\hat{\tau}_{2j+1}(\mu)\hat{\tau}_{2j}(\eta)\Bigr]+\epsilon(\mu,\eta)+\theta(\mu,\eta),
\end{align}
where $\epsilon(\mu,\eta)$ is from Definition \ref{def:GinOE_kernel} and
\begin{align}
\nonumber \hat{p}_j(\mu)&=p_j(\mu)-\frac{\bar{\nu}_{j+1}}{\bar{\nu}_N}p_{N-1}(\mu),\\
\nonumber \hat{q}_j(\mu) &= e^{-\mu^2/2}\hspace{2pt}\sqrt{\mathrm{erfc}(\sqrt{2}|\mathrm{Im}(\mu)|)}\hspace{2pt}\hat{p}_j(\mu),\\
\nonumber \hat{\tau}_j(\mu) &= 
\left\{ 
\begin{array}{ll}
-\frac{1}{2}\int_{-\infty}^{\infty}\mathrm{sgn}(\mu-z)\hspace{3pt}\hat{q}_j(z)\hspace{3pt}dz, & \mu\in \mathbb{R},\\
i\hat{q}_j(\bar{\mu}),  & \mu\in \mathbb{R}_2^+,
\end{array}
\right.\\
\nonumber \kappa(\mu,\eta) &= 
\left\{ 
\begin{array}{lll}
q_{N-1}(\mu)/\bar{\nu}_N, & \eta\in \mathbb{R},\\
0,  & \mathrm{otherwise},\\
\end{array}
\right.\\
\nonumber \theta(\mu,\eta)&=
\big(\chi_{(\eta\in\mathbb{R})}\tau_{N-1}(\mu)-\chi_{(\mu\in\mathbb{R})}\tau_{N-1}(\eta)\big)/\bar{\nu}_N,
\end{align}
with the indicator function $\chi_{(A)}=1$ for $A$ true and zero for $A$ false. Then, let
\begin{align}
\nonumber \bK_{\odd}(\mu,\eta)=\left[
\begin{array}{cc}
S^{\odd}(\mu,\eta) & -D^{\odd}(\mu,\eta)\\
\tilde{I}^{\odd}(\mu,\eta) & S^{\odd}(\eta,\mu)
\end{array}
\right].
\end{align}
\end{definition}
In Appendix \ref{app:GinOE_kernel_elts_odd} the kernel elements for $N$ odd are written out explicitly. As with the even kernel elements in Definition \ref{def:GinOE_kernel} the odd kernel elements also satisfy the inter-relationships in Lemma \ref{lem:Gin_s=d=i}.

\subsubsection{Functional differentiation method}
\label{sec:Gin_odd_fdiff}

As may be expected from the previous, both the functional differentiation methods, using either the $4\times 4$ or $2\times 2$ kernels, can be adapted to the odd case. We find for both that with $N$ odd the proof for the Fredholm operator form of the Pfaffian differs from the even case in exactly the way the odd case of the GOE differed from its respective even case. Explicitly, in order to find a Fredholm operator form of the Pfaffian in (\ref{eqn:odd_gpfsum}), we are led to consider an odd skew-diagonal matrix of the form (\ref{eqn:skew_diag_mat_odd}) and its inverse, which takes us outside the space of matrices that can be decomposed as $\bD\bZ_{2N}^{-1}$, with $\bD$ diagonal, and thus, outside the realm of Pfaffians and quaternion determinants. However, as in the GOE odd case, this involves only technical minutiae, and instead, the real difference between even and odd cases boils down to simply the inclusion of an additional column in the matrices $\bF$ and $\bE_{\varphi}$ (for the $4\times 4$ and $2\times 2$ methods respectively). Since the technique has already been demonstrated in the GOE case, we will only briefly consider these modified methods, starting with the $4\times 4$ kernel.

\begin{proposition}
\label{prop:GinOE_fred_op_odd}
With $x,y\in\mathbb{R}$ and $w,z\in\mathbb{C}\backslash\mathbb{R}$ then, using Definition \ref{def:GinOE_kernel_odd}, the $N$ odd analogue of Proposition \ref{prop:4x4_fred} is
\begin{align}
\label{eqn:Ginoddfred} \mathrm{Pf}\left[\begin{array}{cc}
[\alpha_{j,l}+\beta_{j,l}] & [\nu_j]\\
\left[-\nu_l\right]& 0\\
\end{array}\right]_{j,l=1,...,N}=\bar{\nu}_N\prod_{j=0}^{(N-1)/2-1}r_j \;\qdet [\1_4+\bK_{\mathrm{odd}}(\bt-\1_4)],
\end{align}
where $\bK_{\odd}(\bt-\1_4)$ is an integral operator with kernel
\nonumber \begin{align}
\left[\begin{array}{cc}
\bK_{\odd}(x,y) & \bK_{\odd}(x,z)\\
\bK_{\odd}(w,y) & \bK_{\odd}(w,z)
\end{array}\right] \left[\begin{array}{cccc}
u(y)-1 & 0 & 0 & 0\\
0 & u(y)-1 & 0 & 0\\
0 & 0 & v(z)-1 & 0\\
0 & 0 & 0 & v(z)-1\\
\end{array}\right].
\end{align}
\end{proposition}

\textit{Proof}: If we let
\begin{align}
\nonumber \mathcal{C}:=\left[\begin{array}{cc}
[\alpha_{j,l}+\beta_{j,l}] & [\nu_j]\\
\left[-\nu_l\right]& 0\\
\end{array}\right]_{j,l=1,...,N}
\end{align}
then we note that $\mathcal{C}^{(1)}:=\mathcal{C}\big|_{u=v=1}$ is of the form (\ref{eqn:skew_diag_mat_odd}) (although with $b_j=\bar{\nu}_{j}=0$ for $j$ even), and thus Corollary \ref{cor:qdet=pf} does not apply. So, as in the case of the GOE, we work with $(\Pf\:\mathcal{C})^2=\det\: \mathcal{C}$ instead of $\Pf\:\mathcal{C}$ itself. Note that
\begin{align}
\nonumber \mathcal{C}=\mathcal{C}^{(1)}-\mathcal{C}^{(\tau)},
\end{align}
where
 $\mathcal{C}^{(\tau)}_{j,l}:=(\alpha_{j,l}+\beta_{j,l})^{(\tau)}$ of (\ref{def:abtau}) for $1\leq j < k \leq N$, and $\mathcal{C}^{(\tau)}_{j,N+1}:=\bC_{j,N+1}^{(\tau)}$ of (\ref{eqn:oddCnu}). Now substitute
\begin{align}
\nonumber \bF_{\odd}:=\left[
\begin{array}{cccc}
\psi_1(y) & \cdot \cdot\cdot & \psi_N(y) & 0\\
\epsilon\psi_1[y] & \cdot \cdot\cdot & \epsilon\psi_N[y] & -1\\
\phi_1(z) & \cdot\cdot\cdot & \phi_N(z) & 0\\
\epsilon\phi_1[z] & \cdot\cdot\cdot & \epsilon\phi_N[z] & 0
\end{array}
\right]
\end{align}
for $\bF$ in (\ref{def:GinOE_om_F}) and define $\bA$ to be the $(N+1)\times 4$ integral operator with kernel\\$2(\mathcal{C}^{(1)})^{-1} (\tau(\sigma,\eta)\Omega_G\bF_{\odd})^T$, where the remaining notation is from Proposition \ref{prop:4x4_fred}. Then, with $\bB=\bF_{\odd}$, we apply (\ref{eqn:1+AB}) and obtain
\begin{align}
\label{eqn:GinOE_odd_ktilde} \1_{4}+\bB\bA=\left[\begin{array}{cc}
\tilde{\kappa}^{\odd}_{r,r} & \tilde{\kappa}^{\odd}_{r,c}\\
\tilde{\kappa}^{\odd}_{c,r} & \tilde{\kappa}^{\odd}_{c,c}
\end{array}\right],
\end{align}
where
{\small
\begin{align}
\nonumber \tilde{\kappa}_{r,r}&=\left[\begin{array}{c}
1-\sum_{j=1}^N\frac{2}{r_{\lfloor (j-1)/2 \rfloor}}\Big(\hat{\psi}_j\otimes \sigma \epsilon \hat{G}_j + \hat{\psi}_j \otimes \sigma\epsilon(\sigma \hat{G}_j) \Big)+\frac{1}{\bar{\nu}_N}\psi_N\otimes \sigma\\
\mho
\end{array}\right.\\
\nonumber &\qquad\qquad\qquad\qquad\left.\begin{array}{l}
\sum_{j=1}^N\frac{2}{r_{\lfloor (j-1)/2 \rfloor}}\hat{\psi}_j\otimes \sigma \hat{G}_j\\
1+\sum_{j=1}^N\frac{2}{r_{\lfloor (j-1)/2 \rfloor}}\epsilon \hat{\psi}_j\otimes \sigma \hat{G}_j+\frac{1}{\bar{\nu}_N}1\otimes\sigma\psi_N
\end{array}\right],\\
\nonumber \tilde{\kappa}_{r,c}&=\left[\begin{array}{lr}
-\sum_{j=1}^N\frac{2}{r_{\lfloor (j-1)/2 \rfloor}} \hat{\psi}_j\otimes \eta \epsilon \hat{H}_j & \sum_{j=1}^N\frac{2}{r_{\lfloor (j-1)/2 \rfloor}}\hat{\psi}_j\otimes \eta \hat{H}_j\\
-\sum_{j=1}^N\frac{2}{r_{\lfloor (j-1)/2 \rfloor}} \epsilon \hat{\psi}_j\otimes \eta\epsilon \hat{H}_j- \frac{1}{\bar{\nu}_N}1\otimes\eta\epsilon\phi_N & \sum_{j=1}^N\frac{2}{r_{\lfloor (j-1)/2 \rfloor}}\epsilon \hat{\psi}_j\otimes \eta \hat{H}_j
\end{array}\right],\\
\nonumber \tilde{\kappa}_{c,r}&=\left[\begin{array}{cc}
-\sum_{j=1}^N\frac{2}{r_{\lfloor (j-1)/2 \rfloor}}\hat{\phi}_j\otimes \sigma \epsilon \hat{G}_j +\frac{1}{\bar{\nu}_N}\psi_N\otimes \sigma & \sum_{j=1}^N\frac{2}{r_{\lfloor (j-1)/2 \rfloor}}\hat{\phi}_j\otimes \sigma \hat{G}_j\\
-\sum_{j=1}^N\frac{2}{r_{\lfloor (j-1)/2 \rfloor}}\epsilon \hat{\phi}_j\otimes \sigma\epsilon \hat{G}_j +\frac{1}{\bar{\nu}_N}\epsilon\phi_N\otimes\sigma & \sum_{j=1}^N\frac{2}{r_{\lfloor (j-1)/2 \rfloor}}\epsilon \hat{\phi}_j\otimes \sigma \hat{G}_j
\end{array}\right],\\
\nonumber \tilde{\kappa}_{c,c}&=\left[\begin{array}{cc}
1-\sum_{j=1}^N\frac{2}{r_{\lfloor (j-1)/2 \rfloor}}\hat{\phi}_j\otimes \eta \epsilon \hat{H}_j & \sum_{j=1}^N\frac{2}{r_{\lfloor (j-1)/2 \rfloor}}\hat{\phi}_j\otimes \eta \hat{H}_j\\
-\sum_{j=1}^N\frac{2}{r_{\lfloor (j-1)/2 \rfloor}}\epsilon \hat{\phi}_j\otimes \eta\epsilon \hat{H}_j & 1+\sum_{j=1}^N\frac{2}{r_{\lfloor (j-1)/2 \rfloor}}\epsilon \hat{\phi}_j\otimes \eta \hat{H}_j
\end{array}\right],
\end{align}
}with
\begin{align}
\nonumber \mho&:=-\sum_{j=1}^N\frac{2}{r_{\lfloor (j-1)/2 \rfloor}}\Big(\epsilon\hat{\psi}_j\otimes \sigma \epsilon \hat{G}_j + \epsilon\hat{\psi}_j \otimes \sigma\epsilon(\sigma \hat{G}_j) \Big)\\
\nonumber &+\frac{1}{\bar{\nu}_N}\Big(\epsilon\psi_N\otimes \sigma -1\otimes\sigma\epsilon(\sigma\psi_N) - 1\otimes\sigma\psi_N \Big),
\end{align}
and $\hat{\psi}_j,\hat{\phi}_j,\hat{G}_j$ and $\hat{H}_j$ are given by the `non-hat' versions from Proposition \ref{prop:4x4_fred} with $p_j$ replaced by $\hat{p}_j$ of Definition \ref{def:GinOE_kernel_odd}. Now we use (\ref{eqn:omg_decomp}) to factorise (\ref{eqn:GinOE_odd_ktilde}), and then apply Corollary \ref{cor:sqrtFreds} to obtain the result.

\hfill $\Box$

\begin{remark}
By comparing Propositions \ref{prop:pf_integ_op}, \ref{prop:pf_integ_op_odd}, \ref{prop:4x4_fred} and \ref{prop:GinOE_fred_op_odd} one can see why this method has been used: with minor variations it can be applied to both the even and odd cases of the GOE and the real, Ginibre ensemble, which has a pleasing symmetry. This also points to the future possibilities of easily generalising this method to systems with an arbitrary number of distinct particle species. But that, as they say, is another story (and ahead of known applications). (See Chapter \ref{sec:FW} for possible uses.)
\end{remark}

Substitution of (\ref{eqn:Ginoddfred}) into (\ref{eqn:odd_gpfsum}) yields
\begin{align}
\label{eqn:Z_NGinOE_mod_odd} Z_N^{\odd}[u,v]= \frac{2^{-N(N+1)/4}}{\prod_{l=1}^N \Gamma(l/2)} \; \bar{\nu}_N\prod_{j=0}^{(N-1)/2-1}r_j \;\qdet [\1_4+\bK_{\mathrm{odd}}(\bt-\1_4)],
\end{align}
and, as is now routine, the correlation functions are given upon substitution of (\ref{eqn:Z_NGinOE_mod_odd}) into (\ref{eqn:GinOEfnal_diff_correln}).

\begin{proposition}[\cite{sommers_and_w2008, FM09, Sinc09}]
\label{prop:Gin_oddcorrelns}
For the real Ginibre ensemble of odd-sized matrices the correlation functions for $n_1$ real eigenvalues and $n_2$ non-real complex conjugate pairs of eigenvalues are
\begin{align}
\nonumber &\rho_{(n_1,n_2)}(x_1,...,x_{n_1},w_1,...,w_{n_2})=\mathrm{qdet}\left[\begin{array}{cc}
\bK_{\odd}(x_i,x_j) & \bK_{\odd}(x_i,w_m)\\
\bK_{\odd}(w_l,x_j) & \bK_{\odd}(w_l,w_m)\\
\end{array}\right]_{i,j=1,...,n_1 \atop l,m=1,...,n_2}\\
\nonumber &=\Pf\left(\left[\begin{array}{cc}
\bK_{\odd}(x_i,x_j) & \bK_{\odd}(x_i,w_m)\\
\bK_{\odd}(w_l,x_j) & \bK_{\odd}(w_l,w_m)\\
\end{array}\right]_{i,j=1,...,n_1 \atop l,m=1,...,n_2}\bZ^{-1}_{2(n_1+n_2)}  \right), x_i\in \mathbb{R}, w_i \in \mathbb{R}_2^+.
\end{align}
\end{proposition}

\begin{remark}
To use the $2\times 2$ kernel method, the essential step is to replace $\bE_{\varphi}$ with
\begin{align}
\nonumber \bE_{\varphi,\odd}:=\left[
\begin{array}{cccc}
\varphi_1(Y) & \cdot \cdot\cdot & \varphi_N(Y)&0\\
\epsilon\varphi_1[Y] & \cdot \cdot\cdot & \epsilon\varphi_N[Y]&-1
\end{array}
\right],
\end{align}
in Proposition \ref{prop:GinOE_fred_op}, then Proposition \ref{prop:Gin_oddcorrelns} follows by adapting the odd method used in Proposition \ref{prop:pf_integ_op_odd}.
\end{remark}

\subsubsection{Odd from even}
\label{sec:Gin_oddfromeven}

Here we can, more or less, directly apply the method of Chapter \ref{sec:odd_from_even} to precipitate the odd case from the even case for the real Ginibre ensemble. In fact, the method was originally presented in \cite{FM09} with the real Ginibre case in mind, only using the simpler GOE case to illustrate the technique.

To ensure success, we look for a factorisation analogous to (\ref{eqn:jpdfOEfactorisation}). Using (\ref{eqn:GinOEjpdf}) we separate out the dependence on the eigenvalue $\lambda_1$ to obtain
\begin{align}
\nonumber Q_{N,k}(\Lambda,W)&=C_{N,k}\;e^{-\lambda^2_1/2} \prod^{k}_{j=2} |\lambda_1- \lambda_j|\prod^{(N-k)/2}_{j=1}|\lambda_1-w_j|| \lambda_1-\bar{w}_j|\\
\nonumber &\times \prod_{j=2}^{k}e^{-\lambda_j^2/2} \prod_{j=1}^{(N-k)/2} e^{-(w_j^2+\bar{w}_j^2)/2} \mathrm{erfc}(\sqrt{2}|\mathrm{Im}(w_j)|)|\Delta(\tilde{\Lambda}_1,W)|,
\end{align}
where $\tilde{\Lambda}_j$ is the $\Lambda$ of Proposition \ref{prop:GinOE_eval_jpdf} without $\lambda_j$. Now we let $\lambda_1$ tend to infinity to find
\begin{align}
\label{eqn:sortfactor} \fullsub{Q_{N,k}(\Lambda,W)}{\sim} {\lambda_1\to\infty}{\frac{C_{N,k}} {C_{N-1,k-1}} e^{-\lambda_1^2/2}\lambda_1^{N-1} Q_{N-1,k-1}(\tilde{\Lambda}_1,W)},
\end{align}
which is the required analogue of (\ref{eqn:jpdfOEfactorisation}). Define
\begin{align}
\nonumber g_{N,k}(\lambda_1):=\frac{C_{N,k}}{C_{N-1,k-1}}e^{-\lambda_1^2/2} \lambda_1^{N-1}= \frac{e^{-\lambda_1^2/2} \lambda_1^{N-1}}{k\:2^{N/2}\Gamma(N/2)},
\end{align}
then
\begin{align}
\nonumber \frac{\delta}{\delta u(\lambda_1)}Z_{k,(N-k)/2}[u,v] \mathop{\sim}\limits_{\lambda_1 \to \infty} k\; g_{N,k}(\lambda_1) \; Z_{k-1,(N-k)/2}[u,v],
\end{align}
recalling $Z_{k,(N-k)/2}[u,v]$ from (\ref{eqn:GinOE_gpf_even}). So with $Z_N[u,v]$ from (\ref{eqn:summedup}) we have
\begin{align}
\nonumber \frac{\delta}{\delta u (\lambda_1)} Z_N[u,v] &= \sum_{k=0}^N {}^{\sharp} \frac{\delta}{\delta u(\lambda_1)}Z_{k,(N-k)/2}[u,v]\\
\nonumber &\mathop{\sim}\limits_{\lambda_1 \to \infty} \sum_{k=1}^{N-1} {}^{\sharp}\; k\; g_{N,k}(\lambda_1)\; Z_{k-1,(N-k)/2}[u,v]\\
\nonumber &=\frac{e^{-\lambda_1^2/2}\lambda_1^{N-1}}{2^{N/2}\Gamma(N/2)} \sum_{k=1}^{N-1} {}^{\sharp}\; Z_{k-1,(N-k)/2}[u,v]\\
\label{eqn:ZkNsim} &=\frac{e^{-\lambda_1^2/2} \lambda_1^{N-1}}{2^{N/2} \Gamma(N/2)} Z_{N-1}[u,v],
\end{align}
where the $\sharp$ indicates that the sum is only over those values $k$ with the same parity as the upper terminal. 

Recalling from (\ref{eqn:GinOEfnal_diff_correln}) that
\begin{align}
\rho_{(1,0)}(\lambda_1)=\frac{1}{Z_N[u,v]} \frac{\delta} {\delta u (\lambda_1)} Z_N[u,v]\Big|_{u=v=1}
\end{align}
and that $Z_N[1,1]=1$ for all $N$, it follows from (\ref{eqn:ZkNsim}) that with $g_N(\lambda) :=
k g_{N,k}(\lambda)$
\begin{align}
\nonumber \rho_{(1,0)}^N(\lambda_1) \sim g_N(\lambda_1),
\end{align}
which is the analogue of Lemma \ref{lem:xm_to_infty}, and so we have
\begin{align}
\nonumber Q_{N,k}(\Lambda,W) \mathop{\sim}\limits_{\lambda_1 \to \infty}
\rho_{(1,0)}(\lambda_1) Q_{N-1,k-1}(\tilde{\Lambda}_1,W),
\end{align}
which is the sought factorisation. Now, making use of (\ref{eqn:GinOEfnal_diff_correln}), in the general case we obtain
\begin{align}
\nonumber \rho_{(n_1,n_2)}^{(N)}(x_1,...,x_{n_1}, w_1,...,w_{n_2}) &\mathop{\sim}\limits_{x_1 \to \infty} \rho_{(1,0)}^{(N)}(x_1)\;\rho_{(n_1-1,n_2)}^{(N-1)}(x_2,...,x_{n_1},w_1,...,w_{n_2}),
\end{align}
the analogue of (\ref{eqn:finalOEfactorisation}). So, as with the GOE, knowledge of  the correlation functions for $N$ even enables us to find the correlation functions for a system of $N-1$ eigenvalues by factoring out the density of the largest real eigenvalue and taking the limit.

Using the correlation function in Pfaffian form (recall the Pfaffian kernel (\ref{def:GinPfK}), which we call $K_N$ here), we shift the rows and columns corresponding to the eigenvalue $x_k$ to the far right and bottom of the matrix as so
\begin{align}
\label{eqn:GinHRCR} \mathrm{Pf}\left[\begin{array}{ccc}
K_N(x_i,x_j) & K_N(x_i,w_m) & K_N(x_i,x_k)\\
K_N(w_l,x_j) & K_N(w_l,w_m) & K_N(w_l,x_k)\\
K_N(x_k,x_j) & K_N(x_k,w_m) & K_N(x_k,x_k)\\
\end{array}\right]_{\begin{subarray}{l} i,j=1,...,n_1-1 \\ l,m=1,...,n_2 \end{subarray}}.
\end{align}
Since this involves shifting 2 rows and 2 columns an even number of times the Pfaffian is unchanged. The sizes of the submatrices in (\ref{eqn:GinHRCR}) are:
\begin{itemize}
\item{top left: $2(n_1 -1)\times 2(n_1 -1)$; top centre: $2(n_1 -1)\times 2(n_2)$;\\top right: $2(n_1 -1)\times 2$.}
\item{centre left: $2(n_2)\times 2(n_1-1)$; centre: $2(n_2)\times 2(n_2)$;\\centre right: $2(n_2)\times 2$.}
\item{bottom left: $2\times 2(n_1 -1)$; bottom centre: $2\times 2(n_2)$;\\bottom right: $2\times 2$.}
\end{itemize}
We now perform the same row and column reduction as in (\ref{eqn:Pf3}), finding that the `starred' kernel elements are slightly complicated by the existence of both real and complex eigenvalues. Then, on taking the limit $x_k\to\infty$ the starred kernel elements reduce to their odd counterparts, with $N$ replaced by $N-1$, and we recover Proposition \ref{prop:Gin_oddcorrelns}. The details are ommitted as the procedure is a straightforward modification of that described in Chapter \ref{sec:odd_from_even}.

\subsection{Correlation kernel elements and large $N$ limits}
\label{sec:Ginkernelts}

The summations in the kernel elements of Definitions \ref{def:GinOE_kernel} and \ref{def:GinOE_kernel_odd} can be performed explicitly on substitution of the skew-orthogonal polynomials (\ref{eqn:GinOE_sopolys}) \cite{FN08, b&s2009}. We list the results for $S(\mu,\eta)$ for each of the four combinations of real and complex eigenvalues, but first we recall the definitions
\begin{align}
\nonumber \Gamma(N,x):=&\int_{x}^{\infty}t^{N-1}e^{-t}dt&\gamma(N,x):=&\int_{0}^{x}t^{N-1}e^{-t}dt\\
\label{def:Gammas} =&\Gamma (N) e^{-x} \sum_{j=1}^N \frac{x^{j-1}}{(j-1)!}, &=&\Gamma(N)-\Gamma(N,x),
\end{align}
which are called the \textit{upper} and \textit{lower incomplete gamma functions} respectively. By substituting the polynomials (\ref{eqn:GinOE_sopolys}) into $S(\mu,\eta)$ (with $\mu=x,\eta=y\in\mathbb{R}$) and performing some manipulations involving the formulae
\begin{align}
\nonumber \int_{-\infty}^x e^{-u^2/2}u^{2k+1}du&=-(2k)!!\; e^{-x^2/2}\sum_{l=0}^k \frac{x^{2l}}{(2l)!!},\\
\nonumber \int_{-\infty}^x e^{-u^2/2}u^{2k}du&=(2k-1)!!\int_{-\infty}^x e^{-u^2/2}du\\
\nonumber &-(2k-1)!! \; e^{-x^2/2}\sum_{l=0}^k \frac{x^{2l-1}}{(2l-1)!!},
\end{align}
we obtain a closed form of $S_{r,r}(x,y)$. The other kernel elements can be similarly summed and we have
{\small
\begin{align}
\label{eqn:Ginsummed}
\begin{split}
S_{r,r}(x,y)&=\frac{1}{\sqrt{2\pi}}\left[e^{-(x-y)^2/2}\frac{\Gamma(N-1,xy)}{\Gamma(N-1)}+2^{(N-3)/2}e^{-x^2/2}x^{N-1}\sgn(y)\frac{\gamma(\frac{N-1}{2},y^2/2)}{\Gamma(N-1)}\right],\\
S_{r,c}(x,w)&=\frac{ie^{-(x-\bar{w})^2/2}}{\sqrt{2\pi}}(\bar{w}-x)\frac{\Gamma(N-1,x\bar{w})}{\Gamma(N-1)}\sqrt{\mathrm{erfc}(\sqrt{2}|\mathrm{Im}(w)|)},\\
S_{c,r}(w,x)&=\frac{1}{\sqrt{2\pi}}\left[e^{-(w-x)^2/2}\frac{\Gamma(N-1,wx)}{\Gamma(N-1)}+2^{(N-3)/2}e^{-w^2/2}w^{N-1}\sgn(x)\frac{\gamma(\frac{N-1}{2},x^2/2)}{\Gamma(N-1)}\right]\\
&\times\sqrt{\mathrm{erfc}(\sqrt{2}|\mathrm{Im}(w)|)},\\
S_{c,c}(w,z)&=\frac{ie^{-(w-\bar{z})^2/2}}{\sqrt{2\pi}}(\bar{z}-w)\frac{\Gamma(N-1,w\bar{z})}{\Gamma(N-1)}\sqrt{\mathrm{erfc}(\sqrt{2}|\mathrm{Im}(w)|)}\sqrt{\mathrm{erfc}(\sqrt{2}|\mathrm{Im}(z)|)},
\end{split}
\end{align}
}where we again use the convention that $x,y$ are real and $w,z$ are non-real complex. The remaining kernel elements $D(\mu,\eta)$ and $\tilde{I}(\mu,\eta)$ can also be written in such a form by direct summation or by using Lemma \ref{lem:Gin_s=d=i}. (The one caveat to the previous statement is that there does not seem to be a closed form of $I_{r,r}(x,y)$.) Note that the equations in $(\ref{eqn:Ginsummed})$ are independent of the parity of $N$, and so we suspect that they hold in both the even and odd cases. This can be checked by explicitly performing the sums in Definition \ref{def:GinOE_kernel_odd} using the skew-orthogonal polynomials; one then obtains the same set of equations explicitly. See Appendix \ref{app:Ginsummed} for the full set of summed kernel elements.

Recall from (\ref{eqn:1pt_correln}) that for the GOE the density --- which is identical to the $1$-point correlation function --- of real eigenvalues was given by $S(x,x)$; similarly for the real Ginibre ensemble we see from (\ref{eqn:Ginsummed}) that the density of real eigenvalues is given by \cite{eks1994}
\begin{align}
\nonumber \rho^r_{(1)}(x)&=S_{r,r}(x,x)\\
\label{eqn:Ginreal_density}&=\frac{1}{\sqrt{2\pi}}\left[\frac{\Gamma(N-1,x^2)}{\Gamma(N-1)} +2^{(N-3)/2} e^{-x^2/2} |x|^{N-1}\frac{\gamma(\frac{N-1}{2},x^2/2)} {\Gamma(N-1)}\right],
\end{align}
and the density of complex eigenvalues is given by \cite{Ed97}
\begin{align}
\nonumber \rho^c_{(1)}(w)&=S_{c,c}(w,w)\\
\label{eqn:Gincompx_density}&=\sqrt{\frac{2}{\pi}}\;ve^{2v^2}\frac{\Gamma(N-1,|w|^2)}{\Gamma(N-1)}\erfc(\sqrt{2}v),
\end{align}
where $w=u+iv$.

In Chapter \ref{sec:pnk} we discussed the probability $p_{N,k}$ of obtaining $k$ real eigenvalues from an $N\times N$ real Ginibre matrix, and we mentioned that an interesting related quantity is $E_N$, the expected number of real eigenvalues. By integrating (\ref{eqn:Ginreal_density}) over the real line we have a simpler method of calculating $E_N$ than using $(\ref{eqn:Ginxnreals})$, and, with a result from \cite[3.196.1]{GraRyz2000}, we find
\begin{align}
\nonumber E_N&=\frac{1}{2}+\sqrt{\frac{2}{\pi}}\frac{\Gamma(N+1/2)}{\Gamma(N)}{}_2F_1(1,-1/2;N;1/2)\\
\label{eqn:Ginxnreals2} &=\frac{1}{2} + \sqrt{\frac{2N}{\pi}}\left(1-\frac{3}{8N}-\frac{3}{128N^2}+\frac{27}{1024 N^3}+\frac{499}{32768 N^4}+O(1/N^5) \right),
\end{align}
which was first identified in \cite[Corollaries 5.1 and 5.2]{eks1994}. Clearly, from (\ref{eqn:Ginxnreals2}) we see that
\begin{align}
\label{eqn:bigNEN} E_N\sim\sqrt{\frac{2N}{\pi}}
\end{align}
for large $N$. In \cite{FN07} the authors calculated the large $N$ variance in the number of real eigenvalues
\begin{align}
\label{eqn:Gvar} \sigma_N^2=(2-\sqrt{2})E_N,
\end{align}
which, in Chapter \ref{sec:Ssops}, we will compare to the analogous result for the real spherical ensemble, finding that they are identical.

We can also find the large $N$ limits of the kernel elements. Firstly we look for the limit in the bulk: let $N\to\infty$ with $x,y,w,z$ fixed. Noting from (\ref{def:Gammas}) that $\Gamma(N,x) \to \Gamma(N)$ for large $N$ (and so $\gamma(N,x)\to 0$) we have \cite{FN07,b&s2009}
\begin{align}
\nonumber S^{\mathrm{bulk}}_{r,r}(x,y)&=\frac{1}{\sqrt{2\pi}}e^{-(x-y)^2/2},\\
\nonumber S^{\mathrm{bulk}}_{r,c}(x,w)&=\frac{i}{\sqrt{2\pi}}\sqrt{\mathrm{erfc}(\sqrt{2}|\mathrm{Im}(w)|)}\;(\bar{w}-x)e^{-(x-\bar{w})^2/2},\\
\nonumber S^{\mathrm{bulk}}_{c,r}(w,x)&=\frac{1}{\sqrt{2\pi}}\sqrt{\mathrm{erfc}(\sqrt{2}|\mathrm{Im}(w)|)}\;e^{-(w-x)^2/2},\\
\label{eqn:Ginbulk} S^{\mathrm{bulk}}_{c,c}(w,z)&=\frac{i}{\sqrt{2\pi}}\sqrt{\mathrm{erfc}(\sqrt{2}|\mathrm{Im}(w)|)}\sqrt{\mathrm{erfc}(\sqrt{2}|\mathrm{Im}(z)|)}\;(\bar{z}-w)e^{-(\bar{z}-w)^2/2}.
\end{align}
We will see in Chapter \ref{sec:circlaw} that the eigenvalue support tends to a disk centred at the origin with radius $\sqrt{N}$ and so we can calculate the limiting kernel elements at the real edge (the edge of the support on the real line) by taking $X=x-\sqrt{N}$ (and similarly for $Y,W,Z$). Then, with the following asymptotic forms for large $N$
\begin{align}
\nonumber \gamma(N-j+1,N)&\sim\frac{\Gamma(N-j+1)}{2}\Big( 1+\erf(j/\sqrt{2N}) \Big),\\
\nonumber \Gamma(N-j+1)&\sim \sqrt{2\pi}N^{N-j+1/2}e^{-N}e^{-j^2/2N},\\
\nonumber \Big( 1+\frac{x}{\sqrt{N}}\Big)^{N-1}&\sim e^{x\sqrt{N}-x^2/2},
\end{align}
we have \cite{FN07, b&s2009}
\begin{align}
\nonumber S^{\mathrm{edge}}_{r,r}(X,Y)&=\frac{1}{\sqrt{2\pi}}\Big[ \frac{e^{-(X-Y)^2/2}}{2} \erfc\Big(\frac{X+Y}{\sqrt{2}}\Big)+\frac{e^{-X^2}}{2\sqrt{2}}(1+\erf \:Y)\Big],\\
\nonumber S^{\mathrm{edge}}_{r,c}(X,W)&=\frac{i}{2\sqrt{2\pi}}\sqrt{\mathrm{erfc}(\sqrt{2}|\mathrm{Im}(W)|)}\;(\overline{W}-X)e^{-(X-\overline{W})^2/2}\erfc \Big( \frac{X+\overline{W}}{\sqrt{2}}\Big),\\
\nonumber S^{\mathrm{edge}}_{c,r}(W,X)&=\frac{1}{\sqrt{2\pi}}\Big[\frac{e^{-(W-X)^2/2}}{2}\sqrt{\mathrm{erfc}(\sqrt{2}|\mathrm{Im}(W)|)}\;\erfc \Big( \frac{X+W}{\sqrt{2}}\Big)\\
\nonumber &+\frac{e^{-W^2}}{2\sqrt{2}}(1+\erf\:X)\Big],\\
\nonumber S^{\mathrm{edge}}_{c,c}(W,Z)&=\frac{i}{2\sqrt{2\pi}}\sqrt{\mathrm{erfc}(\sqrt{2}|\mathrm{Im}(W)|)}\sqrt{\mathrm{erfc}(\sqrt{2}|\mathrm{Im}(Z)|)}\\
\label{eqn:Ginedge} &\times(\overline{Z}-W)e^{-(\overline{Z}-W)^2/2}\erfc \Big( \frac{W+\overline{Z}}{\sqrt{2}}\Big).
\end{align}
For a full list of the limiting kernel elements in the bulk and at the edge see (\ref{eqn:Ginbulkall}) and (\ref{eqn:Ginedgeall}) in Appendix \ref{app:Ginsummed}.

These edge and bulk results have been taken somewhere near the real line where the effect of the non-zero density on the real line can still be felt. If one were to look at the limits away from the real line, then one expects this effect to vanish. Indeed this is what happens; from \cite{b&s2009} the limiting bulk correlations for the complex eigenvalues away from the real line are given by
\begin{align}
\label{eqn:Gincbulk} \lim_{N\to\infty}\rho_{(n)}^{c}(w_1,...,w_n)=\det\left[\frac{1}{\pi}e^{-(w_j-\bar{w}_l)^2/2}\right]_{j,l=1,...,n},
\end{align}
and for the complex edge
\begin{align}
\label{eqn:Gincedge} \lim_{N\to\infty}\rho_{(n)}^{c}(W_1,...,W_n)=\det\left[\frac{1}{\pi}e^{-(W_j-\overline{W}_l)^2/2}\erfc\Big(\frac{W_j\bar{u}+\overline{W}_l u}{\sqrt{2}}\Big)\right]_{j,l=1,...,n},
\end{align}
where $W_j=w_j-u\sqrt{N}$ with $u\in\mathbb{C}$ and $|u|=1$ so that $u$ is just rotation around the edge. These correlations are identical to those of the complex Ginibre ensemble, which we expect since (\ref{eqn:Gincbulk}) and (\ref{eqn:Gincedge}) represent the eigenvalue correlations away from the effect of any real eigenvalues, and the complex Ginibre ensemble has no real eigenvalues at all. This naturally leads us onto the topic of universality and the circular law, which we review in the next section.

\subsubsection{Circular law}
\label{sec:circlaw}

Recall that in Chapter \ref{sec:GOE_sums} we discussed the semi-circle law, Proposition \ref{prop:wssl}, which states that for the Hermitian ensembles, with entries drawn from any mean zero, finite variance probability distribution, the eigenvalue density tends towards a semi-circle. There is an analogous result for non-Hermitian matrices called the \textit{circular law}.

If one normalises the eigenvalues by dividing by $\sqrt{N}$ (label these eigenvalues as $\tilde{w}$) then it turns out that the density of complex eigenvalues tends to uniformity on the unit circle. Further, when we recall (\ref{eqn:bigNEN}), which shows the expected number of real eigenvalues goes only as $\sqrt{N}$, then we see that the distribution of general eigenvalues for the real Ginibre ensemble tends to uniformity on the unit disk.

\begin{proposition}[\cite{Ed97}]
\label{prop:Gincirclaw}
The limiting density of eigenvalues, scaled by $1/\sqrt{N}$, in the real Ginibre ensemble is
\begin{align}
\label{eqn:Gcirclaw} \rho^c_{(1)}(\tilde{w})=\pi^{-1}\chi_{|\tilde{w}|<1},
\end{align}
where $\chi_{x}$ is the indicator function.
\end{proposition}

\textit{Proof}: With $w=u+iv$ we change variables $\tilde{u}=u/\sqrt{N}, \tilde{v}=v/\sqrt{N}$ in (\ref{eqn:Gincompx_density}) giving 
\begin{align}
\nonumber \tilde{\rho}_{(1)}(\tilde{w})=N\sqrt{\frac{2N}{\pi}}\; \tilde{v} e^{2N\tilde{v}^2}\frac{\Gamma(N-1,N(\tilde{u}^2+\tilde{v}^2))}{\Gamma(N-1)}\erfc(\sqrt{2N}|\tilde{v}|),
\end{align}
where we have multiplied by $N$ (the Jacobian of the change of variables). We are therefore looking to calculate
\begin{align}
\label{eqn:normdens1} \rho^c_{(1)}(\tilde{w})=\lim_{N\to\infty} \frac{\tilde{\rho}_{(1)}(\tilde{w})}{N}.
\end{align}

Writing out the incomplete Gamma function using the definition
\begin{align}
\nonumber \Gamma(N-1,N\alpha)=\int_{N\alpha}^{\infty} t^{N-2} e^{-t}dt,
\end{align}
we see that the integral will be dominated by the maximum of the integrand in the large $N$ limit. Rewriting this integrand as
\begin{align}
\nonumber e^{(N-2) (\log t)-t}
\end{align}
we maximise this exponent by differentiating to find $t_{\max}=N-2\sim N$. So if $\alpha<1$, the maximum falls inside the interval of integration and the incomplete Gamma function tends to the complete Gamma function for large $N$. If $\alpha>1$, then it will be of lower order than the complete function. This gives us
\begin{align}
\label{def:uGa} \frac{\Gamma(N-1,N\alpha)}{\Gamma(N-1)}\to \left\{\begin{array}{cc}
1, &\alpha<1,\\
0, &\alpha>1.
\end{array}\right.
\end{align}
Lastly, using \cite[7.1.13]{AS1965} we find that for general $\tilde{v}$
\begin{align}
\label{eqn:bigerfc} \sqrt{N}\tilde{v}e^{2N\tilde{v}^2} \erfc(\tilde{v}\sqrt{2N})\mathop{\sim} \limits_{N\to\infty} \frac{1}{\sqrt{2\pi}}.
\end{align}
Substitution of these results into (\ref{eqn:normdens1}) gives (\ref{eqn:Gcirclaw}).

\hfill $\Box$

One can get an immediate sense of this result by looking at the simulation results in Figure \ref{fig:GinOE_eval_plot}. The complex eigenvalues are contained in a disk of radius roughly $\sqrt{N}$. The uniformity is clearly spoiled by the eigenvalues on the real line, but, as discussed above, these real eigenvalues have diminishing effect as $N$ becomes large.

Proposition \ref{prop:Gincirclaw} is specific to the real Ginibre ensemble, however it forms part of a wider class of results collectively known as the \textit{circular law}. The origins of the circular law can be traced back to (at least) the 1960's (although there are claims that it was being discussed a decade earlier \cite{Bai1997}). In \cite[Chapter 12.1]{Mehta1967} Mehta shows that the eigenvalue density for the complex Ginibre ensemble (complex asymmetric matrices) approaches uniformity inside the disk of radius $\sqrt{N}$ and zero outside the disk. Edelman showed the same is true for the real Ginibre ensemble \cite{Ed97} (Proposition \ref{prop:Gincirclaw}). One specific  version of the circular law then states that for iid Gaussian matrices the support of the normalised eigenvalues $\tilde{w}=w/\sqrt{N}$ approaches the unit disk and the density inside the disk approaches uniformity as $N\to\infty$.

The name of the law is often prefixed by that of Girko \cite{Girko1985a, Girko1985b} who attempted to relax the Gaussian constraint; wanting to show that the eigenvalues of matrices with iid entries drawn from any mean zero, finite variance distribution will display the same density in the large $N$ limit. However, the consensus view seems to be that there were sufficiently many errors in Girko's work that the proof did not withstand scrutiny. By making some assumptions on the moments of the distribution Bai \cite{Bai1997} built on Girko's work and furnished a proof under such restrictions. Further refinements were made by G\"{o}tze and Tikhomirov \cite{GT2010}, Pan and Zhou \cite{PZ2010} and Tao and Vu \cite[Corollary 1.17]{TVK10}. We quote the latter result here.
\begin{proposition}[Circular law]
\label{prop:circlaw}
For an $N\times N$ random matrix with iid entries from a distribution with finite mean and variance $1$ the distribution of the normalised eigenvalues approaches the uniform distribution on the unit disk as $N\to\infty$.
\end{proposition}

The proof of Proposition \ref{prop:circlaw} in \cite{TVK10} relies on first establishing the universality of the limiting distribution of eigenvalues. In this context, universality means that the eigenvalue distribution in the large $N$ limit is independent of the probability distribution of the matrix elements. Having established this universality the authors use the circular result for the case of Gaussian distributed elements, which, as mentioned above, was contained in \cite{Mehta1967}, to prove the general circular law.

Note that if one has the circular law in advance, then it gives us a simple way of finding the asymptotic behaviour of $E_N$. First we see that, with $x=y$, (\ref{eqn:Ginbulk}) implies that the limiting density of real eigenvalues is $1/\sqrt{2\pi}$. Since the circular law tells us that general eigenvalues are only supported on $(-\sqrt{N},\sqrt{N})$ as $N\to \infty$, by directly integrating the limiting density over the real line we obtain (\ref{eqn:bigNEN}).

By similar reasoning to Proposition \ref{prop:Gincirclaw} we can find the limiting density of just the real eigenvalues, scaled into the unit disk by letting $\tilde{x}=x/\sqrt{N}$,
\begin{align}
\label{eqn:Grlimdens} \rho_{(1)}^r(\sqrt{N}\tilde{x}) = \sqrt{\frac{N}{2\pi}}.
\end{align}
Remarkably, we see that the real eigenvalues, despite being a lower-order contribution to the system, are also distributed uniformly. We will compare (\ref{eqn:Grlimdens}) to the analogous results in the real spherical (Chapter \ref{sec:SOE}) and real truncated ensembles (Chapter \ref{sec:truncs}) and find the same behaviour.

\newpage

\section{Partially symmetric real Ginibre ensemble}
\setcounter{figure}{0}
\label{sec:tG}

From the previous chapter we know the limiting density of real eigenvalues for the real Ginibre ensemble is constant on the interval $(-\sqrt{N},\sqrt{N})$ (set $x=y$ in the first equation of (\ref{eqn:Ginbulk})), and zero elsewhere. We also know that in the GOE, the limiting density of (real) eigenvalues is supported on $(-\sqrt{2N},\sqrt{2N})$, but is decidedly not constant (recall (\ref{eqn:wssl})); it is a semi-circle. One can imagine that as the symmetry constraint is relaxed the eigenvalue density transitions between these two regimes. Another transition that we expect to see is the probability of all real eigenvalues $p_{N,N}$ decreasing from $1$ in the GOE to that given in (\ref{eqn:GinOEpNN}) for the real Ginibre ensemble. One of the goals of this chapter is to make these ideas concrete by analysing the \textit{partially symmetric real Ginibre ensemble}.

This ensemble was analysed in \cite{SCSS1988} (building on the work of \cite{CS1987}) where they identified the \textit{elliptical law}, which describes the eigenvalue density as its behaviour changes from that of the circular law (\ref{eqn:Gcirclaw}) to that of the semi-circle law (\ref{eqn:wssl}), for Gaussian real asymmetric matrices. (The elliptical law seems to have been first discussed by Girko \cite{Girko1985c, Girko1986}.) The eigenvalue jpdf of the partially symmetric real ensemble was presented in \cite{LS91} along with the asymmetric (real Ginibre) specialisation. Refinements to the analysis were presented in \cite{Efe97}, where it was shown that the density of eigenvalues contains a singular delta function term corresponding to the non-zero density on the real line. The full correlation functions, in the case of $N$ even were contained in \cite{FN08} by generalising the orthogonal polynomial method used for the real Ginibre ensemble. (See \cite{FS03, KS2009} for reviews.)

\subsection{Element distribution}

Any matrix $\bX$ can be decomposed into a sum of symmetric and anti-symmetric matrices; we choose the decomposition
\begin{align}
\label{def:tau_mats} \bX=\frac{1}{\sqrt{b}}\left( \bS+\sqrt{c}\bA \right),
\end{align}
where $c:=(1-\tau)/(1+\tau)$, $-1<\tau<1$ and $b\in\mathbb{R}$. With $\tau\to 1$ then $c\to 0$ and so we obtain symmetric matrices while with $\tau\to-1$ we see that $c\to\infty$, however by suitably scaling $b$ we can control this behaviour and access anti-symmetric matrices. The matrix $\bX$ will be completely asymmetric when $\tau=0$, meaning $c=1$.

For our purposes in this chapter we will take $\bS$ to be a GOE matrix, that is a symmetric real, Gaussian matrix with iid elements distributed as in (\ref{eqn:GOE_el_dist}). The matrix $\bA$ is the anti-symmetric analogue with iid elements
\begin{align}
\label{def:antisymdist} \frac{1}{\sqrt{\pi}}e^{-x_{jl}^2},\quad j<l,
\end{align}
with the remaining elements determined by anti-symmetry. So an ensemble of the matrices (\ref{def:tau_mats}) with $\tau\to1$ recovers the GOE, while with $\tau=0$ we have the real Ginibre ensemble. We also have an anti-symmetric Gaussian ensemble in the limit $\tau\to-1$; see \cite{mehta2004} where the author finds a semi-circular density in analogue with that of the GOE.

We will see that with the inclusion of this parameter we can access statistics interpolating between the ensembles discussed in Chapters \ref{sec:GOE_steps} and \ref{sec:GinOE} of the present work. We will proceed using the $5$ step method, which says that we first require the matrix or element distribution.

\begin{lemma}
\label{lem:tGindX}
The wedge product of the independent elements of $\bX$ in terms of the wedge products of the independent elements of $\bS$ and $\bA$ is
\begin{align}
\nonumber (d\bX)=(2\sqrt{c})^{N(N-1)/2}\;b^{-N^2/2}\;(d\bS)(d\bA).
\end{align}
\end{lemma}

\textit{Proof}: Each element in $d\bX$ contributes a factor of $\sqrt{b}$ and each of the independent elements of $\bA$ contributes a factor of $\sqrt{c}$; we can now ignore $b$ and $c$. From (\ref{def:tau_mats}) the elements of $d\bX$ are
\begin{align}
\nonumber dx_{jl}=\left\{\begin{array}{ll}
ds_{jl}+ da_{jl},&j<l,\\
ds_{jj},&j=l,\\
ds_{lj}-da_{lj},&l<j.
\end{array}\right.
\end{align}
When wedging together each element in the strict upper triangle (of which there are $N(N-1)/2$), we have two choices: either the symmetric or anti-symmetric element. Picking one then forces the corresponding choice in the lower triangle. Picking the other incurs a factor of $(-1)$, which is cancelled by the anti-symmetry of the wedge product. Taking the absolute value then gives the result.

\hfill $\Box$

\begin{proposition}[\cite{FN08}]
Let $\bS$ be a real, symmetric matrix with iid elements distributed as in (\ref{eqn:GOE_el_dist}), and let $\bA$ be a real, anti-symmetric matrix with iid elements according to (\ref{def:antisymdist}). Then the pdf of the matrices $\bX$ from (\ref{def:tau_mats}) is
\begin{align}
\label{eqn:tGinpdf} P(\bX)_{\tau,b}=\frac{b^{N^2/2}}{(2\pi)^{N^2/2}c^{N(N-1)/4}} \; e^{-b\frac{\Tr \bX\bX^T-\tau\Tr \bX^2}{2(1-\tau)}}.
\end{align}
\end{proposition}

\textit{Proof}:  Since the matrices $\bS=[s_{jl}]$ and $\bA=[a_{jl}]$ are independent, we take the product of their elemental probability densities
\begin{align}
\nonumber \prod_{j=1}^N \frac{1}{\sqrt{2\pi}}e^{-s_{jj}^2/2}\prod_{1\leq j < l\leq N}\frac{1}{\sqrt{\pi}}e^{-s_{jl}^2}\prod_{1\leq j < l\leq N}\frac{1}{\sqrt{\pi}}e^{-a_{jl}^2},
\end{align}
which we can rewrite as
\begin{align}
\label{eqn:SAjpdf1} \frac{1}{2^{N/2}}\frac{1}{\pi^{N^2/2}}\; e^{-(\Tr \:\bS^2 -\Tr\: \bA^2)/2}.
\end{align}
Using (\ref{def:tau_mats}) we can express $\bS$ and $\bA$ in terms of $\bX$ and $\bX^T$ thusly
\begin{align}
\label{eqn:tGincovs} \begin{split}
\bS&=\sqrt{\frac{b}{c}}\; \frac{\bX-\bX^T}{2},\\
\bA&=\frac{\sqrt{b}}{2}\; (\bX+\bX^T),
\end{split}
\end{align}
and so
\begin{align}
\label{eqn:TrSA} \Tr \:\bS^2-\Tr\: \bA^2=\frac{b}{1-\tau}\Big( \Tr\: \bX\bX^T-\tau\Tr\: \bX^2 \Big).
\end{align}
Substituting (\ref{eqn:TrSA}) into (\ref{eqn:SAjpdf1}) and multiplying by the Jacobian for the change of variables (\ref{eqn:tGincovs}), which is the content of Lemma \ref{lem:tGindX}, we have the result.

\hfill $\Box$
 
Up to a constant factor, we see that with $\tau=0$ (\ref{eqn:tGinpdf}) reduces to the real Ginibre pdf (\ref{eqn:GinOE_eldist}) and, with $\tau\to 1$ (recalling that in this limit $\bX^T\to \bX$) we have the GOE (\ref{eqn:GOE_el_jpdf}).

The question of interest is: what happens to the eigenvalue distribution as $\tau$ is varied? We can gain some insight into the answer through numerical simulations of these ensembles.

\begin{figure}[htp]
\begin{center}
\includegraphics[scale=0.4]{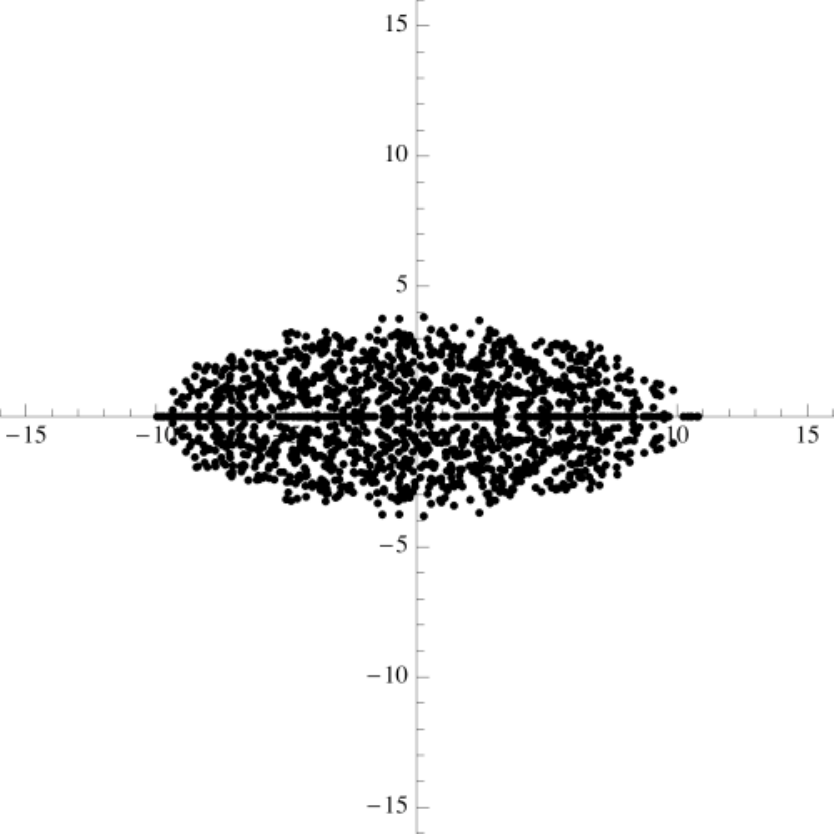}
\hspace{3ex}\includegraphics[scale=0.4]{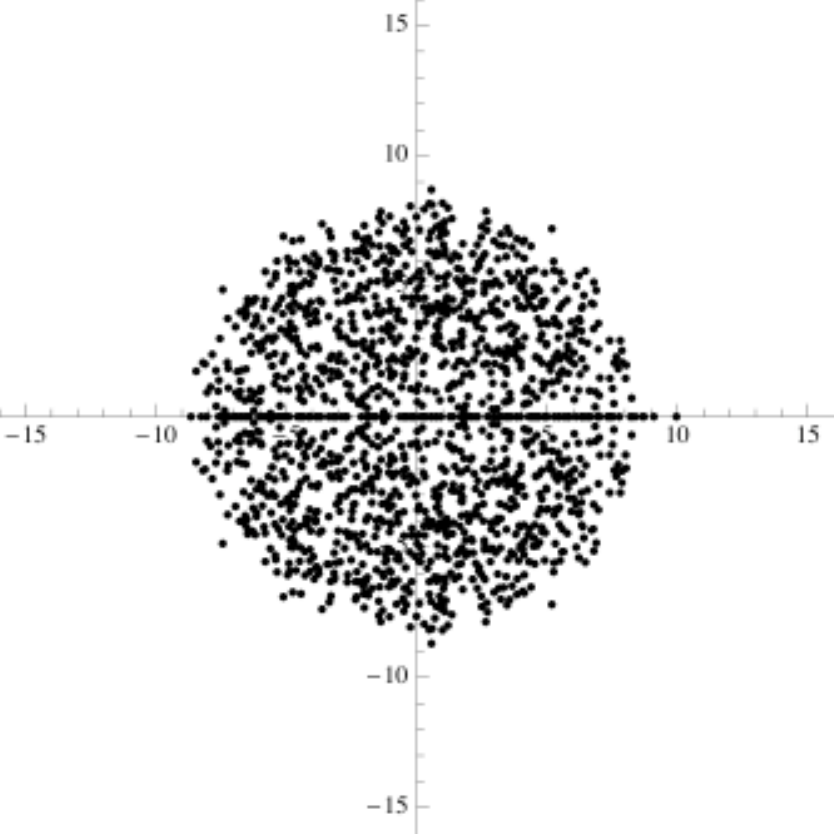}
\hspace{3ex}\includegraphics[scale=0.4]{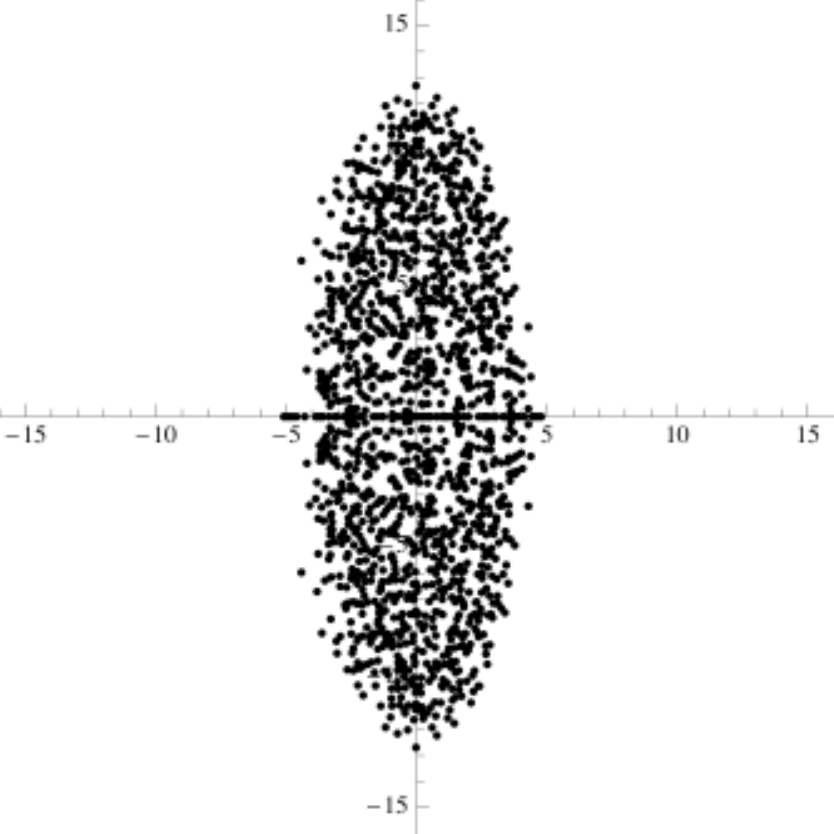}
\caption[Plot of simulated eigenvalues from partially symmetric real Ginibre ensembles, with $\tau=1/2, 0, 1/2$.]{Eigenvalues for 25 independent $64\times64$ matrices as defined in (\ref{def:tau_mats}). The left, middle and right plots correspond to $\tau=\frac{1}{2}, 0,-\frac{1}{2}$ respectively. In all three simulations $b=1$.}
\label{fig:tau_sims}
\end{center}
\end{figure}
\noindent As Figure \ref{fig:tau_sims} illustrates, the eigenvalue distribution for $\tau=0$ is circular with a distinct non-zero density of eigenvalues on the real line (this we of course knew from Chapter \ref{sec:GinOE}). As $\tau\to 1$ the matrices in the ensemble become more symmetric and the eigenvalues tend to congregate near the real line, forming an ellipse with major axis in the real direction. Conversely, with $\tau\to -1$ the eigenvalues collapse onto the imaginary axis as the ensemble approaches anti-symmetry, and the major axis of the ellipse is in the imaginary direction. We will not discuss anti-symmetric ensembles any further but we direct the interested reader to \cite{mehta2004}, and one can also find related self-dual matrices in \cite{Hast1999}. Note that since all these partially symmetric matrices are consistently real, we have a non-zero density of real eigenvalues for all $\tau\in (-1,1)$.

The distributions in Figure \ref{fig:tau_sims} may lead one to conjecture that there is an \textit{elliptical law}, which degenerates to the circular law when $\tau=0$. Indeed this is the case and we will analyse the situation further in Chapter \ref{sec:tGkernelts}. Another interesting point is that the ellipse must collapse onto the real axis in the limit $\tau\to 1$, since we must end up in the GOE in this limit. This leads to singular behaviour in the eigenvalue density as it shifts from being uniform in the ellipse, to semi-circular on the real line. The details of the analysis involve the \textit{strongly symmetric} (or \textit{weakly non-symmetric}) limit \cite{FKS97}, which we discuss in Chapter \ref{sec:tGsslim}.

\subsection{Eigenvalue jpdf}

As mentioned above, (\ref{eqn:tGinpdf}) reduces to (\ref{eqn:GinOE_eldist}) when $\tau=0$. By a simple scaling argument we can use this fact to deduce the eigenvalue jpdf for the partly symmetric ensemble from Proposition \ref{prop:GinOE_eval_jpdf}, which is the corresponding result for $\tau=0$.

\begin{proposition}[\cite{LS91}]
The eigenvalue jpdf for the partially symmetric real Ginibre matrices $\bX$ from (\ref{def:tau_mats}) is
\begin{align}
\nonumber &Q_{N,k,\tau,b}(\Lambda,W)=C_{N,k,\tau,b}\prod_{i=1}^{k}e^{-b\lambda_i^2/2}\\
\label{eqn:tGinejpdf} &\times\prod_{j=1}^{(N-k)/2} e^{-b(w_j^2+\bar{w}_j^2)/2}\;\mathrm{erfc}\left(\sqrt{\frac{2b}{1-\tau}}\; |\mathrm{Im}(w_j)|\right) \big|\Delta (\Lambda\cup W)\big|,
\end{align}
where $\Lambda=\{ \lambda_i \}_{i=1,...,k}$ and $W=\{ w_i,\bar{w}_i \}_{i=1,...,(N-k)/2}$ are the sets of real and non-real complex eigenvalues respectively, and
\begin{align}
\nonumber C_{N,k,\tau,b}:=\frac{b^{N(N+1)/4}(1+\tau)^{N(N-1)/4} 2^{-N(N-1)/4-k/2}}{k!((N-k)/2)!\prod_{l=1}^N\Gamma(l/2)}.
\end{align}
\end{proposition}

\textit{Proof}: Note that with
\begin{align}
\label{def:tGinscaledeval} \bX\to \bY = \sqrt{\frac{b}{1-\tau}}\: \bX,
\end{align}
we have
\begin{align}
\nonumber (d\bX) \to (d\bY) = \left(\frac{b}{(1-\tau)}\right)^{N^2/2} (d\bX),
\end{align}
and so the pdf for the real Ginibre matrices (\ref{eqn:GinOE_eldist}) becomes 
\begin{align}
\nonumber P(\bY) (d\bY) &= P\left(\sqrt{\frac{b} {1-\tau}} \bX \right) (d\bY) \\
\label{eqn:tGintoGincovs}& = \left(\frac{b}{2\pi(1-\tau)}\right)^{N^2/2} e^{-b\Tr\: \bX\bX^T/2(1-\tau)} (d\bX) =: \hat{P}(\bX) (d\bX).
\end{align}
We use (\ref{eqn:tGintoGincovs}) to rewrite (\ref{eqn:tGinpdf}) as
\begin{align}
\nonumber P(\bX)_{\tau,b} = \frac{(1+\tau)^{N(N-1)/4}}{(1-\tau)^{-N(N+ 1)/4}}\; e^{\tau b\: (\Tr\: \bX^2)/2(1-\tau)}\hat{P}(\bX) .
\end{align}
Now from Proposition \ref{prop:GinOE_eval_jpdf} we see that the eigenvalue jpdf corresponding to $\hat{P}(\bX)$, for the scaled matrices, is given by applying (\ref{def:tGinscaledeval}) to (\ref{eqn:GinOEjpdf}) and is
\begin{align}
\nonumber \left(\frac{b}{1-\tau}\right)^{N/2}Q_{N,k}\left(\sqrt{\frac{b}{1-\tau}} \Lambda,\sqrt{\frac{b}{1-\tau}} W \right).
\end{align} 
The exponential factor containing $\Tr\: \bX^2$ can be immediately written in terms of the eigenvalues of $\bX$ and so we have
\begin{align}
\nonumber Q_{N,k,\tau,b}&=\frac{(1+\tau)^{N(N-1)/4}}{(1-\tau)^{-N(N+ 1)/4}}\left(\frac{b}{1-\tau}\right)^{N/2}\prod_{j=1}^k e^{\tau b\: \lambda^2/2(1-\tau)}\\
\nonumber &\times\prod_{j=1}^{(N-k)/2} e^{\tau b\: (w_j^2+\bar{w}_j^2) /2(1-\tau)}\; Q_{N,k}\left(\sqrt{\frac{b}{1-\tau}} \Lambda,\sqrt{\frac{b}{1-\tau}} W \right),
\end{align}
from which the result follows on factoring $\sqrt{b/(1-\tau)}$ out of the Vandermonde product.

\hfill $\Box$

\subsection{Generalised partition function}

Since the structure of the eigenvalue jpdf (\ref{eqn:tGinejpdf}) for the partially symmetric ensemble is identical to that of the real Ginibre ensemble (\ref{eqn:GinOEjpdf}) we can immediately write down the generalised partition function for the former by substituting it into (\ref{def:multi_gen_part_fn}), setting $m=2$.

\begin{proposition}[\cite{FN08}]
With $-1<\tau<1$ the generalised partition function for the partially symmetric real Ginibre ensemble, with $k,N$ even, can be written
\begin{align}
\label{eqn:tGingpfe} Z_{k,(N-k)/2}[u,v]_{\tau}=\frac{b^{N(N+1)/4} (1+\tau)^{N(N-1)/4} }{2^{N(N+1)/4}\prod_{l=1}^N\Gamma(l/2)}[\zeta^{k/2}]\Pf\left[\zeta\: \alpha_{j,l}^{(\tau)}+\beta_{j,l}^{(\tau)}\right]_{j,l =1,...,N},
\end{align}
where $[\zeta^n]$ means the coefficient of $\zeta^n$ and, with monic polynomials $\{ p_{i}(x)\}$ of degree i,
\begin{align}
\label{eqn:tGin_alphabeta} \begin{split} \alpha_{j,l}^{(\tau)} & = \int_{-\infty}^{\infty}dx\; u(x)\int_{-\infty}^{\infty}dy\; u(y)e^{-b(x^2+y^2)/2}p_{j-1}(x)p_{l-1}(y)\; \mathrm{sgn}(y-x),\\
\beta_{j,l}^{(\tau)} & = 2i\int_{\mathbb{R}_+^2}dw\;v(w)\; \mathrm{erfc}\left(\sqrt{\frac{2b}{1-\tau}}\; |\mathrm{Im}(w)|\right)\; e^{-b(w^2+\bar{w}^2)/2}\\
&\times \Bigl(p_{j-1}(w)p_{l-1}(\bar{w})-p_{l-1}(w)p_{j-1}(\bar{w})\Bigr).
\end{split}
\end{align}
\end{proposition}

\begin{proposition}
With $-1<\tau<1$ and $\alpha_{j,l}^{(\tau)},\beta_{j,l}^{(\tau)}$ as in (\ref{eqn:tGin_alphabeta}) the generalised partition function for $k,N$ odd can be written
\begin{align}
\nonumber Z_{k,(N-k)/2}^{\odd}[u,v]_{\tau}&=\frac{b^{N(N+1)/4} (1+\tau)^{N(N-1)/4} }{2^{N(N+1)/4}\prod_{l=1}^N\Gamma(l/2)}\\
\label{eqn:tGingpfo} &\times [\zeta^{(k-1)/2}]\mathrm{Pf}\left[\begin{array}{cc}
\left[\zeta \alpha_{j,l}^{(\tau)}+\beta_{j,l}^{(\tau)}\right] & \left[\nu_j^{(\tau)}\right]\\
\left[-\nu_l^{(\tau)}\right]& 0\\
\end{array}\right]_{j,l=1,...,N},
\end{align}
where
\begin{align}
\label{def:tGinnu} \nu_l^{(\tau)}=\int_{-\infty}^{\infty} e^{-b x^2/2}u(x)p_{l-1}(x)\; dx.
\end{align}
\end{proposition}

The corresponding summed partition functions come from substituting (\ref{eqn:tGingpfe}) and (\ref{eqn:tGingpfo}) into (\ref{eqn:summedup}) and its odd equivalent, resulting in
\begin{align}
\label{eqn:tGinsume} Z_N[u,v]_{\tau}=\frac{b^{N(N+1)/4} (1+\tau)^{N(N-1)/4} }{2^{N(N+1)/4} \prod_{l=1}^N \Gamma(l/2)} \Pf\left[\alpha_{j,l}^{(\tau)}+ \beta_{j,l}^{(\tau)} \right]_{j,l=1,...,N},
\end{align}
and
\begin{align}
\label{eqn:tGinsumo} Z_{N}^{\odd}[u,v]_{\tau}&=\frac{b^{N(N+1)/4} (1+\tau)^{N(N-1)/4} }{2^{N(N+1)/4} \prod_{l=1}^N\Gamma(l/2)} \; \mathrm{Pf}\left[\begin{array}{cc}
\left[\alpha_{j,l}^{(\tau)} +\beta_{j,l}^{(\tau)} \right] & \left[\nu_j^{(\tau)}\right]\\
\left[-\nu_l^{(\tau)}\right]& 0\\
\end{array}\right]_{j,l=1,...,N}.
\end{align}

\subsubsection{Probability of $k$ real eigenvalues}
\label{sec:tGprobs}

As for the real Ginibre ensemble the probability of obtaining $k$ real eigenvalues from an $N\times N$ partially symmetric matrix is given by integrating $Q_{N,k,\tau,b}$ from (\ref{eqn:tGinejpdf}) over all $k$ real and $(N-k)/2$ complex conjugate pairs of eigenvalues. Note that by changing variables $\lambda_j\to \sqrt{b}\lambda_j$ and $w_j\to \sqrt{b}w_j$ the parameter $b$ scales out of this integral and so the probability, which we call $p_{N,k, \tau}$ in this chapter, is independent of $b$, and so $b$ can therefore be set arbitrarily. We choose $b=1$ for convenience in this section.

Recall from Chapter \ref{sec:Gsops} that we put off the discussion of the probabilities until we had obtained the skew-orthogonal polynomials (\ref{eqn:GinOE_sopolys}) relevant to the real Ginibre ensemble. In that case the polynomials were quite simple and the calculation of $\alpha_{j,l}$, $\beta_{j,l}$ and $\bar{\nu}_j$ could be performed. However, in the present setting, we find (in Chapter \ref{sec:tGinsops}) that the polynomials are not as simple as for real Ginibre (they interpolate between the real Ginibre polynomials and the Hermite polynomials of the GOE) and instead the calculations can be done more easily by (following \cite{FN08}) assuming a different (non-skew-orthogonal) form of the polynomials. The polynomials we use are $p_j(x)=x^j$ and, applying integration by parts, we have
\begin{align}
\label{eqn:tGalpha} \alpha_{2j-1,2l}^{(\tau)}\Big|_{u=1}=2^l(l-1)!\sum_{p=1}^l \frac{\Gamma(j+p-3/2)}{2^{p-1}(p-1)!},
\end{align}
\begin{align}
\nonumber \beta_{2j-1,2l}^{(\tau)}\Big|_{v=1} &=-4\mathop{\sum_{s=0}^{2j-2}\sum_{t=0}^{2l-1}}\limits_{s+t \; \odd}(-1)^t \left(\frac{2j-2}{s}\right)\left(\frac{2l-1}{t}\right)\\
\label{eqn:tGbeta} &\times\Gamma(j+l-1-(s+t)/2))\; I_{s+t},
\end{align}
where
\begin{align}
\label{eqn:tGeye} I_{j}=\frac{(-1)^{(j-1)/2}((j-1)/2)!}{2}\left( \sqrt{\frac{2}{1+\tau}}\sum_{p=0}^{(j-1)/2} (-1)^p \left( \frac{1-\tau}{1+\tau}\right)^p \frac{(1/2)_p}{p!}-1\right)
\end{align}
(see \cite{FN08} for the intermediate steps). For the odd case we also need the evaluation of $\nu_j^{(\tau)}\big|_{u=1}$, but with $b=1$ and our current choice of polynomial the evaluation is identical to that in the real Ginibre ensemble using the skew-orthogonal polynomials applicable to that case, and so $\nu_j^{(\tau)}\big|_{u=1}$ is given by (\ref{eqn:Gnu}). We can then calculate the probabilities in terms of the generalised partition functions (\ref{eqn:tGingpfe}) and (\ref{eqn:tGingpfo})
\begin{align}
\nonumber p_{N,k,\tau}=\left\{
\begin{array}{cc}
Z_{k,(N-k)/2}[1,1]_{\tau},&N \mbox{ even},\\
Z_{k,(N-k)/2}^{\odd}[1,1]_{\tau},&N \mbox{ odd}.
\end{array}
\right.
\end{align}

For the case that $k=N$, that is, when all eigenvalues are real, we can write down the evaluation of the probabilities by first noting from (\ref{eqn:tGingpfe}), recalling (\ref{eqn:GinOE_gpf_even}), that
\begin{align}
\nonumber p_{N,N,\tau}=Z_{N,0}[1,1]_{\tau}&=(1+\tau)^{N(N-1)/4}Z_{N,0}[1,1],
\end{align}
since the $\alpha_{j,l}$, with polynomials $p_j(x)=x^j$ do not depend on $\tau$. Then using (\ref{eqn:GinOEpNN}) we see
\begin{align}
\label{eqn:tGinpNN} p_{N,N,\tau}=\left(\frac{1+\tau}{2}\right)^{N(N-1)/4}.
\end{align}
Of course, since (\ref{eqn:tGinpNN}) is independent of the parity of $N$, we obtain the same result if we use the odd analogues. Note that we now see $p_{N,N,\tau}\to 1$ for $\tau\to 1$ and $p_{N,N,\tau}\to 2^{-N(N-1)/4}$ for $\tau\to 0$, facts that we anticipated in the discussion at the beginning of this chapter by considering the GOE and real Ginibre ensemble.

The values of $p_{N,k, \tau}$ for $\tau=1/2$ and $\tau=-1/2$ are contained in Appendix \ref{app:tGinsimpnk}, Tables \ref{tab:pnkxact_simp} and \ref{tab:pnkxact_simm} respectively, where we compare them to some simulated results. From looking at the data in this table we notice that for certain $\tau$ all the probabilities are rational.

\begin{proposition}
\label{prop:tGirr}
The probabilities $p_{N,k, \tau}$, given by (\ref{eqn:tGingpfe}) for $N$ even and (\ref{eqn:tGingpfo}) for $N$ odd, are all rational if $\tau=2r^2-1$, where $-1<r<1$ is a rational number.
\end{proposition}

\textit{Proof}: First note that the gamma functions do not contribute any irrational factors for any value of $\tau$. To see this, first we focus on $N$ even. In this case the Pfaffian in (\ref{eqn:tGingpfe}) will always be a product of $\alpha_{j,l}^{(\tau)}$s and $\beta_{j,l}^{(\tau)}$s, containing $N/2$ factors in total. Each of these factors contributes $\sqrt{\pi}$ from the gamma functions in (\ref{eqn:tGalpha}) and (\ref{eqn:tGbeta}), which are cancelled by $\prod_{l=1}^N \Gamma(l/2)$ in the denominator of the pre-factor. When $N$ is odd, the Pfaffian has $\pi^{(N+1)/4}$, including the contribution from $\nu_j^{(\tau)}\big|_{u=1}$. Again these are cancelled by the product of gamma functions in the pre-factor.

We now deal with the remaining sources of irrationality. With $\tau=2r^2-1$ then we see from (\ref{eqn:tGalpha}), (\ref{eqn:tGbeta}) and (\ref{eqn:tGeye}) that $\alpha_{j,l}$ and $\beta_{j,l}$ contain only rational factors (other than the gamma functions, which we have already dealt with). Since we have an integer number of factors of $\alpha_{j,l}$ and $\beta_{j,l}$ in the expansion of the Pfaffian from (\ref{eqn:tGingpfe}) it follows that the Pfaffian does not contribute to the irrationality of $p_{N,k, \tau}$ when $N$ is even. For $N$ odd each term in the Pfaffian contains a factor of $\nu_j^{(\tau)}\big|_{u=1}$ and, from (\ref{eqn:Gnu}), this contributes $\sqrt{2}$.

The prefactor in (\ref{eqn:tGingpfe}) has the potentially irrational factor
\begin{align}
\nonumber \frac{(1+\tau)^{N(N-1)/4}}{2^{N(N+1)/4}},
\end{align}
however we substitute for $\tau$ as specified above and find
\begin{align}
\label{eqn:r2irr} \frac{(2r^2)^{N(N-1)/4}}{2^{N(N+1)/4}}=r^{N(N-1)/2}\; 2^{-N/2},
\end{align}
which is rational for all even $N$, and for $N$ odd the $\sqrt{2}$ from the Pfaffian cancels the irrational factor in (\ref{eqn:r2irr}).

\hfill $\Box$

\subsection{Skew-orthogonal polynomials}
\label{sec:tGinsops}

As we know from the previous chapters, skew-orthogonalising the Pfaffians in (\ref{eqn:tGinsume}) and (\ref{eqn:tGinsumo}) allows us to calculate the correlation functions. In analogue with Definition \ref{def:Ginip1} we would like to define an inner product based upon $\alpha_{j,l}^{(\tau)}$ and $\beta_{j,l}^{(\tau)}$ from (\ref{eqn:tGin_alphabeta}). Before we do so, however, recall that (\ref{eqn:tGingpfe}) and (\ref{eqn:tGingpfo}) are independent of $b$ and so we can set it to any (positive real) value for our convenience. This convenient value turns out to be
\begin{align}
\label{eqn:tGinb} b=\frac{1}{1+\tau},
\end{align}
and so from here forth we will assume (\ref{eqn:tGinb}).

\begin{definition}
\label{def:tauGinip1}
Define the inner product
\begin{align}
\nonumber &\langle p,q \rangle_{\tau} :=\; \int_{-\infty}^{\infty}dx\int_{-\infty}^{\infty}dy\; e^{-\frac{x^2 + y^2}{2(1+\tau)}} p_{j}(x)p_{l}(y)\hspace{3pt}\mathrm{sgn}(y-x)\\
\nonumber &+2i\int_{\mathbb{R}_+^2}dw\; \mathrm{erfc}\Big(\sqrt{\frac{2}{1- \tau^2}}\;|\mathrm{Im}(w)|\Big) \; e^{-\frac{w^2+\bar{w}^2}{2(1+\tau)}} \Bigl(p_{j}(w)p_{l} (\bar{w})-p_{l}(w)p_{j}(\bar{w}) \Bigr)\\
\label{def:tauGinip} &=\alpha_{j+1,l+1}^{(\tau)}+\beta_{j+1,l+1}^{(\tau)}\big|_{u=v=1, \: b=1/(1+\tau)}.
\end{align}
\end{definition}

\begin{proposition}[\cite{FN08}]
\label{prop:tGinsops}
With $H_j(z)$ the Hermite polynomials from (\ref{eqn:herm_polys}), let $C_j(z)$ be the scaled monic Hermite polynomials
\begin{align}
\label{def:tGsops} C_j(z) :=\left(\frac{\tau}{2}\right)^{j/2}\: H_j\left(\frac{z}{\sqrt{2\tau}} \right).
\end{align}
The polynomials skew-orthogonal with respect to the inner-product (\ref{def:tauGinip}) are
\begin{align}
\nonumber R_{2j}(z)=C_{2j}(z)&,&R_{2j+1}(z)=C_{2j+1}(z)-2j\:C_{2j-1}(z),
\end{align} 
with normalisation
\begin{align}
\label{eqn:tGinnorms} r^{(\tau)}_j=\Gamma (2j+1) \;2 \sqrt{2\pi}\;(1+\tau).
\end{align}
\end{proposition}

The derivation of the polynomials in Proposition \ref{prop:tGinsops} is the general $\tau$ version of that used to obtain Proposition \ref{prop:Ginsops}; the reader is referred to \cite{FN08} for the details, or to \cite{AkePhilSom2010} for a method using an average over a characteristic polynomial.

Note that all but the first term in $H_j(z/(\sqrt{2\tau}))$ has $\tau^l$, where $-j/2<l\leq 0$ and so only the leading term of $C_j(z)$ is non-vanishing as $\tau\to 0$. The leading term has unit coefficient and so Definition \ref{def:tauGinip1} and Proposition \ref{prop:tGinsops} reduce to their real Ginibre counterparts (Definition \ref{def:Ginip1} and Proposition \ref{prop:Ginsops} respectively) when $\tau\to 0$. With $\tau \to 1$, and changing variables $x\to x/\sqrt{2},y\to y/\sqrt{2}$, we reclaim the GOE inner product of Definition \ref{def:GOE_soip} and the skew-orthogonal polynomials of Proposition \ref{prop:GOE_soip}.

\subsection{Correlation functions}

Since the structure of the generalised partition function (\ref{eqn:tGingpfe}) is identical to that of the real Ginibre ensemble (\ref{eqn:GinOE_gpf_even}), we can apply the machinery of Chapter \ref{sec:Gincorrlnse} to find the correlations. Similarly, in \cite{APS2009} the authors adapt the method of averaging over characteristic polynomials from the real Ginibre to the partially symmetric real Ginibre case, although we will not pursue their method here.

We first define a correlation kernel analogous to those in the real Ginibre ensemble (Definition \ref{def:GinOE_kernel}) and the GOE (Definition \ref{def:GOE_correln_kernel}).

\begin{definition}
\label{def:tGine_kernel}
Let $N$ be even. With $R_0,R_1,...$ the skew-orthogonal polynomials of Proposition \ref{prop:tGinsops} and $r^{(\tau)}_0,r^{(\tau)}_1,...$ the corresponding normalisations, define
\begin{align}
\nonumber S(\mu,\eta)_{\tau}&=2\sum_{j=0}^{\frac{N}{2}-1}\frac{1}{r_j^{(\tau)}}\Bigl[q_{2j}(\mu)\varphi_{2j+1}(\eta)-q_{2j+1}(\mu)\varphi_{2j}(\eta)\Bigr],\\
\nonumber D(\mu,\eta)_{\tau}&=2\sum_{j=0}^{\frac{N}{2}-1}\frac{1}{r_j^{(\tau)}}\Bigl[q_{2j}(\mu)q_{2j+1}(\eta)-q_{2j+1}(\mu)q_{2j}(\eta)\Bigr],\\
\nonumber \tilde{I}(\mu,\eta)_{\tau}&=2\sum_{j=0}^{\frac{N}{2}-1}\frac{1}{r_j^{(\tau)}}\Bigl[\varphi_{2j}(\mu)\varphi_{2j+1}(\eta)-\varphi_{2j+1}(\mu)\varphi_{2j}(\eta)\Bigr]+\epsilon(\mu,\eta)\\
\nonumber &=: I(\mu,\eta)_{\tau} +\epsilon(\mu,\eta),
\end{align}
where
\begin{align}
\nonumber h(\mu) &=e^{-\mu^2/2(1+\tau)}\sqrt{\erfc\left(\sqrt{\frac{2}{1-\tau^2}}\: |\mathrm{Im}(\mu)|\right)},\\
\nonumber q_j(\mu) &= h(\mu) R_j(\mu),\\
\nonumber \varphi_j(\mu) &= 
\left\{ 
\begin{array}{ll}
-\frac{1}{2}\int_{-\infty}^{\infty}\mathrm{sgn}(\mu-z)\hspace{3pt}q_j(z)\hspace{3pt}dz, & \mu\in \mathbb{R},\\
iq_j(\bar{\mu}),  & \mu\in \mathbb{R}_2^+,
\end{array}
\right.
\end{align}
and $\epsilon(\mu,\eta)$ is from Definition \ref{def:GinOE_kernel}.

In terms of these quantities, define
\begin{align}
\label{def:tGin_K} \bK^{(\tau)}(\mu,\eta)=\left[
\begin{array}{cc}
S(\mu,\eta)_{\tau} & - D(\mu,\eta)_{\tau}\\
\tilde{I}(\mu,\eta)_{\tau} & S(\eta,\mu)_{\tau}
\end{array}
\right].
\end{align}
\end{definition}

By undertaking either the $4\times 4$ or $2\times 2$ kernel method of Chapter \ref{sec:Gincorrlnse} we find the correlation functions for $N$ even.

\begin{proposition}[\cite{FN08}]
Let $N$ be even. Then, with $\bK^{(\tau)}(\mu,\eta)$ from (\ref{def:tGin_K}), the correlation functions for $n_1$ real and $n_2$ non-real, complex conjugate pairs of eigenvalues in the partially symmetric real Ginibre ensemble are
{\small
\begin{align}
\nonumber &\rho_{(n_1,n_2)}(x_1,...,x_{n_1},w_1,...,w_{n_2})_{\tau}=\qdet\left[\begin{array}{cc}
\bK^{(\tau)}(x_i,x_j) & \bK^{(\tau)}(x_i,w_m)\\
\bK^{(\tau)}(w_l,x_j) & \bK^{(\tau)}(w_l,w_m)
\end{array}\right]_{i,j=1,...,n_1, \atop l,m=1,...,n_2}\\
\nonumber &=\Pf\left(\left[\begin{array}{cc}
\bK^{(\tau)}(x_i,x_j) & \bK^{(\tau)}(x_i,w_m)\\
\bK^{(\tau)}(w_l,x_j) & \bK^{(\tau)}(w_l,w_m)
\end{array}\right]\bZ_{2(n_1+n_2)}^{-1}\right)_{i,j=1,...,n_1, \atop l,m=1,...,n_2}, \quad x_i\in \mathbb{R}, w_i \in \mathbb{R}_2^+.
\end{align}
}
\end{proposition}

We can likewise apply the functional differentiation methods of Chapter \ref{sec:Gin_odd_fdiff} or use the `odd-from-even' approach of Chapter \ref{sec:Gin_oddfromeven} to obtain the $N$ odd case.

\begin{definition}
Let $N$ be odd. With $R_0,R_1,...$ the skew-orthogonal polynomials in Proposition \ref{prop:tGinsops}, $r^{(\tau)}_0,r^{(\tau)}_1,...$ the corresponding normalisations (\ref{eqn:tGinnorms}), and $\nu_j^{(\tau)}$ as in (\ref{def:tGinnu}) (with $\bar{\nu}_j^{(\tau)}:= \nu_j^{(\tau)}\big|_{u=1}$), define
\begin{align}
\nonumber S^{\odd}(\mu,\eta)_{\tau}&=2\sum_{j=0}^{\frac{N-1}{2}-1}\frac{1}{r_j^{(\tau)}}\Bigl[\hat{q}_{2j}(\mu)\hat{\varphi}_{2j+1}(\eta)-\hat{q}_{2j+1}(\mu) \hat{\varphi}_{2j}(\eta)\Bigr]+ \kappa(\mu,\eta),\\
\nonumber D^{\odd}(\mu,\eta)_{\tau}& =2\sum_{j=0}^{\frac{N-1}{2}-1}\frac{1}{r_j^{(\tau)}}\Bigl[\hat{q}_{2j}(\mu)\hat{q}_{2j+1}(\eta)- \hat{q}_{2j+1}(\mu) \hat{q}_{2j}(\eta)\Bigr],\\
\nonumber \tilde{I}^{\odd}(\mu,\eta)_{\tau}& =2\sum_{j=0}^{\frac{N-1}{2}-1}\frac{1}{r_j^{(\tau)}}\Bigl[\hat{\varphi}_{2j} (\mu)\hat{\varphi}_{2j+1}(\eta)- \hat{\varphi}_{2j+1}(\mu) \hat{\varphi}_{2j}(\eta)\Bigr]\\
\nonumber &+\epsilon(\mu,\eta)+\theta(\mu,\eta),
\end{align}
where $\epsilon(\mu,\eta)$ is from Definition \ref{def:GinOE_kernel} and
\begin{align}
\nonumber \hat{R}_j(\mu)&=R_j(\mu)- \frac{\bar{\nu}_{j+1}^{(\tau)}} {\bar{\nu}_N^{(\tau)}} R_{N-1}(\mu),\\
\nonumber \hat{q}_j(\mu) &= h(\mu)\: \hat{R}_j(\mu),\\
\nonumber \hat{\varphi}_j(\mu) &= 
\left\{ 
\begin{array}{ll}
-\frac{1}{2}\int_{-\infty}^{\infty}\mathrm{sgn}(\mu-z)\hspace{3pt}\hat{q}_j(z)\hspace{3pt}dz, & \mu\in \mathbb{R},\\
i\hat{q}_j(\bar{\mu}),  & \mu\in \mathbb{R}_2^+,
\end{array}
\right.\\
\nonumber \kappa(\mu,\eta) &= 
\left\{ 
\begin{array}{lll}
q_{N-1}(\mu)/ \bar{\nu}_N^{(\tau)}, & \eta\in \mathbb{R},\\
0,  & \mathrm{otherwise},\\
\end{array}
\right.\\
\nonumber \theta(\mu,\eta)&=
\big(\chi_{(\eta\in\mathbb{R})}\varphi_{N-1}(\mu)- \chi_{(\mu\in\mathbb{R})}\varphi_{N-1}(\eta)\big)/  \bar{\nu}_N^{(\tau)},
\end{align}
with the indicator function $\chi_{(A)}=1$ for $A$ true and zero for $A$ false. Then, let
\begin{align}
\label{def:tGin_Ko} \bK^{(\tau)}_{\odd}(\mu,\eta)=\left[
\begin{array}{cc}
S^{\odd}(\mu,\eta)_{\tau} & -D^{\odd}(\mu,\eta)_{\tau}\\
\tilde{I}^{\odd}(\mu,\eta)_{\tau} & S^{\odd}(\eta,\mu)_{\tau}
\end{array}
\right].
\end{align}
\end{definition}

\begin{proposition}
Let $N$ be odd. Then with $\bK_{\odd}^{(\tau)}(\mu,\eta)$ from (\ref{def:tGin_Ko}), the $(n_1,n_2)$-point correlation functions for the partially symmetric real Ginibre ensemble are
{\small
\begin{align}
\nonumber &\rho_{(n_1,n_2)}(x_1,...,x_{n_1},w_1,...,w_{n_2})_{\tau}=\qdet\left[\begin{array}{cc}
\bK_{\odd}^{(\tau)}(x_i,x_j) & \bK_{\odd}^{(\tau)}(x_i,w_m)\\
\bK_{\odd}^{(\tau)}(w_l,x_j) & \bK_{\odd}^{(\tau)}(w_l,w_m)
\end{array}\right]_{i,j=1,...,n_1, \atop l,m=1,...,n_2}\\
\nonumber &=\Pf\left(\left[\begin{array}{cc}
\bK_{\odd}^{(\tau)}(x_i,x_j) & \bK_{\odd}^{(\tau)}(x_i,w_m)\\
\bK_{\odd}^{(\tau)}(w_l,x_j) & \bK_{\odd}^{(\tau)}(w_l,w_m)
\end{array}\right]\bZ_{2(n_1+n_2)}^{-1}\right)_{i,j=1,...,n_1, \atop l,m=1,...,n_2}, \quad x_i\in \mathbb{R}, w_i \in \mathbb{R}_2^+.
\end{align}
}
\end{proposition}

\subsection{Correlation kernel elements}
\label{sec:tGkernelts}

We expect that the correlation kernel elements will be deformations of those in the real Ginibre case of Chapter \ref{sec:Ginkernelts}. To establish this claim we use an integral representation of the Hermite polynomials to write (\ref{def:tGsops}) as
\begin{align}
\label{eqn:integHerm} C_n (z)=\frac{1}{\sqrt{\pi}}\int_{-\infty}^{\infty}e^{-t^2}(z+i\sqrt{2\tau}t)^n\;dt,
\end{align}
which we can verify by expanding the integrand using the binomial theorem and comparing the result to (\ref{eqn:herm_polys}). With (\ref{eqn:integHerm}) we can express $D(\mu,\eta)_{\tau}$ in terms of the $D(\mu,\eta)_0:=D(\mu,\eta)$ from Definition \ref{def:GinOE_kernel} thusly
\begin{align}
\nonumber &D(\mu,\eta)_{\tau}=\frac{h(\mu)h(\eta)}{\pi (1+\tau)} \int_{-\infty}^{\infty}dt_1\; e^{-t_1^2}\int_{-\infty}^{\infty}dt_2\; e^{-t_2^2}\\
\label{eqn:DDttrans} & \times \left( w(\mu+ i\sqrt{2\tau}t_1)w(\eta +i\sqrt{2\tau}t_2)\right)^{-1} D(\mu+i\sqrt{2\tau}t_1,\eta+i\sqrt{2\tau}t_2)_0,
\end{align}
where $h(x)$ is from Definition \ref{def:tGine_kernel} and $w(x)=h(x)\Big|_{\tau=0}=e^{-x^2/2}\sqrt{\erfc(\sqrt{2}|\mathrm{Im}(x)|)}$. Similar transformations also hold for $S_{c,c}(\mu,\eta)_{\tau}$ and $S_{r,c}(\mu,\eta)_{\tau}$ (where the second variable is complex conjugated) and $\tilde{I}_{c,c}(\mu,\eta)$ (where both variables are conjugated).

\begin{remark}
\label{rem:tGtrans}
The transformation (\ref{eqn:DDttrans}) does not hold for the remaining kernel elements since they all contain factors of $\varphi_j(x)$.
\end{remark}

To obtain the limiting complex correlation kernel in the bulk (in the strongly non-symmetric limit where $\tau$ is bounded away from $1$) we can directly apply the transform (\ref{eqn:DDttrans}) using the real Ginibre result from (\ref{eqn:Ginbulk}). This yields
\begin{align}
\nonumber S^{\mathrm{bulk}}_{c,c}(w,z)_{\tau}&=\frac{i}{\sqrt{2\pi}}\sqrt{\mathrm{erfc}\left(\sqrt{\frac{2}{1-\tau^2}}|\mathrm{Im}(w)|\right)}\sqrt{\mathrm{erfc}\left(\sqrt{\frac{2}{1-\tau^2}}|\mathrm{Im}(z)|\right)}\\
\nonumber &\times \frac{\bar{z}-w}{(1-\tau^2)} \:e^{-(\bar{z}-w)^2/2(1-\tau^2)},
\end{align}
and so the bulk limiting complex density is given by \cite{FN08}
\begin{align}
\label{eqn:tGbulkcdens} \rho_{(1)}^{\bulk}(w)_{\tau}=S^{\mathrm{bulk}}_{c,c}(w,w)_{\tau}=\sqrt{\frac{2}{\pi}}\: \erfc\left(\sqrt{\frac{2}{1-\tau^2}}\: v\right)\: \frac{v \: e^{2 v^2/(1-\tau^2)}}{1-\tau^2},
\end{align}
where $w=u+iv \in \mathbb{R}_2^{+}$.

For the real case we cannot simply apply (\ref{eqn:DDttrans}) to (\ref{eqn:Ginbulk}) since, as mentioned in Remark \ref{rem:tGtrans} the appearance of $\varphi_j$ factors invalidates its use. Instead, we substitute the skew-orthogonal polynomials of Proposition \ref{prop:tGinsops} and perform similar manipulations to those leading to (\ref{eqn:Ginsummed}) and find \cite{FN08}
\begin{align}
\nonumber S_{r,r}(x,y)_{\tau}&=\frac{e^{-(x^2+y^2)/2(1+\tau)}}{\sqrt{2 \pi}}\sum_{k=0}^{N-2}\frac{C_k(x)C_k(y)}{k!}\\
\label{eqn:Srrsum} &+\frac{e^{-x^2/2(1+\tau)}}{\sqrt{2\pi}(1+\tau)}\frac{C_{N-1}(x)\Phi_{N-2}(y)}{(N-2)!},
\end{align}
giving the large $N$ limit
\begin{align}
\label{eqn:tGSrrlim} \lim_{N\to \infty}S_{r,r}(x,y)_{\tau}=\frac{e^{-(x-y)^2/2(1-\tau^2)}}{\sqrt{2\pi (1-\tau^2)}}=:S_{r,r}^{\bulk}(x,y)_{\tau}.
\end{align}
The limiting bulk density is given by (\ref{eqn:tGSrrlim}) with $x=y$.

By comparing (\ref{eqn:tGbulkcdens}) and (\ref{eqn:tGSrrlim}) to (\ref{eqn:Ginbulk}) we see that we can obtain the general $\tau$ bulk densities for the complex eigenvalues by changing variables
\begin{align}
\label{def:tcovw} u\to u/(\sqrt{1-\tau^2}), v\to v/(\sqrt{1-\tau^2})
\end{align}
in $S_{c,c}^{\bulk}(u+iv, u+iv) dudv$, and for the real case by using
\begin{align}
\label{def:tcovx} x\to x/(\sqrt{1-\tau^2})
\end{align}
in $S_{r,r}^{\bulk}(x,x) dx$. Indeed, through the use of the inter-relationships (\ref{eqn:Gin_s=d=i}) we obtain the bulk limiting form of the general correlation functions from
\begin{align}
\rho_{n_1,n_2}^{(\tau)}(x_1,...,x_{n_1},w_1,...,w_{n_2})dx_1...dx_{n_1}dw_1,...,dw_{n_2}
\end{align}
by applying the scaling (\ref{def:tcovw}) and (\ref{def:tcovx}) to (\ref{eqn:Ginbulk}) and (\ref{eqn:Ginbulkall}).

In \cite{DGIL1994} the authors describe the boundary of a Coulomb gas (which from \cite{forrester?} we know is analogous to a system of eigenvalues of a random matrix) by looking for the point $w=x+iy$ that maximises the difference between the densities of systems with $N$ and $N+1$ particles; to wit, to maximise
\begin{align}
\nonumber \rho_{(1)}(w)\Big|_{N\mapsto N+1}-\rho_{(1)}(w).
\end{align}
Substituting the real Ginibre result (\ref{eqn:Ginsummed}) for $D(s,t)_0$ in (\ref{eqn:DDttrans}), it is shown in \cite{FN08} that
\begin{align}
\label{eqn:tGbdry} \rho_{(1)}(w)\Big|_{N\mapsto N+1}-\rho_{(1)}(w)\sim\frac{\sqrt{2}\;|C_{N-2}(w)|^2}{\pi(1+\tau)\Gamma(N-1)}e^{-\left(2|w|^2- \tau(w^2+ \bar{w}^2)\right) /2(1-\tau^2)}.
\end{align}
By maximising this difference with respect to $w$ the working in \cite{DGIL1994} shows that (\ref{eqn:tGbdry}) implies the boundary is an ellipse with semi-axes $(1+\tau) \sqrt{N}$ and $(1-\tau) \sqrt{N}$; a fact we illustrated in the plots of Figure \ref{fig:tau_sims}. We can then find the partially symmetric analogue to Proposition \ref{prop:Gincirclaw}.

\begin{proposition}[\cite{SCSS1988}]
With $\hat{z}=z/\sqrt{N}$ the limiting distribution of complex eigenvalues $\hat{z}$ in the partially symmetric real Ginibre ensemble is
\begin{align}
\label{eqn:ellaw1} \rho_{(1)}^c(\hat{z})_{\tau}=\frac{\chi_{\hat{z}\in E}}{\pi (1-\tau^2)},
\end{align}
where $E$ is the ellipse centred at the origin with semi-axes $(1+\tau)$ and $(1-\tau)$.
\end{proposition}

\textit{Proof}: We already know from (\ref{eqn:tGbdry}) that the boundary of the support is the ellipse with semi-axes $(1+\tau)$ and $(1-\tau)$. So all that remains is to show that the density is uniform on the ellipse as stated.

Making the change of variables $\hat{w} = w/\sqrt{N}$ and $\hat{z}= z/\sqrt{z}$ in (\ref{eqn:DDttrans}) we have
\begin{align}
\nonumber &D_{c,c}(\sqrt{N} \hat{w}, \sqrt{N} \hat{z})_{\tau}=\frac{h(\sqrt{N} \hat{w})h(\sqrt{N} \hat{z})}{\sqrt{2 \pi} (1+\tau) \pi} \int_{-\infty}^{\infty}dt_1\; e^{-t_1^2}\int_{-\infty}^{\infty}dt_2\; e^{-t_2^2}\\
\label{eqn:tDrtN} & \times (\sqrt{N}\hat{z}+i\sqrt{2\tau} t_2 -\sqrt{N}\hat{w}-i\sqrt{2\tau} t_1) \sum_{j=1}^{N-1} \frac{((\sqrt{N}\hat{w}+i\sqrt{2\tau} t_1)(\sqrt{N}\hat{z}+i\sqrt{2\tau} t_2))^{j-1}} {(j-1)!}.
\end{align}
Using the knowledge that for large $N$
\begin{align}
\nonumber \sum_{j=1}^{N-1} \frac{x^j}{j!} \sim e^{x},
\end{align}
we can perform the resulting integrals in (\ref{eqn:tDrtN}) to obtain
\begin{align}
\nonumber D_{c,c}(\sqrt{N} \hat{w}, \sqrt{N} \hat{z})_{\tau} &\sim \sqrt{\frac{N}{2\pi(1-\tau^2)}} \frac{(\hat{z}-\hat{w})}{1-\tau^2} e^{N(\hat{w}-\hat{z})^2/2(1-\tau^2)}\\
\nonumber & \times \left( \erfc\left(\sqrt{\frac{2N}{1-\tau^2}}\: \mathrm{Im}(\hat{w})\right) \erfc\left(\sqrt{\frac{2N}{1-\tau^2}}\: \mathrm{Im}(\hat{z})\right) \right)^{1/2}.
\end{align}
Using (\ref{eqn:Gin_s=d=i}) we have, by following the same reasoning in Proposition \ref{prop:Gincirclaw},
\begin{align}
\nonumber \rho_{(1)}^c (\sqrt{N} \hat{z})= S_{c,c}(\sqrt{N} \hat{z}, \sqrt{N} \hat{z})_{\tau}\sim \sqrt{\frac{2}{\pi}} \sqrt{\frac{N}{1-\tau^2}} \frac{\hat{y}} {1-\tau^2} e^{2N\hat{y}^2/(1-\tau^2)} \erfc\left(\sqrt{\frac{2N}{1-\tau^2}}\: \hat{y} \right),
\end{align}
where $\hat{z}=\hat{x}+i\hat{y}$. Noting the asymptotic behaviour (\ref{eqn:bigerfc}), we have the result.

\hfill $\Box$

In \cite{SCSS1988} the authors point out that the projection of (\ref{eqn:ellaw1}) gives a generalised semi-circle, which reduces to Wigner's semi-circle in the limit $\tau\to 1$.

\begin{remark}
Note that the analysis of the partially symmetric ensemble neatly explains why there is superficially a discrepancy between the radius of support in the GOE limit ($(-\sqrt{2N}, \sqrt{2N})$ from (\ref{eqn:wssl})) and the real Ginibre limit ($(-\sqrt{N},\sqrt{N})$ from (\ref{eqn:Gcirclaw})); with fixed $b=1/(1+\tau)$ then as $\tau\to 1$ we see from (\ref{def:tau_mats}) that $\bX\to \bS/\sqrt{2}$.
\end{remark}

From the universality established in \cite{TVK10}, we can conclude that since Proposition \ref{prop:circlaw} holds in the case $\tau=0$ for all distributions with finite mean and variance $1$, we also have, by scaling, the equivalent general elliptical law, for general $\tau$.

\subsubsection{Strongly symmetric limit}
\label{sec:tGsslim}

To analyse the regime of cross-over between symmetric and asymmetric matrices we follow \cite{FN08} and use the idea of \cite{FKS97, Efe97} to let
\begin{align}
\label{def:tssla} \tau=1-\alpha^2/N
\end{align}
and allow $N\to\infty$. Since we know that in the large $N$ limit we will recover the semi-circular density (\ref{eqn:wssl}), the eigenvalues will be supported on $[-2\sqrt{N},2\sqrt{N}]$ and the average spacing between them must be on the order of $1/\sqrt{N}$. With this in mind we scale the eigenvalues by
\begin{align}
\label{def:tGascale} x\to x\pi/\sqrt{N},
\end{align}
so that we have unit (real) density, and (\ref{eqn:Srrsum}) becomes
\begin{align}
\nonumber \frac{\pi}{\sqrt{N}} \; S_{rr}\left(\frac{\pi x}{\sqrt{N}}, \frac{\pi y}{\sqrt{N}}\right)_{1-\frac{\alpha^2}{N}} &\sim \sqrt{\frac{\pi} {2N}} e^{-\pi^2 \frac{x^2+y^2} {2N-\alpha^2}} \sum_{k=0}^{N-2} \frac{(1-\alpha^2/N)^k} {2^k k!}\\
\nonumber &\times H_k \left(\frac{\pi x}{\sqrt{2N-2\alpha^2}} \right) H_k \left(\frac{\pi y}{\sqrt{2N-2\alpha^2}} \right),
\end{align}
where we have ignored the second term in (\ref{eqn:Srrsum}) since it tends to zero because of the factorial denominator. Applying the asymptotic formula \cite[18.15.26]{NIST2010}
\begin{align}
\nonumber \frac{\Gamma(n/2+1)}{\Gamma(n+1)}e^{-x^2/2}H_n(x)=\cos(\sqrt{2n+1}\; x-n\pi/2) +O(n^{-1/2}),
\end{align}
and noting the asymptotic behaviour
\begin{align}
\nonumber & (1-\alpha^2/N)^k \sim e^{-k\alpha^2/N},
\end{align}
we have \cite{FN08}
\begin{align}
\nonumber \frac{\pi}{\sqrt{N}} \; S_{rr}\left(\frac{\pi x}{\sqrt{N}}, \frac{\pi y}{\sqrt{N}}\right)_{1-\frac{\alpha^2}{N}} &\sim \sum_{k= 1}^{N-2} \sqrt{\frac{1} {k N}} \; e^{-k\alpha^2/N}\\
\label{eqn:tGSrr1} &\times \cos \left(\sqrt{\frac{2k+1}{2N}} \pi x -\frac{k\pi}{2} \right) \cos \left(\sqrt{\frac{2k+1}{2N}} \pi y -\frac{k\pi}{2} \right),
\end{align}
where we have used the identity
\begin{align}
\nonumber \frac{\Gamma (k+1)}{\Gamma (k/2+1)} = \frac{2^{k} \Gamma ((k+1)/2)} {\sqrt{\pi}},
\end{align}
and then the large $x$ behaviour $\Gamma(x+a)/\Gamma (x) \sim x^a$.

The cosine multiple angle formulae tell us that $\cos x \cos y = (\cos (x+y)+\cos (x-y))/2$ and so the cosine product in (\ref{eqn:tGSrr1}) becomes 
\begin{align}
\nonumber & \frac{1}{2} \cos \left(\sqrt{\frac{2k+1}{2N}} \pi (x+ y) -k\pi \right) + \frac{1}{2} \cos \left(\sqrt{\frac{2k+1}{2N}} \pi (x- y) \right).
\end{align}
To leading order, the contributions from the first term are cancelled because of the $(\pm 1)$ introduced by the $-k\pi$ term, and so we are left with
\begin{align}
\nonumber S_{r,r}(x,y)_{1-\frac{\alpha^2}{N}} &\sim \sum_{k=1}^{N-2} \sqrt{\frac{1} {k N}} \; e^{-k\alpha^2/N} \cos \left( \sqrt{\frac{k} {N}} \pi (x-y) \right).
\end{align}
Letting $t_k :=k /N \in (0,1)$ we rewrite this as
\begin{align}
\label{eqn:tGRSapp} \sum_{k=1}^{N-2} \frac{e^{-\alpha^2 t_k}} {N\sqrt{t_k}}  \cos \left(\pi (x-y) \sqrt{t_k} \right) = \sum_{k=1}^{N-2} \frac{e^{-\alpha^2 t_k}} {\sqrt{t_k}}  \cos \left(\pi (x-y) \sqrt{t_k} \right) \Delta_k,
\end{align}
where $\Delta_k:= t_k-t_{k-1}=1/N$ is the eigenvalue spacing. The right hand side of (\ref{eqn:tGRSapp}) is a Riemann sum approximation to a definite integral and so in the large $N$ limit we have \cite{FN08}
\begin{align}
\label{eqn:tGSrr2}  \frac{\pi}{\sqrt{N}} \; S_{rr}\left(\frac{\pi x}{\sqrt{N}}, \frac{\pi y}{\sqrt{N}}\right)_{1-\frac{\alpha^2}{N}} &\sim \int_0^1 e^{-\alpha^2 t} \cos \left(\pi (x-y) t \right) dt,
\end{align}
where we have changed variables $t\to t^2$. By taking the limit $\alpha\to 0$ in (\ref{eqn:tGSrr2}) we see that we reclaim $S^{\bulk}(x_j,x_k)$ from (\ref{eqn:GOESbulk}) and so we have indeed obtained GOE behaviour as the $\tau \to 1$ limit of the partially symmetric real Ginibre ensemble, as we expected.

We can apply the same procedure to the complex case, recalling that both the real and imaginary parts of each eigenvalue must be scaled according to (\ref{def:tGascale}), that is we make the replacement $w_j \to \pi w_j/\sqrt{N} = \pi(u_j+ iv_j)/\sqrt{N}$. So, with $\tau$ specified by (\ref{def:tssla}), we have
\begin{align}
\nonumber &e^{-(w_j^2+\bar{w}_j^2)/2(1+\tau)} \sqrt{\erfc\left( \sqrt{\frac{2}{1-\tau^2}}\; v_j \right)}\\
\nonumber &\to e^{-\pi^2 (w_j^2+\bar{w}_j^2)/ (4N-2\alpha^2)} \sqrt{\erfc\left( \sqrt{\frac{2}{2\alpha^2/N -\alpha^4/N^2}}\; \frac{\pi v_j}{\sqrt{N}} \right)} \sim \sqrt{\erfc\left( \frac{\pi v_j}{\alpha}\right)},
\end{align}
and we obtain \cite{FN08}
\begin{align}
\nonumber \lim_{N\to\infty}\left(\frac{\pi}{\sqrt{N}}\right)^2 S_{cc} \left( \frac{\pi w_1} {\sqrt{N}}, \frac{\pi w_2} {\sqrt{N}}\right)_{\tau} &=i\pi \; \sqrt{\erfc \left( \frac{\pi v_1}{\alpha}\right) \erfc \left( \frac{\pi v_2}{\alpha}\right)}\\
\label{eqn:tGlimcc}&\times \int_0^1 t e^{-\alpha^2 t^2} \sin (\pi t (\bar{w}_1-w_2)) dt.
\end{align}
By summing (\ref{eqn:tGSrr2}) and (\ref{eqn:tGlimcc}) we reclaim the result of Efetov \cite[(5.30)]{Efe97}.

\newpage

\section{Real Spherical Ensemble}
\setcounter{figure}{0}
\label{sec:SOE}

In Chapter \ref{sec:tG} we generalised the real Ginibre ensemble with the inclusion of the parameter $\tau$, which controlled the degree of symmetry in the matrices. Here we consider a different generalisation that can be viewed from either a geometric viewpoint or as a problem in generalised eigenvalues. Recall that the real Ginibre ensemble possessed an eigenvalue distribution that `naturally' lay in the plane --- by that we mean that in the large $N$ limit the circular law takes effect and the distribution is uniform on the unit (planar) disk. The ensemble of this chapter has eigenvalues that `naturally' live on the sphere; the second of the triumvirate of surfaces of constant curvature discussed in the Introduction.

As it happens, the ensemble that we discuss here is intimately related to a question raised in \cite{eks1994} concerning the distribution of generalised eigenvalues of a pair of real matrices. Recall from the introduction that a generalised eigenvalue of the set $\{ \bA,\bB \}$, where $\bA$ and $\bB$ are $N\times N$ matrices, is a value of $\lambda$ satisfying the equation (\ref{eqn:genEvals}), $\det (\bB -\lambda \bA) =0$. The authors of \cite{eks1994} provide a geometric interpretation of the problem: regard the pair of matrices $\bA,\bB$ as two vectors in $\mathbb{R}^{N^2}$. The corresponding plane spanned by these vectors then intersects the sphere $S^{N^2-1}$ to give a great circle. The real generalised eigenvalues relate to the intersection  of this great circle with the set $\Delta_N$ of all $N \times N$ singular matrices $\bX$ such that $\mathrm{Tr}\: \bX\bX^T=1$ (thus choose $\bX=c(\bA- \lambda \bB)$ for suitable $c$). With $\bA, \bB$ having standard Gaussian entries, the great circle has uniform measure, so the expected number of real eigenvalues is equal to the expected number of intersections of $\Delta_N$ with a random great circle.

Another feature of the random generalised eigenvalue problem studied in \cite{eks1994} is the density $\rho_{(1)}(\lambda)$ of real generalised eigenvalues. By writing $\lambda =\mathrm{tan} \hspace{2pt}\theta$ the generalised eigenvalue equation reads $\det (\mathrm{cos}\theta \bB- \mathrm{sin}\theta \bA)=0$. Using the fact that a pair of standard Gaussians $(x_1,x_2)$ is, as a distribution in the plane, invariant under rotation, it was noted that $(\mathrm{cos} \theta,\mathrm{sin} \theta)$ must be distributed uniformly on the unit circle, and so, by $d\theta = d\lambda /(1+\lambda^2)$,
\begin{align}
\label{eqn:rho}
\rho_{(1)}(\lambda)=\frac{1}{\pi}\frac{E_N}{1+\lambda^2}.
\end{align}

The appearance of circles and spheres can be anticipated. First recall that a Cauchy random variable $\mathcal{C}$ can be defined as the ratio of two Gaussian random variables, and its density is
\begin{align}
\label{eqn:Caudist} \frac{1}{\pi (1+\mathcal{C}^2)}.
\end{align}
(Note that we are only interested in the standard Cauchy distribution here, which is centred at zero and with scale parameter equal to one.) In other words, a Cauchy distributed real variable has uniform distribution on a great circle of a sphere when stereographically projected. Second, note that by factoring out $\bA$ from the determinant in (\ref{eqn:genEvals}) the generalised eigenvalue problem is equivalent to the standard eigenvalue problem for the matrix $\bA^{-1}\bB$. Since $\bA$ and $\bB$ are both random matrices with Gaussian entries, $\bA^{-1}\bB$ is the matrix equivalent of a Cauchy random variable. Indeed, in \cite{eks1994} the authors refer to these matrices as `Cauchy matrices'.

\begin{remark}
Note that by Remark \ref{rem:Ginsing} we can assume that $\bA$ is invertible.
\end{remark}

This leads us to define the \textit{real, spherical ensemble}, consisting of matrices
\begin{align}
\label{def:ainvb} \bY=\bA^{-1}\bB,
\end{align}
where $\bA,\bB$ are $N\times N$ real, Ginibre matrices with elements specified by (\ref{def:Gin_eldist}). In this chapter we shall investigate the statistics of its eigenvalues, particularly in regards to stereographic projection onto the sphere. An analogous ensemble has been studied in \cite{Krish2009,HKPV2009}, where $\bA$ and $\bB$ are complex Ginibre matrices. The real spherical and complex spherical ensembles are analogous in the same way that the real Ginibre and complex Ginibre ensembles are analogous, and the GOE and GUE are analogous.

\begin{figure}[ht]
\begin{center}
\includegraphics[scale=0.6]{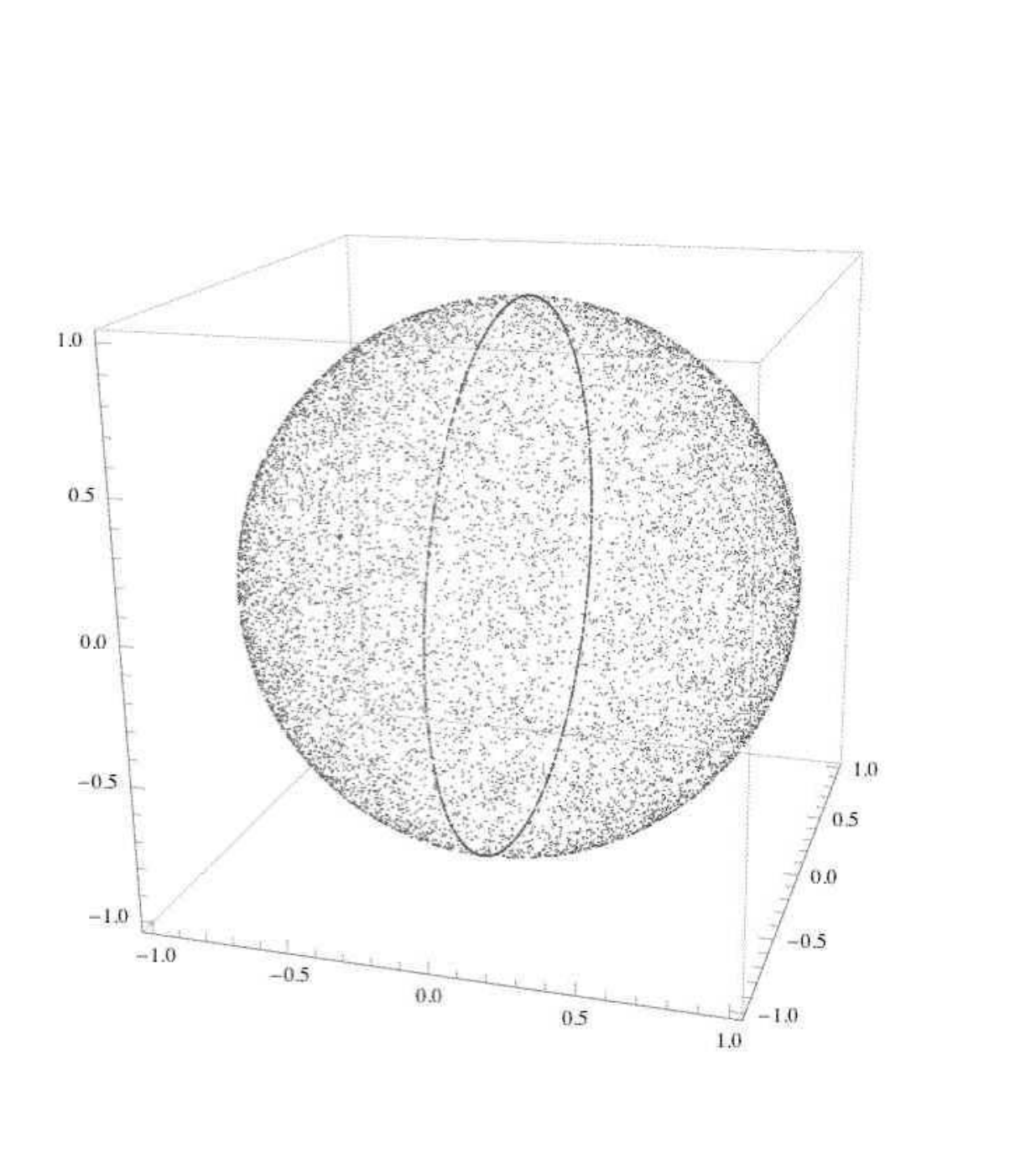}
\end{center}
\caption[Plot of simulated eigenvalues from the real spherical ensemble.]{A stereographic projection of the eigenvalues from 120 independent $100\times 100$ matrices of the form (\ref{def:ainvb}). The great circle of real eigenvalues can be clearly seen.}
\label{fig:dstar}
\end{figure}

As with the real Ginibre matrices, the real spherical matrices differ from their complex comrades is that they exhibit a finite probability of having real eigenvalues, which, in the present case, corresponds to a great circle of uniform non-zero density of eigenvalues. To illustrate this point we have simulated an eigenvalue distribution and stereographically projected it in Figure \ref{fig:dstar}. As with the real Ginibre case, in the large $N$ limit the effect of the real eigenvalues becomes negligible. In \cite{FM11} the authors show that the large $N$ distribution matches that of the complex spherical ensemble, leading them to conjecture that there exists a spherical law that is analogous to the circular law of Proposition \ref{prop:circlaw}. This has since been established \cite{Bord2010}.

A major difference in the analysis of this ensemble compared to that of Chapter \ref{sec:GinOE} is that we will not study the eigenvalues themselves directly. Given that we expect a more or less uniform distribution on the sphere (with the exception of the great circle corresponding to the real line) we use the fractional linear transformation
\begin{align}
\label{7'} z = {1 \over i} {w -1 \over w+1},
\end{align}
which maps the upper half-plane into the unit disk, with the real axis on the circumference. In the case that $z=\lambda \in\mathbb{R}$ then $w$ lies on the unit circle and we can write
\begin{align}
\label{14.2}
\lambda&=\frac{1} {i}\left(1 -\frac{2}{e+1} \right)=\mathrm{tan}\frac{\theta}{2},
\end{align}
with the convention $e:=e^{i\theta}$. In terms of the sphere, this is a projection of one hemisphere into the unit disk. The transformation (\ref{7'}) allows us to take advantage of the rotational symmetry of the problem, enabling us to compute an otherwise intractable integral.

One of the technical consequences of this choice of co-ordinates is that the Pfaffian in the generalised partition function (the real spherical analogue of (\ref{eqn:GinOE_gpf_even})) can be skew-diagonalised for general $\zeta$, which was not possible in the real Ginibre case. This results in probabilities $p_{N,k}$ that are products over the $\alpha$ and $\beta$, which are computationally easier than the Pfaffian or determinant structures heretofore encountered. We will discuss this further in Chapter \ref{sec:Ssops}.

Analysis similar to our study of the real spherical ensemble has been undertaken in \cite{caillol81, FJM1992}, where a Coulomb gas confined to a sphere is examined (although the authors of the latter paper use a system consisting of two oppositely charged species of particle). This viewpoint was exploited in \cite{FM11} to obtain two sum rules for our system here. Also, we remark that there is an analogy with the random polynomials
\begin{align}
\label{eqn:rand_polys}
p_{n}(z)=\sum_{p=0}^n {n \choose p}^{1/2}a_p z^p&,&a_p\sim N[0,1].
\end{align}
When stereographically projected onto the sphere there is of order $\sqrt{N}$ zeros on a great circle corresponding to the real axis \cite{EK95}, but for $N$ large the density is asymptotically uniform on the sphere \cite{Mc09}, which is what we find for the generalised eigenvalues.

\subsection{Element distribution}

As for the other ensembles already considered in this work, we must first establish the elemental distribution. For $N\times N$ matrices $\bA$ and $\bB$ taken from the real Ginibre ensemble, with distributions given by (\ref{eqn:GinOE_eldist}), we wish to write down the probability density function $\mathcal{P}(\bY)$ of $\bY=\bA^{-1}\bB$. This requires changing variables in the joint density of $\bA$ and $\bB$,
\begin{align}
\label{eqn:YjpdfAB} (2\pi)^{-N^2}e^{-\Tr(\bA\bA^T+\bB\bB^T)/2}(d\bA)(d\bB),
\end{align}
to those of $\bY$ and then integrating out the remaining $N^2$ independent variables. For this we will need the following pieces of theory.

\begin{lemma}[\cite{muirhead1982} Theorem 2.1.5]
\label{lem:alpha_tensor_beta}
For $\bX=\alpha \bY \beta$, where $\alpha_{P\times P}$ and $\beta_{Q\times Q}$ are arbitrary real matrices and $\bY_{P\times Q}$ has $PQ$ independent entries (ie. the wedge product $(d\bY)$ has $PQ$ factors) then
\begin{align}
\nonumber (d\bX)&=\left|\mathrm{det}(\alpha \otimes\beta^T)\right|(d\bY)\\
\nonumber &=\left|\mathrm{det}(\alpha)^Q\mathrm{det}(\beta)^P\right|(d\bY).
\end{align} 
\end{lemma}

\begin{lemma}[\cite{muirhead1982} Theorem 2.1.14]
\label{lem:tilde_const_covarbs}
For an $N\times M$ matrix (with $M\geq N$) $\bX$, if $\bM=\bX\bX^T$ then
\begin{align}
\label{eqn:tilde_const_covarbs} (d\bX)= \tilde{c} \; \mathrm{det}\bM^{(N-M-1)/2}(d\bM),
\end{align}
where $\tilde{c}$ is independent of $\bM$.
\end{lemma}

The following is another corollary of the Selberg integral (\ref{def:Selb}).
\begin{corollary}[\cite{forrester?} Proposition 4.7.3]
\label{cor:varselb}
By making the replacements $ t_l \mapsto x_l/L, 2\lambda=\beta, \lambda_1=\beta L/2$ and $\lambda_2=\beta L/2$ in (\ref{def:Selb}) we have the limit
\begin{align}
\nonumber &\lim_{L\to\infty} L^{N+N a \beta/2+\beta N(N-1)/2}S_N(\beta a/2,\beta L/2, \beta/2)\\
\nonumber &=\int_0^{\infty}dx_1\cdot\cdot\cdot \int_0^{\infty}dx_N \prod_{l=1}^N x_l^{\beta a/2} e^{-\beta x_l/2}\prod_{1\leq j < l \leq N}|x_l -x_j|^{\beta}\\
\label{cor:varselb1} &=(\beta /2)^{-N(a\beta /2+1+(N-1)\beta /2)}\prod_{j=0}^{N-1}\frac{\Gamma \left(1+(j+1)\beta /2\right)\Gamma (a\beta/2+1+j\beta/2)}{\Gamma (1+\beta/2)}.
\end{align}
\end{corollary}

We may now establish the jpdf for the elements of the matrix $\bY$.
\begin{proposition}
\label{prop:elementjpdf}
Let $\bA,\bB$ be $N\times N$ real Ginibre matrices, having elements distributed according to (\ref{def:Gin_eldist}), and let $\bY=\bA^{-1}\bB$. The probability density function of $\bY$ is
\begin{align}
\label{eqn:elementjpdf} \mathcal{P}(\bY)&=\pi^{-N^2/2}\prod_{j=0}^{N-1} \frac{\Gamma((N+1)/2+j/2)}{\Gamma((j+1)/2)}\; \mathrm{det}(\mathbf{1}_N+\bY\bY^T)^{-N}.
\end{align}
\end{proposition}

\textit{Proof:} Writing $\bB=\bA\bY$ we let $\alpha=\bA$ and $\beta=\mathbf{1}_{N}$ in Lemma \ref{lem:alpha_tensor_beta} to see that
\begin{align}
\label{eqn:dBi} (d\bB)&=|\mathrm{det} \bA|^N(d\bY).
\end{align}
Using (\ref{eqn:dBi}) we change variables in (\ref{eqn:YjpdfAB}) to obtain
\begin{align}
\nonumber (2\pi)^{-N^2}e^{-\frac{1}{2}\mathrm{Tr}\left(\bA\bA^T(\mathbf{1}_N+\bY\bY^T)\right)}|\mathrm{det}\bA\bA^T|^{N/2}(d\bA)(d\bY).
\end{align}
Setting $\bC:=\bA\bA^T$, Lemma \ref{lem:tilde_const_covarbs} tells us that $(d\bA)=\tilde{c}\: (\det \bC)^{-1/2}(d\bC)$ for some $\tilde{c}$ to be determined. Integrating over $\bC$ (noting that $\bC$ is positive definite, denoted $\bC>0$) we have
\begin{align}
\nonumber &\mathcal{P}(\bY)(d\bY)=(2\pi)^{-N^2}\tilde{c}\int_{\bC>0}(\mathrm{det}\: \bC)^{(N-1)/2}e^{-\frac{1}{2} \mathrm{Tr}\left(\bC (\mathbf{1}_N+\bY\bY^T)\right)} (d\bC)(d\bY)\\
\nonumber &=(2\pi)^{-N^2}\tilde{c}\int_{\bC>0}(\mathrm{det}\: \bC)^{(N-1)/2} e^{-\frac{1}{2}\mathrm{Tr}\left((\mathbf{1}_N+\bY\bY^T)^{1/2}\bC (\mathbf{1}_N+\bY\bY^T)^{1/2}\right)} (d\bC)(d\bY).
\end{align}
Carrying out the change of variables $\bC\mapsto  (\mathbf{1}_N+\bY\bY^T)^{1/2}\bC(\mathbf{1}_N+\bY\bY^T)^{1/2}$ we use Lemma \ref{lem:adma} to find
{\small
\begin{align} 
\nonumber \mathcal{P}(\bY)(d\bY)=(2\pi)^{-N^2}\tilde{c}\; \mathrm{det}(\mathbf{1}_N+\bY\bY^T)^{-N}\int_{\bC>0} (\mathrm{det}\: \bC)^{(N-1)/2} e^{-\frac{1}{2} \mathrm{Tr}\: \bC}(d\bC)(d\bY).
\end{align}
}Taking an integral transform of both sides of (\ref{eqn:tilde_const_covarbs}) we can calculate $\tilde{c}$ as
\begin{align}
\nonumber \int e^{-\mathrm{Tr}(\bA\bA^T)/2}(d\bA)&=\tilde{c}\int_{\bC>0}e^{-(\mathrm{Tr}\: \bC)/2}\frac{(d\bC)}{(\mathrm{det}\: \bC)^{1/2}},
\end{align}
where, on the LHS, we have $N^2$ standard Gaussian integrals and so
\begin{align}
\nonumber \tilde{c}&=\frac{(2\pi)^{N^2/2}}{\int_{\bC>0} (\mathrm{det}\: \bC)^{-1/2} e^{-(\mathrm{Tr}\: \bC)/2}(d\bC)}.
\end{align}
Substituting for $\tilde{c}$ in the above formula for $\mathcal{P}(\bY) (d\bY)$ gives
\begin{align}
\nonumber \mathcal{P}(\bY)(d\bY)&= (2\pi)^{-N^2/2}\mathrm{det}(\mathbf{1}_N+\bY\bY^T)^{-N}\frac{\int_{\bC>0}\mathrm{det}(\bC)^{(N-1)/2}e^{-\mathrm{Tr}(\bC)/2}(d\bC)}{\int_{\bC>0}\mathrm{det}(\bC)^{-1/2}e^{-\mathrm{Tr}(\bC)/2}(d\bC)}(d\bY).
\end{align}
Since $\bC$ is symmetric we may use Proposition \ref{prop:GOE_J} to rewrite the ratio of integrals as
\begin{align}
\nonumber &\frac{\int_{\bC>0}(\mathrm{det}\: \bC)^{(N-1)/2} e^{-(\mathrm{Tr}\: \bC)/2}(d\bC)} {\int_{\bC>0} (\mathrm{det}\: \bC)^{-1/2} e^{-(\mathrm{Tr}\: \bC)/2}(d\bC)}\\
\nonumber & =  \frac{\int_{(0,\infty)^N} \prod_{l=1}^Nx_l^{(N-1)/2}e^{-x_l/2} \prod_{j<k}^N|x_k-x_j|dx_1 \cdot\cdot\cdot dx_N}{\int_{(0,\infty)^N}\prod_{l=1}^Nx_l^{-1/2} e^{-x_l/2} \prod_{j<k}^N|x_k-x_j| dx_1\cdot\cdot\cdot dx_N},
\end{align}
which is seen to be a ratio of the Selberg-type integrals of Corollary \ref{cor:varselb}. The result follows on using the formula (\ref{cor:varselb1}).

\hfill $\Box$

From the discussion at the beginning of this chapter we know that a Cauchy variable has distribution (\ref{eqn:Caudist}) and we see that (\ref{eqn:elementjpdf}) is a matrix analogue of this distribution, which we anticipated since $\bA^{-1}\bB$ is the matrix analogue of a Cauchy variable. Note that when $N=1$ (\ref{eqn:elementjpdf}) is exactly (\ref{eqn:Caudist}).

\subsection{Eigenvalue distribution}
\label{sec:Sevaldist} 

As we know from the study of the real Ginibre ensemble, for a general $N \times N$ non-symmetric real matrix, we will have have $0 \leq k \leq N$ real eigenvalues, where $N$ has the same parity as $k$. From knowledge of (\ref{eqn:elementjpdf}), by a suitable change of variables, we can extract the eigenvalue distribution for each allowed $k$. In this task we are motivated by the work on the analogous complex spherical ensemble of Hough \textit{et al} \cite{HKPV2009} (see also \cite{FK2009}). In particular, we again work with the real Schur decomposition (\ref{eqn:GinOE_decomp}), yielding $\bY=\bQ\bR_N\bQ^T$, where $\bQ$ is real orthogonal (each column is an eigenvector of $\bY$, with the restriction that the entry in the first row is positive) and $\bR_N$ is the same as $\bR$ in (\ref{def:triangular_mat}) (where we have introduced a subscript denoting the number of rows and columns for later convenience). For a unique decomposition we impose the ordering (\ref{9'}).

Since we are looking to change variables from the elements of $\bY$ to the eigenvalues of $\bY$ as implied by the real Schur decomposition, before proceeding we first need knowledge of the corresponding Jacobian. With $\tilde{\bR}_N$ the strictly upper triangular part of $\bR_N$, we know from Proposition \ref{prop:GinJ} that
\begin{align}
\nonumber (d\bY)&=2^{(N-k)/2}\prod_{j<p}|\lambda(R_{pp})-\lambda(R_{jj})| \prod_{l=k+1}^{(N+k)/2} |b_l-c_l|\\
\nonumber &\times(d\tilde{\bR}_N)(\bQ^Td\bQ) \prod_{s=1}^{k}d\lambda_s \prod_{l=k+1}^{(N+k)/2}dx_ldb_ldc_j,
\end{align}
using the notation of (\ref{def:GinOEpods}). Note that the dependence on $\bQ$ can be immediately dispensed with by integrating over $(\bQ^Td\bQ)$ using (\ref{eqn:RTdR_integ}).

So far the procedure is exactly the same as for the real Ginibre ensemble, but the integration over the elements of $\tilde{\bR}_N$ is no longer straightforward. Indeed, following \cite{HKPV2009}, we integrate over the columns in $\tilde{\bR}_N$ corresponding to each of the eigenvalues (or pair of complex conjugate eigenvalues) in turn, starting with the two columns on the far right (corresponding to the complex conjugate pair with largest real part). This process is then iterated from right to left, until the complex eigenvalue columns are exhausted. We then move on to iteratively integrate over the single columns above the real eigenvalues, which will then leave us with the eigenvalue jpdf. This procedure can be found in the work of Hua \cite{Hua1963}.

\subsubsection{Complex eigenvalue columns}
\label{sec:iitc}

In the region $j>k$ of $\bR_N$ we can isolate the last two rows and columns to write
\begin{align}
\label{eqn:RNdecomp} \bR_N=\left[\begin{array}{cc}
\bR_{N-2} & u\\
\mathbf{0}^T & z_m
\end{array}\right],
\end{align}
where $u$ is of size $(N-2)\times 2$ and $\0^T$ is of size $2\times (N-2)$. So then
\begin{align}
\nonumber &\1_N +\bR_N \bR_N^T\\
\nonumber &=\left[\begin{array}{cc}
\1_{N-2}+\bR_{N-2} \bR_{N-2}^T+ uu^T & uz_m^T\\ 
z_mu^T & \1_2+z_m z_m^T
\end{array}\right]\\
\nonumber &=\left[\begin{array}{cc}
\1_{N-2}+ \bR_{N-2}\bR_{N-2}^T+ uu^T -uz_m^T (\1_2+z_mz_m^T)^{-1} z_mu^T & \0\\ 
z_mu^T & \1_2 +z_mz_m^T
\end{array}\right],
\end{align}
where we have used elementary row operations to obtain the second equality. Before proceeding, note the identity
\begin{align}
\nonumber &z_m^T(\1_2 +z_mz_m^T)^{-1}z_m\\
\nonumber &= z_m^T (\1_2 -z_mz_m^T+ z_mz_m^Tz_m z_m^T-z_mz_m^Tz_mz_m^Tz_mz_m^T+...)z_m\\
\nonumber &=(\1_2 -z_m^Tz_m+z_m^Tz_mz_m^T z_m-...) z_m^Tz_m\\
\nonumber &=(\1_2 +z_m^Tz_m)^{-1} z_m^T z_m\\
\label{eqn:zm1} &=\1_2-(\1_2 +z_m^Tz_m)^{-1}.
\end{align}
This enables us to expand the determinant as so
\begin{align}
\nonumber &\det \left(\1_N+\bR_N \bR_N^T \right)\\
\nonumber &= \det \left(\1_2+z_mz_m^T \right)\det \left(\1_{N-2}+\bR_{N-2} \bR_{N-2}^T+ u(\1_2+z_m z_m^T)^{-1} u^T \right)\\
\nonumber &=\det \left(\1_2+z_m z_m^T)\det(\1_{N-2}+\bR_{N-2} \bR_{N-2}^T \right)\\
\nonumber &\times \det \left(\1_{N-2}+(\1_{N-2}+\bR_{N-2} \bR_{N-2}^T)^{-1} u(\1_2 + z_m z_m^T)^{-1} u^T \right)\\
\nonumber &=\det \left(\1_2+z_mz_m^T) \det(\1_{N-2}+\bR_{N-2} \bR_{N-2}^T \right)\\
\label{eqn:det1RR} &\times\det \left(\1_2+(\1_2+ z_mz_m^T)^{-1/2}u^T (\1_{N-2}+ \bR_{N-2}\bR_{N-2}^T)^{-1}u (\1_2+ z_mz_m^T)^{-1/2}\right),
\end{align}
where we have used (\ref{eqn:zm1}) to obtain the first quality and (\ref{eqn:1+AB}) to obtain the third.

We are now in a position to integrate over the elements of the matrix $u$
{\small
\begin{align}
\nonumber &\int\frac{(du)}{\det(\1_N+\bR_N \bR_N^T)^N}=\frac{1}{\det(\1_2+z_m z_m^T)^N \det(\1_{N-2}+\bR_{N-2} \bR_{N-2}^T)^N}\\
\nonumber &\times \int\frac{(du)}{ \det(\1_{2}+ (\1_2+z_m z_m^T)^{-1/2} u^T(\1_{N-2}+\bR_{N-2} \bR_{N-2}^T)^{-1} u(\1_2+z_m z_m^T)^{-1/2})^N},
\end{align}
}where the integral for each independent real component of $u$ is over the real line. Changing variables $ v=(\1_{N-2}+\bR_{N-2} \bR_{N-2}^T)^{-1/2} \: u\: (\1_2+z_m z_m^T)^{-1/2}$ we use Lemma \ref{lem:alpha_tensor_beta} to find
\begin{align}
\nonumber \int\frac{(du)}{\det(\1_N+\bR_N \bR_N^T)^N}&=\frac{1}{\det(\1_2+z_mz_m^T)^{N/2+1}\det(\1_{N-2}+\bR_{N-2} \bR_{N-2}^T)^{N-1}}\\
\nonumber &\times\int\frac{(dv)}{\det(\1_{2}+v^Tv)^N}.
\end{align}
Iterating over all columns corresponding to complex eigenvalues we have
\begin{align}
\nonumber &\int\frac{(du_{N-2})\cdot\cdot\cdot (du_{k+1})}{\det(\1_N+\bR_N \bR_N^T)^N}= \frac{1}{\det(\1_k+\bR_k \bR_k^T)^{(N+k)/2}}\\
\label{eqn:with_subs} &\times \prod_{s=k+1}^{(N+k)/2} \frac{1}{\det(\1_2+z_sz_s^T)^{N/2+1}}\prod_{s=0}^{(N-k)/2-1}\int\frac{(dv_{N-2-2s})}{\det(\1_2+v_{N-2-2s}^Tv_{N-2-2s})^{N-s}},
\end{align}
where the subscripts $*$ on the matrices $v_*,du_*,dv_*$ denote their number of
rows.

To evaluate each of the $(N-k)/2$ integrals we use a method similar to that used in Proposition \ref{prop:elementjpdf}. Firstly, for each $v_{N-2-2s}$, we let $v_{N-2-2s}^Tv_{N-2-2s}=\bC$ and apply Lemma \ref{lem:tilde_const_covarbs} to get 
\begin{equation}\label{11'}
(dv_{N-2-2s})=\tilde{c}\: (\det \bC)^{(N-2s-5)/2}(d\bC),
\end{equation}
 and 
\begin{eqnarray}
\label{eqn:Sctilde} \tilde{c}\int (\det \bC)^{(N-2s-5)/2}e^{-\Tr \: \bC}(d\bC)=\int e^{-\Tr(v^Tv)}(dv)=\pi^{N-2s-2}.
\end{eqnarray}
So, with $\kappa:=(N-2s-5)/2$,
\begin{align}
\nonumber &\int\frac{(dv_{N-2-2s})}{\det(\1_2+v_{N-2-2s}^Tv_{N-2-2s})^{N-s}}=\pi^{N-2s-2}\frac{\int(\det \bC)^{\kappa}\det(\1_2+ \bC)^{s-N} (d\bC)}{\int(\det \bC)^{\kappa}e^{-\Tr \: \bC}(d\bC)}\\
\nonumber & = \pi^{N-2s-2}\int_0^{\infty}\int_0^{\infty}\frac{x_1^{\kappa}}{(1+x_1)^{N-s}}\frac{x_2^{\kappa}}{(1+x_2)^{N-s}}|x_1-x_2|dx_1dx_2\\
\nonumber & \quad \times\left( \int_0^{\infty}\int_0^{\infty}x_1^{\kappa}x_2^{\kappa}e^{-x_1}e^{-x_2} |x_1-x_2|dx_1dx_2  \right)^{-1}\\
\nonumber & =\pi^{N-2s-2}\int_0^1\int_0^1y_1^{\kappa} y_2^{\kappa}(1-y_1)^{(N-1)/2}(1-y_2)^{(N-1)/2} |y_1-y_2|dy_1dy_2\\
\label{eqn:compx_selberg} & \quad \times\left( \int_0^{\infty}\int_0^{\infty}x_1^{\kappa}x_2^{\kappa}e^{-x_1}e^{-x_2} |x_1-x_2|dx_1dx_2  \right)^{-1},
\end{align}
where use was made of Proposition \ref{prop:GOE_J} for the second equality, and the change of variables $y=x/(1+x)$ for the third. We now have a ratio of Selberg-type integrals which can be evaluated using (\ref{cor:varselb1}) as
\begin{align}
\nonumber &\int\det(\1_2+v_{N-2-2s}^T v_{N-2-2s})^{-(N-s)}(dv)\\
\label{12'} &=\pi^{N-2s-2}\frac{\Gamma((N+1)/2)}{\Gamma(N-s-1/2)} \frac{\Gamma(N/2+1)} {\Gamma(N-2)}.
\end{align}
The case $N$ odd, $k=1$, corresponding to $s=(N-1)/2-1$ is special since then $v_{N-2-2s}$ consists of 1 row and 2 columns, and thus is the only case in which the number of rows is less than the number of columns and so Lemma \ref{lem:tilde_const_covarbs} does not apply. We must then write
$$
\det ({\bf 1}_2 + v^T_{N-2-2s} v_{N-2-2s})^{-p} = (1 + v_{N-2-2s} v_{N-2-2s}^T)^{-p},
$$
using (\ref{eqn:1+AB}). However, it turns out that the change this implies to 
(\ref{eqn:compx_selberg}) does not effect the evaluation (\ref{12'}). So in all cases, after having integrated over the columns corresponding to complex eigenvalues we are left with
\begin{align}
\nonumber &\int\frac{(du_{N-2})\cdot\cdot\cdot(du_{k+1})}{\det(\1_N+\bR_N \bR_N^T)^N} =\prod_{s=k+1}^{(N+k)/2}\frac{1}{\det(\1_2+z_sz_s^T)^{N/2+1}}\\
\nonumber &\qquad \times\prod_{s=0}^{(N-k)/2-1}\pi^{N-2s-2}\frac{\Gamma((N+1)/2)}{\Gamma(N-s-1/2)}\frac{\Gamma(N/2+1)}{\Gamma(N-2)} \hspace{3pt}\frac{1}{\det(\mathbf{1}_k+\bR_k \bR_k^T)^{(N+k)/2}}.
\end{align}

It remains to compute the integrals over the columns corresponding to the real eigenvalues.

\subsubsection{Real eigenvalue columns}
\label{sec:iitr}

We see that we are left with a function of $\bR_k$, which is the upper-left sub-block of $\bR_N$. Similar to the process in the previous section, we isolate the last row and column
\begin{eqnarray}
\nonumber \bR_k=\left[\begin{array}{cc}
\bR_{k-1} & u_{k-1}\\
\0^T & \lambda_k
\end{array}\right],
\end{eqnarray}
where now $u_{k-1}$ is of size $(k-1)\times 1$ and $\0^T$ is of size $1\times (k-1)$. Following the same procedure that led to (\ref{eqn:det1RR}) for the columns corresponding to the complex eigenvalues, we find
\begin{align}
\nonumber \det(\1_k+\bR_k \bR_k^T)&= (1+ \lambda_k^2)\det(\1_{k-1}+\bR_{k-1} \bR_{k-1}^T)\\
\nonumber &\times \left(1+(1+ \lambda_k^2)^{-1} u_{k-1}^T (\1_{k-1}+\bR_{k-1} \bR_{k-1}^T)^{-1} u_{k-1} \right).
\end{align}
Setting $v_{k-1}=(\1_{j-1}+ \bR_{j-1} \bR_{j-1}^T)^{-1/2}u_{k-1} (1+ \lambda_j^2)^{-1/2}$ and again making use of Lemma \ref{lem:alpha_tensor_beta} we have
\begin{align}
\nonumber &\int\frac{(du_{k-1})}{\det(\1_k+\bR_k \bR_k^T)^{(N+k)/2}}\\
\nonumber &=\frac{1}{(1+\lambda_k^2)^{(N+1)/2}\det \left(\1_{k-1}+\bR_{k-1} \bR_{k-1}^T \right)^{(N+k-1)/2}} \int\frac{(dv_{k-1})}{(1+ v_{k-1}^T v_{k-1})^{(N+k)/2}}.
\end{align}
Iterating over the remaining columns of $\bR_k$ gives
\begin{align}
\nonumber &\int\frac{(du_{k-1})\cdot\cdot\cdot(du_1)}{\det(\1+\bR_k \bR_k^T)^{(N+k)/2}}\\
\nonumber &=\prod_{s=1}^k\frac{1}{(1+\lambda_s^2)^{(N+1)/2}} \prod_{s=1}^{k-1} \int\frac{(dv_{k-s})}{(1+ v_{k-s}^T v_{k-s})^{(N+k)/2-(s-1)/2}}
\end{align}
(cf.~(\ref{eqn:with_subs})). To evaluate the integrals, we use the same method as for the integrals in (\ref{eqn:with_subs}) --- involving Lemma \ref{lem:tilde_const_covarbs} and a now one-dimensional case of the Selberg integral, which is the beta integral. This gives
\begin{align}
\nonumber \int\frac{(dv_{k-s})}{(1+v_{k-s}^Tv_{k-s})^{(N+k)/2-(s-1)/2}}=\pi^{(k-s)/2}\frac{\Gamma((N+1)/2)}{\Gamma((N+k-s+1)/2)},
\end{align}
and so
\begin{align}
\nonumber &\int\frac{(du_{k-1})\cdot\cdot\cdot(du_1)}{\det(\1_k+\bR_k \bR_k^T)^{(N+k)/2}}\\
\nonumber &=\prod_{s=1}^k\frac{1}{(1+\lambda_s^2)^{(N+1)/2}}\prod_{s=1}^{k-1}\pi^{(k-s)/2}\frac{\Gamma((N+1)/2)}{\Gamma((N+k-s+1)/2)}.
\end{align}

\subsubsection{Eigenvalue jpdf and fractional linear transformation}

According to the working in the preceding section (\ref{eqn:elementjpdf}) has been reduced to the following distribution of $\{ \lambda_i,x_i,b_i,c_i\}$ (with $x_i,b_i$ and $c_i$ from (\ref{eqn:Schurz}))
\begin{align}
\nonumber &\pi^{-(N-k)/4}\Gamma((N+1)/2)^{N/2}\Gamma(N/2+1)^{N/2} \left( \frac{\Gamma((N+1)/2)}{\Gamma(N/2+1)}\right)^{k/2} \prod_{j=1}^{N-1} \frac{1} {\Gamma(j/2)^2}\\
&\nonumber \quad  \times\prod_{s=k+1}^{(N+k)/2} \frac{1}{\det(\1+z_sz_s^T)^{N/2+1}} \prod_{s=1}^{k} \frac{1}{(1+ \lambda^2_s)^{(N+1)/2}} 2^{(N-k)/2}\prod_{l=k+1}^{(N+k)/2} |b_l-c_l|\\
\label{eqn:reduced_dist}& \quad \times \quad \prod_{j<p}|\lambda(R_{pp})-\lambda(R_{jj})|,
\end{align}
where use has been made of the simplification
{\small
\begin{align}
\nonumber &\prod_{s=1}^{k-1}\pi^{(k-s)/2}\frac{ \Gamma((N+1)/2)}{ \Gamma((N+k-s+1)/2)}\prod_{s=0}^{(N-k)/2-1}\pi^{N-2s-2}\frac{ \Gamma((N+1)/2)}{ \Gamma(N-s-1/2)}\frac{\Gamma(N/2+1)}{ \Gamma(N-s)}\\
\nonumber &= \pi^{(k+N^2-2N)/4} \Gamma((N+1)/2)^{N/2} \Gamma(N/2+1)^{N/2} \left(\frac{\Gamma((N+1)/2)}{\Gamma(N/2+1)} \right)^{k/2}\\
\nonumber & \times \prod_{s=0}^{N-1}\frac{1} {\Gamma((N+1+s)/2)}.
\end{align}
}

With $\epsilon_s=1+x_s^2+y_s^2$ and $\delta_s=b_s-c_s$ we see that $\det (\mathbf{1}_2+z_sz_s^T) = \epsilon_s^2+\delta_s^2$. We can use (\ref{eqn:GinOE_covs}) to change variables from $b_l,c_l$ to $y_l,\delta_l$
recalling the correction that $-\infty < \delta < \infty$. Now we integrate over $\delta$
\begin{align}
\nonumber &\int_{\delta=-\infty}^{\delta=\infty} \frac{|b_s-c_s|}{\det(\1+z_sz_s^T)^{N/2+1}}\: dx_s db_s dc_s\\
\nonumber &=4y_s\int_{\delta=0}^{\delta=\infty} \frac{\delta \; d\delta}{(\epsilon_s^2+\delta^2)^{N/2+1} \sqrt{4y^2+\delta^2}}\: dx_sdy_s\\
\label{eqn:delta_integ} &=4y_s\int_{t=2y_s}^{t=\infty}\frac{dt}{(\epsilon_s^2-4y_s^2+t^2)^{N/2+1}}\: dx_s dy_s.
\end{align}
Substituting (\ref{eqn:delta_integ}) in (\ref{eqn:reduced_dist}) as appropriate
gives the reduced jpdf, but (\ref{eqn:delta_integ}) as written appears intractable for
further analysis. On the other hand the discussion at the beginning of this chapter suggests that when projected on to the sphere the eigenvalue density is unchanged by rotation in the $X$--$Z$ plane, where $X,Y,Z$ are the co-ordinates after stereographic projection. This suggests that simplifications can be achieved by an appropriate mapping of the half-plane that contains the rotational symmetry of the half sphere.

We therefore introduce the fractional linear transformation (\ref{7'}) mapping the upper half-plane to the interior of the unit disk, with (\ref{14.2}) (recalling the definition of $e$) mapping the real line to a great circle through the poles. In particular, the complicated dependence on $x_s$ and $y_s$ in (\ref{eqn:delta_integ}) is now unravelled.

\begin{lemma}
Let $\epsilon_s = 1 + x_s^2 + y_s^2$. With the change of co-ordinates (\ref{7'}) we have
\begin{align}
\nonumber &y_s\int_{t=2y_s}^{t=\infty}\frac{dt}{(\epsilon_s^2- 4y_s^2+ t^2)^{N/2+1}} \: dx_s dy_s\\
\label{14.3} & \qquad = \frac{(1-|w_s|^2)\: |1+w_s|^{2N-4}} {2^{2N-2} \: |w_s|^{N+1}} \int_{\frac{|w_s|^{-1} -|w_s|}{2}}^{\infty} \frac{dt}{\left(1+t^2\right)^{N/2+1}} \: du_s dv_s.
\end{align}
\end{lemma}

\textit{Proof}: Noting that
\begin{align}
\nonumber y_s &= {1 - |w_s|^2 \over |1 + w_s|^2}, \\
\nonumber \epsilon_s^2 - 4y_s^2 &= {16 |w_s|^2 \over |1 + w_s|^4}, \\
\nonumber dx_s dy_s &= \Big | {dz_s \over dw_s} \Big |^2 du_s dv_s =
{4 \over |1 + w_s|^4} du_s dv_s,
\end{align}
we reduce the given expression to
\begin{align}
\label{eqn:14.3a} \frac{16} {|1 + w_s|^4} \frac{1 - |w_s|^2} {|1 + w_s|^2}
\int_{\frac{2(1 - |w_s|^2)} {|1+w_s|^2}}^{\infty}
{dt \over \left( \frac{16 |w_s|^2} {|1 + w_s|^4} + t^2 \right)^{N/2 + 1}} \, du_s dv_s.
\end{align}
The RHS of (\ref{14.3}) results from (\ref{eqn:14.3a}) after the change of variables $t \mapsto 4|w_s| t/|1 + w_s|^2$.

\hfill $\square$

For the product of differences in (\ref{eqn:reduced_dist}), the substitutions (\ref{7'}) and (\ref{14.2}) give
{\small
\begin{align}
\nonumber \prod_{j<p}^k|\lambda_p-\lambda_j|&=(-2i)^{k(k-1)/2}\prod_{s=1}^k\frac{(\bar{e}_s)^{(k-1)/2}}{|e_s+1|^{k-1}}\prod_{j<p}^k(e_p-e_j)
\end{align}
for the real-real factors,
\begin{align}
\nonumber &\prod_{j=1}^k\prod_{s=k+1}^{(N+k)/2}|\lambda_j-z_s||\lambda_j-\bar{z}_s|=(-1)^{k(N-k)/2}2^{(N-k)k}\prod_{j=1}^k(\bar{e}_j)^{(N-k)/2}\left| \frac{1}{e_j+1}\right|^{(N-k)}\\
\nonumber &\times \prod_{s=k+1}^{(N+k)/2}(\bar{w}_s)^k\left| \frac{1}{w_s+1}\right|^{2k}\prod_{j=1}^k\prod_{s=k+1}^{(N+k)/2}(w_s-e_j)\left(\frac{1}{\bar{w}_s}-e_j\right)
\end{align}
for the real-complex factors, and
\begin{align}
\nonumber &\prod_{k+1\leq a < b \leq (N+k)/2}\hspace{-26pt} |z_a-z_b||\bar{z}_a-\bar{z}_b| \hspace{-6pt} \mathop{\prod_{c,d=k+1}^{(N+k)/2}}_{c \neq d} \hspace{-6pt} |z_c-\bar{z}_d|=(-2)^{2\left(\frac{N-k}{2}\frac{N-k-2}{2}\right)} \prod_{j=k+1}^{(N+k)/2} \hspace{-6pt} (1-|w_j|^2)^{-1}\\
\nonumber &\times \hspace{-6pt} \prod_{s=k+1}^{(N+k)/2}(\bar{w})^{N-k-1}\left| \frac{1}{w_s+1}\right|^{2(N-k-2)}\prod_{a < b}(w_b-w_a)\left( \frac{1}{\bar{w}_b}-\frac{1}{\bar{w}_a} \right)\prod_{c,d=k+1}^{(N+k)/2}\left( \frac{1}{\bar{w}_d}-w_c\right)
\end{align}
}for complex-complex. An essential feature is that, apart from the creation of some one-body terms, the product of difference structure is conserved by the substitutions. Combining all this, and using the identity
\begin{align}
\nonumber \left| \frac{1}{e_j+1}\right|^{N-1}&= \left(\frac{1} {2\: \mathrm{cos} (\theta_j/2)}\right)^{N-1},
\end{align}
we have the explicit form of the eigenvalue jpdf in the variables (\ref{7'}) and (\ref{14.2}).

\begin{proposition}
\label{thm:ainvb_jpdf}
Let $\bA,\bB$ be $N\times N$ real Ginibre matrices having elements distributed according to (\ref{def:Gin_eldist}), and let $\bY=\bA^{-1} \bB$. In the variables (\ref{7'}) and (\ref{14.2}) the eigenvalue jpdf of $\bY$, conditioned to have $k$ real eigenvalues ($k$ being of the same parity as $N$), is 
\begin{align}
\label{eqn:q(y)} \mathcal{Q}(\bY)= A_{k,N} \prod_{j=1}^k \tau(e_j) \prod_{s=k+1}^{(N+k)/2} \frac{1} {|w_s|^2}\: \tau(w_s)\tau \left(\frac{1}{\bar{w}_s}\right) \Delta\left(\mathbf{e}, \mathbf{w}, \mathbf{\frac{1}{\bar{w}}} \right),
\end{align}
with $\mathbf{e}=\{e_1,...,e_k \}$, $\mathbf{w}=\{w_1,\bar{w}_1,...,w_{(N-k)/2},\bar{w}_{(N-k)/2} \}$ and
\begin{align}
\nonumber A_{k,N}&= \quad \frac{(-1)^{(N-k)/2((N-k)/2-1)/2+(N-k)k/2-k(k-1)/4}}{2^{(N(N-1)+k)/2}}\prod_{j=1}^N\frac{1}{\Gamma(j/2)^2}\\
&\nonumber \times\Gamma((N+1)/2)^{N/2}\Gamma(N/2+1)^{N/2},\\
\nonumber \tau(x)&=\left( \frac{1}{x}\right)^{(N-1)/2}\left[\frac{1}{\sqrt{\pi}}\int_{\frac{|x|^{-1}-|x|}{2}}^{\infty}\frac{dt}{\left(1+t^2\right)^{N/2+1}}\right]^{1/2},
\end{align}
\begin{align}
\nonumber \Delta\left(\mathbf{e},\mathbf{w},\mathbf{\frac{1}{\bar{w}}}\right)&=\prod_{j<p}(e_p-e_j)\prod_{j=1}^k\prod_{s=k+1}^{(N+k)/2}(w_s-e_j) \left(\frac{1} {\bar{w}_s}-e_j \right)\\
\nonumber &\quad \times \prod_{a<b}(w_b-w_a)\left(\frac{1}{\bar{w}_s}-\frac{1}{\bar{w}_s}\right) \prod_{c,d=k+1}^{(N+k)/2}\left( \frac{1}{\bar{w}_d}-w_c \right).
\end{align}
\end{proposition}

Note that when $x\in\mathbb{R}$ we have
\begin{align}
\label{eqn:taureal} \tau(x)=\left( \frac{1}{x}\right)^{(N-1)/2} \left( \frac{\Gamma((N+1)/2)} {2\: \Gamma(N/2)} \right)^{1/2}.
\end{align}

\subsection{Generalised partition function}

Conceptually, the procedure in this section is identical to that of Chapter \ref{sec:Ggpf} for both even and odd. To find a Pfaffian form of the generalised partition function (or ensemble average) we integrate over the eigenvalue jpdf (\ref{eqn:q(y)}), introducing some indeterminants $u,v$. Writing the Vandermonde product as an $N\times N$ determinant we apply the method of integration over alternate variables ultimately resulting in an $N/2 \times N/2$ Pfaffian. However, the key technical difference from Chapter \ref{sec:Ggpf} is a re-ordering of the matrix columns during the alternate variable integration, which is slightly more complicated in the odd case (\ref{eqn:poly_order_odd}) than the even (\ref{eqn:poly_ordering}). It will turn out that these re-orderings result in separate even and odd skew-orthogonal polynomials (see Propositions \ref{prop:skew_polys_even} and \ref{prop:odd_polys}), which have the benefit of skew-orthogonalising the generalised partition function Pfaffian for general $\zeta$, unlike in the real Ginibre ensemble.

With $\mathcal{Q}(\bY)$ from (\ref{eqn:q(y)}) we use (\ref{def:multi_gen_part_fn}) (with $m=2$) to define the generalised partition function of the spherical ensemble
\begin{align}
\nonumber Z_{k,(N-k)/2}[u,v]_S&=\frac{1}{((N-k)/2)!}\int_0^{\theta_2}d\theta_1\int_{\theta_1}^{\theta_3}d\theta_2 \cdot\cdot\cdot\int_{\theta_{k-1}}^{2\pi}d\theta_k\prod_{l=1}^ku(e_l)\\
\label{eqn:ZkN-k} &\times\int_{\Omega}dw_{k+1} \cdot\cdot\cdot \int_{\Omega} dw_{(N+k)/2} \prod_{s=k+1}^{(N+k)/2}v(w_s) \: \mathcal{Q}(\bY)
\end{align}
for fixed $k$, where $\Omega$ is the unit disk, and the factor of $1/((N-k)/2)!$ comes from relaxing the ordering constraint on the complex eigenvalues. Note the ordering of the angles $\theta_j$ corresponding to the real eigenvalues is in accord with the ordering (\ref{9'}) of $\{\lambda_i\}$. As for the GOE and the real Ginibre ensemble it is at this point that parity considerations become important.

\subsubsection{$N$ even}

\begin{proposition}
\label{prop:gen_part_fn}
Let $\{ p_{j}(x)\}$ be a set of monic polynomials of degree $j$, and define 
\begin{align}
\label{eqn:q=p} q_{2j}(x)=p_{2j}(x)&,& q_{2j+1}(x)=p_{N-1-2j}(x).
\end{align}
The generalised partition function for the real spherical ensemble, with $N$ even, is 
\begin{align}
\nonumber Z_{k,(N-k)/2}[u,v]_S&=\frac{(-1)^{(N/2)(N/2-1)/2}}{2^{N(N-1)/2}} \Gamma((N+1)/2)^{N/2} \Gamma(N/2+1)^{N/2}\\
\label{eqn:genpartfn} &\times\prod_{s=1}^{N}\frac{1} {\Gamma(s/2)^2}\; [\zeta^{k}] \mathrm{Pf} \left[\zeta^2 \alpha_{j,l}+ \beta_{j,l} \right],
\end{align}
with $[\zeta^k ]$ denoting the coefficient of $\zeta^k$, and where
{\small
\begin{align}
\nonumber \alpha_{j,k} &=-\frac{i}{2}\int_0^{2\pi}d\theta_1\: u(e_1)\tau(e_1)\int_0^{2\pi}d\theta_2\:  u(e_2)\tau(e_2)q_{j-1}(e_1)q_{k-1}(e_2)\: \mathrm{sgn}(\theta_2-\theta_1),\\
\label{15'} \beta_{j,k} &=\int_{\Omega} dw\hspace{3pt}v(w)\tau(w) \tau\left(\frac{1}{\bar{w}}\right)\frac{1}{|w|^2}\left(q_{j-1}(w)q_{k-1}\left(\frac{1}{\bar{w}}\right) - q_{k-1}(w)q_{j-1}\left(\frac{1}{\bar{w}}\right) \right).
\end{align}
}
\end{proposition}

\textit{Proof:} With $p_l(x)$ an arbitrary monic polynomial of degree $l$, using (\ref{eqn:vandermonde_polys}) the Vandermonde product in $\mathcal{Q}(\bY)$ can be written
{\small
\begin{align}
\nonumber &\Delta\left(\mathbf{e},\mathbf{w},\mathbf{\frac{1}{\bar{w}}}\right) =\mathrm{det}\left[\begin{array}{c}
[p_{l-1}(e_j)]_{j=1,...,k}\vspace{3pt}\\
\left[p_{l-1}(w_s)\right]_{s=k+1,...,(N+k)/2}\\
\left[p_{l-1}(1/\bar{w}_s)\right]_{s=k+1,...,(N+k)/2}
\end{array}\right]_{l=1,...,N}\\
\nonumber &=(-1)^{(N-k)/2((N-k)/2-1)/2}\mathrm{det}\left[\begin{array}{c}
[p_{l-1}(e_j)]_{j=1,...,k}\vspace{3pt}\\
\left[\begin{array}{c}
p_{l-1}(w_s)\\
p_{l-1}(1/\bar{w}_{s})
\end{array}
\right]_{s=k+1,...,(N+k)/2}
\end{array}\right]_{l=1,...,N},
\end{align}
}where, for the second equality, we have interlaced the rows corresponding to complex conjugate pairs; this will be convenient later. 

Next, as in Proposition \ref{prop:GinOE_gpf_even}, we apply the method of integration over alternate variables to the $e_j$, which correspond to the real eigenvalues,
{\small
\begin{align}
\nonumber &Z_{k,(N-k)/2}[u,v]_S= (-1)^{(N-k)/2((N-k)/2-1)/2} \frac{A_{k,N}}{(k/2)!((N-k)/2)!}\\
\nonumber &\times \int_0^{2\pi} d\theta_2\int_{0}^{2\pi} d\theta_4 \cdot\cdot\cdot \int_{0}^{2\pi} d\theta_k \int_{\Omega}dw_{k+1} \cdot\cdot\cdot \int_{\Omega} dw_{(N+k)/2} \prod_{s=k+1}^{(N+k)/2}v(w_s)\\
\nonumber &\times \hspace{-6pt} \prod_{s=k+1}^{(N+k)/2}\hspace{-4pt}\frac{1}{|w_s|^2} \tau(w_s) \tau \left(\frac{1} {\bar{w}_s} \right) \mathrm{det}\left[\begin{array}{c}
\left[\begin{array}{c}
\int_{0}^{\theta_{2j}}u(\theta)\tau(e)p_{l-1}(e)d\theta \\
u(\theta_{2j})\tau(e_{2j})p_{l-1}(e_{2j})\end{array}\right]_{j=1,...k/2} \vspace{6pt}\\
\left[\begin{array}{c}
p_{l-1}(w_s)\\
p_{l-1}(1/\bar{w}_s)
\end{array}
\right]_{s=k+1,...,(N+k)/2}
\end{array}\right]_{l=1,...,N}.
\end{align}
}Re-order columns in the determinant according to
\begin{align}
\label{eqn:poly_ordering}
p_0,p_{N-1},p_2,p_{N-3},\cdot\cdot\cdot ,p_{N-2},p_1,
\end{align}
which introduces a factor of $(-1)^{(N/2)(N/2-1)/2}$. For labeling purposes define
\begin{align}
\nonumber q_{2j}(x)=p_{2j}(x)&,&q_{2j+1}(x)=p_{N-1-2j}(x).
\end{align}
Expanding the determinant according to its definition as a signed sum over permutations, then performing the remaining integrations gives
\begin{align}
\nonumber & Z_{k,(N-k)/2}[u,v]_S = (-1)^{(N-k)/2((N-k)/2-1)/2}\frac{A_{k,N}}{(k/2)!((N-k)/2)!} \\
\nonumber & \qquad \times \sum_{P \in S_N} \varepsilon(P) \prod_{l=1}^{k/2}
a_{P(2l-1),P(2l)} \prod_{l=k/2+1}^{N/2} b_{P(2l-1),P(2l)},
\end{align}
where
\begin{align}
\nonumber a_{j,k} &= \int_0^{2\pi} d \theta_1 \, u(\theta_1) \tau(e_1) q_{j-1}(e_1)
\int_0^{\theta_1} d \theta_2 \, u(\theta_2) \tau(e_2) q_{k-1}(e_2),\\
\nonumber b_{j,k} &= \int_{\Omega} dw \: v(w) \tau(w) \tau\left(\frac{1}{\bar{w}}\right) \frac{1}{|w|^2}\hspace{2pt}q_{j-1}(w) \: q_{k-1}\left(\frac{1}{\bar{w}}\right).
\end{align}
If we now impose the restriction $P(2l) > P(2l-1)$, ($l=1,\dots,N/2$) this can be rewritten as
\begin{align}
\nonumber Z_{k,(N-k)/2}[u,v]_S & = (-1)^{(N-k)/2((N-k)/2-1)/2} (2i)^{k/2} A_{k,N}\\
\label{15.1} & \times \sum_{P \in S_N \atop P(2l) > P(2l-1)} \hspace{-12pt}\varepsilon(P) \hspace{6pt}\prod_{l=1}^{k/2} \alpha_{P(2l-1),P(2l)}
 \prod_{l=k/2+1}^{N/2} \beta_{P(2l-1),P(2l)},
\end{align}
with $\alpha_{j,k}, \beta_{j,k}$ given by (\ref{15'}). With Definition \ref{def:pfaff} we can write (\ref{15.1}) in terms of a Pfaffian, and (\ref{eqn:genpartfn}) follows.

\hfill $\Box$

The summed up partition function for the real symmetric ensemble we define as
\begin{align}
\label{ZS} Z_N[u,v]_S:=  \sum_{k=0 \atop k \: {\rm even}}^N  Z_{k,(N-k)/2}[u,v]_S
\end{align}
analogously to (\ref{eqn:summedup}), and substitution of (\ref{eqn:genpartfn}) yields
\begin{align}
\nonumber Z_N[u,v]_S & =  \frac{(-1)^{(N/2)(N/2-1)/2}}{2^{N(N-1)/2}} \Gamma((N+1)/2)^{N/2} \Gamma(N/2+1)^{N/2}\\
\label{21} &\times\prod_{s=1}^{N}\frac{1}{\Gamma(s/2)^2} \mathrm{Pf} \left[\alpha_{j,l}+ \beta_{j,l} \right].
\end{align}

Recall that $Z_{k,(N-k)/2}[1,1]_S$ is the probability of finding $k$ real eigenvalues and $(N-k)/2$ complex eigenvalues, and so the generating function for these probabilities $p_{N,k}$ is
\begin{align}
\label{def:ZNxi} Z_N(\zeta)_S&:= \sum_{k=0 \atop k \: {\rm even}}^N \zeta^k p_{N,k} \: = \: \sum_{k=0}^{N/2}\zeta^{2k} Z_{2k,(N-2k)/2}[1,1]_S,
\end{align}
which becomes
\begin{align}
\nonumber Z_N(\zeta)_S &=\frac{(-1)^{(N/2)(N/2-1)/2}}{2^{N(N-1)/2}} \Gamma((N+1)/2)^{N/2} \Gamma(N/2+1)^{N/2}\\
\label{eqn:ZNxi} &\times\prod_{j=1}^{N}\frac{1} {\Gamma(j/2)^2}\; \mathrm{Pf} \left[\zeta^2 \alpha_{j,l}+ \beta_{j,l} \right]\Bigg|_{u=v=1}.
\end{align}

\subsubsection{$N$ odd}

The calculation of the generalised partition function for $N$ odd proceeds along the same lines as Proposition \ref{prop:gen_part_fn} for $N$ even, but with a more complicated ordering of the columns during the integration over alternate variables, the purpose of which is to aid in finding the skew-orthogonal polynomials.

\begin{proposition}
\label{prop:gen_part_fn_odd}
With monic polynomials $\{ p_j(x)\}$ of degree $j$ and $\alpha_{j,l},\beta_{j,l}$ as in (\ref{15'}), the generalised partition function for the real spherical ensemble, with $N$ odd, is
\begin{align}
\nonumber Z_{k,(N-k)/2}^{\odd}[u,v]_S &=\frac{(-1)^{(N-1)/4((N-1)/2)-1)}} {2^{N(N-1)/2}} \Gamma((N+1)/2)^{N/2} \Gamma(N/2+1)^{N/2}\\
\label{eqn:gen_fn_odd} &\times\prod_{s=1}^{N}\frac{1}{\Gamma(s/2)^2}\; [\zeta^{k-1}]\mathrm{Pf}\left[\begin{array}{cc}
\left[\zeta^2\alpha_{i,j} +\beta_{i,j}\right] & \left[\nu_i\right]\\
 \left[-\nu_j\right] & 0\\
\end{array}\right]_{i,j=1,...,N},
\end{align}
where,
\begin{align}
\label{def:nu} \nu_l:=\frac{1}{\sqrt{2}}\int_0^{2\pi}u(\theta)\tau(e)q_{l-1}(e) d\theta,
\end{align}
and
\begin{align}
\nonumber &\left.
\begin{array}{l}
q_{2j}=p_{2j},\\
q_{2j+1}=p_{N-1-2j},
\end{array}
\right\}0 \leq 2j<(N-1)/2,\\
\nonumber &\left.
\begin{array}{l}
q_{2j}=p_{2j+1},\\
q_{2j+1}=p_{N-1-(2j+1)},
\end{array}
\right\}(N-1)/2 \leq 2j<N-1,\\
\label{eqn:odd_polys} &\left.
\begin{array}{l}
q_{N-1}=p_{(N-1)/2}.
\end{array}
\right.
\end{align}
\end{proposition}

\textit{Proof:} As for the even case write the Vandermonde product of $\mathcal{Q}(\bY)$ as
\begin{align}
\nonumber &\Delta\left(\mathbf{e},\mathbf{w},\mathbf{\frac{1}{\bar{w}}}\right)=\mathrm{det}\left[\begin{array}{c}
[p_{l-1}(e_j)]_{j=1,...,k-1}\vspace{3pt}\\
\left[p_{l-1}(w_s)\right]_{s=k+1,...,(N+k)/2}\\
\left[p_{l-1}(1/\bar{w}_s)\right]_{s=k+1,...,(N+k)/2}
\end{array}\right]_{l=0,...,N}\\
\label{eqn:odd_vand} &=(-1)^{(N-k)/2((N-k)/2-1)/2} \mathrm{det}\left[\begin{array}{c}
[p_{l-1}(e_j)]_{j=1,...,k}\vspace{3pt}\\
\left[\begin{array}{c}
p_{l-1}(w_s)\\
p_{l-1}(1/\bar{w}_{s})
\end{array}
\right]_{s=k+1,...,(N+k)/2}\\
\left[ p_{l-1}(e_k)\right]
\end{array}\right]_{l=1,...,N},
\end{align}
where we have moved the row corresponding to the $k$th real eigenvalue to the bottom of the matrix. (It proves to be more convenient to convert this latter matrix to Pfaffian form than the equivalent matrix where the $k$th row is not moved.) This always involves an even number of transpositions so no overall factor is required. The shifted row corresponds to the single unpaired real eigenvalue that must exist in any odd-sized real matrix, a fact that is guaranteed by $N$ and $k$ being of the same parity. The factors of $-1$ come from the reordering of complex eigenvalue rows, exactly as in the even case. Now we substitute (\ref{eqn:odd_vand}) into (\ref{eqn:ZkN-k}) and apply integration over alternate variables, as in Proposition \ref{prop:gen_part_fn}, to find
{\small
\begin{align}
\nonumber &Z_{k,(N-k)/2}^{\odd}[u,v]_S=(-1)^{(N-k)/2((N-k)/2-1)/2}\frac{A_{k,N}}{((k-1)/2)!((N-k)/2)!}\\
\nonumber &\times \int_0^{2\pi}d\theta_2\int_{0}^{2\pi} d\theta_4 \cdot\cdot\cdot \int_{0}^{2\pi}d\theta_{k-1}\int_{\Omega}dw_{k+1} \cdot\cdot\cdot \int_{\Omega} dw_{(N+k)/2}\prod_{s=k+1}^{(N+k)/2}v(w_s)\\
\nonumber &\times \hspace{-6pt}\prod_{s=k+1}^{(N+k)/2} \hspace{-4pt} \frac{1}{|w_s|^2} \tau(w_s)\tau\left(\frac{1}{\bar{w}_s}\right) \mathrm{det}\left[\begin{array}{c}
\left[\begin{array}{c}
\int_{0}^{\theta_{2j}}u(\theta)\tau(e)p_{l-1}(e)d\theta \\
u(\theta_{2j})\tau(e_{2j})p_{l-1}(e_{2j})\end{array}\right]_{j=1,...(k-1)/2}\\
\left[\begin{array}{c}
p_{l-1}(w_s)\\
p_{l-1}(1/\bar{w}_s)
\end{array}
\right]_{s=k+1,...,(N+k)/2}\\
\left[ \int_{0}^{2\pi}u(\theta)\tau(e)p_{l-1}(e)\hspace{2pt}d\theta \right]
\end{array}\right]_{l=1,...,N}.
\end{align}
}

We need to reorder the columns of the determinant in a similar way to that of (\ref{eqn:poly_ordering}), although with the key difference of shifting the middle column to the end. The re-ordering is then
\begin{align}
\nonumber &p_0,p_{N-1},p_2,p_{N-3},... ,p_{(N-1)/2-\epsilon_{1,2}}, p_{(N-1)/2+ \epsilon_{1,2}},\\
\label{eqn:poly_order_odd} &p_{(N-1)/2+\epsilon_{2,1}}, p_{(N-1)/2- \epsilon_{2,1}}, ..., p_{N-4},p_{3},p_{N-2},p_{1},p_{(N-1)/2},
\end{align}
where
\begin{align}
\nonumber &\epsilon_{1,2}=\left\{
\begin{array}{ll}
1, & \mbox{for $(N-1)/2$ even},\\
2, & \mbox{for $(N-1)/2$ odd},
\end{array}
\right.\\
\nonumber &\epsilon_{2,1}=\left\{
\begin{array}{ll}
2, & \mbox{for $(N-1)/2$ even},\\
1, & \mbox{for $(N-1)/2$ odd}.
\end{array}
\right.
\end{align} 
This introduces a factor of $(-1)^{(N-1)/2+(N-1)/2((N-1)/2-1)/2}$. Also, for $N$ odd, the factors of $-1$ in $A_{k,N}$ can be re-written by noting
\begin{align}
\nonumber (-1)^{(N-k)k/2-k(k-1)/4}=(-1)^{(N-1)/2-(k-1)/4},
\end{align}
which gives us an overall factor of
\begin{align}
\nonumber &(-1)^{(N-k)/2((N-k)/2-1)/2}\; (-1)^{(N-1)/2+(N-1)/2((N-1)/2-1)/2}\; A_{k,N}\\
\nonumber &= \frac{(-1)^{(N-1)/4((N-1)/2)-1)-(k-1)/4}}{2^{(N(N-1)+k)/2}} \Gamma((N+1)/2)^{N/2} \Gamma(N/2+1)^{N/2}\\
\nonumber &\quad\times \prod_{s=1}^{N}\frac{1}{\Gamma(s/2)^2}.
\end{align}
Now we again expand the determinant as a signed sum over permutations and impose the restriction $P(2l)>P(2l-1)$, which gives us the odd analogue of (\ref{15.1}),
{\small
\begin{align}
\nonumber &Z_{k,(N-k)/2}^{\odd}[u,v]_S  =  \frac{(-1)^{(N-1)/4((N-1)/2)-1)}}{2^{N(N-1)/2}}\Gamma((N+1)/2)^{N/2} \Gamma(N/2+1)^{N/2}\\
\nonumber &\times \prod_{j=1}^{N} \frac{1}{\Gamma(j/2)^2} \sum_{P \in S_N \atop P(2l) > P(2l-1)} \hspace{-12pt}\varepsilon(P)\; \nu_{P(N),N+1} \prod_{l=1}^{(k-1)/2} \hspace{-6pt} \alpha_{P(2l-1),P(2l)} \prod_{l=(k+1)/2}^{(N-1)/2} \beta_{P(2l-1),P(2l)},
\end{align}
}where $\nu_{P(N)}:=\nu_{P(N),N+1}$ is given by (\ref{def:nu}). Using the Pfaffian definition (\ref{def:Pf}), (\ref{eqn:gen_fn_odd}) now follows.

\hfill $\Box$

The analogous definition to (\ref{ZS}) for $N$ odd is
\begin{align}
\label{eqn:gen_fn_odd1} Z_N^{\odd}[u,v]_S:=  \sum_{k=1 \atop k \: {\rm odd}}^N  Z_{k, (N-k)/2}^{\odd} [u,v]_S,
\end{align}
and substituting (\ref{eqn:gen_fn_odd}) we have
\begin{align}
\nonumber Z_N^{\odd}[u,v]_S &=\frac{(-1)^{(N-1)/4((N-1)/2)-1)}}{2^{N(N-1)/2}}\Gamma((N+1)/2)^{N/2}\Gamma(N/2+1)^{N/2}\\
\nonumber &\times\prod_{s=1}^{N}\frac{1}{\Gamma(s/2)^2}\; \mathrm{Pf}\left[\begin{array}{cc}
\left[\alpha_{j,l} +\beta_{j,l} \right] & \left[\nu_j \right]\\
 \left[-\nu_l \right] & 0\\
\end{array}\right]_{j,l=1,...,N}.
\end{align}
The generating function for the probabilities with $N$ odd is
\begin{align}
\label{def:SZNo} Z_N^{\odd}(\zeta)_S&:= \sum_{k=1 \atop k \: {\odd}}^N \zeta^k p_{N,k} \: = \: \sum_{k=0}^{(N-1)/2}\zeta^{2k+1} Z_{2k+1,(N-2k-1)/2}^{\odd}[1,1]_S,
\end{align}
and
\begin{align}
\nonumber Z_N^{\odd}(\zeta)_S &=\frac{(-1)^{(N-1)/4((N-1)/2)-1)}} {2^{N(N-1)/2}} \Gamma((N+1)/2)^{N/2} \Gamma(N/2+1)^{N/2}\\
\label{eqn:ZNxio} &\times \zeta \prod_{s=1}^{N}\frac{1}{\Gamma(s/2)^2}\;  \mathrm{Pf}\left[\begin{array}{cc}
\left[\zeta^2\alpha_{j,l} +\beta_{j,l}\right] & \left[\nu_j \right]\\
 \left[-\nu_l \right] & 0\\
\end{array}\right]_{j,l=1,...,N}\Bigg|_{u=v=1}.
\end{align}

In the next section we will see that with the relevant skew-orthogonal polynomials, the probabilities for both even and odd can be calculated in a straightforward manner.

\subsection{Skew-orthogonal polynomials}
\label{sec:Ssops}

Recall that the Pfaffian in the generating function (\ref{eqn:GinOE_probsGF_pf}) for the even case of the real Ginibre ensemble was brought to skew-diagonal form (\ref{eqn:skew_diag_mat}) using the polynomials (\ref{eqn:GinOE_sopolys}) only when $\zeta=1$. This means that the calculation of the probabilities seems to inevitably involve the calculation of a Pfaffian or determinant --- recall that we obtained a chequerboard $N\times N$ Pfaffian, which can be rewritten as an $N/2 \times N/2$ determinant according to (\ref{eqn:chequer}) --- and so is computationally intensive when using exact arithmetic. Separately, when discussing the probabilities of the partially symmetric real Ginibre ensemble in Chapter \ref{sec:tGprobs}, we found that we could use the monomials $p_j(x)=x^j$ to calculate the $\alpha_{2j-1,2l}^{(\tau)}\big|_{u=1}$ and $\beta_{2j-1,2l}^{(\tau)}\big|_{v=1}$ individually, however these did not skew-diagonalise the matrix.

For the problem at hand we can obtain both of these benefits simultaneously: we explicitly construct polynomials $q_i(x)$ that, for general $\zeta$, skew-diagonalise the matrix in (\ref{eqn:ZNxi}). The polynomials turn out to be quite simple, and it was knowledge of these polynomials that motivated the definition of the $\{q_i(x)\}$ in terms of the $\{p_i(x)\}$ in (\ref{eqn:q=p}). Further, we find that with some modification, we can also use a similar set of polynomials to bring the matrix in (\ref{eqn:ZNxio}) to a form which, for our purposes, is equivalent to the odd skew-diagonal form (\ref{eqn:skew_diag_mat_odd}), for all $\zeta$.

\begin{definition}
For monic polynomials $p_j(x),p_l(x)$ of degree $j$ and $l$ respectively, define the inner product
\begin{align}
\nonumber &\langle p_j,p_l \rangle_S:=-\frac{i}{2}\int_0^{2\pi}d\theta_1\: \tau(e_1) \int_0^{2\pi} d\theta_2 \: \tau(e_2) p_j(e_1) p_l(e_2)\: \mathrm{sgn}(\theta_2-\theta_1)\\
\nonumber &+\int_{\Omega} dw\; \tau(w) \tau \left(\frac{1}{\bar{w}}\right) \frac{1} {|w|^2} \left(p_j(w) p_l\left(\frac{1}{\bar{w}}\right) - p_l(w) p_j\left(\frac{1} {\bar{w}}\right) \right)\\
\label{def:Sip} &=\alpha_{j+1,l+1}+ \beta_{j+1,l+1}\big|_{u=v=1},
\end{align}
where $\alpha_{j,l}, \beta_{j,l}$ are from (\ref{15'}).
\end{definition}
We will also find it convenient to define
\begin{align}
\nonumber \hat{\alpha}_{j,l}&:=\alpha_{j,l}\big|_{u=1},\\
\label{def:hatab1} \hat{\beta}_{j,l}&:=\beta_{j,l}\big|_{v=1},
\end{align}
and
\begin{align}
\nonumber \hat{\alpha}_l&:=\hat{\alpha}_{2l+1,2l+2},\\
\label{def:hatab} \hat{\beta}_l&:=\hat{\beta}_{2l+1,2l+2}.
\end{align}

\begin{proposition}
\label{prop:skew_polys_even}
The inner product (\ref{def:Sip}) satisfies the skew-orthogonality conditions (\ref{eqn:GinOE_soprops}) using the polynomials $p_j(x)=x^j$ and thus, according to (\ref{eqn:q=p}),
\begin{eqnarray}
\label{eqn:skew_polys} q_{2j}(x)=x^{2j}, \qquad q_{2j+1}(x)=x^{N-1-2j}.
\end{eqnarray}
The normalisations $\hat{\alpha}_l, \hat{\beta}_l$ from (\ref{def:hatab}) are
\begin{align}
\nonumber \hat{\alpha}_l&=\frac{2\pi}{N-1-4l} \frac{\Gamma((N+1)/2)} {\Gamma(N/2+1)},\\
\label{eqn:hatabeval} \hat{\beta}_l&=\frac{2\sqrt{\pi}}{N-1-4l} \left( 2^N \frac{\Gamma(2l+1) \Gamma(N-2l)} {\Gamma(N+1)}-\sqrt{\pi}\: \frac{\Gamma((N+1)/2)} {\Gamma(N/2+1)}\right).
\end{align}
\end{proposition}

\textit{Proof}: The skew-symmetry property $\hat{\alpha}_{j,l}= -\hat{\alpha}_{l,j}$, $\hat{\beta}_{j,l}= -\hat{\beta}_{l,j}$ can be checked by observation, so to establish the result we must show that each of $\hat{\alpha}_{j,l}$ and $\hat{\beta}_{j,l}$ are non-zero only for $j=2t+1,l=2t+2$ in which case they have the evaluations stated.

From (\ref{15'}), we have
\begin{align}
\nonumber \hat{\alpha}_{j+1,l+1}&=c \: \frac{i}{2}\int_0^{2\pi}d\theta_1\: e^{i\theta_1 (\tilde{j} - (N-1)/2)}\int_{\theta_1}^{2\pi}d\theta_2\:  e^{i\theta_2(\tilde{l}- (N-1)/2}\\
\nonumber &-c \: \frac{i}{2}\int_0^{2\pi}d\theta_1\: e^{i\theta_1 (\tilde{j} - (N-1)/2)}\int_0^{\theta_1}d\theta_2\:  e^{i\theta_2(\tilde{l}- (N-1)/2},
\end{align}
where $c$ is a constant factor and $\tilde{j}=2j$ or $\tilde{j}=N-1-2j$ for $j$ even or odd respectively. Performing the inner integrals over $\theta_2$, using the fact that $\tilde{j}\neq (N-1)/2$ since $j$ is an integer and $N$ is even, we find
\begin{align}
\label{eqn:sopsa1} \hat{\alpha}_{j+1,l+1}&=c\: \frac{2}{2\tilde{l}-N+1} \int_0^{2\pi}d\theta_1\: e^{i\theta_1 (\tilde{j} +\tilde{l} - N+1)},
\end{align}
which is non-zero only in the case that $\tilde{l}=N-1-\tilde{j}$, which implies $j=2t+1,l=2t+2$ (for $\hat{\alpha}_{j+1,l+1}$ positive). The evaluation of (\ref{eqn:sopsa1}) in this case is straightforward. To obtain the conditions on $s$ and $t$ where $\hat{\beta}_{s,t}\neq 0$ we repeat the procedure used above by writing out $\tau(w),\tau(\bar{w}^{-1}),q_{2j}$ and $q_{2j+1}$. The fact that $s=2j+1$ and $t=2j+2$ for a non-zero evaluation is then immediate.

It remains to evaluate $\hat{\beta}_{2j+1,2j+2}=: \hat{\beta}_{j}$, which turns out to require knowledge of a non-standard form of the beta integral. Thus, after converting to polar co-ordinates, setting $c:=|w|^2$ and integrating by parts, one obtains
\begin{align}
\label{eqn:beta1} \hat{\beta}_{j}=-\frac{2\pi^{3/2}}{N-1-4j}\frac{\Gamma((N+1)/2)}{\Gamma(N/2+1)}+\frac{2^{N+1}\pi}{N-1-4j}\int_0^1\frac{c^{2j}+c^{N-2j-1}}{(1+c)^{N+1}}dc.
\end{align}
According to \cite[Equation 3.216 (1)]{GraRyz2000}, for general $a,b$ such that Re$\, b>0$,
Re$\, (a-b) > 0$,
\begin{align}
\nonumber \int_0^1(t^{b-1} + t^{a-b-1})(1 + t)^{-a} \, dt = \frac{\Gamma(b)\Gamma(a-b)}{\Gamma(a)},
\end{align}
and so, with $b = y$, $a-b = x$, we have a non-standard form of the beta integral
\begin{align}
\nonumber \int_0^1 t^{x-1}(1-t)^{y-1} \, dt = \frac{\Gamma(x)\Gamma(y)}{\Gamma(x+y)}.
\end{align}
The stated formula for $\hat{\beta}_{j}$ now follows.

Alternatively, by some manipulations the integral in (\ref{eqn:beta1}) can be transformed to
\begin{align}
\nonumber \int_0^1 \frac{c^{2j}+c^{N-2j-1}} {(1+c)^{N+1}}dc&=\frac{1} {2j+1}{}_2 F_1 (2j+1,N+1,2j+2;-1)\\
\label{eqn:beta2} &+\frac{1}{N-2j} {}_2 F_1 (N-2j,N+1,N+1-2j;-1).
\end{align}
We see that the RHS of (\ref{eqn:beta2}) must equal $\Gamma(2j+1) \Gamma(N-2j)/ \Gamma(N+1)$ for the result to be obtained, a condition which can be shown to be equivalent to the statement
\begin{align}
\label{eqn:beta3} 1=\sum_{s=0}^{2j}\frac{2^{s-N} \Gamma(N-s)}{\Gamma(2j+1-s) \Gamma(N-2j)} +\sum_{s=0}^{N-2j-1}\frac{2^{s-N} \Gamma(N-2j-s)}{\Gamma(2j+1) \Gamma(N-2j)}.
\end{align}
With $j=0$ the RHS of (\ref{eqn:beta3}) can be evaluated as
\begin{align}
\nonumber \frac{1}{2^N}+\sum_{s=0}^{N-1}\frac{1}{2^{N-s}}=1,
\end{align}
where we have used the formula $\sum_{s=1}^j 2^{-j}=1-2^{-j}$. We now establish (\ref{eqn:beta3}) inductively for all integer $j\in [0, N/2-1]$. Through the use of Zeilberger's algorithm \cite{Zeil1990a,Zeil1990b} (in Mathematica form \cite{PaSc1994}) we obtain a proof that
\begin{align}
\nonumber &\sum_{s=0}^{2j}\frac{2^{s-N} \Gamma(N-s)}{\Gamma(2j+1-s) \Gamma(N-2j)} +\sum_{s=0}^{N-2j-1}\frac{2^{s-N} \Gamma(N-2j-s)}{\Gamma(2j+1) \Gamma(N-2j)}\\
\nonumber &=\sum_{s=0}^{2j+1}\frac{2^{s-N} \Gamma(N-s)}{\Gamma(2j+2-s) \Gamma(N-2j-1)} +\sum_{s=0}^{N-2j-2}\frac{2^{s-N} \Gamma(N-2j-1-s)}{\Gamma(2j+2) \Gamma(N-2j)},
\end{align}
so the RHS of (\ref{eqn:beta3}) is unchanged by $2j\mapsto 2j+1$ for all $j\in [0,N/2-1]$, and so this establishes (\ref{eqn:hatabeval}).

\hfill $\square$

Since Proposition \ref{prop:skew_polys_even} tells us that both $\hat{\alpha}_{j,l}$ and $\hat{\beta}_{j,l}$ are independently skew-\\orthogonalised by the polynomials (\ref{eqn:skew_polys}) for general $\zeta$, we have that
\begin{align}
\label{eqn:soPf} \mathrm{Pf}\left[\zeta^2\alpha_{j,l}+ \beta_{j,l} \right]_{j,l=1,2,...,N} \Big|_{u=v=1} &=\prod_{l=0}^{N/2-1} \zeta^2\hat{\alpha}_l+ \hat{\beta}_l.
\end{align}
Substitution of (\ref{eqn:soPf}) into (\ref{eqn:ZNxi}) gives us the generating function for the probabilities $p_{N,k}$
\begin{align}
\nonumber Z_N(\zeta)_S&=\frac{(-1)^{(N/2)(N/2-1)/2}}{2^{N(N-1)/2}} \Gamma((N+1)/2)^{N/2} \Gamma(N/2+1)^{N/2}\\
\label{eqn:Sprobs} &\times\prod_{s=1}^{N}\frac{1} {\Gamma(s/2)^2} \prod_{l=0}^{N/2-1} \zeta^2\hat{\alpha}_l+ \hat{\beta}_l.
\end{align}
The probability $p_{N,N}$ that all eigenvalues are real is the coefficient of $\zeta^N$ in (\ref{eqn:Sprobs}) and is
\begin{align}
\nonumber p_{N,N}&=\pi^{N/2}\frac{\Gamma((N+1)/2)^{N/2}}{2^{N(N-2)/2}} \prod_{s=1}^{N}\frac{1} {\Gamma(s/2)^2} \prod_{l=0}^{N/2-1}\frac{1}{N-1-4l},
\end{align}
where we have used the fact that $\lfloor N/4 \rfloor$ (the floor function of $N/4$) is of the same parity as $N/2(N/2-1)/2$.

We can also use the polynomials (\ref{eqn:skew_polys}) to calculate the expected number of real eigenvalues $E_N$ and the variance $\sigma^2_N$.

\begin{corollary}
\label{3.5}
With $N$ even the expected number of real eigenvalues of an $N \times N$ matrix $\bY=\bA^{-1}\bB$ is \cite{eks1994}
\begin{align}
\label{29} E_N=\left. \frac{\partial}{\partial \zeta}Z_N(\zeta) \right|_{\zeta=1}=\sum_{l=0}^{N/2-1}\frac{2\hspace{1pt}\hat{\alpha}_l} {\hat{\alpha}_l+\hat{\beta}_l} =
\frac{\sqrt{\pi} \Gamma((N+1)/2)}{\Gamma(N/2)}.
\end{align}
The variance in the number of real eigenvalues is
\begin{align}
\nonumber \sigma_N^2 &= {\partial^2 \over \partial \zeta^2} Z_N(\zeta) \Big |_{\zeta = 1} + E_N - E_N^2 \: = \: 2E_N - 4 \sum_{l=0}^{N/2 - 1} {\hat{\alpha}_l^2 \over (\hat{\alpha}_l + \hat{\beta}_l)^2} \nonumber\\
\label{30} &= 2E_N - 2 \sqrt{\pi}  {\Gamma((N+1)/2)^2 \Gamma(N-1/2) \over \Gamma(N/2)^2 \Gamma(N)}.
\end{align}
\end{corollary}

\textit{Proof}: The second equalities follow from the first and (\ref{eqn:Sprobs}) through simple differentiation, while for the third equalities use has been made of the summations
\begin{align}
\nonumber \sum_{j=0}^{N/2 - 1} \Big ( {N - 1 \atop 2j} \Big ) = 2^{N-2}
\end{align}
and
\begin{align}
\nonumber \sum_{j=0}^{N/2 - 1} \Big ( {N - 1 \atop 2j} \Big )^2 &= \frac{1}{2}\sum_{j=0}^{N-1} \Big ( {N - 1 \atop j} \Big )^2 = \frac{1}{2} \Big ( {2N - 2 \atop N-1} \Big ) = \frac{\Gamma(2N-1)} {2(\Gamma(N))^2}\\
\nonumber &= {2^{2N-3} \; \Gamma ((2N-1)/2) \over \sqrt{\pi} \; \Gamma(N)},
\end{align}
where use was made of \cite[0.157 (1)]{GraRyz2000}.

\hfill $\Box$

The result (\ref{29}) was first derived by Edelman \textit{et al.}~\cite{eks1994} using ideas from integral geometry. (This statement is generalised to one concerning eigenvalues of a matrix polynomial in \cite{EK95}.) A corollary, also noted in \cite{eks1994}, is that for $N \to \infty$
\begin{align}
\label{eqn:SENasy} E_N \: \sim \: \sqrt{\pi N \over 2} \bigg ( 1 - {1 \over 4N} + {1 \over 32 N^2} + {5 \over 128 N^3} - {21 \over 2048 N^4} + {\rm O} \Big ( {1 \over N^5} \Big ) \bigg ),
\end{align}
which gives the leading order behaviour for large $N$. We also note that, to leading order, (\ref{30}) implies the variance is related to the mean by $\sigma_N^2 \sim (2 - \sqrt{2} )E_N$, which coincidentally (?) is the same asymptotic relation as (\ref{eqn:Gvar}), the analogous result for the real Ginibre ensemble.

The explicit form of the generating function (\ref{eqn:Sprobs}) allows for the computation of the large $N$ limiting form of the probability density of the scaled number of real eigenvalues.

\begin{proposition}
\label{pAB}
Let $\sigma_N^2$ and $E_N$ be as in Corollary \ref{3.5}, and let $p_{N,k}$ be the probability of obtaining $k$ eigenvalues from $N$ total eigenvalues. We have
\begin{align}
\nonumber \lim_{N \to \infty} \: {\rm sup}_{x \in (-\infty,\infty)} \Big | \sigma_N p_{N,\lfloor \sigma_N x + E_N \rfloor} - {1 \over \sqrt{2 \pi}} e^{- x^2/2} \Big | = 0,
\end{align}
where $\lfloor \cdot \rfloor$ denotes the floor function.
\end{proposition}

\noindent
\textit{Proof}: For a given $n$, let $\{p_n(k)\}_{k=0,1,\dots,n}$ be a sequence such that
\begin{align}
\nonumber P_N(x) = \sum_{k=0}^n p_n(k) x^k
\end{align}
has the properties that the zeros of $P_N(x)$ are all on the real axis, and $P_N(1)=1$.
Let
\begin{align}
\nonumber \mu_n = \sum_{k=0}^n k p_n(k)&,& \sigma_n^2 = \sum_{k=0}^n k^2 p_n(k) - \mu_n^2,
\end{align}
and suppose $\sigma_n \to \infty$ as $n \to \infty$. A local limit theorem due to Bender 
\cite{Be73} gives
\begin{align}
\nonumber \lim_{n \to \infty} \: \mathrm{sup}_{x \in (-\infty,\infty)} \Big | \sigma_n p_n(\lfloor \sigma_n x + \mu_n \rfloor) - \frac{1} {\sqrt{2 \pi}} e^{-x^2/2} \Big | &= 0.
\end{align}
Application of this general theorem to $Z_N(\zeta)$ in (\ref{def:ZNxi}), with $\zeta^2 = x$, gives the stated result.
\hfill $\square$
\newline

\begin{figure}
\begin{center}
\includegraphics[scale=0.6,trim=0 0mm 80mm 540, clip=true]{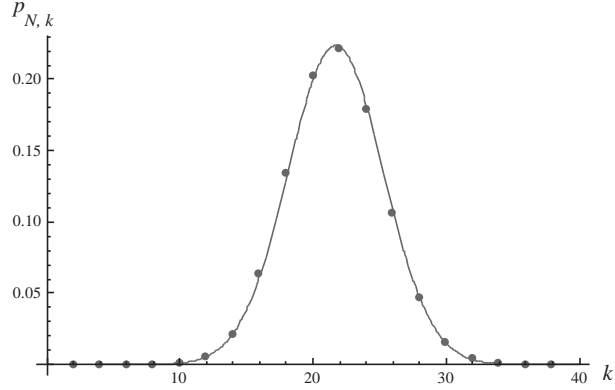}
\end{center}
\caption[Comparison of real spherical $p_{N,k}$ to the analytic prediction.]{A plot of $p_{300,k}$, that is, the probability of finding $k$ real eigenvalues from a $300\times 300$ real matrix $\bY=\bA^{-1}\bB$, where $\bA,\bB$ are real matrices with iid Gaussian elements. The points were calculated using (\ref{eqn:Sprobs}), while the solid line is the Gaussian curve implied by Proposition \ref{pAB} (with a normalising factor of $2$ since $N$ and $k$ must be of the same parity).}
\label{fig:pnk}
\end{figure}

The implication of Proposition \ref{pAB} is that our probabilities $p_{N,k}$ (suitably scaled) will tend to lie on a Gaussian curve as $N$ becomes large. In Figure \ref{fig:pnk} we have calculated the value of $p_{N,k}$ for $N=300,k={2,4,...,38}$ and overlaid it with the Gaussian curve given by Proposition \ref{pAB}; the agreement is clear. (We do not, as yet, have any further results to explain the slight systematic shift of the points relative to the curve.)

\subsubsection{$N$ odd}

As discussed at the beginning of Chapter \ref{sec:Ssops} we can find polynomials that will reduce the matrix in (\ref{eqn:ZNxio}) to an easily computable form for all $\zeta$. The column reordering (\ref{eqn:poly_order_odd}) means while these polynomials are still monomials, compared to the even case the labeling is more complicated, since there was the additional movement of the middle column to the end. The first half of the polynomials are the same as the even case, while the second half are modified by $j\rightarrow j+1/2$. The middle polynomial must be singled out for special treatment.

\begin{proposition}
\label{prop:odd_polys}
Let $N$ be odd. The skew-orthogonal polynomials with respect to the inner product (\ref{def:Sip}) are
\begin{align}
\nonumber &\left.
\begin{array}{l}
q_{2j}(x)=x^{2j},\\
q_{2j+1}(x)=x^{N-1-2j},
\end{array}
\right\}0 \leq 2j<(N-1)/2,\\
\nonumber &\left.
\begin{array}{l}
q_{2j}(x)=x^{2j+1},\\
q_{2j+1}(x)=x^{N-1-(2j+1)},
\end{array}
\right\}(N-1)/2 \leq 2j<N-1,
\end{align}
and
\begin{align}
\nonumber &q_{N-1}(x)=x^{(N-1)/2}\\
\nonumber &+ \sum_{j=0}^{ (N-1)/2-1}\left(\frac{ \langle q_{2j+1}, x^{(N-1)/2}\rangle_S} {\hat{\alpha}_{j}+ \hat{\beta}_{j}} q_{2j}(x)- \frac{\langle q_{2j}, x^{(N-1)/2}\rangle_S} {\hat{\alpha}_{j}+ \hat{\beta}_{j}} q_{2j+1}(x)\right).
\end{align}
With these polynomials
\begin{align}
\nonumber &\left.
\begin{array}{l}
\hat{\alpha}_j^{\odd}=\hat{\alpha}_j,\\
\hat{\beta}_j^{\odd}=\hat{\beta}_j,
\end{array}
\right\}0 \leq 2j<(N-1)/2,\\
\nonumber &\left.
\begin{array}{l}
\hat{\alpha}_j^{\odd}=\hat{\alpha}_{j+1/2},\\
\hat{\beta}_j^{\odd}=\hat{\beta}_{j+1/2},
\end{array}
\right\}(N-1)/2 \leq 2j<N-1,\\
\nonumber &\left.
\begin{array}{l}
\nonumber \hat{\alpha}_{s,N}+\hat{\beta}_{s,N} =\hat{\alpha}_{N,s} +\hat{\beta}_{N,s}=0, \qquad s \leq N,
\end{array}
\right.\\
\nonumber &\left.
\begin{array}{l}
\nonumber \bar{\nu}_N :=\nu \big|_{u=1}=\pi \sqrt{\frac{\Gamma((N+1)/2)}{\Gamma(N/2+1)}},
\end{array}
\right.
\end{align}
where $\hat{\alpha}_j, \hat{\beta}_j$ are as in (\ref{eqn:hatabeval}) and
\begin{align}
\nonumber \hat{\alpha}_{j+1/2}&=\frac{2\pi}{N-3-4j}\frac{\Gamma((N+1)/2)}{\Gamma(N/2+1)},\\
\nonumber \hat{\beta}_{j+1/2}&=\frac{2\sqrt{\pi}}{N-3-4j}\left( 2^N\frac{\Gamma(2j+2)\Gamma(N-2j-1)}{\Gamma(N+1)}-\sqrt{\pi}\frac{\Gamma((N+1)/2)}{\Gamma(N/2+1)}\right).
\end{align}
\end{proposition}

\textit{Proof:} For $0\leq 2j<(N-1)/2$ we have the result by Proposition \ref{prop:skew_polys_even} and replacing $j\mapsto j+1/2$ we have result for $(N-1)/2\leq 2j < N-1$. By the construction of $q_{N-1}(x)$, we see that $\hat{\alpha}_{s,N}+\hat{\beta}_{s,N}=0$ for $1\leq s\leq N$. Writing out $\hat{\nu}_l$ using the definition (\ref{def:nu}), the fact that it is non-zero only for $l=N$ is clear, that is, only when $l=N$ does the angular dependence cancel from the integral, in which case the evaluation is straightforward.

\hfill $\Box$

With the polynomials of Proposition \ref{prop:odd_polys} the matrix in (\ref{eqn:ZNxio}) for general $\zeta$ (and $u=v=1$) has the structure
\begin{align}
\label{def:Ssod} \left[\begin{array}{ccc}
\bP & \bg_{N-1} & \0_{N-1}\\
-\bg_{N-1}^T & 0 & h_{N}\\
\0_{N-1}^T & -h_N & 0
\end{array}\right],
\end{align}
where $\bP$ is skew-diagonal and $\bg_{N-1}$ is an $N-1$ dimensional non-zero vector. This structure happens to be identical to that of (\ref{eqn:skew_inverse_odd}) with $\bA^{-1}=\bP,g_j=c_j$ and $h_N=-b_N^{-1}$, and we therefore know from (\ref{eqn:Pfinvo}) that the Pfaffian of the matrix in (\ref{def:Ssod}) is equal to
\begin{align}
\nonumber h_N\prod_{j=1}^{(N-1)/2}p_j.
\end{align}
This means the odd analogue of (\ref{eqn:soPf}) is
\begin{align}
\nonumber &\mathrm{Pf}\left.\left[\begin{array}{cc}
\left[\zeta^2\alpha_{i,j} +\beta_{i,j}\right] & \left[\nu_i\right]\\
 \left[-\nu_j\right] & 0\\
\end{array}\right]_{i,j=1,...,N}\right|_{u=v=1}\\
\nonumber &= \bar{\nu}_N \prod_{l=0}^{\lceil (N-1)/4\rceil-1} \left( \zeta^2\hat{\alpha}_l +\hat{\beta}_l \right) \prod_{l=\lceil (N-1)/4\rceil}^{(N-1)/2-1} \left( \zeta^2\hat{\alpha}_{l+1/2}+ \hat{\beta}_{l+1/2} \right).
\end{align}
where $\lceil x\rceil$ is the ceiling function on $x$. Substituting this into (\ref{eqn:ZNxio}) gives
{\small
\begin{align}
\nonumber Z_N^{\odd}(\zeta)&=\frac{(-1)^{(N-1)/4((N-1)/2)-1)}}{2^{N(N-1)/2}} \Gamma((N+1)/2)^{N/2} \Gamma(N/2+1)^{N/2} \prod_{s=1}^{N}\frac{1}{\Gamma(s/2)^2}\\
\label{eqn:gfprobso} &\times \zeta \: \bar{\nu}_N \prod_{l=0}^{\lceil (N-1)/4\rceil-1} \left( \zeta^2\hat{\alpha}_l+ \hat{\beta}_l \right) \prod_{l= \lceil (N-1)/4\rceil}^{(N-1)/2-1} \left( \zeta^2\hat{\alpha}_{l+1/2} + \hat{\beta}_{l+1/2} \right).
\end{align}
}

From (\ref{eqn:gfprobso}) we can calculate the expected number of real eigenvalues in the case of $N$ odd, which we know from \cite{eks1994} is given by (\ref{29}) independent of the parity of $N$. Similarly, we can check that the formula (\ref{30}) for the variance also holds independent of the parity of $N$.

\begin{corollary}
\label{prop:odd_EN_var}
For $N$ odd, the expected number of real eigenvalues of $\bY$ can be written
\begin{align}
\nonumber E_N^{\odd} &= \frac{\partial}{\partial \zeta}Z_N^{\odd}(\zeta)\bigg|_{\zeta=1}\\
\nonumber &= 1+\sum_{l=0}^{\lceil (N-1)/4\rceil-1}\frac{2\: \hat{\alpha}_l} {\hat{\alpha}_l+ \hat{\beta}_l}+\sum_{l=\lceil (N-1)/4\rceil}^{(N-1)/2-1}\frac{2\: \hat{\alpha}_{l+1/2}} {\hat{\alpha}_{l+1/2}+ \hat{\beta}_{l+1/2}},
\end{align}
which has evaluation (\ref{29}). The variance for $N$ odd is
{\small
\begin{align}
\nonumber (\sigma_N^{\odd})^2&=\frac{\partial^2} {\partial\zeta^2}Z_N^{\odd}(\zeta)\bigg|_{\zeta=1}\\
\nonumber &= 2(E_N-1)-\sum_{l=0}^{\lceil (N-1)/4\rceil-1} \frac{4\: \hat{\alpha}_l^2} {(\hat{\alpha}_l+ \hat{\beta}_l)^2}+\sum_{l=\lceil (N-1)/4\rceil}^{(N-1)/2-1}\frac{4\: \hat{\alpha}_{l+1/2}^2}{(\hat{\alpha}_{l+1/2} +\hat{\beta}_{l+1/2})^2},
\end{align}
}
which has evaluation (\ref{30}).
\end{corollary}

\textit{Proof:} The formulae in terms of $\hat{\alpha}_l$ and $\hat{\beta}_l$ follow from
(\ref{eqn:gfprobso}) recalling the definition of $Z_N^{\odd}$ from (\ref{def:SZNo}). For the summations we use the identity
\begin{align}
\nonumber \sum_{l=0}^{\lceil (N-1)/4\rceil-1}{N-1 \choose 2l}^p+\sum_{l=\lceil (N-1)/4\rceil}^{(N-1)/2-1}{N-1 \choose 2l+1}^p&=\sum_{l=0}^{(N-1)/2-1}{N-1 \choose l}^p,
\end{align}
which holds for integer $p$ and for both $(N-1)/4\in\mathbb{Z}$ and $(N-1)/4\in\mathbb{Z}+1/2$. We obtain this identity by checking these two cases.

\hfill $\Box$

The values of $p_{N,k}$ for $N=2,...,7$, calculated using (\ref{eqn:Sprobs}) and (\ref{eqn:gfprobso}), are listed in Table \ref{table:Spnk} of Appendix \ref{app:Sprobs}, along with the results of a simulation of 100,000 matrices. A remarkable fact can be immediately seen in the table: the probabilities for even $N$ are polynomials in $\pi$ of degree $N/2$, while for odd $N$ they are rational numbers. The key difference is that $(N+1)/2$ and $N/2+1$ alternate as integers and half integers, depending on whether $N$ is even or odd. These values introduce factors of $\sqrt{\pi}$ through the gamma functions.

\begin{proposition}\label{p310}
Let $p_{N,k}$ be the probability of finding $k$ real eigenvalues in a matrix $\bY=\bA^{-1}\bB$, where $\bA,\bB$ are real Ginibre matrices. Then for $N$ even, $p_{N,k}$ is a polynomial in $\pi$ of degree $N/2$. For $N$ odd, $p_{N,k}$ is a rational number. 
\end{proposition}

\textit{Proof}: For $N$ even, from (\ref{eqn:hatabeval}) $\hat{\alpha}_l$ and the second term in $\hat{\beta}_l$ both yield factors of $\pi^{3/2}$. The pre-factor in (\ref{eqn:genpartfn}) yields $\pi^{-N/4}$. Combining these two facts we find the highest power of $\pi$ is $N/2$. Noting that the first term in $\hat{\beta}_l$ has a factor of $\pi^{1/2}$ and expanding the product in (\ref{eqn:genpartfn}) gives terms of lower order in $\pi$.

For the odd case, the pre-factor in (\ref{eqn:gen_fn_odd}) gives $\pi^{-N/4-1/2}$. Then by noting that 
$(\zeta^2\hat{\alpha}_l+\hat{\beta}_{l})$ and $(\zeta^2 \hat{\alpha}_{l+1/2}+\hat{\beta}_{l+1/2})$ both give factors of $\pi^{1/2}$ and $\nu_N$ gives $\pi^{3/4}$ we see that the end result is a rational number.

\hfill $\Box$

\subsection{Correlation functions}
\label{sec:Scorrelns}

As we did for the GOE and for the real Ginibre ensemble we would like to make use of knowledge of the Pfaffian form of the generating function (\ref{21}), and the skew-orthogonal polynomials (\ref{eqn:skew_polys}), to compute the ($N$ even) correlation functions $\rho_{(k_1,k_2)}$. In the present setting, the latter specifies the probability density for $k_1$ eigenvalues occurring at specific points on the unit circle, and $k_2$ eigenvalues occurring at specific points in the unit disk. Analogous to (\ref{eqn:GinOEfnal_diff_correln}), the $(k_1,k_2)$-point correlation function can be calculated in terms of the summed up generalised partition function (\ref{ZS}) by functional differentiation,
\begin{equation}\label{ZS1}
\rho_{(k_1,k_2)}(\mathbf{e},\mathbf{w})
= {1 \over Z_N[u,v]} {\delta^{k_1+k_2} \over \delta u(e_1) \cdots \delta u(e_{k_1})
\delta v(w_1) \cdots \delta v(w_{k_2}) } Z_N[u,v] \Big |_{u=v=1}.
\end{equation}

Comparing (\ref{eqn:q(y)}) to (\ref{eqn:GinOEjpdf}) and (\ref{eqn:genpartfn}) to (\ref{eqn:GinOE_gpf_even}), we see that the equations governing the eigenvalue statistics in the real Ginibre and real spherical ensembles are strikingly similar. As such, we expect that the correlation functions for the spherical ensemble will display similar characteristics to those of the real Ginibre ensemble. Indeed, this is what we find.

\begin{definition}
\label{def:Scorrelne}
Let $N$ be even, and $\{q_j(x) \}$ be the set of monic skew-orthogonal polynomials (\ref{eqn:skew_polys}). Define
\begin{align}
\nonumber D(x_i,x_j)_S&=\sum_{l=0}^{\frac{N}{2}-1}\frac{1}{r_l}\Bigl[a_{2l}(x_i)a_{2l+1}(x_j)-a_{2l+1}(x_i)a_{2l}(x_j)\Bigr],\\
\nonumber S(x_i,x_j)_S&=\sum_{l=0}^{\frac{N}{2}-1}\frac{1}{r_l}\Bigl[a_{2l}(x_i)b_{2l+1}(x_j)-a_{2l+1}(x_i)b_{2l}(x_j)\Bigr],\\
\nonumber \tilde{I}(x_i,x_j)_S&=\sum_{l=0}^{\frac{N}{2}-1}\frac{1}{r_l}\Bigl[b_{2l}(x_i)b_{2l+1}(x_j)-b_{2l+1}(x_i)b_{2l}(x_j)\Bigr]+\epsilon(x_i,x_j),
\end{align}
where
\begin{align}
\nonumber a_j(x) &= 
\left\{ 
\begin{array}{ll}
|x|^{-1}\tau(x)\hspace{2pt}q_j(x),  & x\in  \mathbb{D},\\
\sqrt{-i/2}\hspace{2pt}\tau(x) q_j(x),  & x\in \partial \mathbb{D},\\
\end{array}
\right.\\
\nonumber b_j(x) &= 
\left\{ 
\begin{array}{ll}
|x|^{-1}\tau(\bar{x}^{-1})\hspace{2pt}q_j(\bar{x}^{-1}),  & x\in \mathbb{D},\\
\sqrt{-i/2}\int_{0}^{2\pi}\tau(e^{i\theta})q_j(e^{i\theta})\mathrm{sgn}(\theta-\mathrm{arg}(x))d\theta, & x\in \partial \mathbb{D},\\
\end{array}
\right.\\
\nonumber \epsilon(x_i,x_j) &= 
\left\{ 
\begin{array}{ll}
\mathrm{sgn}(\mathrm{arg}(x_i)-\mathrm{arg}(x_j)),  & x_i,x_j\in \partial \mathbb{D},\\
0,  & \mathrm{otherwise},\\
\end{array}
\right.\\
\nonumber r_l&=\hat{\alpha}_l+\hat{\beta}_l,
\end{align}
and $\mathbb{D}$ is the unit disk, with $\partial\mathbb{D}$ its boundary. Also define
\begin{align}
\label{eqn:kernel} \bK_N(s,t)_S=\left[\begin{array}{cc}
S(s,t)_S & -D(s,t)_S \\
\tilde{I}(s,t)_S& S(t,s)_S \\
\end{array}\right].
\end{align}
\end{definition}

From Proposition \ref{prop:gen_part_fn} we see that the generalised partition function for the real spherical ensemble ($N$ even) is structurally identical to the analogous quantity for the real Ginibre ensemble from Proposition \ref{prop:GinOE_gpf_even} upon the identifications $p_j(w) \leftrightarrow q_j(w)$, $p_j(\bar{w}) \leftrightarrow q_j(1/ \bar{w})$. So, by the working in Chapter \ref{sec:Gincorrlnse} we obtain the correlation functions for the real spherical ensemble.

\begin{proposition}
\label{thm:correlns}
Let $N$ be even, $\bY$ be an $N \times N$ matrix as in (\ref{def:ainvb}). The $(k_1,k_2)$-point correlation function is
\begin{align}
\label{eqn:correlns} &\rho_{(k_1,k_2)}(\mathbf{e},\mathbf{w})=\mathrm{qdet}\left[\begin{array}{cc}
\bK_N(e_i,e_j)_S & \bK_N(e_i,w_m)_S\\
\bK_N(w_l,e_j)_S & \bK_N(w_l,w_m)_S\\
\end{array}\right]\\
\nonumber &=\Pf\left( \left[\begin{array}{cc}
\bK_N(e_i,e_j)_S & \bK_N(e_i,w_m)_S\\
\bK_N(w_l,e_j)_S & \bK_N(w_l,w_m)_S\\
\end{array}\right]\bZ^{-1}_{2(k_1+k_2)}\right),
e_i\in \partial \mathbb{D}, \; w_i \in \mathbb{D},
\end{align}
where $\be=\{e_1,...,e_{k_1} \}$ and $\{w_1,...,w_{k_2} \}$, and $\bZ_{2n}$ is from (\ref{def:Z2N}).
\end{proposition}

Similar to the even case, the generalised partition functions (\ref{eqn:GinOE_gpf_odd}) and (\ref{eqn:gen_fn_odd}) are structurally identical under the replacement mentioned above, but note that we cannot use the odd-from-even method of Chapters \ref{sec:odd_from_even} and \ref{sec:Gin_oddfromeven}. Recall that the method relies on removing the variable corresponding to the largest real eigenvalue off to infinity. However, in the current situation we have transformed our real eigenvalues $\lambda_j$ into the complex exponentials $e_j$ according to (\ref{14.2}) (recalling the definition of $e_j$), and so the variables in (\ref{eqn:correlns}) corresponding to the real eigenvalues are constrained to lie on the unit circle. So the question of whether the method applies to the spherical ensemble is equivalent to asking whether it can be applied to the circular ensembles. It may be possible to project the eigenvalues from the circle onto the real line, remove one to infinity and then project back, although this is, as yet, unknown. (In \cite{FM09}, only the Gaussian and Ginibre ensembles were considered.) However, the method of functional differentiation still remains viable, and by applying the procedures of Chapter \ref{sec:Gin_odd_fdiff} to (\ref{eqn:gen_fn_odd}) we can obtain the $N$ odd correlations.

\begin{definition}
Let $N$ be odd, and $\{q_j(x) \}$ be the set of monic skew-orthogonal polynomials of Proposition \ref{prop:odd_polys}. Define
\begin{align}
\nonumber &D^{\odd}(x_i,x_j)_S=\sum_{l=0}^{\lceil (N-1)/4\rceil -1}\frac{1}{r_l}\Bigl[a_{2l}(x_i)a_{2l+1}(x_j)-a_{2l+1}(x_i)a_{2l}(x_j)\Bigr]\\
\nonumber &+\sum_{l=\lceil (N-1)/4\rceil}^{(N-1)/2 -1}\frac{1}{r_{l+1/2}}\Bigl[a_{2l}(x_i)a_{2l+1}(x_j)-a_{2l+1}(x_i)a_{2l}(x_j)\Bigr],\\
\nonumber &S^{\odd}(x_i,x_j)_S=\sum_{l=0}^{\lceil (N-1)/4\rceil -1}\frac{1}{r_l}\Bigl[a_{2l}(x_i)b_{2l+1}(x_j)-a_{2l+1}(x_i)b_{2l}(x_j)\Bigr]\\
\nonumber &+\sum_{l=\lceil (N-1)/4\rceil}^{(N-1)/2 -1}\frac{1}{r_{l+1/2}}\Bigl[a_{2l}(x_i)b_{2l+1}(x_j)-a_{2l+1}(x_i)b_{2l}(x_j)\Bigr]+\kappa(x_i,x_j),\\
\nonumber &\tilde{I}^{\odd}(x_i,x_j)_S=\sum_{l=0}^{\lceil (N-1)/4\rceil -1}\frac{1}{r_l}\Bigl[b_{2l}(x_i)b_{2l+1}(x_j)-b_{2l+1}(x_i)b_{2l}(x_j)\Bigr]\\
\nonumber &+\sum_{l=\lceil (N-1)/4\rceil}^{(N-1)/2 -1}\frac{1}{r_{l+1/2}}\Bigl[b_{2l}(x_i)b_{2l+1}(x_j)-b_{2l+1}(x_i)b_{2l}(x_j)\Bigr]+\epsilon(x_i,x_j)+\sigma(x_i,x_j),
\end{align}
where $a_j(x),b_j(x)$ and $\epsilon(x_i,x_j)$ are as in Definition \ref{def:Scorrelne}, and
\begin{align}
\nonumber \kappa(x_i,x_j) &= 
\left\{ 
\begin{array}{ll}
\frac{\tau(x_i)}{\nu_N}q_{N-1}(x_i),  & x_j\in \partial \mathbb{D},\\
0,  & \mathrm{otherwise},\\
\end{array}
\right.\\
\nonumber \sigma(x_i,x_j) &= 
\left\{ 
\begin{array}{ll}
\frac{1}{\nu_N}(b_{N-1}(x_i)-b_{N-1}(x_j)), & x_i,x_j\in \partial \mathbb{D},\\
-\frac{1}{\nu_N}b_{N-1}(x_j), &x_i\in\partial \mathbb{D},x_j\in \mathbb{D},\\
\frac{1}{\nu_N}b_{N-1}(x_i), &x_i\in \mathbb{D},x_j\in\partial \mathbb{D},\\
0, & \mathrm{otherwise},\\
\end{array}
\right.\\
\nonumber r_{l+1/2}&=\hat{\alpha}_{l+1/2}+\hat{\beta}_{l+1/2}.
\end{align}
Also define
\begin{align}
\label{eqn:Skernelo} \bK_N^{\odd}(s,t)_S=\left[\begin{array}{cc}
S^{\odd}(s,t)_S & -D^{\odd}(s,t)_S \\
\tilde{I}^{\odd}(s,t)_S& S^{\odd}(t,s)_S \\
\end{array}\right].
\end{align}
\end{definition}

\begin{theorem}
With $N$ odd, the correlation functions for $\bY=\bA^{-1}\bB$ are given by
\begin{align}
\nonumber &\rho_{(k_1,k_2)}(\mathbf{e},\mathbf{w})=\mathrm{qdet}\left[\begin{array}{cc}
\bK_N^{\odd}(e_i,e_j)_S & \bK_N^{\odd}(e_i,w_m)_S\\
\bK_N^{\odd}(w_l,e_j)_S & \bK_N^{\odd}(w_l,w_m)_S\\
\end{array}\right]\\
\label{eqn:Scorrelnso} &=\Pf\left( \left[\begin{array}{cc}
\bK_N^{\odd}(e_i,e_j)_S & \bK_N^{\odd}(e_i,w_m)_S\\
\bK_N^{\odd}(w_l,e_j)_S & \bK_N^{\odd}(w_l,w_m)_S\\
\end{array}\right]\bZ^{-1}_{2(k_1+k_2)}\right), \qquad
e_i\in \partial \mathbb{D}, \; w_i \in \mathbb{D}.
\end{align}
\end{theorem}

\subsection{Kernel element evaluations and scaled limits}
\label{sec:SOElims}

As for the GOE and real Ginibre ensembles, the correlations in (\ref{eqn:correlns}) and (\ref{eqn:Scorrelnso}) are completely determined by the $2\times 2$ kernels (\ref{eqn:kernel}) and (\ref{eqn:Skernelo}). The elements of these kernels countenance relations analogous to those of Lemma \ref{lem:Gin_s=d=i}. We only list the relations for the even kernel; those for the odd kernel are similar.
\begin{lemma}
\label{lem:S_s=d=i}
The elements of the correlation kernel (\ref{eqn:kernel}) obey the relations
\begin{align}
\nonumber \tilde{I}_{r,r}(e_1,e_2)_S&=\int_{\theta_1}^{\theta_2}S_{r,r}(e,e_2)_S \; d\theta +\mathrm{sgn} (\theta_1-\theta_2),\\
\nonumber \tilde{I}_{r,c}(e,w)_S&=-\tilde{I}_{c,r}(w,e)_S=\frac{1}{|w|^2}S_{c,r}(\bar{w}^{-1},e)_S,\\
\nonumber \tilde{I}_{c,c}(w_1,w_2)_S&=\frac{1}{|w_1|^2} S_{c,c} (\bar{w}_1^{-1}, w_2)_S,\\
\nonumber D_{r,r}(e_1,e_2)_S&=\frac{\partial}{\partial \theta_2} S_{r,r}(e_1, e_2)_S,\\
\nonumber D_{c,r}(w,e)_S&=D_{r,c}(e,w)_S=\frac{1} {|w|^2} S_{r,c}(e, \bar{w}^{-1})_S,\\
\label{eqn:kernelrelations} D_{c,c}(w_1,w_2)_S&= \frac{1} {|w_2|^2} S_{c,c}(w_1,\bar{w}_2^{-1})_S.
\end{align}
\end{lemma}

We can write the $S_{*,*}(s,t)$ in a summed-up form, which, remarkably, holds for both even and odd.

\begin{proposition}
\label{prop:summedupS}
The elements $S_{r,r}(s,t)_S$, $S_{r,c}(s,t)_S$, $S_{c,r}(s,t)_S$ and $S_{c,c}(s,t)_S$ of the correlation kernel (\ref{eqn:kernel}), corresponding to real-real, real-complex, complex-real and complex-complex eigenvalue pairs respectively, can be evaluated as
{\small
\begin{align}
\nonumber S_{r,r}(e_1,e_2)_S&=\frac{\Gamma((N+1)/2)}{2\sqrt{\pi}\Gamma(N/2)}\; \mathrm{cos}\left(\frac{\theta_2-\theta_1}{2}\right)^{N-1},
\end{align}
\begin{align}
\nonumber S_{c,r}(w,e_1)_S&=\left(\frac{-i}{\sqrt{\pi}}\right)^{1/2}\frac{1}{r_w}\frac{iN}{2^N\sqrt{\pi}}\sqrt{\frac{\Gamma((N+1)/2)}{\Gamma(N/2+1)}} \left[ \int_{\frac{r_w^{-1}-r_w}{2}}^{\infty} \frac{dt} {(1+t^2)^{N/2+1}}\right]^{1/2}\\
\nonumber &\times\left( \frac{e^{-i(\theta_w-\theta_1)/2}}{r_w^{1/2}}+\frac{e^{i(\theta_w-\theta_1)/2}}{r_w^{-1/2}} \right)^{N-1},\\
\nonumber S_{r,c}(e_1,w)_S&=\left( \frac{-i}{\sqrt{\pi}} \right)^{1/2}\frac{1}{r_w}\frac{N(N-1)}{2^{N+2}\sqrt{\pi}}\left[ \int_{\frac{r_w^{-1}-r_w}{2}}^{\infty}\frac{dt}{(1+t^2)^{N/2+1}}\right]^{1/2}\sqrt{\frac{\Gamma((N+1)/2)}{\Gamma(N/2+1)}}\\
\nonumber &\times \left( \frac{e^{-i(\theta_1-\theta_w)/2}}{r_w^{1/2}} +\frac{e^{i(\theta_1-\theta_w)/2}}{r_w^{-1/2}}\right)^{N-2}\left( \frac{e^{-i(\theta_1-\theta_w)/2}}{r_w^{1/2}} -\frac{e^{i(\theta_1-\theta_w)/2}}{r_w^{-1/2}}\right),\\
\nonumber S_{c,c}(w,z)_S&=\frac{N(N-1)}{2^{N+1}\pi r_w r_z} \left[\int_{\frac{r_w^{-1}-r_w}{2}}^{\infty} \frac{dt}{\left(1+t^2\right)^{N/2+1}}\right]^{1/2} \left[\int_{\frac{r_z^{-1}-r_z}{2}}^{\infty} \frac{dt}{\left(1+t^2\right)^{N/2+1}}\right]^{1/2}\\
\nonumber &\times \left( \frac{e^{i(\theta_z-\theta_w)/2}}{(r_wr_z)^{1/2}}+\frac{e^{-i(\theta_z-\theta_w)/2}}{(r_wr_z)^{-1/2}} \right)^{N-2} \left( \frac{e^{i(\theta_z-\theta_w)/2}}{(r_wr_z)^{1/2}}-\frac{e^{-i(\theta_z-\theta_w)/2}}{(r_wr_z)^{-1/2}} \right),
\end{align}
}
for $N$ even or odd, where $w,z:=r_w e^{i\theta_w}, r_z e^{i\theta_z}$.
\end{proposition}

\textit{Proof:} Using the binomial theorem we can establish the identity
\begin{eqnarray}
\nonumber \frac{1}{2}\left(\frac{d}{dx}(1+x)^{2n-1}+\frac{d}{dx}(1-x)^{2n-1}\right)=\sum_{p=0}^{n-1}2p {2n-1 \choose 2p}x^{2p-1}
\end{eqnarray}
With this identity and the polynomials of Proposition \ref{prop:skew_polys_even} (for the even case) and Proposition \ref{prop:odd_polys} (for the odd case) the respective sums can be performed.

\hfill $\Box$

\noindent From Proposition \ref{prop:summedupS} we can use the equations (\ref{eqn:kernelrelations}) to obtain the other kernel elements.

According to Propositions \ref{thm:correlns} and \ref{prop:summedupS}, we know that the real and complex densities are given by
\begin{align}
\label{eqn:SOElimdensr} \rho_{(1)}^{r}(\theta)&:=\rho_{(1,0)}(e)=S_{r,r}(e,e)_S =\frac{\Gamma((N+1)/2)} {2\sqrt{\pi} \Gamma(N/2)},
\end{align}
and
\begin{align}
\nonumber \rho_{(1)}^{c}(w)&:=\rho_{(0,1)}(w)=S_{c,c}(w,w)_S\\
\label{eqn:rhoc} &= \frac{N(N-1)} {2^{N+1} \pi r^2} \left( {1 \over r} + r \right)^{N-2} \left( \frac{1} {r} - r \right) \int_{\frac{r^{-1} - r} {2}}^\infty \frac{dt} {(1 + t^2)^{N/2 + 1}},
\end{align}
respectively. Note that $\int_{0}^{2\pi} \rho_{(1)}^{r}(\theta) d\theta= E_N$ and so the evaluation (\ref{29}) is immediate from (\ref{eqn:SOElimdensr}). Since $\rho_{(1)}^{r}(\theta)=E_N/2\pi$ we can use (\ref{eqn:SENasy}) to give us the large $N$ form of the density
\begin{align}
\label{eqn:Srlimdens} \rho_{(1)}^{r}(\theta)  \sim {1 \over 2 \pi} \sqrt{{\pi N \over 2}},
\end{align}
while integration by parts of (\ref{eqn:rhoc}) shows
\begin{align}
\label{eqn:Sclimdens} \rho_{(1)}^{\rm c}(w) \sim {(N-1) \over \pi} {1 \over (1 + r^2)^2} - {N - 1 \over N - 2}
{1 \over \pi} {1 \over (1 - r^2)^2} + \rm{O}\Big ( {1 \over N} \Big ),
\end{align}
valid for $r \in [0, 1 - {\rm O}(1/\sqrt{N})]$. Note that the real eigenvalue density is only proportional to $\sqrt{N}$, and so to leading order in $N$ the density of general eigenvalues is
\begin{align}
\label{df}
{N \over \pi} {1 \over (1 + r^2)^2},
\end{align}
for all $r \in [0,1]$. This is a Cauchy distribution and, projected stereographically onto the half sphere, gives a uniform distribution. Note also that we have a uniform density of real eigenvalues in (\ref{eqn:Srlimdens}), which we recall was manifested in the analogous real Ginibre result (\ref{eqn:Grlimdens}).
\begin{remark}
This $1/N$ convergence  in (\ref{eqn:Sclimdens}) should be contrasted with the exponential convergence in the case of the polynomials (\ref{eqn:rand_polys}) (see \cite{Mc09}).
\end{remark}

The uniform spherical distribution implied by (\ref{df}) led to a conjecture in \cite{FM11} that there is a \textit{spherical law} analogous to the circular law of Proposition \ref{prop:circlaw}. This conjecture has since been proven by adapting the method of \cite{TVK10} used to prove the circular law. Specifically, in \cite{Bord2010} the author shows that the eigenvalue density for a matrix $\bY=\bA^{-1}\bB$, where $\bA,\bB$ have general iid mean zero, variance one distributed elements, converges to that where $\bA,\bB$ are (complex) Gaussian distributed. Then using (\ref{df}) the spherical law follows.

\begin{proposition}[Spherical law, \cite{Bord2010}]
\label{prop:sphlaw}
Let $\bY=\bA^{-1}\bB$, where $\bA,\bB$ are $N\times N$ matrices with iid elements from a distribution with mean $0$ and variance $1$. Then the density of eigenvalues of $\bY$ approaches the uniform distribution on the unit sphere under stereographic projection as $N \to \infty$.
\end{proposition}

\subsubsection{Scaled limit}
\label{sec:SOEsclims}

Before  implementing the fractional linear transformations (\ref{7'}) and (\ref{14.2}), we have from (\ref{eqn:rho}) that the density of real eigenvalues near the origin is proportional to $E_N$, and thus $\sqrt{N}$. Taking a scaled limit --- involving changing the variables so that the real and complex densities are of order unity --- is of interest in this case because the resulting correlations are expected to be the same as for the eigenvalues of the real Ginibre ensemble, scaled near the origin, which we calculated in Chapter \ref{sec:Ginkernelts} and listed the results in (\ref{eqn:Ginbulkall}) of Appendix \ref{app:Ginsummed}. This expectation is based upon the geometrical knowledge that locally a sphere resembles a plane.
\begin{remark}
In the complex case, an analogy between the eigenvalue jpdf of the generalised eigenvalue problem and the Boltzmann factor for the two-dimensional one-component plasma on a sphere \cite{caillol81}, together with the analogy between the eigenvalue jpdf for the Ginibre matrices and the two-dimensional one-component plasma in the plane \cite{AJ80} allow this latter point (the similarity to the bulk real Ginibre correlations) to be anticipated from a Coulomb gas perspective.
\end{remark}

In the present problem, with our use of the transformed variables (\ref{7'}) and (\ref{14.2}), the original origin has been mapped to $(1,0)$. We must choose scaled co-ordinates so that in the vicinity of this point the real and complex eigenvalues have a density of order unity. For the real eigenvalues, from the knowledge that their expected value is of order $\sqrt{N}$ and that they are uniform on the unit circle, with $e_j:=e^{ix_j}$, we scale
\begin{eqnarray}
\label{eqn:largeNreal} x_j\mapsto \frac{2X_j}{\sqrt{N}}.
\end{eqnarray}
For the complex eigenvalues, which total of order $N$ in the unit disk, an order one density will result by writing
\begin{eqnarray}
\label{eqn:largeNcomplex} w_j\mapsto 1+\frac{2i}{\sqrt{N}}W_j
\end{eqnarray}
Note that the real and imaginary parts have been interchanged to match the geometry of the problem in the Ginibre ensemble, that is so the eigenvalues are again distributed in the upper half-plane, including the real line. The factors of $2$ in (\ref{eqn:largeNreal}) and (\ref{eqn:largeNcomplex}) are included so an exact correspondence with the results of (\ref{eqn:Ginbulk}) and (\ref{eqn:Ginbulkall}) can be obtained.

\begin{remark}
Since the eigenvalue density is rotationally invariant under (\ref{7'}), we can choose any point on the unit circle in (\ref{eqn:largeNcomplex}); we have chosen the image of the original origin for convenience.
\end{remark}

Since $S_{r,r}(x,x)_S$ is interpreted as a density, the normalised quantity is $S_{r,r}(x,x)_Sdx$. It then follows that in the more general case we must look at the scaled limit of $S_{r,r}(x,y)_S \sqrt{dx dy}$ and $S_{c,c}(w_1, w_2)_S\sqrt{d^2w_1 d^2w_2}$. For $S_{r,c}(x, w)_S$ and $S_{c,r}(w,x)_S$ we require that the product $S_{r,c}(x,w)_S S_{c,r}(w,x)_S dx d^2w$ has a well defined limit. From (\ref{eqn:largeNreal}) and (\ref{eqn:largeNcomplex}) we see
\begin{align}
\nonumber \sqrt{dxdy}&\mapsto\frac{2} {\sqrt{N}} \sqrt{dXdY},\\
\nonumber dxd^2w&\mapsto\left( \frac{4} {N}\right)^{3/2} dXd^2W,\\
\nonumber \sqrt{d^2w_1d^2w_2}&\mapsto\frac{4} {N} \sqrt{d^2W_1d^2W_2}.
\end{align}
With this change of variables the large $N$ form of the correlation kernel for the spherical ensemble matches that of the Ginibre ensemble.

\begin{proposition}
\label{prop:scaled_limit}
Recall $\bK_{N}(s,t)_S$ from (\ref{eqn:kernel}). Replacing $x_j$ and $w_j$ according to (\ref{eqn:largeNreal}) and (\ref{eqn:largeNcomplex}) then taking $N\to\infty$ gives
\begin{align}
\nonumber \frac{2}{\sqrt{N}}\; \bK_N(e_i,e_j)_S&\sim \bK_{r,r}^{\bulk}(X_i,X_j),\\
\nonumber \frac{2^{5/2}}{N}\; \bK_N(e_i,w_j)_S&\sim \bK_{r,c}^{\bulk}(X_i,W_j),\\
\nonumber \frac{\sqrt{2}}{\sqrt{N}}\; \bK_N(w_i,e_j)_S&\sim -\left(\bK_{r,c}^{\bulk} (W_i,X_j)\right)^{T},\\
\nonumber \frac{4}{N}\; \bK_N(w_i,w_j)_S&\sim \bK_{c,c}^{\bulk}(W_i,W_j),
\end{align}
with $\bK_{r,r}^{\bulk}(\mu,\eta),\bK_{r,c}^{\bulk}(\mu,\eta)$ and $\bK_{c,c}^{\bulk}(\mu,\eta)$ as in (\ref{eqn:Ginbulkall}).
\end{proposition}

\textit{Proof:} From the explicit functional forms of Proposition \ref{prop:summedupS}, we see that elementary limits suffice. For example, changing variables $t\mapsto 2t/\sqrt{N}$ shows
\begin{align}
\nonumber \int_{\frac{r_w^{-1}-r_w}{2}}^{\infty} \frac{dt} {(1+t^2)^{N/2+1}} \sim\sqrt{\frac{\pi}{2N}}\: \mathrm{erfc}(\sqrt{2}\mathrm{Im}W).
\end{align} 
Combining such calculations we obtain
\begin{align}
\nonumber \frac{2}{\sqrt{N}}\; S_{r,r}(e_i,e_j)_S&\sim\frac{1}{\sqrt{2\pi}}e^{-(X_i-X_j)^2/2},\\
\nonumber \frac{2^{5/2}}{N}\; S_{r,c}(e_i,w_j)_S&\sim\frac{\sqrt{-i}} {\sqrt{2\pi}} \sqrt{\mathrm{erfc}(\sqrt{2} \mathrm{Im}W_j)}\: e^{-(X_i-\overline{W}_j)^2/2}\; i(\overline{W}_j-X_i),\\
\nonumber \frac{\sqrt{2}} {\sqrt{N}}\; S_{c,r}(w_i,e_j)_S&\sim\frac{\sqrt{-i}}{\sqrt{2\pi}} \sqrt{\mathrm{erfc} (\sqrt{2}\mathrm{Im}W_i)}\: ie^{-(W_i-X_j)^2/2},\\
\nonumber \frac{4}{N}\; S_{c,c}(w_i,w_j)_S&\sim\frac{1}{\sqrt{2\pi}}\sqrt{\mathrm{erfc}(\sqrt{2} \mathrm{Im}W_i)} \sqrt{\mathrm{erfc}(\sqrt{2}\mathrm{Im}W_j)}\\
\nonumber &\times i(\overline{W}_j-W_i)e^{-(W_i-\overline{W}_j)^2/2},
\end{align}
which is in agreement with the diagonal entries on the RHS of the present proposition, as implied by (\ref{eqn:Ginbulk}) (when one recalls that $S_{r,c}$ and $S_{c,r}$ never appear individually; only as the product $S_{r,c}S_{c,r}$).

Performing the remaining limits, or by recalling the inter-relationships (\ref{eqn:kernelrelations}), the other kernel elements $D$ and $I$ can be obtained from $S$, giving the off-diagonal entries required in (\ref{eqn:Ginbulkall}).

\hfill $\Box$

\subsection{Averages over characteristic polynomials}
\label{sec:SOEcharpolys}

In \cite{sommers2007, sommers_and_w2008} the author/s used an average over characteristic polynomials to obtain the correlation kernels. This method is again successfully applied in \cite{KSZ2010} to the real truncated ensemble (which we will discuss in Chapter \ref{sec:truncs}). As emphasised in \cite{FK07,APS2009} there is a large class of eigenvalue jpdfs such that the eigenvalue density is given in terms of an average over the corresponding characteristic polynomials. This is true of the one-point function (density) for the complex eigenvalues, with $N\to N+2$ (for convenience), in the present generalised eigenvalue problem for which the jpdf is given by (\ref{eqn:q(y)}). To establish such, write
\begin{align}
\label{eqn:charpoly} C_N(z)=\prod_{j=1}^k(z-e_j)\prod_{s=k+1}^{(N+k)/2}(z-w_s)(z-1/\bar{w}_s)
\end{align}
for the characteristic polynomial in the $N\times N$ case of $\bY=\bA^{-1}\bB$ conditioned to have $k$ real eigenvalues, with eigenvalues transformed according to (\ref{7'}) and (\ref{14.2}). Letting
\begin{align}
\nonumber \tilde{G}_{N}:=\frac{(-1)^{(N/2)(N/2-1)/2}}{2^{N(N-1)/2}} \Gamma((N+1)/2)^{N/2} \Gamma(N/2+1)^{N/2} \prod_{s=1}^{N}\frac{1}{\Gamma(s/2)^2},
\end{align}
which is the pre-factor in (\ref{eqn:ZNxi}), then it follows from (\ref{eqn:q(y)}) and the definition of the density that
\begin{eqnarray}
\label{eqn:charpolydensity} \rho_{(1)}^{(N+2, c)}(z)=\frac{\tilde{G}_{N+2}}{\tilde{G}_N} \frac{1}{|z|^2}\tau(z)\tau(1/\bar{z})(1/\bar{z}-z)\langle C_N(z)C_N(1/\bar{z})\rangle,
\end{eqnarray}
where the superscript $N+2$ denotes the number of eigenvalues in the system. Of course we can therefore read off from (\ref{eqn:rhoc}) the exact form of the average in (\ref{eqn:charpolydensity}). Moreover, in keeping with the development in \cite{APS2009}, we can use our integration methods to compute the more general average $\langle C_N(z_1)C_N(1/\bar{z}_2) \rangle$, which we expect to be closely related to $S_{c,c}(z_1,z_2)_S$, in accordance with known results from the real, complex and real quaternion Ginibre ensembles \cite{kanzieper2001, AV2003, AB07, APS2009, sommers_and_w2008}.

Note that we will also introduce a superscript on $\alpha$ and $\beta$ to indicate the size of system that they relate to, that is $\alpha_{j,l}^{(t)}$ has $j,l=1,...,t$, and $\hat{\alpha}_{s}^{(t)}$ are the corresponding normalisations.

\begin{proposition}
\label{prop:SOEcharpoly}
With the characteristic polynomial $C_N(z)$ given by (\ref{eqn:charpoly}) and $\langle\cdot\rangle$ an average with respect to (\ref{eqn:q(y)}) summed over $k$, one has
\begin{align}
\nonumber &\langle C_N(z_1)C_N(1/\bar{z}_2) \rangle=\frac{\tilde{G}_{N}}{\tilde{G}_{N+2}} (1/\bar{z}_2-z_1)^{-1}\\
\nonumber &\times \sum_{s=0}^{N/2}\frac{1}{\alpha_s^{(N+2)}+ \beta_s^{(N+2)}} \left( z_1^{2s}(1/\bar{z}_2)^{2s+1}-z_1^{2s+1}(1/\bar{z}_2)^{2s}\right)\\
\label{eqn:charpolyD} &=\frac{\tilde{G}_{N}}{\tilde{G}_{N+2}} (1/\bar{z}_2 -z_1)^{-1} \left(\frac{ \tau(z_1)}{|z_1|} \frac{\tau(1/\bar{z}_2)} {|z_2|}\right)^{-1} \left.D_{c,c}(z_1,1/\bar{z}_2) \right|_{N\to N+2},
\end{align}
where it is assumed $N$ is even. Furthermore
\begin{align}
\nonumber \left.S_{c,c}(z_1,z_2)_S \right|_{N \to N+2} &=  \frac{\tilde{G}_{N+2}}{\tilde{G}_N} \frac{\tau(z_1)}{|z_1|} \frac{ \tau(1/\bar{z}_2)} {|z_2|}(1/\bar{z}_2-z_1)\langle C_N(z_1)C_N(1/\bar{z}_2) \rangle,
\end{align}
from which we reclaim (\ref{eqn:charpolydensity}).
\end{proposition}

\textit{Proof:} From (\ref{eqn:q(y)}) we see that
\begin{align}
\nonumber  C_N(z_1)C_N(1/\bar{z}_2)\mathcal{Q}(Y) =A_{k,N}\prod_{j=1}^k\tau(e_j)\prod_{s=k+1}^{(N+k)/2}\frac{1}{|w_s|^2}\tau(w_s)\tau\left(\frac{1}{\bar{w}_s}\right)\\
\nonumber \times(1/\bar{z}_2-z_1)^{-1}\Delta \left(\mathbf{e}, \mathbf{w}, \mathbf{ \frac{1}{\bar{w}}}, z_1,\frac{1}{\bar{z}_2}\right).
\end{align}
Integrating over $\mathbf{e}$ and $\mathbf{w}$ gives
\begin{align}
\nonumber \langle C_N(z_1)C_N(1/\bar{z}_2\rangle\: \rule[-8pt]{0.5pt}{18pt}_{\: k \: {\rm fixed}}&= \tilde{G}_N(1/ \bar{z}_2-z_1)^{-1}\\
\nonumber &\times [\varkappa^{N/2}] [\zeta^{k}]\mathrm{Pf}\left[\varkappa \left(\zeta^2 \hat{\alpha}_{j,l}^{(N+2)}+ \hat{\beta}_{j,l}^{(N+2)} \right)+ \gamma_{j,l}^{(N+2)}\right],
\end{align}
where $\gamma_{j,l}^{(t)}=q_{j-1}(z_1)q_{l-1}(1/\bar{z}_2)-q_{l-1}(z_1)q_{j-1}(1/\bar{z}_2)$ $(j,l=1,...,t)$, and recalling from (\ref{def:hatab1}) that the `hat' on $\alpha,\beta$ indicates that $u=v=1$. Note that the matrix
\begin{align}
\nonumber \left[\varkappa \left(\zeta^2 \hat{\alpha}_{j,l}^{(N+2)}+ \hat{\beta }_{j,l}^{(N+2)} \right)+ \gamma_{j,l}^{(N+2)}\right]
\end{align}
is of dimension $(N+2) \times (N+2)$. Summing over $k$ leads to
\begin{align}
\nonumber &\langle C_N(z_1)C_N(1/\bar{z}_2\rangle:=  \sum_{k=0 \atop k \: {\rm even}}^N  \langle C_N(z_1)C_N(1/\bar{z}_2\rangle\: \rule[-8pt]{0.5pt}{18pt}_{\: k \: {\rm fixed}}\\
\nonumber &=\tilde{G}_N (1/\bar{z}_2-z_1)^{-1}[\varkappa^{N/2}]\mathrm{Pf} \left[\varkappa \left(\hat{\alpha}_{j,l}^{(N+2)} + \hat{\beta}_{j,l}^{(N+2)} \right)+ \gamma_{j,l}^{(N+2)}\right].
\end{align}
Using the skew-orthogonal polynomials (\ref{eqn:skew_polys}), we find
\begin{align}
\nonumber &[\varkappa^{N/2}]\mathrm{Pf} \left[\varkappa \left(\hat{\alpha}_{j,l}^{(N+2)} + \hat{\beta}_{j,l}^{(N+2)} \right)+ \gamma_{j,l}^{(N+2)}\right]\\
\nonumber &\qquad =\sum_{s=0}^{N/2}\gamma_{2s+1,2s+2}\prod_{j=0 \atop j\neq s}^{N/2} \left( \hat{\alpha}_{j}^{(N+2)} +\hat{\beta}_{j}^{(N+2)} \right).
\end{align}
The evaluation now follows upon recalling the form of
\begin{align}
\nonumber Z_{N+2}(1)_S =1=\tilde{G}_{N+2} \prod_{l=0}^{N/2} \hat{\alpha}_l^{(N+2)} +\hat{\beta}_l^{(N+2)}
\end{align}
implied by (\ref{eqn:Sprobs}).

The expression for $\left.S_{c,c}(z_1,z_2)_S\right|_{N\to N+2}$ is a simple manipulation of (\ref{eqn:charpolyD}).

\hfill $\Box$

\newpage

\section{Truncations of orthogonal matrices}
\setcounter{figure}{0}
\label{sec:truncs}

As discussed in the Introduction, the three homogeneous two-dimensional manifolds with constant Gaussian curvature $\kappa>0,\kappa=0$, or $\kappa<0$ (being the sphere, plane and pseudo- or anti-sphere respectively) correspond to certain random matrix ensembles. We have identified the real Ginibre ensemble (Chapter \ref{sec:GinOE}) with the plane and the real spherical ensemble (Chapter \ref{sec:SOE}) with the sphere. In this chapter we consider the last of the triumvirate: the anti-sphere, which corresponds to ensembles of truncated unitary or orthogonal matrices.

The \textit{real truncated ensemble} consists of sub-blocks of random, Haar-distributed orthogonal matrices (recall (\ref{eqn:Haar})). Enter, stage left, the following definition.
\begin{definition}
With $N=L+M$, let $\bR$ be an $N\times N$ Haar distributed random orthogonal matrix. Decompose $\bR$ as follows
\begin{align}
\label{def:Rdecomp} \bR=\left[\begin{array}{cc}
\bA & \bB\\
\bC & \bD
\end{array}\right],
\end{align}
where $\bA$ is of size $L\times L$ and $\bD$ is of size $M\times M$, with the dimensions of $\bB$ and $\bC$ implicit.
\end{definition}
We will concern ourselves with the analysis of the bottom-right block, labeled here as $\bD$.

\begin{remark}
Since the matrix $\bR$ is Haar distributed, it is invariant under linear transformations and so every sub-block of size $M\times M$ manifests the same statistics. We have chosen the bottom-right sub-block to keep our notation consistent with that in \cite{Forrester2010a}, which is, in turn, consistent with \cite{Krish2009}.
\end{remark}

An important distinction between this ensemble and those of the previous chapters in this work is that the eigenvalues of $\bD$ (with $L>0$) must lie strictly inside the unit circle. (This consideration has implications for the calculation of the odd correlations, which we will discuss in Chapter \ref{sec:TOEcorrelno}.) This can be seen by considering the induced $2$-norm on $N\times N$ matrices
\begin{align}
\label{def:i2n} ||\bM||_2 = \max \{ ||\bM x||_2: x\in \mathbb{R}^N, ||x||_2=1\},
\end{align}
where $||x||_2=\sqrt{x_1^2+x_2^2+...+x_N^2}$ is the usual $2$-norm on (real) vectors; (\ref{def:i2n}) is equivalent to the \textit{spectral norm} $\sqrt{\lambda_{\max} (\bM^T \bM)}$, which in our case gives us the eigenvalue of largest absolute value. We can think of the matrix $\bD$ as
\begin{align}
\nonumber \hat{\bR}=\left[\begin{array}{cc}
\0 & \0 \\
\0 & \bD
\end{array}\right],
\end{align}
and from the definition of (\ref{def:i2n}) we see that $||\hat{\bR}||_2 =||\bD||_2$. By writing $\hat{\bR}=\bP \bR$, where $\bP$ is some matrix, we apply the sub-multiplicative property ($||\bA\bB||\leq ||\bA|| ||\bB||$) of the spectral norm to find $||\hat{\bR}||_2 \leq ||\bP||_2 ||\bR||_2=||\bP||_2$, where the equality follows since orthogonal matrices have norm one. Note that $\bP$ is the projection operator on any vector $x$ (it sends some number of elements to zero), and so $||\bP x||_2\leq ||x||_2$, which implies $||\bP||_2\leq 1$. Therefore the eigenvalues of $\hat{\bR}$ must lie in the unit disk, and consequently, so must those of $\bD$.

\begin{remark}
\label{rem:evals<1}
Note that we can assume that all eigenvalues satisfy $|\lambda|<1$ since, in the space of all Haar distributed random orthogonal matrices, matrices whose truncation yields eigenvalues on the unit circle form a set of measure zero. 
\end{remark}

As happened with the Gaussian and spherical ensembles, truncated matrices with complex entries (that is, truncations of unitary matrices) were more successfully dealt with before those with real entries (see \cite{Z&S2000,P&R2003} for the details relating to the complex truncated ensemble). In \cite{Z&S2000}, while focussing primarily on the complex case, the authors briefly discuss the real case and complete some preliminary calculations. Yet it was not until \cite{KSZ2010} the distribution of the sub-blocks and their eigenvalues were established. The authors of the latter paper go on to find a Pfaffian form of the correlations and compute the correlation kernel, which largely completes the study; although in \cite{Forrester2010a} the author finds the skew-orthogonal polynomials and establishes a correspondence between the eigenvalues of $\bD$ and Kac random polynomials in the particular case when $M=N-1$. The fact that the correlation functions were found before the skew-orthogonal polynomials means (naturally) that the so-called \textit{orthogonal polynomial} method, which heretofore in this work we have been using, was not the method by which the correlations were established. Indeed, the method of \cite{KSZ2010} is based upon earlier work \cite{AV2003,APS2009} using averages over characteristic polynomials, a method we discussed in Chapter \ref{sec:SOEcharpolys}. In this chapter we proceed with the same five-step procedure as used in earlier chapters, with the caveat that it is not yet known how to derive the skew-orthogonal polynomials independently of the correlation kernel. (In \cite{ForNagRai2006} and \cite{AkeKiePhi2010} the authors present determinantal and Pfaffian (respectively) formulae for finding skew-orthogonal polynomials in the basis of the monomials, however it is not known how to perform the integrals this method entails; see also \cite[Ex. 6.1 Q4]{forrester?} and \cite{Eyn2001}.)

We expect that there are three limiting regimes of interest, which correspond to the behaviour of $\alpha:=M/N$: \textit{i}) $\alpha\to 1$, where $L$ is fixed or grows only subdominantly; \textit{ii}) $0<\alpha<1$, where both $L$ and $M$ grow proportionately to $N$, that is,
\begin{align}
\label{def:alphaML} M=\alpha N &&\mbox{and}&& L=\gamma N=(1-\alpha)N;
\end{align}
\textit{iii}) $\alpha\to 0$, where $M$ is fixed or grows subdominantly.  If $\alpha$ is restricted to be less than $1$ (cases (\textit{ii}) and (\textit{iii})), then the ensemble will exhibit \textit{weakly orthogonal} behaviour, which matches that of the real Ginibre ensemble. This is anticipated, since with a relatively large number of rows and columns deleted then the elements in $\bD$ will only weakly feel the orthogonality constraint imposed on $\bR$, and so the elements should be more or less independent Gaussians. There is a well known theorem of Jiang \cite{Jiang2006} to this effect. On the other hand, if $\alpha\to 1$ then the orthogonality of $\bR$ is keenly felt and so we call this case the \textit{strongly orthogonal} regime. In this limit, the behaviour is analogous to that of the real spherical ensemble, with the spherical geometry replaced by that of anti-spherical geometry. This will lead us to an anti-spherical conjecture, analogous to Proposition \ref{prop:sphlaw}, at the end of this chapter.

\subsection{Matrix distribution}
\label{sec:TOEmjpdf}

Since the matrix $\bR$ is orthogonal, we must have $\bC \bC^T +\bD \bD^T=\1_M$ giving that the joint distribution of $\bC$ and $\bD$ is
\begin{align}
\label{eqn:TOEmatdist2} P(\bC, \bD)\propto \delta (\bC \bC^T +\bD \bD^T- \1_M).
\end{align}
To furnish the normalisation we begin by noting that this constraint defines the manifold $V_{N,M}=\{ (\bC,\bD)\in \mathbb{R}^{N\times M} : \bC\bC^T +\bD\bD^T =\1 \}$ in the set of all $N\times M$ matrices. We think of $V_{N,M}$ as a set of $M$ orthonormal vectors of length $N$; this manifold is called the \textit{Stiefel manifold}. A well-known algebraic result tells us that the Stiefel manifold can be written as a quotient group.
\begin{lemma}
\label{lem:orbstab1} Let $X$ be a set, $G$ a group that acts on $X$ and $G_x$ the stabilizer of $x\in X$ by $G$, that is $G_x=\{g\in G : g * x=x \}$ where $*$ is the group operation. Then
\begin{align}
\label{eqn:orbstab1} G/G_x \cong G * x.
\end{align}
\end{lemma}

\textit{Proof}: Define the map $\psi : G/G_x \to G*x$ by $\psi(g_x)=g*x$, where $g$ is a representative from the equivalence class $g_x\in G/G_x$. Since any element of $g*x\in G*x$ has a corresponding class in $G/G_x$, we have that $\psi$ is surjective. To establish injectivity, let $g_x,h_x\in G/G_x$, then we have $\psi(g_x)=\psi(h_x) \Leftrightarrow g*x=h*x \Leftrightarrow h^{-1}g*x=x \Leftrightarrow h^{-1}g \in G_x \Leftrightarrow h \equiv g \mod G_x \Leftrightarrow g_x=h_x$. Note that we have used the fact that $G_x$ is a subgroup of $G$ to establish the \textit{only if} direction in the third relation. $\psi$ is therefore a bijection, and the result follows.

\hfill $\Box$

\noindent We call $G*x$ the \textit{orbit} of the element $x\in X$ under the action of $G$.

If $X$ is finite then we can deduce the \textit{orbit-stabliser theorem}.

\begin{proposition}[Orbit-Stabliser Theorem]
\label{prop:orbstab}
Let $X$ be a finite set, $G$ a finite group and $G_x$ the stabilizer of $x\in X$ by $G$. Then, with $\Orb (x):= G*x$ the orbit of $x$ under $G$, we have
\begin{align}
\nonumber |G|/|G_x|=|\Orb (x)|,
\end{align}
where $|H|$ is the order (or cardinality) of the group $H$.
\end{proposition}

\textit{Proof}: By Lagrange's (Group) Theorem we have that $|G|=[G:H]\cdot |H|$, where $[G:H]$ is defined as the cardinality of the coset space $G/H$, called the \textit{index} of $H$ in $G$. Since $\Orb (x):= G*x$, the result follows from the isomorphic relation (\ref{eqn:orbstab1}).

\hfill $\Box$

Lagrange's theorem is a statement about how many cosets of $H$ there are in $G$, however if one is interested in the volume of an infinite set, then we need to make precise the statement ``$G$ is $\gamma$ times larger $H$". By carefully defining suitable measures, taking limits and viewing the isomorphism in (\ref{eqn:orbstab1}) as a topological isomorphism, one establishes that \cite{nachbin1965}
\begin{align}
\label{eqn:vol/vol} \vol (G/G_x) = \frac{\vol (G)} {\vol (G_x)} = \vol (\Orb (x)),
\end{align}
where the volume is with respect to Haar measure (\ref{def:volON}).

To apply these results to our business we take $G=O(N)$, the orthogonal group of degree $N$, which acts on $V_{N,M}$. Note that any set of orthonormal vectors can be transformed into any other under the operation of $O(N)$. This is the same as saying that the action of $O(N)$ on $V_{N,M}$ is transitive, that is $O(N)*f=V_{N,M}$, for any $f\in V_{N,M}$. To pick a specific stabliser, we may choose $f_0=\{ e_1,...,e_M\}$, where the $e_j$ are the standard basis vectors in $\mathbb{R}^N$. Then
\begin{align}
\nonumber G_{f_0}=\left\{ \left[
\begin{array}{cc}
\1_{M} & \0\\
\0 & \bQ_{L\times L}
\end{array} \right] : \bQ_{L\times L} \mbox{ is an orthogonal $L\times L$ matrix} \right\} \cong O(L).
\end{align}
Combining this information, we have from (\ref{eqn:orbstab1}) that $O(N)/O(L)\cong V_{N,M}$ (where, as noted above, the isomorphism is to be interpreted as a topological isomorphism between locally compact topological groups). From (\ref{eqn:vol/vol}) we then have
\begin{align}
\nonumber \vol(V_{N,M})=\frac{\vol (O(N))}{\vol (O(L))},
\end{align}
where $\vol(O(N))$ is the volume of $O(N)$ with respect to Haar measure, and is given by (\ref{def:volO}). We have therefore found the normalisation in (\ref{eqn:TOEmatdist2}) giving \cite{KSZ2010}
\begin{align}
\label{eqn:TOEmatpdf1} P(\bC, \bD)=\frac{\vol (O(L))}{\vol (O(N))} \; \delta (\bC \bC^T +\bD \bD^T- \1_M).
\end{align}

The distribution of the sub-block $\bD$ itself, however, is a somewhat delicate procedure. For small $M$ (compared to $L$) the elements of the sub-block are roughly independent Gaussians, and the distribution is smooth \cite{Jiang2006}. But when $M$ becomes too large the orthogonality of $\bR$ introduces singular effects to the distribution. This is also seen in the analogous case of truncations of unitary matrices \cite{JN03}. We will proceed to integrate over the matrices $\bC$ in (\ref{eqn:TOEmatpdf1}), although we will find that we must accept the restriction $L\geq M$. Interestingly, it will turn out that the subsequent eigenvalue jpdf is insensitive to these concerns.

\begin{proposition}[\cite{Forrester2006,KSZ2010}]
\label{prop:TOEelpdf}
With $L\geq M$ the probability density function for $\bD$, an $M\times M$ sub-block of $\bR$ is
\begin{align}
\label{eqn:truncpdf} P(\bD)=\frac{(\vol(O(L)))^2}{\vol(O(N))\vol(O(L-M))}\det(\1-\bD^T\bD)^{(L-M-1)/2}.
\end{align}
\end{proposition}

\textit{Proof}: To obtain the matrix pdf of $\bD$ we will integrate over the matrices $\bC$ in (\ref{eqn:TOEmatpdf1}). First note that the delta function therein can be rewritten as the product of delta functions
\begin{align}
\nonumber \prod_{j=1}^M \delta(\hat{c}_{jj}+\hat{d}_{jj}-1) \prod_{1\leq i < j \leq M} \delta (\hat{c}_{ij} +\hat{d}_{ij}),
\end{align}
where $\bC \bC^T=[\hat{c}_{i,j}]_{i,j=1,...,M}$ and $\bD \bD^T=[\hat{d}_{i,j}]_{i,j=1,...,M}$. Using a Fourier integral representation of the delta function we can write
\begin{align}
\nonumber \delta(\hat{c}_{jj}+\hat{d}_{jj}-1) &= \frac{1}{2\pi} \int_{-\infty}^{\infty} e^{i (\hat{c}_{jj} +\hat{d}_{jj}-1)h_{jj}}dh_{jj},\\
\nonumber  \delta (\hat{c}_{ij} +\hat{d}_{ij}) &=\frac{1}{2\pi} \int_{-\infty}^{\infty} e^{i (\hat{c}_{ij} +\hat{d}_{ij})h_{ij}}dh_{ij},
\end{align}
where the $h_{ij}$ are real variables, and so
\begin{align}
\nonumber &\int \delta (\bC \bC^T + \bD \bD^T -\1_M) (d\bC) =\\
\label{eqn:TOEdeltainteg1} &\left( \frac{1}{2\pi} \right)^{M(M+1)/2} \int (d\bC) \int (d\bH) \; e^{i \: \Tr (\bH(\bC\bC^T +\bD\bD^T -\1_M))},
\end{align}
where $\bH$ is an $M\times M$ real, symmetric matrix with diagonal elements $h_{jj}$ and off-diagonal elements $h_{ij}/2$. Perturbing $\bH$ using $\bH - \mu \1_M$, where $\mu>0$, will enable us to use the integral evaluation \cite[Proposition 3.2.8]{forrester?}
\begin{align}
\nonumber &I_{m,n}(\bQ_m):= \int e^{\frac{i}{2} \Tr (\bH_m \bQ_m)} \big(\det (\bH_m-\mu \1_m) \big)^{-n} (d\bH_m)\\
\label{eqn:Imninteg} & =\frac{2^m \pi^{m(m+1)/2} i^{m} (-1)^{m(m-1)/2}} { \prod_{j=n-m+1}^n \Gamma(j) } \left( \det \left( \frac{i}{2} \bQ_m \right) \right)^{n-(m+1)/2} e^{\frac{i}{2} \mu \Tr (\bQ_m)},
\end{align}
with $n\geq m/2$, where $\bH_m$ and $\bQ_m$ are $m\times m$ real, symmetric matrices. (From this point we will only concern ourselves with proportional relations; the normalisation will be calculated \textit{post hoc}.) Introducing the aforementioned perturbation, we rewrite (\ref{eqn:TOEdeltainteg1}) as
\begin{align}
\nonumber &\int \delta (\bC \bC^T + \bD \bD^T -\1_M) (d\bC)\\
\nonumber &\propto \lim_{\mu \to 0^{+}} \int (d\bC) \int (d\bH) \; e^{i \: \Tr ((\bH- i\mu\1_M) (\bC\bC^T +\bD\bD^T -\1_M))}\\
\nonumber &\propto \lim_{\mu \to 0^{+}} \int (d\bC) \int (d\bH) \; \det (\bH -i\mu\1_M)^{-L/2} \; e^{i \: \Tr (\bC\bC^T +(\bH- i\mu\1_M) (\bD\bD^T -\1_M))},
\end{align}
where we have changed variables $\bC \to (\bH-i \mu\1_M)^{1/2}\: \bC$, which introduces a Jacobian of $\det (\bH -i\mu\1_M)^{-L/2}$ using Lemma \ref{lem:alpha_tensor_beta}. We may now interchange the order of integration and perform the integrals over $\bC$ to find that the matrix pdf is proportional to
\begin{align}
\nonumber &\lim_{\mu \to 0^+} \int (d\bH) \; \det (\bH -i\mu\1_M)^{-L/2} \; e^{i \: \Tr \big( (\bH- i\mu\1_M) (\bD\bD^T -\1_M) \big)}\\
\nonumber &= \lim_{\mu \to 0^+} \int (d\bH) \; \det (\bH -i\mu\1_M)^{-L/2} \; e^{i \: \Tr \big( \bH (\bD\bD^T -\1_M) \big)}.
\end{align}
With $m=M, n=L/2$ and $\bQ_m=2(\bD\bD^T -\1_M)$ the preceding limit can be written in terms of the integral $I_{m,n}$ and we may now make use of (\ref{eqn:Imninteg}) to yield
\begin{align}
\nonumber &\lim_{\mu \to 0^+} I_{M,L/2} (2(\bD\bD^T -\1_M))\\
\nonumber &\propto \lim_{\mu \to 0^+} (\det (\1_M -\bD \bD^T) )^{(L-M-1)/2} e^{\mu \Tr (\bD \bD^T -\1_M) }\\
\nonumber &= \det (\1_M -\bD \bD^T)^{(L-M-1)/2}.
\end{align}
Note that, by the condition on $m$ and $n$ in (\ref{eqn:Imninteg}), we must have $L\geq M$.

The proportionality constant $\tilde{C}_{M,L}$ is defined by the condition that
\begin{align}
\label{eqn:TOEmpjdfnorm1} \tilde{C}_{M,L} \int \det (\1_M -\bD^T\bD)^{(L-M-1)/2} (d\bD) =1.
\end{align}
To calculate $\tilde{C}_{M,L}$ we will apply the change of variables $\bD^T\bD = \bG$, and so we first calculate the constant for this transformation as so
\begin{align}
\nonumber 1& =\left(\frac{1}{2\pi} \right)^{M^2/2} \int e^{-\Tr (\bD^T\bD)/2} (d\bD)\\
\nonumber &=b_M \left(\frac{1}{2\pi} \right)^{M^2/2} \int (\det \bG)^{-1/2} e^{-\Tr \bG/2}  (d\bG)\\
\label{eqn:TOEmpjdfnorm2} &=B_M \left(\frac{1}{2\pi} \right)^{M^2/2} \int_0^{\infty}d\lambda_1 \cdot\cdot\cdot \int_0^{\infty} d\lambda_M \prod_{j=1}^M \lambda_{j}^{-1/2} e^{-\lambda_j /2} \prod_{j<l} |\lambda_l -\lambda_j|,
\end{align}
where the $\lambda_j$ are the eigenvalues of $\bG$. To obtain the second equality we have used Lemma \ref{lem:tilde_const_covarbs} and to obtain the third we have used Proposition \ref{prop:GOE_J}; $b_M$ and $B_M$ respectively incorporate the proportionality constants for these transformations. Note that the integral in the second equality is over positive definite matrices. With Corollary \ref{cor:varselb} we can evaluate the integral in (\ref{eqn:TOEmpjdfnorm2}) and find
\begin{align}
\nonumber B_M^{-1}&= \pi^{-M^2/2} \; \prod_{j=0}^{M-1} \frac{\Gamma ((j+3)/2) \Gamma ((j+1)/2)} {\Gamma (3/2)}\\
\label{eqn:BM} &= \pi^{-M(M+1)/2} \; \Gamma (M+1) \prod_{j=1}^{M-1} (\Gamma (j/2))^2.
\end{align}

Now we change variables in (\ref{eqn:TOEmpjdfnorm1}) to obtain
\begin{align}
\nonumber \left(\tilde{C}_{M,L} \right)^{-1} &= b_M \int (\det \bG)^{-1/2} \det (\1_M -\bG)^{(L-M-1)/2} (d \bG)\\
\nonumber &= B_M \int_0^1 d\lambda_1 \cdot\cdot\cdot \int_0^1 d\lambda_M \prod_{j=1}^M \lambda_j^{-1/2} (1-\lambda_j)^{(L-M-1)/2} \prod_{j<l} |\lambda_l- \lambda_j|.
\end{align}
With $B_M$ from (\ref{eqn:BM}) and using (\ref{eqn:Selbint}) we then have
\begin{align}
\nonumber \left(\tilde{C}_{M,L} \right)^{-1} &= \frac{\pi^{M(M+1)/2}} {\Gamma(M+1) \prod_{j=1}^M (\Gamma(j/2))^2}\\
\nonumber &\times  \prod_{j=0}^{M-1} \frac{\Gamma ((j+1)/2) \Gamma ((L-M+1+j)/2) \Gamma ((j+3)/2)} {\Gamma ((L+1+j)/2) \Gamma (3/2)}\\
\nonumber &=\pi^{M^2/2} \prod_{j=0}^{M-1} \frac{\Gamma((L-M+1+j)/2)} {\Gamma((L+1+j)/2)}.
\end{align}
By noting that
\begin{align}
\nonumber \prod_{j=0}^{M-1} \Gamma ((L+1+j)/2) &= \frac{\prod_{j=1}^{L+M} \Gamma (j/2)}{\prod_{j=1}^{L} \Gamma (j/2)},\\
\nonumber \prod_{j=0}^{M-1} \Gamma ((L-M+1-j)/2) &= \frac{\prod_{j=1}^L \Gamma(j/2)}{\prod_{j=1}^{L-M} \Gamma (j/2)}
\end{align}
we can manipulate $\tilde{C}_{M,L}$ to give us the result.

\hfill $\Box$

\begin{figure}[htp]
\begin{center}
\includegraphics[scale=0.4]{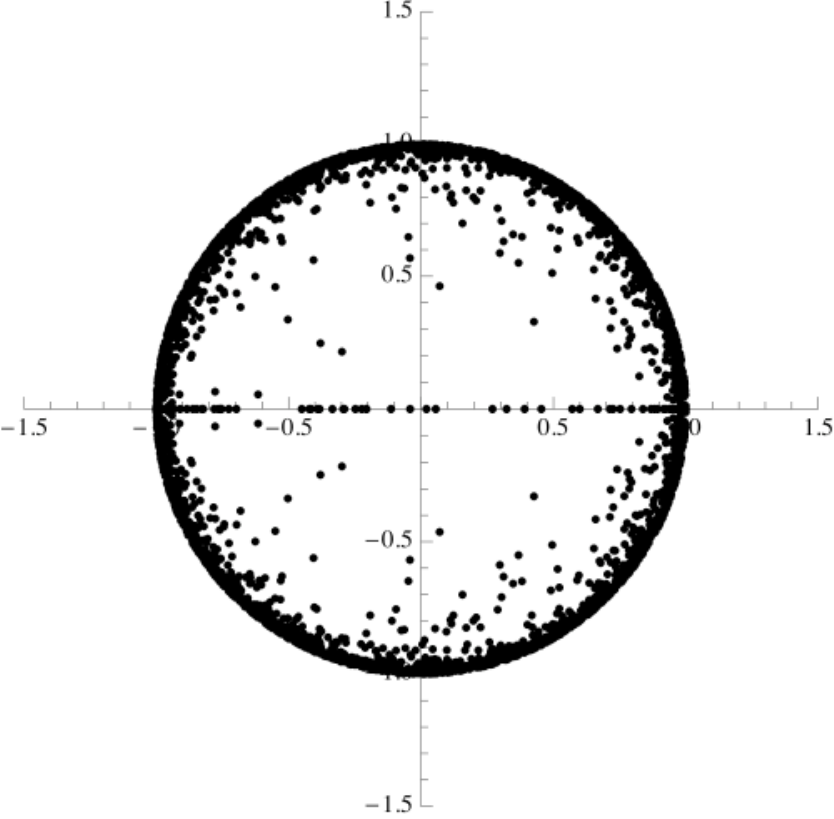}
\hspace{3ex}\includegraphics[scale=0.4]{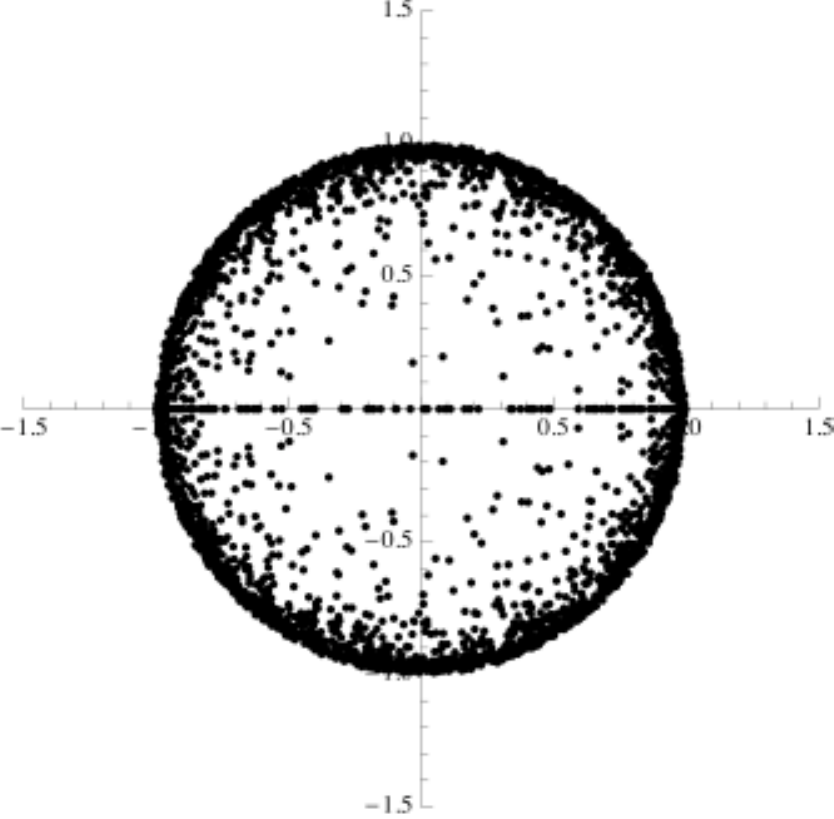}
\hspace{3ex}\includegraphics[scale=0.4]{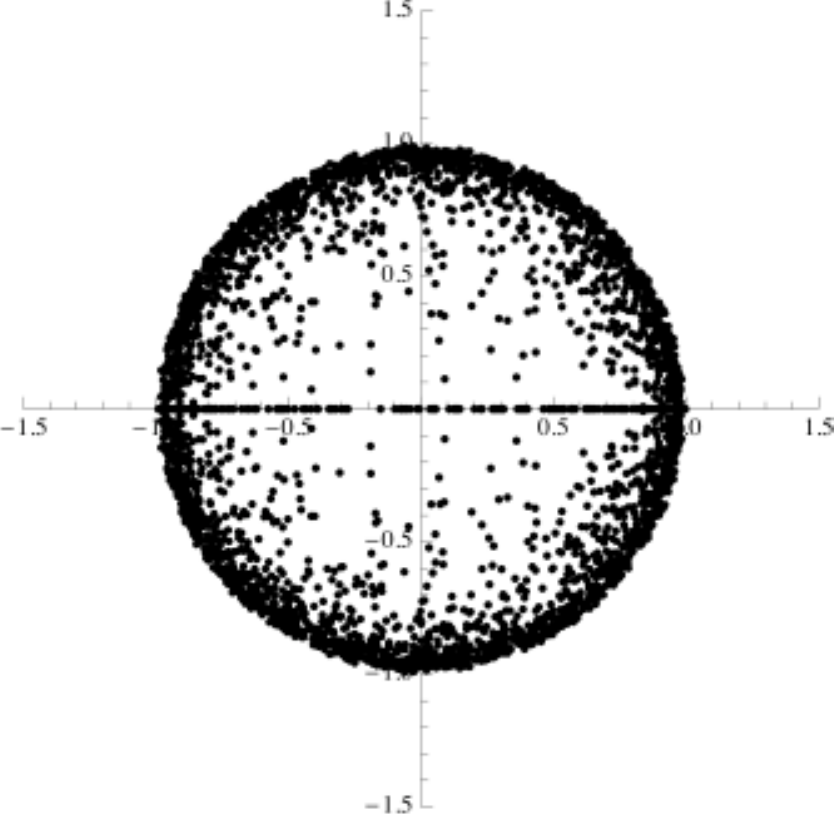}
\hspace{3ex}\includegraphics[scale=0.4]{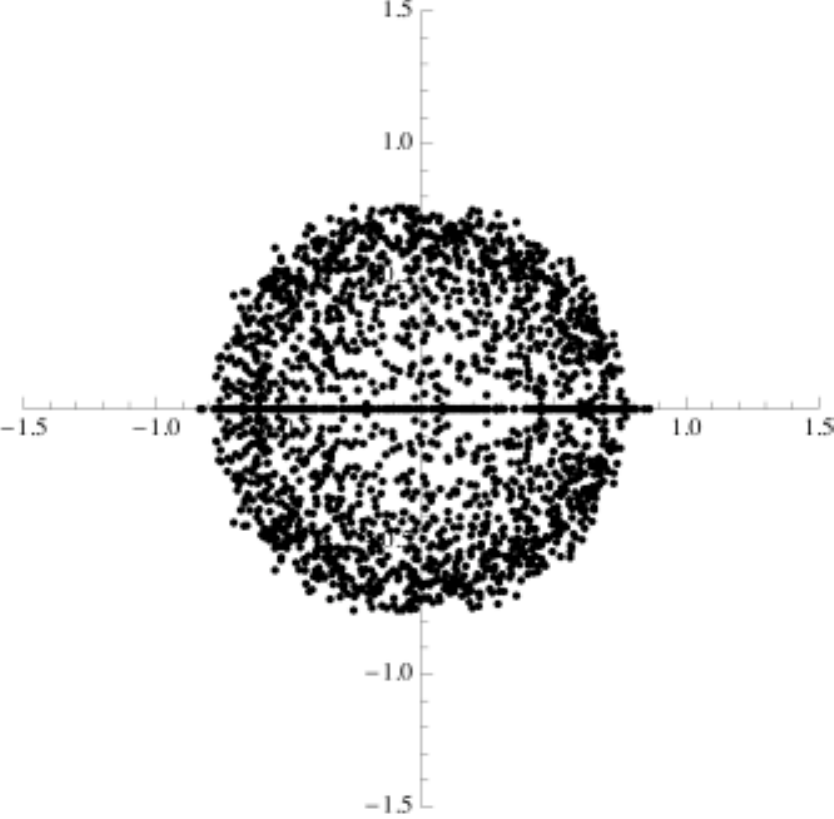}
\hspace{3ex}\includegraphics[scale=0.4]{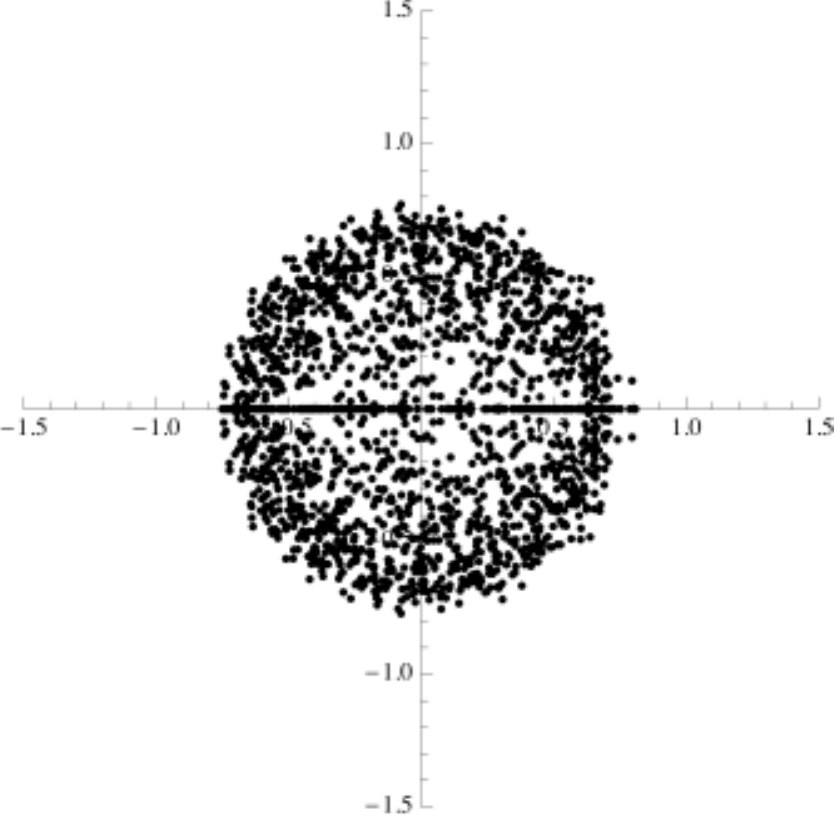}
\hspace{3ex}\includegraphics[scale=0.4]{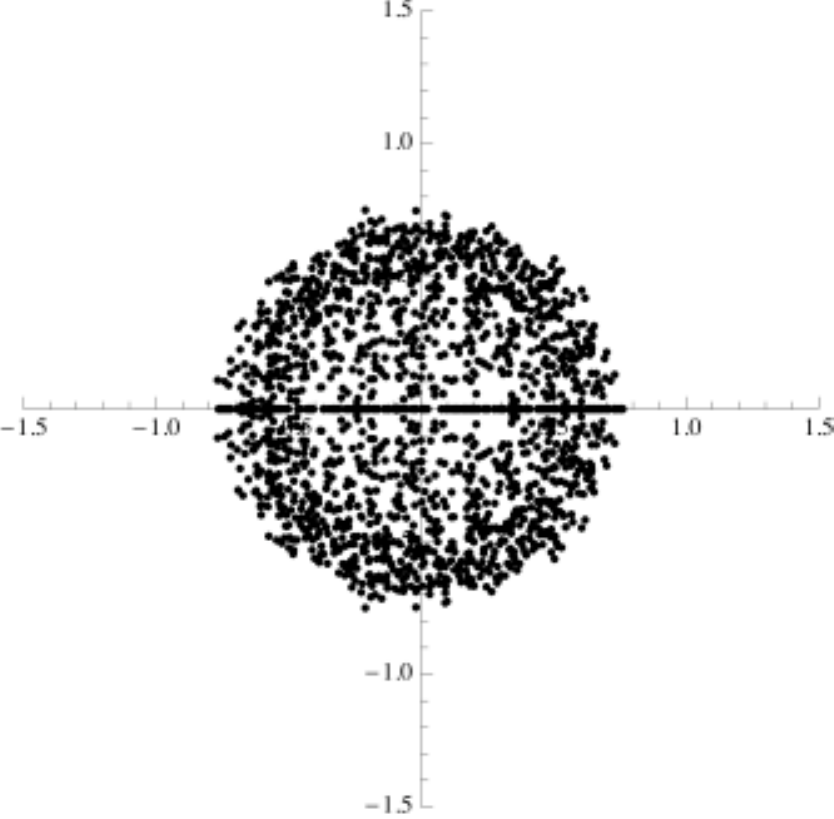}
\hspace{3ex}\includegraphics[scale=0.4]{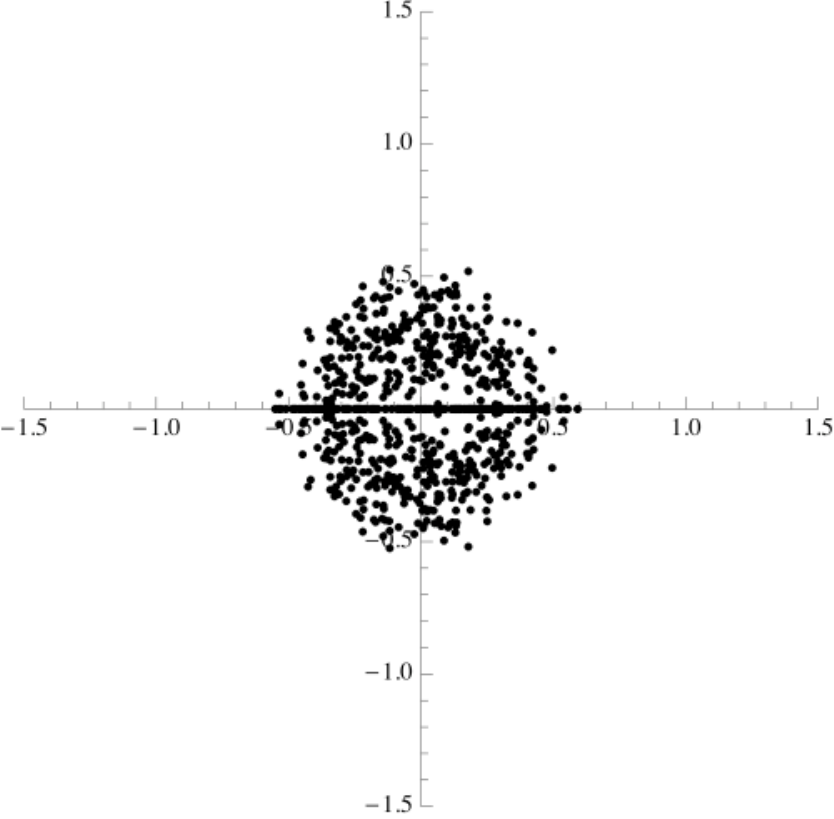}
\hspace{3ex}\includegraphics[scale=0.4]{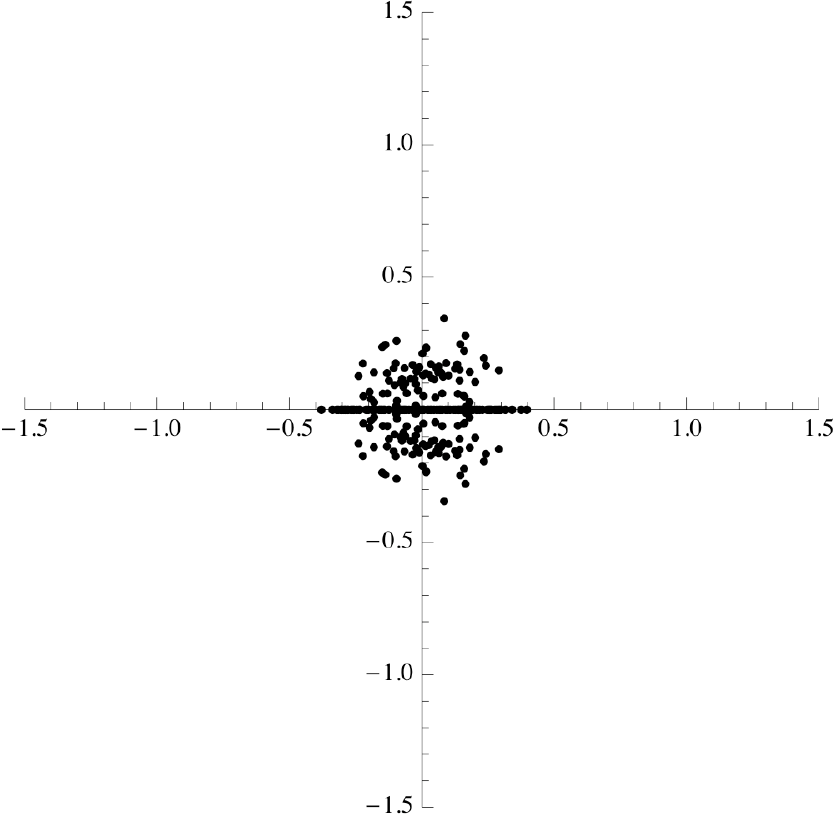}
\caption[Simulated eigenvalue plots of real truncated ensembles with $N=75$ and $M=74, 72, 70, 42, 38, 36, 15, 5$.]{Eigenvalue plots for truncations of $50$ independent $75\times 75$ random Haar distributed orthogonal matrices, with (from left to right, top to bottom) $M=74,72,70,42,38,36,15,5$.}
\label{fig:TOE_sims}
\end{center}
\end{figure}

As with the ensembles in our previous chapters, we find that computer simulations are easily generated and provide an immediate guide to the results we can expect. In Figure \ref{fig:TOE_sims} we have plotted eigenvalues from various truncations of 50 independent $75\times 75$ Haar distributed random orthogonal matrices. Note that for small $L$ (large $M$) the eigenvalues are largely congested on the unit circle, and as $L\to N$ they are restricted to a smaller disk, becoming more uniformly spread. Also note that, in all cases, there is clearly a non-zero density of real eigenvalues, which, as we have noted in previous chapters, is a distinctive feature of ensembles of real matrices.

\begin{remark}
The generation of Haar distributed random matrices was done by first generating a random Gaussian matrix (that is, a real Ginibre matrix) and then applying Gram-Schmidt orthogonalisation to obtain a Haar distributed orthogonal matrix. For a very readable introductory description of this see \cite{Diaconis2005}, or for a more technical treatment see \cite{P&R2003}.
\end{remark}

\subsection{Eigenvalue distribution}

In order to write the eigenvalue distribution in a concise way we will use the following definition.

\begin{definition}
With $\mu\in\mathbb{C}$ let
\begin{align}
\label{def:truncw} \omega(\mu)=\left\{\begin{array}{ll}
\left(\frac{L(L-1)}{2\pi}|1-\mu^2|^{L-2}\int_{2|\mathrm{Im}(\mu)|/|1-\mu^2|}^{1}(1-t^2)^{(L-3)/2}dt\right)^{1/2},& L>1,\\
\left(\frac{1}{2\pi}\right)^{1/2}|1-\mu^2|^{-1/2},&L=1.
\end{array}
\right.
\end{align}
\end{definition}

Note that the integral in (\ref{def:truncw}) (for $L>1$) will play the same role here as the $\erfc$ function did in (\ref{eqn:GinOEjpdf}) for the real Ginibre ensemble, and as the integral in the function $\tau(x)$ did in (\ref{eqn:q(y)}) for the spherical ensemble. If $\mu=x\in\mathbb{R}$ then the integral in the first case of (\ref{def:truncw}) becomes the Beta function
\begin{align}
\nonumber \int_{0}^{1} (1-t^2)^{(L-3)/2}dt &= \frac{1}{2} \int_{0}^{1} t^{-1/2} (1-t)^{(L-3)/2}dt \\
\label{eqn:Bxy1} &=: \frac{1}{2} \; B\left( \frac{1} {2},\frac{L-1}{2} \right) = \frac{\sqrt{\pi}}{2} \frac{\Gamma((L-1)/2)}{\Gamma (L/2)}
\end{align}
and so
\begin{align}
\label{eqn:TOErweight} \omega(x)= \frac{(1-x^2)^{L/2-1}}{\sqrt{2} \; \pi^{1/4}}\left(L\frac{\Gamma((L+1)/2)}{\Gamma(L/2)}\right)^{1/2},
\end{align}
which, with $L=1$, gives us the second case in (\ref{def:truncw}) immediately.

At the end of the previous chapter we found the explicit distribution of the sub-block matrices $\bD$, under the restriction that $M \leq L$, by integrating out the dependence on the sub-block $\bC$. This result, (\ref{eqn:truncpdf}), gives us a convenient way of finding the eigenvalue distribution using the tools that we successfully employed to obtain the distribution of the eigenvalues in the real Ginibre and real spherical ensembles in earlier chapters. The single caveat is that we have the aforementioned restriction $M \leq L$, which, as discussed in the introduction to this chapter, is a weakly orthogonal regime. The need for this restriction with this approach can be seen in (\ref{eqn:TOEmpjdfnorm1}), which can only be satisfied for $(L-M-1)/2\geq 1/2$. It turns out that the eigenvalue jpdf for this restricted case is identical to that in the general case. We will proceed with the calculation of the jpdf, assuming $M \leq L$, in the same fashion that led to (\ref{eqn:q(y)}). This will be followed by a discussion of how to establish the general case.

\begin{proposition}[\cite{KSZ2010}]
\label{prop:Tejpdf}
Let $\bD$ be an $M\times M$ sub-block of an $(L+M)\times (L+M)$ Haar distributed random orthogonal matrix as in (\ref{def:Rdecomp}). Then $\bD$ has $k$ real eigenvalues $\Lambda =\{\lambda_1,...,\lambda_k\}$ and $M-k$ complex conjugate paired eigenvalues $W =\{z_1,\bar{z}_1 ,... ,z_{(M-k)/2},$\\$ \bar{z}_{(M-k)/2}\}$. With the eigenvalues ordered as in (\ref{9'}) the eigenvalue jpdf is
\begin{align}
\label{eqn:TOEevaljpdf} Q(\Lambda, W)_T&=C_{M,L}\prod_{j=1}^k \omega(\lambda_j)\prod_{j=1}^{(M-k)/2}2i\;\omega(z_j)^2 \; \Delta(\Lambda \cup W),
\end{align}
where $\Delta(\{x_j \})=\prod_{i<j}(x_j-x_i)$ as in Proposition \ref{prop:GinOE_eval_jpdf},
\begin{align}
\label{def:CML} C_{M,L}:=\frac{\vol(O(L))\vol(O(M))}{\vol(O(L+M))}\left(\frac{(2\pi)^L}{L!}\right)^{M/2},
\end{align}
and $\vol(O(X))$ is from (\ref{def:volO}).
\end{proposition}

\begin{remark}
Note that (\ref{def:CML}) is the corrected normalisation; it was incorrectly stated in \cite{KSZ2010}.
\end{remark}

\textit{Proof}: Letting $\tilde{C}_{M,L}$ stand for the prefactor in (\ref{eqn:truncpdf}) we see that
\begin{align}
\nonumber \tilde{C}_{M,L} \det(1-\bD^T\bD)^{(L-M-1)/2}
\end{align}
is structurally similar to (\ref{eqn:elementjpdf}) and so we expect that we may apply a Schur decomposition (\ref{eqn:GinOE_decomp}), followed by the iterated integration technique of Chapter \ref{sec:Sevaldist}, to make progress. We write $\bD=\bQ\bR_M\bQ^T$, where $\bR_M$ is a the block upper-triangular matrix (\ref{def:triangular_mat}) with the diagonal $1\times 1$ and $2\times 2$ blocks corresponding to the real and complex conjugate pairs of eigenvalues respectively, and $\bQ$ is the real orthogonal matrix of corresponding eigenvectors (with the restriction that the first row of $\bQ$ is positive). With the ordering (\ref{9'}) the decomposition is unique.

The Jacobian of the change of variables from the elements of $\bD$ to the eigenvalues of $\bD$ is given by Proposition \ref{prop:GinJ}, 
\begin{align}
\nonumber (d\bD)&=2^{(M-k)/2}\prod_{j<p}|\lambda(R_{pp})-\lambda(R_{jj})| \prod_{l=k+1}^{(M+k)/2} |b_l-c_l|\\
\nonumber &\times(d\tilde{\bR}_M)(\bQ^Td\bQ) \prod_{s=1}^{k}d\lambda_s \prod_{l=k+1}^{(M+k)/2}dx_ldb_ldc_l,
\end{align}
and we recall the integral over $(\bQ^T\bQ)$ from (\ref{eqn:RTdR_integ}). By decomposing
\begin{align}
\nonumber \bR_M=\left[\begin{array}{cc}
\bR_{M-2} & u_{M-2}\\
\0^T & z_m
\end{array}\right]
\end{align}
as in (\ref{eqn:RNdecomp}), we may now directly apply the iterated integration technique (for the complex eigenvalue columns) of Chapter \ref{sec:iitc} by replacing $\det(\1_N+ \bR_N\bR_N^T)^{-N}$ therein with $\det(\1_M- \bR_M\bR_M^T)^{(L-M-1)/2}$, giving
\begin{align}
\nonumber &\int\det (\1_M- \bR_M\bR_M^T)^{(L-M-1)/2} (du_{M-2})\cdot\cdot\cdot (du_{k+2})\\
\nonumber &=\det (\1_k-\bR_k\bR_k^T)^{(L-k-1)/2} \prod_{s= k+1}^{(M+k)/2} \det (\1_2-z_sz_s^T )^{(L-3)/2}\\
\label{eqn:Tdet1-RMRMT} &\times\prod_{s=0}^{(M-k)/2-1}\int \det (\1_2 -v_{M-2s-2}^Tv_{M-2s-2})^{s+(L-M-1)/2} (dv_{M-2s-2}),
\end{align}
where $v_j=(\1_j- \bR_j\bR_j^T)^{-1/2} u_{j} (\1_2- z_{m-(M-2-j)/2}z_{m-(M-2-j)/2}^T)^{-1/2}$. Note that so far we have only changed variables and have not performed any of the integrals on the LHS of (\ref{eqn:Tdet1-RMRMT}). The task now is to perform the integrals on the RHS.

Defining the $2\times 2$ matrix $\bC:= v_{M-2s-2}^T v_{M-2s-2}$ we use Lemma \ref{lem:tilde_const_covarbs} to write
\begin{align}
\label{eqn:Tctildecov} (dv_{M-2s-2}) = \tilde{c} \: \det \bC ^{(M-2s-5)/2} (d\bC).
\end{align}
In order to calculate $\tilde{c}$ (since it is independent of $v_{M-2s-2}$), we can take the elements of $v_{M-2s-2}$ to be independent variables in the interval $(-\infty, \infty)$, as was done in (\ref{eqn:Sctilde}). Then, using the notation $p=M-2s-2, \; q=(M-2s-5)/2$ for ease of expression,
\begin{align}
\nonumber \tilde{c} = \frac{\int e^{-\Tr \: v_{p}^T v_{p}} (d v_{p})} {\int e^{-\Tr \: \bC} \left(\det \bC\right)^{q} (d\bC)} = \frac{\pi^{p}} {\int_0^{\infty}dx_1 \int_0^{\infty}dx_2  \: x_1^q \: e^{-x_1} \: x_2^{q} \: e^{-x_2} |x_2- x_1|},
\end{align}
with $x_1$ and $x_2$ the eigenvalues of $\bC$, having used Proposition \ref{prop:GOE_J} to change variables from the entries of $\bC$ to its eigenvalues in the denominator of the second equality. We can now calculate the integrals on the RHS of (\ref{eqn:Tdet1-RMRMT}) by the change of variables (\ref{eqn:Tctildecov})
\begin{align}
\nonumber &\int \det (\1_2- v_{p}^Tv_{p})^{s+(L-M-1)/2} (dv_{p})\\
\nonumber & = \tilde{c} \int \det \left(\1_2-\bC \right)^{s+(L-M-1)/2} \left(\det \bC\right)^q (d\bC)\\
\nonumber & = \pi^p \frac{\int_0^{1}dy_1 \int_0^{1}dy_2  \: \big((1-y_1) (1-y_2)\big)^{s+(L-M-1)/2} (y_1 y_2)^{q} e^{-x_1} e^{-x_2} |x_2- x_1|} {\int_0^{\infty}dx_1 \int_0^{\infty}dx_2  \: x_1^q \: e^{-x_1} \: x_2^{q}\: e^{-x_2} |x_2- x_1|}\\
\label{eqn:ccoleval} &=\pi^{M-2s-2} \frac{\Gamma((L-M+1)/2+s) \Gamma((L-M)/2+1+s)} {\Gamma((L-1)/2) \Gamma(L/2)},
\end{align}
where we have used the Selberg integrals (\ref{eqn:Selbint}) and (\ref{cor:varselb1}).

If we let
\begin{align}
\nonumber \bR_k=\left[\begin{array}{cc}
\bR_{k-2} & u_{k-2}\\
\0^T & z_m
\end{array}\right]
\end{align}
then we can apply the method of Chapter \ref{sec:iitr} to integrate over the off-diagonal elements in the real eigenvalue columns, and we obtain
\begin{align}
\nonumber &\int \det (\1_k- \bR_k\bR_k^T)^{(L-k-1)/2} (du_k) \cdot\cdot\cdot (du_1)\\
\label{eqn:TOErcol} &= \prod_{s=1}^k (1-\lambda_s^2)^{L/2-1} \prod_{s=1}^{k-1}\int (1 - v_{k-s}^T v_{k-s})^{(L-k+s)/2-1}(dv_{k-s}).
\end{align}
The integral on the right hand side of (\ref{eqn:TOErcol}) can be evaluated analogously to (\ref{eqn:ccoleval}),
\begin{align}
\nonumber \int (1 - v_{k-s}^T v_{k-s})^{(L-k+s)/2-1}(dv_{k-s})=\pi^{(k-s)/2}\frac{\Gamma ((L-k+s)/2)}{\Gamma(L/2)}.
\end{align}

Combining the preceding we find the density of the truncated matrix $\bD$ is
\begin{align}
\nonumber &\tilde{C}_{M,L} 2^{(M-k)/2} \prod_{j<p}|\lambda(R_{pp})-\lambda(R_{jj})| \frac{\pi^{M(M+1)/4}}{\prod_{j=1}^M \Gamma(j/2)} \\
\nonumber &\times \prod_{s=1}^k (1-\lambda_s^2)^{L/2-1} \pi^{(k-s)/2}\frac{\Gamma ((L-k+s)/2)}{\Gamma(L/2)}\\
\nonumber & \times \prod_{l=k+1}^{(M+k)/2} |b_l-c_l| \det (\1_2-z_lz_l^T )^{(L-3)/2}\\
\nonumber & \times  \prod_{l=0}^{(M-k)/2-1} \pi^{M-2l-2} \frac{\Gamma((L-M+1)/2+l) \Gamma((L-M)/2+1+l)} {\Gamma((L-1)/2) \Gamma(L/2)}\\
\label{eqn:TOEjpdf1} & \times   \prod_{s=1}^{k}d\lambda_s \prod_{l=k+1}^{(M+k)/2}dx_ldb_ldc_l.
\end{align}
We now wish to remove the dependence on $b_l$ and $c_l$. With $\epsilon_l:=1-x_l^2-y_l^2$ and $\delta_l:=b_l-c_l$ then by explicit calculation we see that $\det(\1_2-z_lz_l^T) =\epsilon_l^2 -\delta_l^2$. Using (\ref{eqn:GinOE_covs}) (noting again the correction there to the domain of $\delta$) we have
\begin{align}
\nonumber &\int_{\delta=-\epsilon_l}^{\delta=\epsilon_l}|b_l-c_l| \det(\1_2-z_lz_l^T)^q \; dx_ldb_ldc_l = 4y_l\int_{\delta=0}^{\delta=\epsilon_l}\frac{\delta(\epsilon_l^2- \delta^2)^{q}\;d\delta}{\sqrt{\delta^2+4y_l^2}}dx_ldy_l\\
\label{eqn:TOEerfclem} &=4y_l|1- z_l^2|^{2q+1}\int_{t=2|y_l|/|1-z_l^2|}^{t=1}(1-t^2)^q\; dt\; dx_ldy_l,
\end{align}
where the second equality follows after some changes of variable analogous to those of (\ref{eqn:delta_integ}).

By some rearrangement we have the identities
\begin{align}
& \nonumber \prod_{l=0}^{(M-k)/2-1} \Gamma((L-M+1)/2+l) \Gamma((L-M)/2+1+l) \prod_{s=1}^k \Gamma ((L-k+s)/2)\\
\nonumber &= \frac{\vol (O(L-M))}{\vol (O(L))} \frac{2^L \pi^{L(L+1)/4}}{2^{L-M} \pi^{(L-M) (L-M+1)/4}},\\
\nonumber & \prod_{s=1}^k \Gamma(L/2) \prod_{l=0}^{(M-k)/2 -1}\Gamma((L-1)/2) \Gamma(L/2) = \Gamma (L/2)^k \big(2^{2-L}\sqrt{\pi}\; \Gamma(L-1) \big)^{(M-k)/2}
\end{align}
and, recalling that
\begin{align}
\nonumber \prod_{l=k+1}^{(M+k)/2}(-2iy_l)\prod_{j<p}|\lambda(R_{pp}) - \lambda(R_{jj})|=\Delta(\Lambda \cup W),
\end{align}
(\ref{eqn:TOEjpdf1}) becomes
\begin{align}
\nonumber &\tilde{C}_{M,L} \Delta(\Lambda \cup W) \vol (O(M)) \frac{\vol (O(L-M))}{\vol (O(L))} \left( \frac{(2\pi)^{L}}{\Gamma (L+1)} \right)^{M/2} \\
\nonumber &\times \prod_{s=1}^k \left( \frac{L(L-1)}{2\pi}\right)^{1/2} (1-\lambda_s^2)^{L/2-1} \left[ \int_{0}^{1} (1-t^2)^{(L-3)/2}dt \right]^{1/2}\\
\nonumber & \times \prod_{l=k+1}^{(M+k)/2} 2i \frac{L(L-1)}{2\pi} |1- z_l^2|^{L-2}\int_{t=2|y_l|/|1-z_l^2|}^{t=1}(1-t^2)^{(L-3)/2} \; dt\\
\nonumber & \times   \prod_{s=1}^{k}d\lambda_s \prod_{l=k+1}^{(M+k)/2}  dx_ldy_l,
\end{align}
where we have used (\ref{eqn:Bxy1}) and the facts
\begin{align}
\nonumber \prod_{l=0}^{(M-k)/2 -1}\pi^{M-2l-2} = \pi^{-(M-k)(M+k-2)/4},&& \Gamma (z) \Gamma (z+1/2) = \frac{\sqrt{\pi} \; \Gamma (2z)}{2^{2z-1}}.
\end{align}
The result (\ref{eqn:TOEevaljpdf}) now follows by simplifying.

\hfill $\Box$

\begin{remark}
Note the domains of integration in the numerator of the second equality of (\ref{eqn:ccoleval}); this follows since $\bD$ is sub-unitary (that is, all its eigenvalues are $<1$) and so all the $\bR_j$ are sub-unitary, which further implies that the norms of the columns of $u_j$ and $v_j$ are, at most, unity.
\end{remark}

\begin{remark}
\label{rem:TOEana}
The proof of Proposition \ref{prop:Tejpdf} neatly demonstrates the analogy with the earlier ensembles considered in this work. For instance, comparing (\ref{eqn:TOEerfclem}) with the same step in the real Ginibre ensemble (\ref{eqn:GinOE_erfc}), we see that the integral in (\ref{eqn:TOEerfclem}) is the equivalent of the $\erfc$ function for the real truncated ensemble. Likewise, we can see the similarity to (\ref{14.3}) in the real spherical ensemble.
\end{remark}

Recall that the restriction $L\geq M$ in Proposition \ref{prop:Tejpdf} originates from the procedure of integrating over the sub-blocks $\bC$ in (\ref{eqn:TOEmatpdf1}), using (\ref{eqn:Imninteg}), to obtain the distribution of the matrices $\bD$. In \cite{KSZ2010} the restriction on the size of the truncation is side-stepped by using the Schur decomposition of $\bD$ in (\ref{eqn:TOEmatpdf1}) before integrating, and then applying the delta function orthogonality conditions to progressively integrate over $\tilde{\bR}$ and $\bC$. This method involves the introduction of new matrices $\bX_j$ that act on sub-blocks of $\bC$; these matrices $\bX_j$ satisfy a recursive relation that makes the problem tractable. It turns out that when done in this way the Jacobian introduced by the integration over $\tilde{\bR}$ is exactly cancelled by that introduced by integration over $\bC$. The problem is then reduced to a product of integrals over $2\times 2$ blocks, which can each be done explicitly. In performing this calculation the authors of \cite{KSZ2010} were guided by the analogous calculation in the case of truncations of unitary matrices \cite{Z&S2000, FS03, JN03, Reffy2005}. The end result is identical to (\ref{eqn:TOEevaljpdf}), which is why we have dropped the $L\geq M$ condition from the statement of Proposition \ref{prop:Tejpdf}.

With the eigenvalue jpdf specified we can find the probability of all real eigenvalues in the same way that we obtained (\ref{eqn:pNN1}).
\begin{proposition}
With the eigenvalue probability distribution $Q(\Lambda, W)_T$ of (\ref{eqn:TOEevaljpdf}) the probability of obtaining all real eigenvalues is
\begin{align}
\nonumber p_{M,M}&=\frac{2^{M(L-1)+M^2/2}\: C_{M,L}} {\pi^{3M/4} \Gamma(M+1)} \left(L\frac{ \Gamma((L+1)/2)} {\Gamma(L/2)}\right)^{M/2}\\
\label{eqn:TOEpMM} &\times \prod_{j=0}^{M-1} \frac{\Gamma((L+j)/2)^2\Gamma((j+3)/2)}{\Gamma(L+(M+j-1)/2)}.
\end{align}
\end{proposition}

\textit{Proof}: Setting $k=M$ in (\ref{eqn:TOEevaljpdf}) gives
\begin{align}
\nonumber Q(\Lambda ,\emptyset)_T \Big|_{k=M}&=C_{M,L}\left(L\frac{\Gamma((L+1)/2)}{2 \sqrt{\pi} \; \Gamma(L/2)}\right)^{M/2}\\
\label{eqn:TOEk=M} & \times\prod_{j=1}^{M}(1-\lambda_j^2)^{L/2-1} \prod_{j<l}(\lambda_l-\lambda_j).
\end{align}
Letting $A_{M,L}$ stand for the pre-factor which is independent of the $\lambda_j$ in (\ref{eqn:TOEk=M}) and $a:=M(L-1)+M(M-1)/2$ we integrate over the $\lambda_j$
\begin{align}
\nonumber &p_{M,M}=\frac{A_{M,L}}{M!} \int_{-1}^{1}d\lambda_1\cdot\cdot\cdot\int_{-1}^{1} d\lambda_M \prod_{j=1}^M(1-\lambda_j^2)^{L/2-1}\prod_{j<l}|\lambda_l-\lambda_j|\\
\label{eqn:TOEPMM1} &=A_{M,L}\:2^{a} \int_{0}^{1}ds_1\cdot\cdot\cdot\int_{0}^{1}ds_{M}\prod_{j=1}^{M}\Big( s_j(1-s_j)\Big)^{L/2-1}\prod_{j<l}|s_l-s_j|,
\end{align}
where, to obtain the first equality, we removed the ordering on the $\lambda_j$, incurring a factor of $(M!)^{-1}$, and, for the second, we changed variables $s_j=(1+\lambda_j)/2$. We see that (\ref{eqn:TOEPMM1}) is now in the form of a Selberg integral (\ref{def:Selb}). Applying (\ref{eqn:Selbint}) to (\ref{eqn:TOEPMM1}) gives the result.

\hfill $\Box$

The reader will have noted that we can compare the probabilities in (\ref{eqn:TOEpMM}) with simulation, however once we find the generalised partition function and skew-orthogonal polynomials in the following chapters, we will have a formula for the probability of a general number of real eigenvalues, and so we delay further discussion of this point until Chapter \ref{sec:TOEsops}.

\subsection{Generalised partition function}

The eigenvalue jpdf (\ref{eqn:TOEevaljpdf}) is structurally identical to that of the real Ginibre ensemble in (\ref{eqn:GinOEjpdf}), and so we can directly apply the method of Proposition \ref{prop:GinOE_gpf_even} to obtain the generalised partition function for $M$ even.

\begin{proposition}
Let $M$ be even and
\begin{align}
\nonumber \alpha_{j,l}&=\int_{-1}^{1}dx\;u(x)\int_{-1}^{1}dy\;u(y)\: \omega(x) \omega(y)p_{j-1}(x) p_{l-1}(y) \:\sgn(y-x),\\
\label{def:TOEab} \beta_{j,l}&=2i\int_{\mathbb{D}^+}dz\;v(z)\:\omega(z)^2\Big( p_{j-1}(z) p_{l-1}(\bar{z}) -p_{l-1}(z) p_{j-1}(\bar{z}) \Big),
\end{align}
where $\{ p_j(x)\}_{j=1,2,...}$ are monic polynomials of degree $j$, and $\mathbb{D}^+$ is the upper half of the unit disk. The generalised partition function for $k$ real eigenvalues and $M-k$ non-real complex eigenvalues in the real truncated ensemble is
\begin{align}
\label{eqn:TOEgpfe} Z_{k,(M-k)/2}[u,v]_T=C_{M,L}\; [\zeta^{k/2}]\Pf[\zeta\alpha_{j,l}+\beta_{j,l}]_{j,l=1,...,M},
\end{align}
with $C_{M,L}$ from (\ref{def:CML}).
\end{proposition}

Likewise, we can apply the method of Proposition \ref{prop:GinOE_gpf_odd} to find a Pfaffian form for $M$ odd.

\begin{proposition}
Let $M$ be odd, $\alpha_{j,l},\beta_{j,l}$ be as in (\ref{def:TOEab}),
\begin{align}
\label{def:vartheta} \vartheta_j :=\int_{-1}^{1}u(x)\;\omega(x) p_{j-1}(x)dx,
\end{align}
and $C_{M,L}$ as in (\ref{def:CML}). The generalised partition function for the real truncated ensemble with $M$ odd is
\begin{align}
\label{eqn:TOEgpfo} Z^{\odd}_{k,(M-k)/2}[u,v]_T=C_{M,L}\; [\zeta^{(k-1)/2}]\Pf\left[\begin{array}{cc}
\left[\zeta\alpha_{j,l}+\beta_{j,l}\right] & [\vartheta_j]\\
\left[\vartheta_l\right] & 0
\end{array}\right]_{j,l=1,...,M}.
\end{align}
\end{proposition}

The summed-up generalised partition functions are then
\begin{align}
\nonumber Z_M[u,v]_T&:=\sum_{k=0}^M{}^{\sharp}Z_{k,(M-k)/2}[u,v]_T\\
\label{eqn:sumgpfe} &=C_{M,L}\; \Pf[\alpha_{j,l}+\beta_{j,l}]_{j,l=1,...,M},
\end{align}
and
\begin{align}
\nonumber Z_M^{\odd}[u,v]_T&:=\sum_{k=1}^M{}^{\sharp}Z^{\odd}_{k,(M-k)/2}[u,v]_T\\
\label{eqn:sumgpfo} &=C_{M,L}\; \Pf\left[\begin{array}{cc}
\left[\zeta\alpha_{j,l}+\beta_{j,l}\right] & [\vartheta_j]\\
\left[\vartheta_l\right] & 0
\end{array}\right]_{j,l=1,...,M},
\end{align}
where the ${}^{\sharp}$ indicates that the sum is only over those values of $k$ with the same parity as $M$.

A generating function analogous to (\ref{def:GinOE_probsGF}) for the probabilities $p_{M,k}$ of obtaining $k$ real eigenvalues from $\bD$ is
\begin{align}
\label{eqn:TOEpMkse} Z_M(\zeta)_T=C_{M,L}\; \Pf[\zeta\alpha_{j,l}+\beta_{j,l}]_{j,l=1,...,M} \Big|_{u=v=1},
\end{align}
with the probability of all real eigenvalues given by
\begin{align}
\nonumber p_{M,M}=C_{M,L}\; \Pf[\alpha_{j,l}\big|_{u=1}]_{j,l=1,...,M},
\end{align}
from which we can recover (\ref{eqn:TOEpMM}), using the skew-orthogonal polynomials (\ref{eqn:TOEsops}) of the next section. The odd probabilities are found similarly,
\begin{align}
\label{eqn:TOEpMkso} Z_M^{\odd}(\zeta)_T=C_{M,L}\; \Pf\left[\begin{array}{cc}
\left[\zeta\alpha_{j,l}+\beta_{j,l}\right] & [\vartheta_j]\\
\left[\vartheta_l\right] & 0
\end{array}\right]_{j,l=1,...,M} \Bigg|_{u=v=1}.
\end{align}

\subsection{Skew-orthogonal polynomials}
\label{sec:TOEsops}

As with the GOE, real Ginibre and real spherical ensembles, we can use the orthogonal polynomial method to make progress towards the eigenvalue correlation functions. Although this was not historically how the theory progressed, in order to keep the development of the truncated ensemble consistent with the other ensembles in this work, we introduce the polynomials at this point; the derivation of them relies on a procedure involving averages over characteristic polynomials from Chapter \ref{sec:SOEcharpolys}. We will outline its use in the current context in Chapter \ref{sec:TOEkernelts}. As mentioned at the beginning of this chapter, there are formulae for calculating the skew-orthogonal polynomials independently of the correlation kernel however, the application of these methods to the real truncated ensemble has not so far been successful.

\begin{definition}
With $x,y\in\mathbb{R}$ and $z\in\mathbb{C}\backslash\mathbb{R}$ define the inner product for the real truncated ensemble
\begin{align}
\label{def:TOEsoip} \langle p_j,p_l \rangle_T&:=\int_{-1}^1\int_{-1}^1\omega(x)\omega(y)p_j(x)p_l(y)\sgn(y-x)dxdy\\
\nonumber &+2i\int_{\mathbb{D}^{+}}\omega(z)^2\big( p_j(z)p_l(\bar{z})-p_l(z)p_j(\bar{z})\big)dz\\
\nonumber &=\alpha_{j+1,l+1}+\beta_{j+1,l+1}\Big|_{u=v=1},
\end{align}
where $\mathbb{D}^{+}$ is the upper half unit disk.
\end{definition}

We now seek monic polynomials that satisfy analogous conditions to those of (\ref{eqn:GinOE_soprops}), that is, polynomials for which
\begin{align}
\nonumber \langle p_{2j},p_{2l}\rangle_T = \langle p_{2j+1},p_{2l+1}\rangle_T=0 &,&\langle p_{2j},p_{2l+1}\rangle_T=-\langle p_{2l+1},p_{2j}\rangle_T=\delta_{j,l}\:r_j.
\end{align}
The requisite polynomials are similar to those of (\ref{eqn:GinOE_sopolys}) for the real Ginibre ensemble.

\begin{proposition}[\cite{Forrester2010a}]
The polynomials
\begin{align}
\label{eqn:TOEsops} p_{2j}(x)=x^{2j}&,&p_{2j+1}(x)=x^{2j+1}-\frac{2j}{L+2j}x^{2j-1},
\end{align}
are skew-orthogonal with respect to the inner product (\ref{def:TOEsoip}), with
\begin{align}
\label{eqn:TOEsopsr} \langle p_{2j},p_{2j+1}\rangle_T=r_j=\frac{L!(2j)!}{(L+2j)!}.
\end{align}
\end{proposition}

These polynomials were first presented in \cite{Forrester2010a}, and, as mentioned above, the derivation relies on the method of averaging over the characteristic polynomial of sub-blocks $\bD_{(M-2) \times(M-2)}$ of $\bR_{(N-2)\times(N-2)}$. This method was used in \cite{KSZ2010} as a way to calculate the kernel element $D(\mu,\eta)_T$ (to be introduced in (\ref{def:TOESDI1})), and the result of which was used by Forrester to extract the skew-orthogonal polynomials (\ref{eqn:TOEsops}) \cite{Forrester2010a}. As mentioned at the beginning of this chapter, it is not known how to derive the polynomials directly without first evaluating the correlation kernel; formulae such as those in \cite{ForNagRai2006} and \cite{AkeKiePhi2010} lead to integrals that appear intractable. In Chapter \ref{sec:FW} we make some suggestions for how the skew-orthogonal polynomials might be found using the classical Jacobi polynomials.

With the polynomials (\ref{eqn:TOEsops}) we can evaluate the probabilities $p_{M,k}$ implied by (\ref{eqn:TOEpMkse}) and (\ref{eqn:TOEpMkso}) for specific values of $L$. In Appendix \ref{app:TOEpMks} we have tabulated the probabilities for $M=2,...,6$ with $L=1,2,3,8$, and compared them to simulations. Note that as $L$ increases (giving the more weakly orthogonal regimes), the $p_{M,k}$ approach the probabilities $p_{N,k}$ of the real Ginibre ensemble in Table \ref{tab:pnkxact_sim} of Appendix \ref{app:GinOE_kernel_elts} --- for example, compare Table \ref{tab:pnkxact_sim} to Table \ref{tab:TOEpnkL8}.

\subsection{Correlation functions}

\subsubsection{M even}

To again highlight the analogy with the correlations of the real Ginibre and real spherical ensembles, we use similar notation here for the correlation kernel, trusting that no confusion is likely.

\begin{definition}
\label{def:TOESDI}
Let $p_0,p_1,...$ be the skew-orthogonal polynomials (\ref{eqn:TOEsops}) and $r_0,r_1,...$ the normalisations (\ref{eqn:TOEsopsr}). With $\epsilon(\mu,\eta)$ from Definition \ref{def:GinOE_kernel}, let
\begin{align}
\label{def:TOESDI1} \begin{split}
S(\mu,\eta)_T&=2\sum_{j=0}^{\frac{M}{2}-1}\frac{1}{r_j}\Bigl[q_{2j}(\mu)\tau_{2j+1}(\eta)-q_{2j+1}(\mu)\tau_{2j}(\eta)\Bigr],\\
D(\mu,\eta)_T&=2\sum_{j=0}^{\frac{M}{2}-1}\frac{1}{r_j}\Bigl[q_{2j}(\mu)q_{2j+1}(\eta)-q_{2j+1}(\mu)q_{2j}(\eta)\Bigr],\\
\tilde{I}(\mu,\eta)_T&=2\sum_{j=0}^{\frac{M}{2}-1}\frac{1}{r_j}\Bigl[\tau_{2j}(\mu)\tau_{2j+1}(\eta)-\tau_{2j+1}(\mu)\tau_{2j}(\eta)\Bigr]+\epsilon(\mu,\eta),
\end{split}
\end{align}
where
\begin{align}
\nonumber q_j(\mu) &= \omega(\mu)\; p_j(\mu),\\
\label{def:TOEtau} \tau_j(\mu) &= 
\left\{ 
\begin{array}{ll}
-\frac{1}{2}\int_{-1}^{1}\mathrm{sgn}(\mu-z)\hspace{3pt}q_j(z)\hspace{3pt}dz, & \mu\in \mathbb{R},\\
iq_j(\bar{\mu}),  & \mu\in \mathbb{D}^+.
\end{array}
\right.
\end{align}
Define
\begin{align}
\label{def:TOEkernel} \bK(\mu,\eta)_T=\left[
\begin{array}{cc}
S(\mu,\eta)_T & - D(\mu,\eta)_T\\
\tilde{I}(\mu,\eta)_T & S(\eta,\mu)_T
\end{array}
\right].
\end{align}
\end{definition}

Note that the $S,D,\tilde{I}$ of (\ref{def:TOESDI1}) obey the same inter-relationships as in the real Ginibre case, contained in Lemma \ref{lem:Gin_s=d=i}, similar to those for the real spherical ensemble, Lemma \ref{lem:S_s=d=i}.

Since (\ref{eqn:TOEevaljpdf}) and (\ref{eqn:TOEgpfe}) are structurally identical to their real Ginibre counterparts (\ref{eqn:GinOEjpdf}) and (\ref{eqn:GinOE_gpf_even}), with $\omega(\mu)$ replacing $e^{-\mu^2/2}\sqrt{\erfc(\sqrt{2}|\mathrm{Im}(\mu)|)}$, we can apply the methods of Chapter \ref{sec:Gincorrlnse} to obtain the correlation functions, the details of which we omit.

\begin{proposition}
\label{prop:TOEcorrelne}

Let $M$ be even. With $\bK(\mu,\eta)_T$ from (\ref{def:TOEkernel}), the real truncated eigenvalue correlation function for $n_1$ real and $n_2$ non-real, complex conjugate pairs is
{\small
\begin{align}
\nonumber &\rho_{(n_1,n_2)}(x_1,...,x_{n_1},z_1,...,z_{n_2})_T =\qdet\left[\begin{array}{cc}
\bK_{rr}(x_i,x_j)_T & \bK_{rc}(x_i,z_m)_T\\
\bK_{cr}(z_l,x_j)_T & \bK_{cc}(z_l,z_m)_T
\end{array}\right]_{i,j=1,...,n_1 \atop l,m=1,...,n_2}\\
\nonumber &=\Pf\left(\left[\begin{array}{cc}
\bK_{rr}(x_i,x_j)_T & \bK_{rc}(x_i,z_m)_T\\
\bK_{cr}(z_l,x_j)_T & \bK_{cc}(z_l,z_m)_T
\end{array}\right]\bZ_{2(n_1+n_2)}^{-1}\right)_{i,j=1,...,n_1 \atop l,m=1,...,n_2}, x_i\in \mathbb{R}, z_i \in \mathbb{D}^+,
\end{align}
}where $\mathbb{D}^+$ is the upper half of the unit disk.
\end{proposition}

\subsubsection{M odd}
\label{sec:TOEcorrelno}

The $M$ odd correlations use an analogous kernel to that in Definition \ref{def:GinOE_kernel_odd}. 

\begin{definition}
Let $p_0,p_1,...$ be the skew-orthogonal polynomials (\ref{eqn:TOEsops}), $r_0,r_1,...$ the normalisations (\ref{eqn:TOEsopsr}) and $\overline{\vartheta}_j:=\vartheta_j\big|_{u=1}$ from (\ref{def:vartheta}). With $\epsilon(\mu,\eta)$ from Definition \ref{def:GinOE_kernel}, let
\begin{align}
\nonumber S^{\odd}(\mu,\eta)_T&=2\sum_{j=0}^{\frac{M-1}{2}-1}\frac{1}{r_j}\Bigl[\hat{q}_{2j}(\mu)\hat{\tau}_{2j+1}(\eta)-\hat{q}_{2j+1}(\mu)\hat{\tau}_{2j}(\eta)\Bigr]+\kappa(\mu,\eta),\\
\nonumber D^{\odd}(\mu,\eta)_T&=2\sum_{j=0}^{\frac{M-1}{2}-1}\frac{1}{r_j}\Bigl[\hat{q}_{2j}(\mu)\hat{q}_{2j+1}(\eta)-\hat{q}_{2j+1}(\mu)\hat{q}_{2j}(\eta)\Bigr],\\
\nonumber \tilde{I}^{\odd}(\mu,\eta)_T&=2\sum_{j=0}^{\frac{M-1}{2}-1}\frac{1}{r_j}\Bigl[\hat{\tau}_{2j}(\mu)\hat{\tau}_{2j+1}(\eta)-\hat{\tau}_{2j+1}(\mu)\hat{\tau}_{2j}(\eta)\Bigr]+\epsilon(\mu,\eta)+\theta(\mu,\eta),
\end{align}
where
\begin{align}
\nonumber \hat{p}_j(\mu)&=p_j(\mu)-\frac{\overline{\vartheta}_{j+1}}{\overline{\vartheta}_N}p_{N-1}(\mu),\\
\nonumber \hat{q}_j(\mu) &= \omega(\mu)\;\hat{p}_j(\mu),\\
\nonumber \hat{\tau}_j(\mu) &= 
\left\{ 
\begin{array}{ll}
-\frac{1}{2}\int_{-1}^{1}\mathrm{sgn}(\mu-z)\hspace{3pt}\hat{q}_j(z)\hspace{3pt}dz, & \mu\in \mathbb{R},\\
i\hat{q}_j(\bar{\mu}),  & \mu\in \mathbb{D}^+,
\end{array}
\right.\\
\nonumber \kappa(\mu,\eta) &= 
\left\{ 
\begin{array}{lll}
q_{N-1}(\mu)/\overline{\vartheta}_N, & \eta\in \mathbb{R},\\
0,  & \mathrm{otherwise},\\
\end{array}
\right.\\
\nonumber \theta(\mu,\eta)&=
\big(\chi_{(\eta\in\mathbb{R})}\tau_{N-1}(\mu)- \chi_{(\mu\in\mathbb{R})} \tau_{N-1}(\eta)\big)/  \overline{\vartheta}_N,
\end{align}
and $q_j(\mu),\tau_j(\mu)$ are from Definition \ref{def:TOESDI}, with the indicator function $\chi_{(A)}=1$ for $A$ true and zero for $A$ false.
Then, let
\begin{align}
\label{def:TOEkernelo} \bK_{\odd}(\mu,\eta)_T=\left[
\begin{array}{cc}
S^{\odd}(\mu,\eta)_T & -D^{\odd}(\mu,\eta)_T\\
\tilde{I}^{\odd}(\mu,\eta)_T & S^{\odd}(\eta,\mu)_T
\end{array}
\right].
\end{align}
\end{definition}

As for the even case, we see that (\ref{eqn:TOEgpfo}) is identical in structure to its real Ginibre counterpart (\ref{eqn:GinOE_gpf_odd}), and so we were able to immediately write down the correlation functions. However, as for the odd case of the real spherical ensemble in Chapter \ref{sec:Scorrelns}, we cannot apply the `odd-from-even' technique of Chapter \ref{sec:Gin_oddfromeven}. (Recall that for the GOE (Chapter \ref{sec:odd_from_even}) and the real Ginibre ensemble (Chapter \ref{sec:Gin_oddfromeven}), we were able to generate the odd correlations from the known even cases by removing one of the eigenvalues off to infinity, effectively leaving a system of $k-1$ real eigenvalues, and $N-1$ total eigenvalues. Of course, for the GOE, $N=k$.) In the real spherical ensemble, we were unable to use this technique because we had applied the fractional linear transformation (\ref{7'}), since the natural co-ordinate set for analysis was the unit disk.  Hence, all the transformed eigenvalues (from the upper half plane) were constrained to have $|z|\leq 1$.

For the real truncated ensemble, as discussed above Remark \ref{rem:evals<1}, all the eigenvalues are necessarily contained inside the unit disk by the nature of truncations of orthogonal matrices, and so the removal of one eigenvalue off to infinity is meaningless. Of course, we can still use the functional differentiation method of Chapter \ref{sec:Gin_odd_fdiff}, which again highlights the utility of the method of Propositions \ref{prop:pf_integ_op} and \ref{prop:pf_integ_op_odd} (and their real Ginibre counterparts Propositions \ref{prop:4x4_fred}, \ref{prop:GinOE_fred_op} and \ref{prop:GinOE_fred_op_odd}). As for the even case in the previous section, we omit the details.

\begin{proposition}
Let $M$ be odd. With $\bK_{\odd}(\mu,\eta)_T$ from (\ref{def:TOEkernelo}), the real truncated eigenvalue correlation function for $n_1$ real and $n_2$ non-real, complex conjugate pairs is
{\small
\begin{align}
\nonumber &\rho_{(n_1,n_2)}(x_1,...,x_{n_1},z_1,...,z_{n_2})_T=\qdet\left[\begin{array}{cc}
\bK_{rr}^{\odd}(x_i,x_j)_T & \bK_{rc}^{\odd}(x_i,z_m)_T\\
\bK_{cr}^{\odd}(z_l,x_j)_T & \bK_{cc}^{\odd}(z_l,z_m)_T
\end{array}\right]_{i,j=1,...,n_1 \atop l,m=1,...,n_2}\\
\nonumber &=\Pf\left(\left[\begin{array}{cc}
\bK_{rr}^{\odd}(x_i,x_j)_T & \bK_{rc}^{\odd}(x_i,z_m)_T\\
\bK_{cr}^{\odd}(z_l,x_j)_T & \bK_{cc}^{\odd}(z_l,z_m)_T
\end{array}\right]\bZ_{2(n_1+n_2)}^{-1}\right)_{i,j=1,...,n_1 \atop l,m=1,...,n_2}, x_i\in \mathbb{R},z_i \in \mathbb{D}^+.
\end{align}
}
\end{proposition}

\subsection{Correlation kernel elements}
\label{sec:TOEkernelts}

In the previous ensembles considered in this work, we have successfully used the method of orthogonal polynomials to find the correlation kernel elements. However, as already mentioned, in \cite{Forrester2010a} the skew-orthogonal polynomials were found by working backwards from the earlier work of \cite{KSZ2010}, which already identified the correlation kernel using averages over characteristic polynomials. We will follow these developments here, first looking at the evaluation of the characteristic polynomial, then using the known inter-relationships between the kernel elements (Lemma \ref{lem:Gin_s=d=i}) to find most of the kernel. The remaining elements can be obtained from the same inter-relationships after performing a calculation analogous to one in \cite{FN08} pertaining to the partially symmetric real Ginibre ensemble of Chapter \ref{sec:tG}.

Since $\det(t-\bX_{n\times n})=\prod_{j=1}^n(t-x_j)$ where $\{x_j\}$ are the eigenvalues of $\bX_{n\times n}$, we see from Proposition \ref{prop:SOEcharpoly} that \cite{Forrester2010a}
\begin{align}
\nonumber &(\mu-\eta)\big\langle \det(\mu-\bD_{(M-2)\times (M-2)})(\eta-\bD_{(M-2)\times (M-2)})\big\rangle\\
\nonumber&=\sum_{j=0}^{M/2-1}\frac{1}{r_j} \Big( p_{2j+1}(\mu)p_{2j}(\eta) -p_{2j}(\mu)p_{2j+1} (\eta)\Big)\\
\nonumber &=-(2\:\omega(\mu)\omega(\eta))^{-1}D(\mu,\eta)_T.
\end{align}
Combining this with the result of \cite[Eq. (8)]{KSZ2010} we have
\begin{align}
\label{eqn:SOEDsum} D(\mu,\eta)_T=2\:\omega(\mu)\omega(\eta)(\eta-\mu)\sum_{j=0}^{M-2}\frac{(L+j)!}{L!j!}(\mu \eta)^j.
\end{align}
Note from Definition \ref{def:TOESDI} that (\ref{eqn:SOEDsum}) holds for all four combinations of real and complex variables $\mu,\eta$, and further, we see that it is parity independent (so it holds for $M$ odd as well). Using (\ref{eqn:Gin_s=d=i}) we can obtain some of the other kernel elements via the relations
\begin{align}
\nonumber S_{cc}(z_1,z_2)_T&= iD_{cc}(z_1,\bar{z}_2)_T,\\
\nonumber \tilde{I}_{cc}(z_1,z_2)_T&= iS_{cc}(\bar{z}_1,z_2)_T=-D_{cc} (\bar{z}_1,\bar{z}_2)_T,\\
\label{eqn:TOErels1} S_{rc}(x,z)_T&= iD_{cc}(x,\bar{z})_T.
\end{align}
The remaining kernel elements are related by the equations 
\begin{align}
\nonumber S_{cr}(z,x)_T&=S_{rr}(z,x)_T,\\
\nonumber \tilde{I}_{rr}(x,y)_T&=-\int_{x}^{y}S_{rr}(t,y)_Tdt+\frac{1}{2}\sgn(x-y),\\
\label{eqn:TOErels2} \tilde{I}_{cr}(z,x)_T&=-\tilde{I}_{rc}(x,z)_T=iS_{cr}(z,x)_T.
\end{align}
Accordingly, if we find an expression for $S_{rr}(x,y)_T$ then we will have completely specified the kernel. Using techniques from \cite{FN08} the calculation of this quantity is carried out in \cite[Section 3.2]{Forrester2010a}. First note that, with the skew-orthogonal polynomials (\ref{eqn:TOEsops}), and $q_j(x), \tau_j(x)$ from Definition \ref{def:TOESDI}, we have the facts
\begin{align}
\nonumber &\sum_{j=0}^{M/2-1} \frac{q_{2j+1}(x) \tau_{2j}(y)} {r_j} = \frac{x^{M-1} \tau_{M-2}(y)} {r_{M/2-1}}\\
\nonumber & - \sum_{j=0}^{M/2-2} \frac{(L+2j+1)!} {L! (2j+1)!} x^{2j+1} \left( \tau_{2j+2} (y) -\frac{2j+1} {L+2j+1} \tau_{2j}(y) \right),
\end{align}
and
\begin{align}
\nonumber &p_{2j+1}(x)=-\frac{(1-x^2)^{1-L/2}}{L+2j}\;\frac{d}{dx}\Big( x^{2j} (1-x^2)^{L/2}\Big),\\
\nonumber &p_{2j+2}(x)-\frac{2j+1}{L+2j+1}p_{2j}(x)=-\frac{(1-x^2)^{1-L/2}}{L+2j+1}\frac{d}{dx}\Big( x^{2j+1}(1-x^2)^{L/2}\Big),
\end{align}
and, from the definition of $\tau_j(\mu)$ (\ref{def:TOEtau}), the corollaries
\begin{align}
\nonumber &\tau_{2j+1}(x)=\frac{ (1-x^2)^{L/2}}{L+2j}\left(\frac{L}{2 \sqrt{\pi}} \frac{ \Gamma((L+1)/2)}{\Gamma(L/2)}\right)^{1/2} x^{2j},\\
\label{eqn:TOEtaus} &\tau_{2j+2}(x)-\frac{2j+1}{L+2j+1}\tau_{2j}(x)= \frac{ (1-x^2)^{L/2}} {L+2j+1} \left( \frac{L} {2 \sqrt{\pi}} \frac{\Gamma ((L+1)/2)} {\Gamma(L/2)} \right)^{1/2} x^{2j+1},
\end{align}
for $x\in\mathbb{R}$. Substitution of these into the definition of $S_{rr}(x,y)_T$ from (\ref{def:TOESDI1}) yields
\begin{align}
\nonumber &S_{rr}(x,y)_T=-\frac{2\:\omega(x)}{r_{M/2-1}}x^{M-1}\tau_{M-2}(y)\\
\label{eqn:TOEsumS} &+\frac{\Gamma((L+1)/2)}{\sqrt{\pi} \;\Gamma(L/2)}(1-x^2)^{L/2-1}(1-y^2)^{L/2}\sum_{j=0}^{M-2}\frac{(L+j-1)!}{(L-1)!j!}x^j y^j.
\end{align}
So by (\ref{eqn:TOErels2}) we have now specified all kernel elements.

From the preceding chapters in this work we expect that the density of real eigenvalues will be given by $S_{rr}(x,x)_T$, which is the $n_1=1, n_2=0$ case of Proposition \ref{prop:TOEcorrelne}. Using the expression for $S_{rr}(x,x)_T$ given by (\ref{eqn:TOEsumS}), we apply an identity for the regularised Beta function \cite{KS2009},
\begin{align}
\nonumber I_{s}(a,b)&:=\int_{0}^{s}t^{a-1}(1-t)^{b-1}dt/B(a,b)\\
\label{eqn:BetaId} &\;=1-(1-s)^{b}\sum_{j=0}^{a-1}{b-1+j \choose j}s^j,
\end{align}
to write the density as
\begin{align}
\nonumber &\rho_{(1)}^r(x)_T=S_{rr}(x,x)_T=-\frac{2\:\omega(x)}{r_{M/2-1}}x^{M-1}\tau_{M-2}(x)\\
\label{eqn:TOErdens} &+\frac{\Gamma((L+1)/2)}{\sqrt{\pi} \;\Gamma(L/2)}\frac{1-I_{x^2}(M-1,L)}{1-x^2},
\end{align}
which is equivalent to that in \cite{KSZ2010}.

The complex-complex density is more straightforward; from (\ref{eqn:SOEDsum}), using (\ref{eqn:TOErels1}), we obtain
\begin{align}
\nonumber S_{cc}(w,z)_T = 2i \; \omega(w)\omega(\bar{z}) (\bar{z}-w) \sum_{j=0}^{M-2} \frac{(L+j)!}{L! j!} (w\bar{z})^j,
\end{align}
from which we see that the complex density, assuming $L \neq 1$, is
\begin{align}
\nonumber \rho_{(1)}^c (z)_T &=S_{cc}(z,z)_T=4\; \mathrm{Im}(z) \; (\omega(z))^2 \mathrm{Im}(z) \sum_{j=0}^{M-2} \frac{(L+j)!}{L! j!} |z|^{2j}\\
\nonumber & =\frac{2 \: \mathrm{Im}(z) L(L-1)}  {\pi} |1-z^2|^{L-2} \int_{2|\mathrm{Im}(z)|/|1-z^2|}^{1}(1-t^2)^{(L-3)/2}dt \\
\nonumber &\times \sum_{j=0}^{M-2}\frac{(L+j)!}{L! j!} |z|^{2j}\\
\nonumber & = \frac{2\: \mathrm{Im}(z) L(L-1)}  {\pi} \frac{|1-z^2|^{L-2}} {(1-|z|^2)^{L+1}} \int_{2|\mathrm{Im}(z)|/|1-z^2|}^{1}(1-t^2)^{(L-3)/2}dt \\
\label{eqn:TOEcdens} &\times  \big(1-I_{|z|^2}(M-1,L+1)\big),
\end{align}
where we have used (\ref{eqn:BetaId}) to obtain the last equality.

In the case $L= 1$ we require the second definition in (\ref{def:truncw}), so (\ref{eqn:TOErdens}) and (\ref{eqn:TOEcdens}) respectively become
\begin{align}
\nonumber \rho_{(1)}^r(x)_T \Big|_{L=1}&=\frac{\sqrt{2} (M-1) x^{M-1} \tau_{M-2}(x)} {\sqrt{\pi} (1-x^2)^{1/2}} + \frac{(1-x^{2M-2})} {\pi (1-x^2)},\\
\nonumber \rho_{(1)}^c (z)_T \Big|_{L=1}&=\frac{2 \: \mathrm{Im}(z)}{\pi |1-z^2| (1-|z|^2)^2} \left( 1-M|z|^{2M-2}+ (M-1)|z|^{2M} \right),
\end{align}
where the incomplete Beta functions are integrated using elementary techniques. Of course, we could also have found this directly from (\ref{def:TOESDI1}) using the $L=1$ cases of (\ref{eqn:TOEtaus}),
\begin{align}
\nonumber &\tau_{2j+1}(x)\Big|_{L=1}= \frac{(1-x^2)^{1/2}} {\sqrt{2 \pi} (2j+1)}\; x^{2j},\\
\nonumber &\tau_{2j+2}(x)\Big|_{L=1} - \frac{2j+1} {2j+2} \tau_{2j}(x)\Big|_{L=1}= \frac{(1-x^2)^{1/2}}{\sqrt{2 \pi} (2j+2)}\; x^{2j+1}.
\end{align}

\subsubsection{Large $M$ and $L$ limits}

As discussed in the introduction to this chapter, there are three limiting regimes of interest: $\alpha\to 1$, $0<\alpha<1$ and $\alpha\to 0$, where we recall $\alpha =M/N$. We now investigate the real and complex densities in each of these cases.

\noindent\textbf{\underline{Real density}}

We see from (\ref{eqn:TOErels1}) and (\ref{eqn:TOErels2}) that if we have expressions for the limits of $D(\mu,\eta)_T$ and $S_{rr}(x,y)_T$ then we can obtain all the limiting kernel elements by those relations. To that end, note that we can use the classical identity, valid for $|t|<1$,
\begin{align}
\label{eqn:1on1-t} \frac{1}{(1-t)^n}=\sum_{j=0}^{\infty}{n-1 +j \choose j}t^j,
\end{align}
to obtain the large $M$, fixed $L$ limit of (\ref{eqn:SOEDsum})
\begin{align}
\nonumber \lim_{M\to\infty \atop L \; \mathrm{fixed}} D(\mu,\eta)_T= \frac{2\:\omega(\mu)\omega(\eta)\:(\eta-\mu)}{(1-\mu\eta)^{L+1}},
\end{align}
for all combinations of real and complex $\mu$ and $\eta$. For $\mu,\eta\in \mathbb{R}$ the large $M$ limit of (\ref{eqn:TOEsumS}) is
\begin{align}
\label{eqn:TOElimS} \lim_{M\to\infty \atop L \; \mathrm{fixed}} S_{rr}(x,y)_T=\frac{ \Gamma((L+1)/2)} {\sqrt{\pi} \;\Gamma(L/2)}\frac{(1-x^2)^{L/2-1}(1-y^2)^{L/2}}{(1-xy)^L}.
\end{align}
A particularly interesting case of this limit was identified in \cite{Forrester2010a}, where, with $L=1$, the correlation kernel is identical to that encountered when looking at the correlations between the zeroes of the Kac polynomials. In Appendix \ref{app:TOElims} we list these kernels. By letting $x=y$ in (\ref{eqn:TOElimS}) we have the real density in the strongly orthogonal regime, which is the $\alpha\to 1$ limit of (\ref{eqn:TOElimdensL}) below. (It turns out that the limiting densities for $\alpha\to 1$ regime have the same form as those in the $0< \alpha<1$ regime, and so we state them together in Proposition \ref{prop:TOElargedens}.)

Since the Beta function is a key factor in the densities, for the remainder of the chapter we will make repeated use of its asymptotic behaviour
\begin{align}
\label{eqn:betafixed2} B(x,y)\sim \sqrt{2\pi} \frac{x^{x-1/2} y^{y-1/2}}{(x+y)^{x+y-1/2}},\quad x,y \to \infty,
\end{align}
and
\begin{align}
\label{eqn:betafixed} B(x,y) \sim \frac{\Gamma (y)} {x^y},\quad x \to \infty, y \mbox { fixed}.
\end{align}

\begin{proposition}[\cite{KSZ2010}]
\label{prop:TOElargedens}
The limiting real density in the strongly orthogonal regime $(\alpha\to 1)$ is given by
\begin{align}
\label{eqn:TOElimdensL} \rho_{(1)}^r(x)_T \mathop{\sim}\limits_{N \to \infty} \frac{ \Gamma((L+1)/2)}{\sqrt{\pi} \;\Gamma(L/2)} \frac{1}{1-x^2}& ,& -1 <x< 1,
\end{align}
and in the weakly orthogonal regimes $(0<\alpha<1$ or $\alpha\to 1)$ is given by
\begin{align}
\label{eqn:TOElimdensLw} \rho_{(1)}^r(x)_T \mathop{\sim}\limits_{N \to \infty} \sqrt{\frac{L}{2\pi}} \frac{1}{1-x^2}& ,& -\sqrt{\alpha} <x< \sqrt{\alpha}.
\end{align}
\end{proposition}

\textit{Proof}: We will assume that the first term in (\ref{eqn:TOErdens}) tends to zero for the time being, and concentrate on the second term. We will then show that the first term does indeed tend to zero as assumed.

The result (\ref{eqn:TOElimdensL}) in the case $\alpha\to 1$ follows from (\ref{eqn:TOElimS}) with $x=y$, or by substituting (\ref{eqn:1on1-t}) into (\ref{eqn:TOErdens}) (recalling (\ref{eqn:BetaId})). For (\ref{eqn:TOElimdensLw}), that is with $\alpha <1$, we must analyse the behaviour of $I_{x^2} (M-1,L)=\int_{0}^{x^2} t^{M-2} (1-t)^{L-1} dt/B(M-1,L)$, and we use the same approach that led to (\ref{def:uGa}), which entails rewriting the integrand thusly
\begin{align}
\nonumber t^{M-2} (1-t)^{L-1} = \exp\Big( (\alpha N -2)\log t +((1-\alpha)N-1)\log (1-t) \Big).
\end{align}
Note that in the large $N$ limit, the integrand will be dominated by the maximum of $(\alpha N -2)\log t +((1-\alpha)N-1)\log (1-t)$, which, by differentiating, we find to be
\begin{align}
\nonumber t_{\max}=(\alpha N-2)/(N-3) \mathop{\sim}\limits_{N \to \infty} \alpha.
\end{align}
Also recall that $B(a,b)$ is the $s\to 1$ limit of $\int_{0}^{s} t^{a-1} (1-t)^{b-1} dt$, and so
\begin{align}
\label{eqn:Ix^2} I_{x^2} (M-1,L) \mathop{\sim}\limits_{N \to \infty} \left\{\begin{array}{cc}
1, & x^2 >\alpha,\\
0, & x^2 < \alpha,
\end{array} \right.
\end{align}
which gives us the bounds in (\ref{eqn:TOElimdensL}).

It remains to show that the first term in (\ref{eqn:TOErdens}) tends to zero for large $N$. Writing it out we have
\begin{align}
\label{eqn:TOESrr1} \frac{L}{2\pi} \frac{ \Gamma((L+1)/2)}{\sqrt{\pi} \;\Gamma(L/2)} \frac{(1-x^2)^{L/2-1} x^{M-1}}{B(L,M-2)} \int_{-1}^{1} \sgn (x-z) (1-z^2)^{L/2-1} p_{M-2}(z) dz.
\end{align}
Note that 
\begin{align}
\nonumber &\int_{-1}^{1} \sgn (x-z) (1-z^2)^{L/2-1} p_{M-2}(z) dz\\
\nonumber & = \frac{1}{2} \left[ \int_{0}^{x^2} z^{(M-3)/2} (1-z)^{L/2-1} dz - \int_{x^2}^{1} z^{(M-3)/2} (1-z)^{L/2-1} dz\right]\\
\nonumber & = B (x^2,(M-1)/2,L/2) - B(M-1/2,L/2)/2\\
\nonumber & = B((M-1)/2,L/2)\Big(I_{x^2} ((M-1)/2, L/2) -1/2 \Big).
\end{align}
From (\ref{eqn:Ix^2}) we see that this tends to $\pm B((M-1)/2,L/2)/2$ for large $N$ (that is, as either or both of $L$ and $M$ tend to $\infty$); without loss of generality we can take the positive part since if $a$ tends to zero, then so does $-a$. Now assume that $L$ is fixed. Using the asymptotic behaviour of the Beta function (\ref{eqn:betafixed}) we see the large $M$ behaviour of (\ref{eqn:TOESrr1}) is
\begin{align}
\label{eqn:TOESrr2} A_L\; M^{2L} x^{M-1},
\end{align} 
where $A_L$ is the remaining factors independent of $M$. Since $x<1$, (\ref{eqn:TOESrr2}) tends to zero for large $M$.

In the case that $L, M\to \infty$, we need a more involved approach. Using (\ref{eqn:betafixed}) we can write
\begin{align}
\label{eqn:bigLBeta} \frac{\Gamma((L+1)/2)}{\sqrt{\pi} \;\Gamma(L/2)} = \frac{L-1}{2\pi} \;B\left( \frac{1} {2},\frac{L-1} {2} \right) \mathop{\sim}\limits_{L \to \infty} \sqrt{\frac{L}{2\pi}},
\end{align}
and using (\ref{eqn:betafixed2}) we have
\begin{align}
\nonumber \frac{1} {B(L,M-2)} \sim \frac{1}{\sqrt{2\pi}} \frac{(L+M-2)^{L+M-5/2}}{L^{L-1/2}\; (M-2)^{M-5/2}}.
\end{align}
Combining the preceding with the proportionality of $M$ and $L$ to $N$ (according to (\ref{def:alphaML})), we obtain
\begin{align}
\nonumber \frac{2\:\omega(x)}{r_{M/2-1}}x^{M-1}\tau_{M-2}(x) \sim \frac{(1-x^2)^{(1-\alpha)N /2} x^{\alpha N}} {\sqrt{2\pi}} \frac{\Big( (1-\alpha) N+\alpha N\Big)^{N/2}} {\Big( (1-\alpha)N \Big)^{(1-\alpha)N/2} (\alpha N)^{\alpha N/2}}.
\end{align}

By rearranging, we find that for large $N$
\begin{align}
\nonumber \frac{2\:\omega(x)}{r_{M/2-1}}x^{M-1}\tau_{M-2}(x) \to 0 \Leftrightarrow \left( \frac{1-x^2} {1-\alpha} \right)^{(1-\alpha) N/2} \left( \frac{x^2} {\alpha} \right)^{\alpha N/2} \to 0,
\end{align}
or equivalently, by exponentiating, that
\begin{align}
\label{eqn:srr1termz} (1-\alpha) \log\left( \frac{1-x^2} {1-\alpha} \right) + \alpha \log \left( \frac{x^2} {\alpha} \right) < 0.
\end{align}
By differentiating the LHS of (\ref{eqn:srr1termz}) with respect to $\alpha$ and $x$, we find that it is strictly decreasing for $x,\alpha\in (0,1)$ and $\alpha\neq x^2$, and since the LHS is zero for $\alpha = x^2$ (\ref{eqn:srr1termz}) is satisfied for all $x$ and $\alpha$ except for where they are equal. We can see that the first term in (\ref{eqn:TOErdens}) goes to zero for $\alpha = x^2$ when combining the asymptotic Beta function behaviours together under that assumption.

\hfill $\Box$

We see from (\ref{eqn:SOElimdensr}) that (\ref{eqn:TOElimdensL}) is the anti-spherical analogue of the real eigenvalue density for the real spherical ensemble (recalling that the variables in the former case had been transformed according to (\ref{14.2})). Indeed, in the case $L=1$ (in the $\alpha\to 1$ strongly orthogonal regime), (\ref{eqn:TOElimdensL}) obviously reduces to
\begin{align}
\label{eqn:TlimdensL1} \lim_{M\to\infty}\rho_{(1)}^r(x)_T\Big|_{L=1}=\frac{1}{\pi(1-x^2)},
\end{align}
indicating (via the $artanh$ function) that the real eigenvalues are uniformly distributed on what might be called a `great anti-circle' on the anti-sphere. This formula appeared in \cite{Forrester2010a} where it was identified with the asymptotic density of real zeroes of Kac random polynomials (from \cite{BDi1997}).

With (\ref{def:alphaML}), we see that (\ref{eqn:TOElimdensLw}) is
\begin{align}
\label{eqn:TOELlimdens} \rho_{(1)}^r(x)_T \mathop{\sim}\limits_{N \to \infty} \sqrt{\frac{(1-\alpha)N} {2\pi}} \frac{1}{1-x^2}&,&-\sqrt{\alpha}<x<\sqrt{\alpha}.
\end{align}
Comparing (\ref{eqn:TOELlimdens}) to (\ref{eqn:Grlimdens}) we can see that in the (weakly orthogonal) regimes $\alpha< 1$ we have a $\sqrt{N}$ prefactor, which matches that in the Ginibre case.

\noindent\textbf{\underline{Complex density}}

Note that we could have found (\ref{eqn:TlimdensL1}) from (\ref{eqn:TlimL1}) as the 1-point real--real correlation. Likewise, we can read off the 1-point complex--complex correlation for $L=1$,
\begin{align}
\label{eqn:Tlimdensnrl1} \lim_{M\to \infty} \rho_1^{c}(z) \Big|_{L=1} =\frac{2y}{\pi (1-|z|^2)^2|1-z^2|}.
\end{align}
With only marginally more work, we can establish the complex density for arbitrary $L$. As for the real density, the complex densities in the three regimes can all be stated together.

\begin{proposition}[\cite{KSZ2010}]
The limiting complex density in the strongly orthogonal regime $(\alpha\to 1)$ is
\begin{align}
\nonumber \rho_{(1)}^c(z)_T \mathop{\sim}\limits_{N \to \infty} & \frac{2\: \mathrm{Im}(z) L(L-1)}  {\pi} \frac{|1-z^2|^{L-2}} {(1-|z|^2)^{L+1}}\\
\label{eqn:TOElimdensc} &\times \int_{2|\mathrm{Im}(z)|/|1-z^2|}^{1}(1-t^2)^{(L-3)/2}dt,& -1 <z< 1,
\end{align}
and in the weakly  orthogonal regimes $(0<\alpha<1$ or $\alpha\to 1)$ is given by
\begin{align}
\label{eqn:TOEldc} \rho_{(1)}^c (z)_T \sim \frac{(1- \alpha)N} {\pi} \frac{1}{(1- |z|^2)^2}&,& -\sqrt{\alpha} <z< \sqrt{\alpha}.
\end{align}
\end{proposition}

\textit{Proof}: To obtain (\ref{eqn:TOElimdensc}), we begin with (\ref{eqn:TOEcdens}) and apply the same reasoning as that that led to (\ref{eqn:Ix^2}). For (\ref{eqn:TOEldc}) we must consider the large $L$ limit of the integral in (\ref{eqn:TOElimdensc}). We note that the integrand is maximised for $t$ as close to zero as possible. This implies that the dominant contribution to the density will be when $t\to v:= 2|\mathrm{Im}(z)| /|1-z^2|$. We exponentiate the logarithm of the integrand and Taylor expand up to first order about the point $v$ to find
\begin{align}
\nonumber \frac{L-3}{2} \log (1-v^2) -\frac{(L-3)v} {1-v^2} (t-v).
\end{align}
Integrating this from $v$ to $1$ we have
\begin{align}
\nonumber &\int_v^1 (1-t^2)^{(L-3)/2} dt \sim (1-v^2)^{(L-3)/2} e^{\frac{(L-3)v^2} {1-v^2}} \int_v^1e^{-\frac{(L-3)vt} {1-v^2}} dt\\
\nonumber &=\frac{(1-v)^{(L-1)/2}} {(L-3)v} - \frac{(1-v)^{(L-1)/2}} {(L-3)v} e^{\frac{(L-3)v} {1-v^2} (v^2-v)}\\
\label{eqn:TOEintTE} & \sim \frac{(1-v^2)^{(L-1)/2}} {L t}= \frac{(1-|z|^2)^{L-1}} {L \: t \: |1-z^2|^{L-1}},
\end{align}
since $v^2-v<0$, where we have used the fact that $|1-z^2|^2-4y^2 = (1-|z|^2)^2$. Use of (\ref{eqn:TOEintTE}) gives us the result (\ref{eqn:TOEldc}).

\hfill $\Box$

\begin{remark}
Note that we can indeed interpret (\ref{eqn:Tlimdensnrl1}) as the $L=1$ case of (\ref{eqn:TOElimdensc}) since in that case we must have used the weight on the second line of (\ref{def:truncw}) and so the integral and the factors $L(L-1)$ on the RHS of (\ref{eqn:TOElimdensc}) do not appear.
\end{remark}

Comparison of (\ref{eqn:TOEldc}) with (\ref{df}) again shows clearly that we have the anti-spherical analogue of the latter, and on projection to the hyperbolic plane, we have a $\sqrt{L}$ density inside a unit disk, which matches (\ref{eqn:Gcirclaw}) in the real Ginibre ensemble. Also note that (\ref{eqn:TOEldc}) is rotationally invariant, meaning that the symmetry-breaking effect of the real eigenvalues has been overwhelmed by the large number of eigenvalues surrounding any non-real complex point. This implies that this ensemble should be manifesting behaviour similar to that of an ensemble of truncated unitary matrices. Indeed, comparison of (\ref{eqn:TOEldc}) to results in \cite{Z&S2000} confirms this expectation to be true. This same effect was observed in Chapter \ref{sec:Ginkernelts} (where the real Ginibre ensemble degenerates to the complex Ginibre ensemble) and in Chapter \ref{sec:SOElims} (where the real spherical ensemble approaches the complex spherical ensemble) at points away from the real line, in the limit of a large number of eigenvalues. In Chapter \ref{sec:uasc} we will discuss this further.

\noindent\textbf{\underline{Expected number of real eigenvalues}}

Part of the consideration of this unification of the correlation functions for analogous real and complex ensembles is the expected number of real eigenvalues. Since the effect of the real eigenvalues becomes negligible in the limit of large matrix dimension, we expect that the number of real eigenvalues grows more slowly than the number of total eigenvalues. This turns out to be true, however we shall see that the expectation is qualitatively different in the strong and weak orthogonality regimes.

\begin{proposition}[\cite{KSZ2010}]
The expected number of real eigenvalues for fixed $0<\alpha<1$ and large $N$ is
\begin{align}
\label{eqn:ENT} E_{N}^{(T)} \sim 2 \frac{ \Gamma((L+1)/2)}{\sqrt{\pi} \;\Gamma(L/2)} \; \mathrm{artanh}\sqrt{\alpha}.
\end{align}
\end{proposition}

\textit{Proof}: The result is obtained by integrating (\ref{eqn:TOElimdensL}) over $(-\sqrt{\alpha}, \sqrt{\alpha})$.

\hfill $\Box$

\begin{itemize}
\item{\underline{Strongly orthogonal}: To obtain the behaviour in the $\alpha\to 1$ strongly orthogonal regime we use the logarithmic expression for $\mathrm{artanh} \: z$ to write
\begin{align}
\nonumber 2 \; \mathrm{artanh} \: \sqrt{\alpha}=\log \frac{1+\sqrt{\alpha}} {1-\sqrt{\alpha}} &= \log \frac{1+\sqrt{1-\gamma}} {1-\sqrt{1-\gamma}}\\
\nonumber &\sim \log \frac{4-\gamma} {\gamma} = \log (4-\gamma) -\log L +\log N,
\end{align}
with $\gamma$ from (\ref{def:alphaML}). Combining this with (\ref{eqn:ENT}) we have for large $M$ and small $L$
\begin{align}
\label{eqn:ENTb} E_M^{(T)}\sim \frac{ \Gamma((L+1)/2)}{\sqrt{\pi} \;\Gamma(L/2)} \log M.
\end{align}}
\item{\underline{Weakly orthogonal}: Using the expansion $\mathrm{artanh} (x)=x+ x^3/2+ x^5/5+...$ we have, in the case that $\alpha\to 0$,
\begin{align}
\label{eqn:ENTa} E_{N}^{(T)} \sim \sqrt{\frac{2(\alpha-\alpha^2)N} {\pi}} \sim \sqrt{\frac{2 M} {\pi}},
\end{align}
where we have also used (\ref{eqn:bigLBeta}) and recalled from (\ref{def:alphaML}) that $\alpha=M/N$. Comparing (\ref{eqn:ENTa}) to (\ref{eqn:bigNEN}) we see that, in the weakly orthogonal limits, we have correspondence with the real Ginibre ensemble.}
\end{itemize}

\subsubsection{Universality and the `anti-spherical' conjecture}
\label{sec:uasc}

As we saw in the previous section, the expected number of real eigenvalues in both the strongly orthogonal (\ref{eqn:ENTb}) and weakly orthogonal limits (\ref{eqn:ENTa}) grows much more slowly than does the total number of eigenvalues. This same behaviour was manifest in both the real Ginibre and real spherical ensembles (see Chapters \ref{sec:Ginkernelts} and \ref{sec:Ssops} respectively), which led to the observation that, in the limit of large matrix dimension, the general eigenvalue density of the real (Ginibre/spherical) ensemble approached that of the corresponding complex ensemble. Recall from (\ref{eqn:TOEldc}), and the discussion below it, that we have the same phenomenon here. This leads us to conjecture that there is a law analogous to the circular law (Proposition \ref{prop:Gincirclaw}) and spherical law (Proposition \ref{prop:sphlaw}) for the anti-spherical case.
\begin{conjecture}[Anti-spherical law]
\label{con:asl}
Let $\bX$ be an $N\times N$ matrix with iid entries, of zero mean and variance one. If $\hat{\bX}$ is obtained from $\bX$ by Gram-Schmidt orthogonalisation and
\begin{align}
\nonumber \hat{\bX}= \left[\begin{array}{cc}
\bA_{L\times L} & \bB_{L\times M}\\
\bC_{M\times L} & \bD_{M\times M}
\end{array}\right]_{N\times N},
\end{align}
then the eigenvalues of $\bD_{M\times M}$ are uniformly distributed, when projected on the anti-sphere, in the limit of large $N$.
\end{conjecture}

\begin{remark}
In \cite{Z&S2000} the authors present some interesting diagrams showing the eigenvalue density profiles for various values of $M$ and $N$ for truncations of both orthogonal and unitary matrices.
\end{remark}

We are also now in a position to identify a different correspondence. In the discussion at the beginning of this chapter we anticipated that for $L$ large we should obtain some Ginibre-like behaviour. We can make the correspondence exact in the regime $\alpha\to 0$. In effect we will undertake the anti-spherical analogue of the procedure in Chapter \ref{sec:SOEsclims}; we focus our view on a small enough region surrounding the origin such that the curvature of the underlying space (here $\kappa<0$) can be neglected, and so we are approximating the planar case. We must keep the eigenvalues sufficiently close to the real line to preserve the distinctive $\beta=1$ behaviour and so we scale them by $1/\sqrt{L}$, which we will find results in the bulk real Ginibre statistics.

Recall from the discussion at the beginning of Chapter \ref{sec:TOEkernelts} that once the kernel elements $D$ and $S_{rr}$ are specified, we can use the interrelations (\ref{eqn:TOErels1}) and (\ref{eqn:TOErels2}) to obtain the remaining ones. First note that
\begin{align}
\label{eqn:bigLfact} \frac{(L+j)!}{L!} \sim L^{j},
\end{align}
and so
\begin{align}
\nonumber \sum_{j=0}^{M-2} \frac{(L+j)!}{L! j!} \left(\frac{\mu\eta}{L}\right)^j \mathop{\sim}\limits_{L \to \infty} \sum_{j=0}^{M-2} \frac{(\mu\eta)^j }{j!}=e^{\mu\eta} \: \frac{\Gamma (M-1,\mu\eta)}{\Gamma(M-1)}.
\end{align}
We also have
\begin{align}
\nonumber \frac{L-2}{L} \log |1-\mu^2/L| \sim -\frac{\mu^2+\bar{\mu}^2}{4},
\end{align}
and
\begin{align}
\nonumber \int_{\nu}^1 (1-t^2)^{(L-3)/2} dt &\sim \int_{\nu}^1 e^{-L\: t^2/2} dt =\sqrt{\frac{2}{L}} \int_{\nu \sqrt{L/2}}^{\sqrt{L/2}} e^{-t^2} dt\\
\label{eqn:interfclim} &\sim \sqrt{\frac{2}{L}} \int_{\sqrt{2}\: \mathrm{Im}(z)}^{\infty} e^{-t^2} dt= \sqrt{\frac{\pi} {2L}} \; \erfc (\sqrt{2}\: \mathrm{Im}(z)),
\end{align}
where $\nu= 2 \: \mathrm{Im}(z/\sqrt{L})/ |1-z^2/L|$. Using these in (\ref{eqn:SOEDsum}) we have
\begin{align}
\nonumber D\left(\mu/\sqrt{L}, \eta/\sqrt{L} \right)_T &\mathop{\sim}\limits_{L \to \infty} \frac{L (\eta-\mu)}{\sqrt{2\pi}} e^{-(\mu^2+ \bar{\mu}^2)/4} e^{-(\eta^2+ \bar{\eta}^2)/4} \sqrt{\erfc \left( \sqrt{2} \mathrm{Im}(\mu) \right) \erfc \left( \sqrt{2} \mathrm{Im}(\eta) \right)}\\
\label{eqn:bigLD} & \times e^{\mu\eta} \: \frac{\Gamma (M-1,\mu\eta)}{\Gamma (M-1)}.
\end{align}

Next, we apply the same $1/\sqrt{L}$ scaling to (\ref{eqn:TOEsumS}),
\begin{align}
\nonumber &S_{rr}(x/\sqrt{L}, y/\sqrt{L})_T = \frac{L}{2} \frac{ \Gamma((L+1)/2)}{\sqrt{\pi}\; \Gamma(L/2)} \; (x/\sqrt{L})^{M-1} (1-x^2/L)^{L/2-1} \frac{(L+M-2)!} {L! (M-2)!}\\
\nonumber &\times \int_{-1}^{1} \sgn (y/\sqrt{L} -z) z^{M-2} (1-z^2)^{L/2-1}dz\\
\label{eqn:TOESrrbigL} & + \frac{ \Gamma((L+1)/2)}{\sqrt{\pi} \Gamma(L/2)} (1-x^2/L)^{L/2-1} (1-y^2/L)^{L/2} \sum_{j=0}^{M-2} \frac{(L+j-1)!}{(L-1)! j!} \left( \frac{xy}{L} \right)^j,
\end{align}
with $M$ assumed to be even for convenience. The second term in (\ref{eqn:TOESrrbigL}) can be dealt with using exactly the same procedure that led to (\ref{eqn:bigLD}), while for the first term note that
\begin{align}
\nonumber &\int_{-1}^{1} \sgn (y/\sqrt{L} -z) z^{M-2} (1-z^2)^{L/2-1}dz = 2 \; \sgn (y) \int_{0}^{y/\sqrt{L}} z^{M-2} (1-z^2)^{L/2-1}dz\\
\nonumber & \mathop{\sim}\limits_{L \to \infty} 2\; \sgn (y) \int_0^{y/\sqrt{L}} z^{M-2} e^{-Lz^2/2} dz = \sgn (y) \left(\frac{2}{L} \right)^{(M-1)/2} \gamma ((M-1)/2,y^2/2),
\end{align}
which gives
\begin{align}
\nonumber S_{rr}(x/\sqrt{L}, y/\sqrt{L})_T \mathop{\sim}\limits_{L \to \infty} &\sqrt{\frac{L}{2 \pi}} \left( 2^{(M-3)/2} \sgn (y) x^{M-1} e^{-x^2/2} \frac{\gamma((M-1)/2,y^2/2)}{\Gamma(M-1)} \right.\\
\label{eqn:bigLSrr} &\left. + e^{-(x-y)^2/2} \frac{\Gamma (M-1,xy)}{\Gamma(M-1)} \right),
\end{align}
where we have also used (\ref{eqn:bigLfact}).

We can use (\ref{eqn:TOErels1}) and (\ref{eqn:TOErels2}) to compare (\ref{eqn:bigLD}) and (\ref{eqn:bigLSrr}) to the corresponding real Ginibre results in (\ref{eqn:Ginsummed}). Also, by taking $x=y\in \mathbb{R}$ and $\mu=\eta=z\in \mathbb{C}$, we obtain the real and complex densities
\begin{align}
\nonumber \rho_{(1)} ^{r} \left(x/\sqrt{L}\right) = S_{rr}\left(x/\sqrt{L}, x/\sqrt{L} \right)_T \mathop{\sim}\limits_{L \to \infty} &\sqrt{\frac{L}{2 \pi}} \left( 2^{(M-3)/2} |x|^{M-1} e^{-x^2/2} \frac{\gamma((M-1)/2,x^2/2)}{\Gamma(M-1)} \right.\\
\label{eqn:bigLSrd} &\left. + \frac{\Gamma (M-1,x^2)}{\Gamma(M-1)} \right),
\end{align}
\begin{align}
\label{eqn:bigLDcd} \rho_{(1)} ^{c} \left(z/\sqrt{L}\right) = S_{cc}\left(z/\sqrt{L}, z/\sqrt{L} \right)_T \mathop{\sim}\limits_{L \to \infty} \sqrt{\frac{2}{\pi}}\: L\: y\: e^{- (z+\bar{z})^2/2} \frac{\Gamma (M-1,|z|^2)}{\Gamma (M-1)}\: \erfc (\sqrt{2} y),
\end{align}
which agree with (\ref{eqn:Ginreal_density}) and (\ref{eqn:Gincompx_density}), up to a factor of $\sqrt{L}$ in the real case, and $L$ in the complex case. (The same factors appeared in (\ref{eqn:bigLD}) and (\ref{eqn:bigLSrr}).) We anticipate these factors since in the large $N$ limit the complex eigenvalues, for instance, are supported on a disk of radius $\sqrt{\alpha}$ and so
\begin{align}
\nonumber \frac{\# \mathrm{eigenvalues}}{\mathrm{area}} = \frac{M}{\pi \alpha} \mathop{\sim}\limits_{\alpha \to 0} \frac{L}{\pi}.
\end{align}
The real case follows, since those eigenvalues are supported on an interval of radius $\sqrt{L}$.

\begin{figure}[htp]
\begin{center}
\subfloat[]{\label{fig:GinOEn10j200} \includegraphics[scale=0.5]{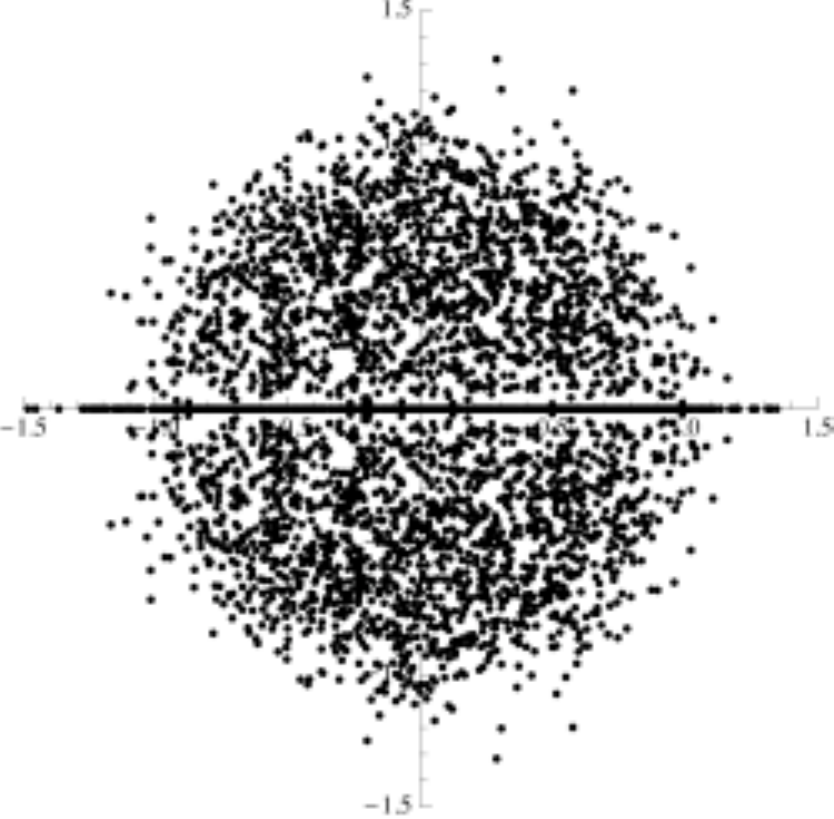}} \hspace{24pt}
\subfloat[]{\label{fig:TOEn1600m20j200} \includegraphics[scale=0.5]{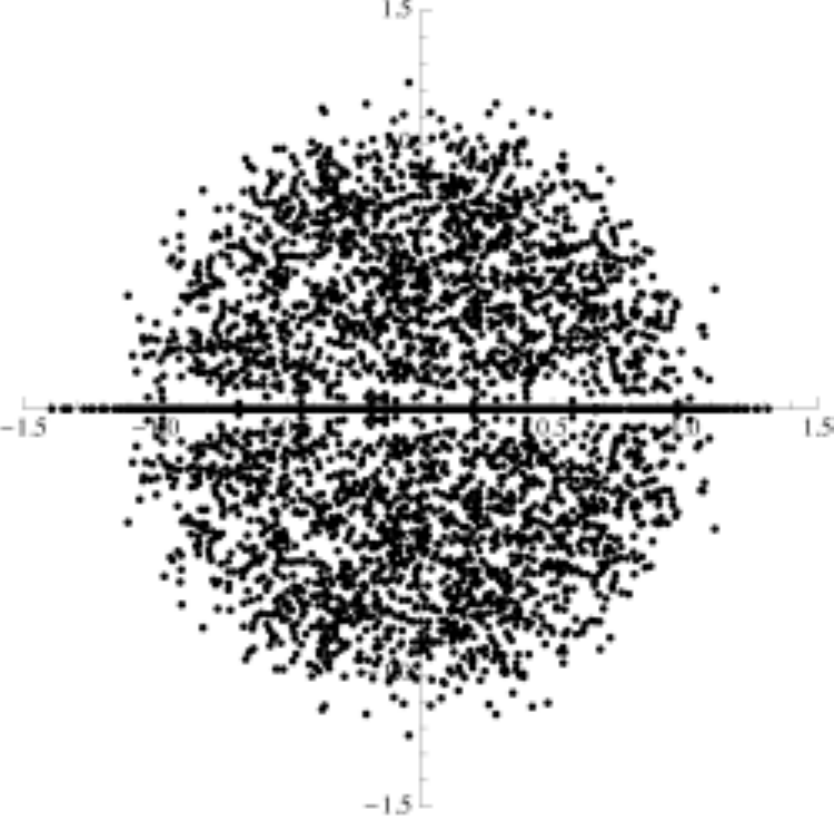}}
\caption[Comparison of simulated eigenvalue plots for the real Ginibre ensemble and the weakly orthogonal limit of the real truncated ensemble.]{Eigenvalues from 200 instances of: (a) $20\times 20$ real Ginibre matrices, scaled by $1/\sqrt{N}$; and (b) $1600\times 1600$ real truncated matrices with $M=20$, scaled by $\sqrt{N/M}$.}
\label{fig:TOEtoGin}
\end{center}
\end{figure}

As expected, we have found correspondence with the real Ginibre ensemble in the weakly orthogonal regimes --- demonstrated by (\ref{eqn:bigLDcd}) and (\ref{eqn:bigLSrd}). A visual demonstration is found in Figure \ref{fig:TOEtoGin}, where \ref{fig:GinOEn10j200} represents the standard real Ginibre ensemble and \ref{fig:TOEn1600m20j200} represents the large $L$ limit of the real truncated ensemble; both sets of eigenvalues have been scaled to the unit disk.
We may also compare the $p_{N,k}$ statistics in Tables \ref{tab:pnkxact_sim} and \ref{tab:TOEpnkL8} where we find some convergence.

\newpage

\section{Further work}
\setcounter{figure}{0}
\label{sec:FW}

The first obvious direction to proceed from the results presented in this work is to establish (one way or the other) Conjecture \ref{con:asl}, the anti-spherical law. It seems that an extension of the method of Tao and Vu \cite{TVK10} (which they used for the circular law), akin to that of Bordenave \cite{Bord2010} for the spherical law, will enable this statement to be proven. We can also ask a deeper question on the eigenvalue distribution of these truncated ensembles. Recall from the discussion in Chapter \ref{sec:TOEmjpdf} that the distribution of the sub-block matrix $\bD$ from (\ref{def:Rdecomp}) contains singular factors when $L<M$ (that is, for a small truncation), but when $L\geq M$ then we have the continuous expression (\ref{eqn:truncpdf}). The question then is: why is it that the eigenvalue distribution (\ref{eqn:TOEevaljpdf}), which we established from (\ref{eqn:truncpdf}), turns out to be identical to that obtained in \cite{KSZ2010}, where the restriction on the relative size of $L$ and $M$ was circumvented (as outlined below Remark \ref{rem:TOEana})? Somehow the singularities in the distribution of the sub-block vanish when we change variables to the distribution of eigenvalues.

Another anomaly in the case of the truncated ensemble is that we have not been able to strictly apply the 5-step method of Chapter \ref{sec:GOE_steps} as we did successfully with the other ensembles in this work; the missing component is that we do not yet know how to calculate the skew-orthogonal polynomials (\ref{eqn:TOEsops}) independently of the correlation kernel (\ref{eqn:SOEDsum}). One way to do this might be by first noting from \cite[Chapter 6]{forrester?} that with $a=b=L-1$ the Jacobi weight function
\begin{align}
\nonumber (1-x)^{(a-1)/2}(1+x)^{(b-1)/2}, \quad x\in (-1,1),
\end{align}
is structurally similar to (\ref{eqn:TOErweight}). The relevant skew-orthogonal polynomials are constructed from the classical Jacobi polynomials as described by Forrester \cite{forrester?}. It may be possible to use these results, perhaps by introducing an interpolating parameter like that used in Chapter \ref{sec:tG}, to obtain the skew-orthogonal polynomials for the real truncated ensemble.

As discussed in the Introduction, Dyson's three-fold way tells us that random matrix ensembles are naturally classified by the parameter $\beta$ in (\ref{eqn:evalbeta}). In this work we have focused on the $\beta=1$ real matrix ensembles; the $\beta=4$ cases of the spherical and truncated ensembles are yet to be investigated, while for the $\beta=4$ Ginibre ensemble see \cite{kanzieper2001}.

\begin{figure}[htp]
\begin{center}
\includegraphics[scale=0.625]{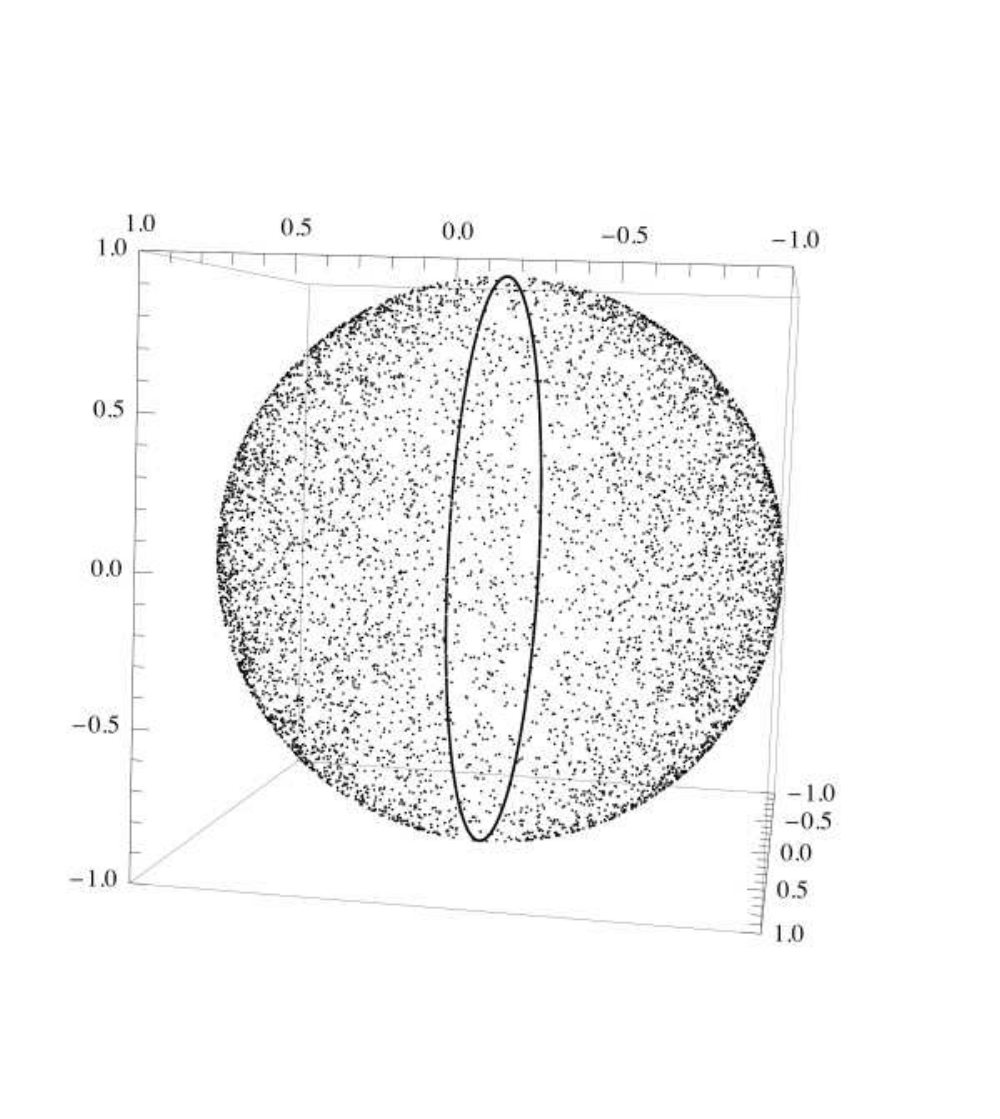}
\caption[Simulated eigenvalue plot for a spherical ensemble of GOE matrices.]{Stereographic projection of the eigenvalues of 1000 instances of the product $\bA^{-1}\bB$, where $\bA,\bB$ are GOE matrices.}
\label{fig:sGOE}
\end{center}
\end{figure}

From the construction of the real spherical ensemble in Chapter \ref{sec:SOE} we see that the product $\bA^{-1}\bB$ (with $\bA$ and $\bB$ having Gaussian entries) is a matrix generalisation of a Cauchy random variable. We may attempt to form a spherical ensemble from any such product of Gaussian matrices, including the Hermitian Gaussian ensembles (GOE, GUE and GSE). Note that while the inverse of a Hermitian matrix is still Hermitian, the product of two Hermitian matrices is not in general Hermitian. This means that we expect the eigenvalues of $\bA,\bB$ to be complex. For instance, take $\bA, \bB$ as GOE matrices (\ref{eqn:GOE_el_dist}), then by simulation we obtain Figure \ref{fig:sGOE}. Since the product $\bA^{-1}\bB$ is a general real matrix we are not surprised to find a ring corresponding to a non-zero density of real eigenvalues.

\begin{figure}[htp]
\begin{center}
\subfloat[]{\label{fig:sGUE} \includegraphics[scale=0.625]{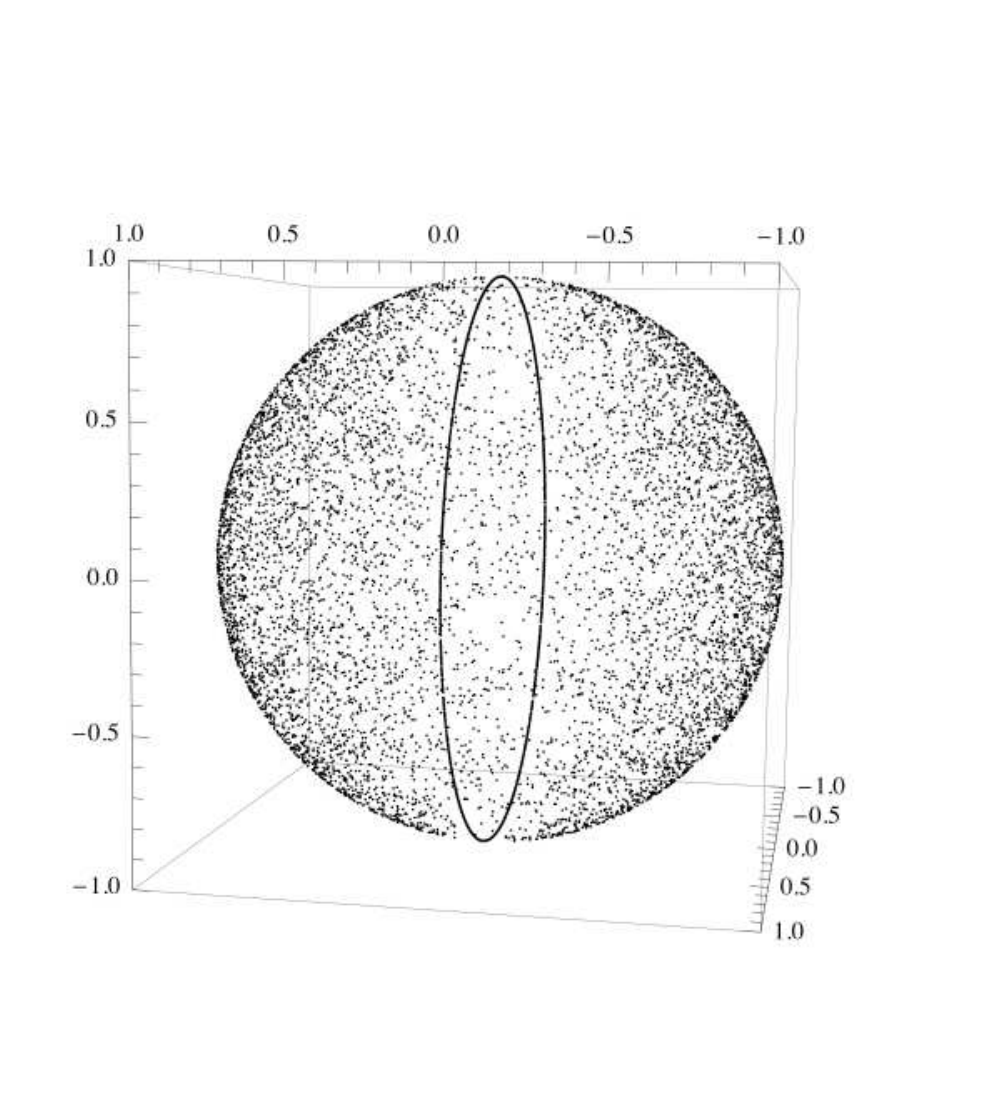}}
\subfloat[]{\label{fig:sGSE} \includegraphics[scale=0.625]{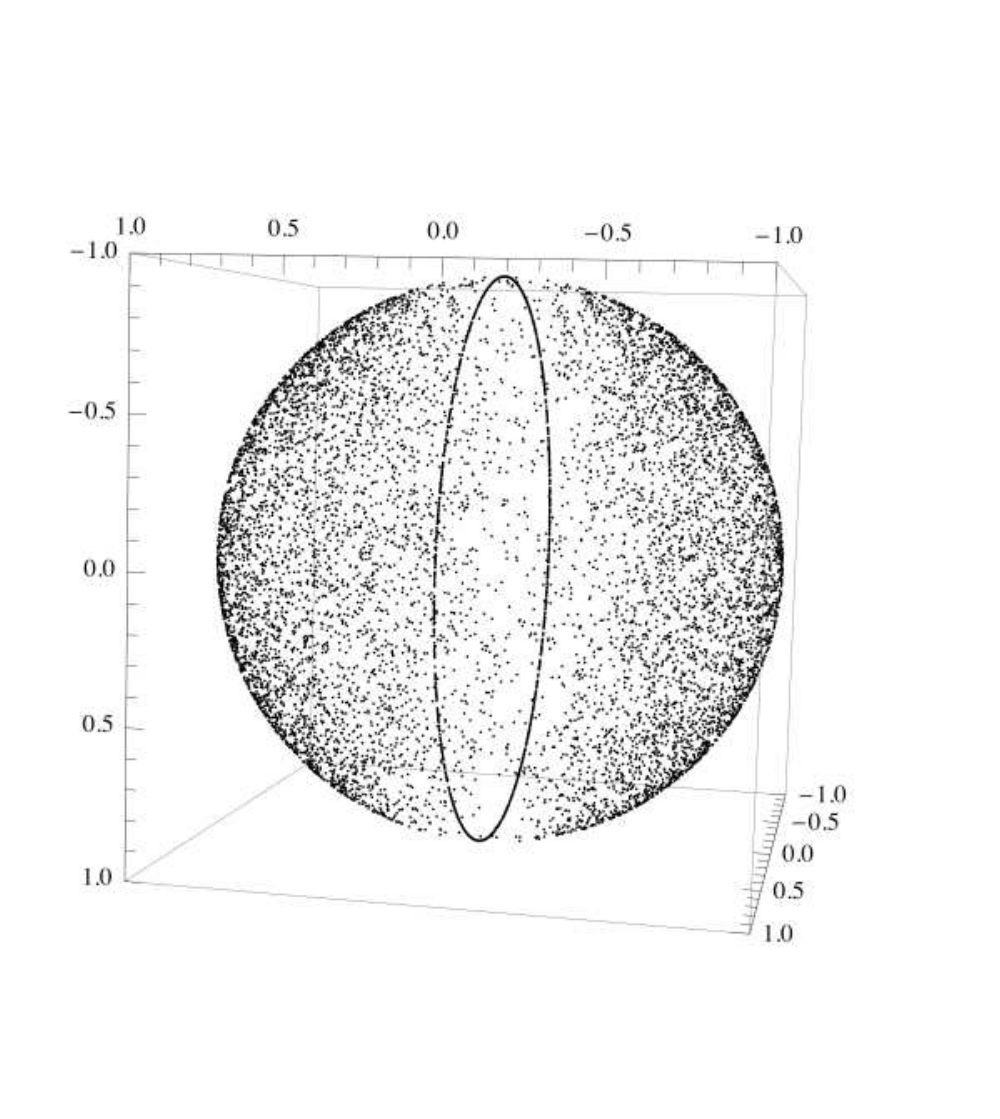}}
\caption[Simulated eigenvalue plots for spherical ensembles of GUE and GSE matrices.]{Stereographic projection of the eigenvalues of 1000 instances of the product $\bA^{-1}\bB$, where $\bA,\bB$ are (a) GUE matrices and (b) GSE matrices.}
\label{fig:sGUEsGSE}
\end{center}
\end{figure}

However, if the matrices are taken from the GUE or the GSE, as in Figure \ref{fig:sGUEsGSE}, then we see the same ring, although these matrices have complex entries, and so \textit{a priori} we do not expect any real eigenvalues at all. Superficially, these distributions resemble those of the real spherical ensemble (Figure \ref{fig:dstar}), although note that for increasing $\beta$ (that is, from the GOE with $\beta=1$ in Figure \ref{fig:sGOE} to the GUE with $\beta=2$ in Figure \ref{fig:sGUE}, to the GSE with $\beta=4$ in Figure \ref{fig:sGSE}) there is stronger repulsion from the great circle. It would be interesting to discover if this density is, firstly integrable, and secondly identical to the analogous spherical ensembles.

\begin{figure}[htp]
\begin{center}
\subfloat[]{\label{fig:cosquarea} \includegraphics[scale=0.625]{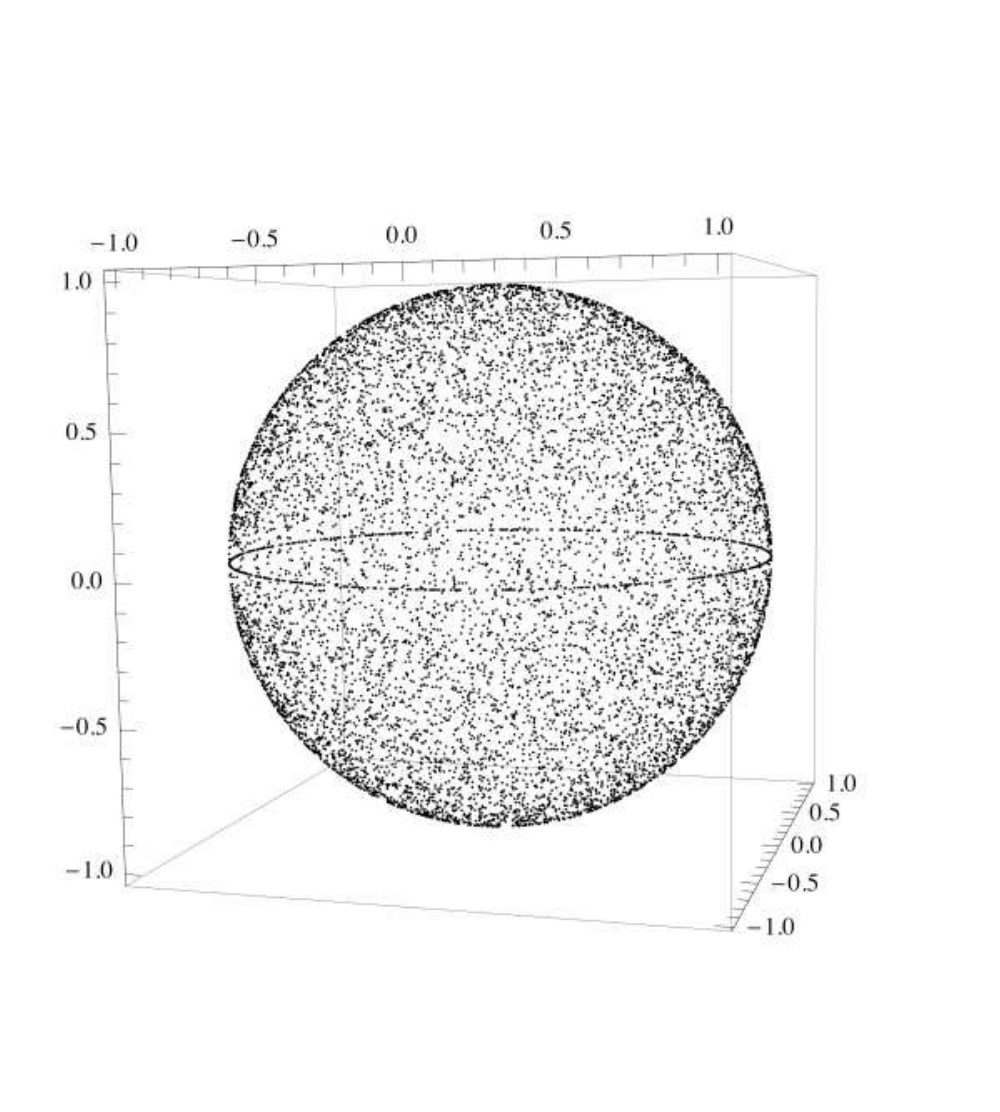}}
\subfloat[]{\label{fig:cosquareb} \includegraphics[scale=0.625]{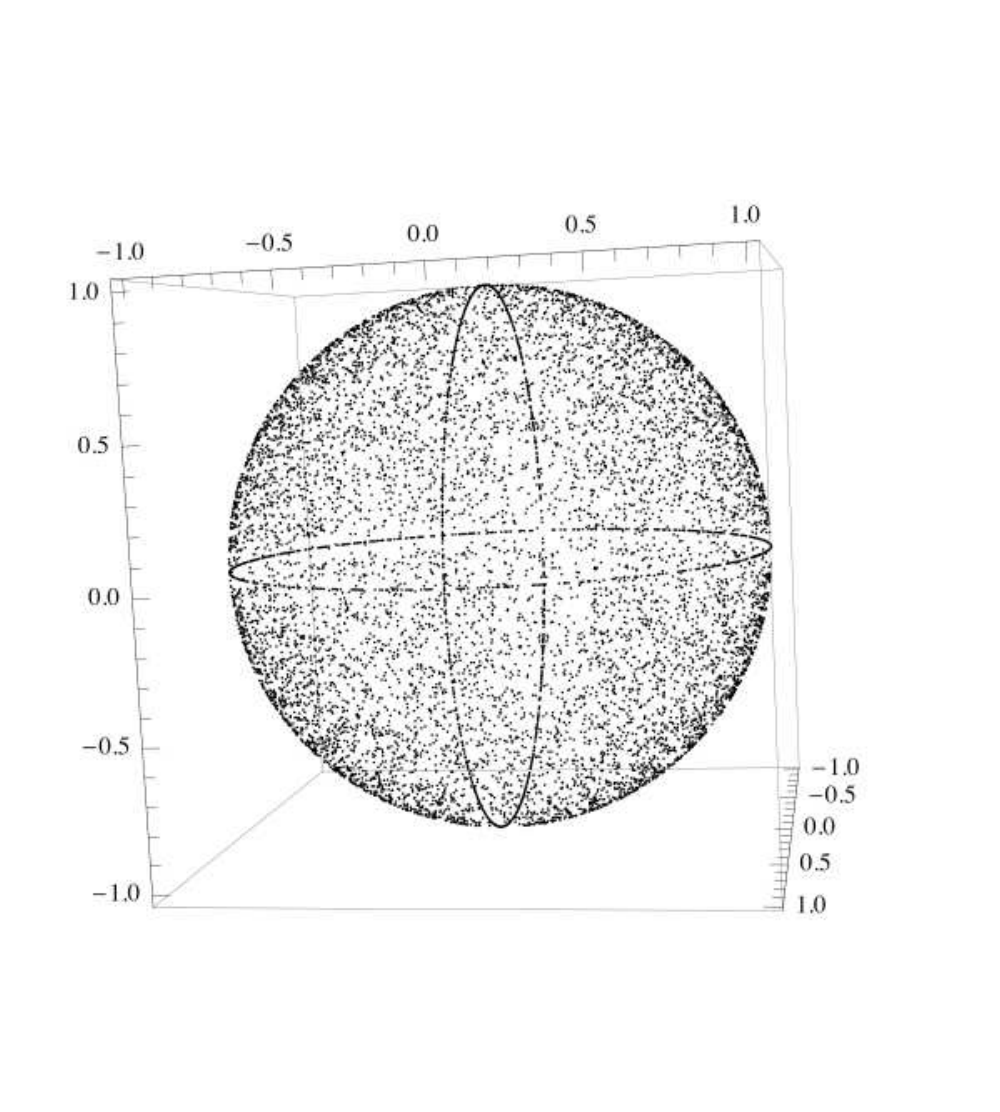}}
\caption[Simulated eigenvalue plots of complex and real $*$-cosquare matrices.]{Stereographic projection of the eigenvalues of 1000 instances of the product $\bA^{*}\bA^{-1}$, where $\bA$ is (a) a complex Ginibre matrix and (b) a real Ginibre matrix.}
\end{center}
\end{figure}

As suggested by Christopher Sinclair, from numerical simulations we can find another construction that seems to reproduce the real spherical distribution: the $*$-cosquare ensemble $\bA^*\bA^{-1}$, where $\bA$ is a complex Ginibre matrix. As we see in Figure \ref{fig:cosquarea}, we have a ring of eigenvalues around the equator instead of through the poles as in Figure \ref{fig:dstar}, but otherwise the distributions seem identical. This equatorial ring comes from the non-zero density of eigenvalues on the unit circle that is an attribute of these $*$-cosquare ensembles; in the $1\times 1$ case this is clear since $r e^{-i\theta}/r e^{i\theta}=e^{-2i\theta}$ obviously lies on the unit circle. One suspects that the distribution of $\bA^*\bA^{-1}$ should match that of the real Ginibre ensemble.
We can further enliven matters by taking $\bA$ a real Ginibre matrix, which results in a spherical distribution such as in Figure \ref{fig:cosquareb}. Note that we now have two rings; one corresponding to the eigenvalues on the unit circle (coming from the  $*$-cosquare construction) and another corresponding to the real eigenvalues that we expect in any real matrix. This represents a set of three species of eigenvalue. In this case the constructions we developed in Chapter \ref{sec:Gincorrlnse} for calculating the correlation functions of multiple disjoint sets of eigenvalues will likely prove useful. For example, we expect to find a $6\times 6$ kernel in the result analogous to (\ref{eqn:4x4_fred}), to which we could apply functional differentiation to calculate the correlation functions.

\begin{figure}[htp]
\begin{center}
\includegraphics[scale=0.625]{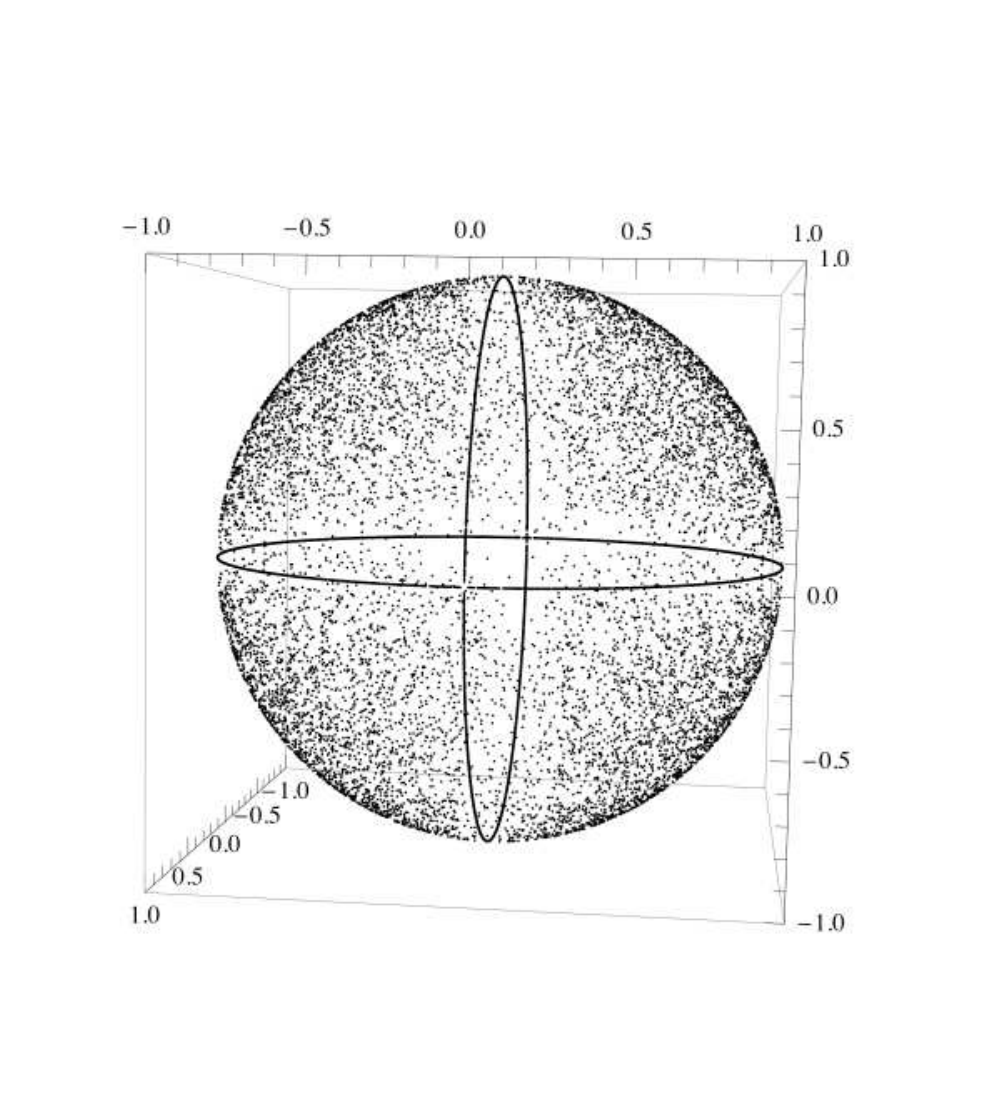}
\caption[Simulated eigenvalue plot of a spherical ensemble formed from a chiral ensemble.]{Stereographic projection of the eigenvalues of 1000 instances of the product $(\bC_{1} \bD_{1}^T)^{-1}\bC_{2}\bD_{2}^T$, where $\bC_1,\bC_2, \bD_1, \bD_2$ are $10\times 10$ real Ginibre matrices.}
\label{fig:sABT}
\end{center}
\end{figure}

We can obtain a similar distribution by taking spherical ($\bA^{-1}\bB$) products of matrices that arise in the study of chiral ensembles. A chiral ensemble contains matrices of the form
\begin{align}
\nonumber \bX = \left[\begin{array}{cc}
\0_{L\times L} & \bC_{L\times M}\\
(\bD_{L\times M})^T & \0_{M\times M}
\end{array} \right];
\end{align}
these ensembles have attracted increasing interest over the last two decades (beginning with \cite{Verb1994a, Verb1994b}) because of their relationship with quantum chromodynamics (QCD). Using the relation
\begin{align}
\nonumber \det \left[ \begin{array}{cc}
\bM_1 & \bM_2\\
\bM_3 & \bM_4
\end{array}\right] = \det (\bM_4) \det (\bM_1-\bM_2(\bM_4)^{-1}\bM_3)
\end{align}
the eigenvalues of $\bX$ are seen to be given by the $\pm$ square roots of the eigenvalues of the product $\bC\bD^T$, which implies that they come in three distinct species: purely real, purely imaginary, and $\pm$ conjugate paired quadruplets. In the case that
\begin{align}
\label{eqn:APSch} \bC=\bP+\mu \bQ,\qquad \bD=\bP^T-\mu \bQ^T,
\end{align}
where $\bP, \bQ$ are iid real Gaussian matrices and $\mu\in (0, 1]$, the eigenvalue distribution and correlation functions are calculated in \cite{AkePhilSom2010}. Since these chiral matrices have three eigenvalue species, to obtain a distribution resembling that in Figure \ref{fig:cosquareb}, we might ask for the eigenvalue distribution of the matrix
\begin{align}
\label{eqn:chProd1} \bX_{1}^{-1}\bX_2 = \left[\begin{array}{cc}
\0 & \bC_{1}\\
\bD_{1}^T & \0
\end{array} \right]^{-1} \left[\begin{array}{cc}
\0 & \bC_{2}\\
\bD_{2}^T & \0
\end{array} \right]= \left[\begin{array}{cc}
(\bD_{1}^T)^{-1}\bD_{2}^T & \0\\
\0 & (\bC_{1})^{-1}\bC_{2}
\end{array} \right],
\end{align}
where the $\bC_i$ and $\bD_j$ are all $N\times N$ iid real Gaussian matrices. However the RHS of (\ref{eqn:chProd1}) implies that the set of eigenvalues of $\bX_{1}^{-1}\bX_2$ is just the union of the eigenvalue sets of the individual real spherical matrices $(\bD_{1}^T)^{-1}\bD_{2}^T$ and $(\bC_{1})^{-1}\bC_{2}$, and so we obtain a distribution like that in Figure \ref{fig:dstar}. Instead, if we take the $\pm$ square roots of the eigenvalues of $\bY=\bA^{-1}\bB$, where
\begin{align}
\label{eqn:AMch} \bA= \bC_{1} \bD_{1}^T, \qquad \bB= \bC_{2}\bD_{2}^T,
\end{align}
with $\bC_1, \bC_2, \bD_1, \bD_2$ each an iid real Gaussian matrix, then we obtain a distribution such as that in Figure \ref{fig:sABT} under stereographic projection. Although the matrix $\bY$ can be written as $(\bC_{1} \bD_{1}^T)^{-1}\bC_{2}\bD_{2}^T = (\bC_{1}^{-T} \bD_{1}^{-1})^{T}\bC_{2}\bD_{2}^T$, this is a quite different ensemble to that studied in \cite{AkePhilSom2010} since the distribution of the factors $\bA^{-T}$ and $\bB$ in (\ref{eqn:AMch}) are not the same as those of (\ref{eqn:APSch}).
Investigation of these ensembles as well as the $*$-cosquare ensembles above (with the tools described in this work) may lead to the hyperpfaffian structures considered in \cite{Sinclair2010}.

As discussed in several parts of the present work, the eigenvalues of random matrices are mutually repulsive, and so have a more uniform distribution than that typically seen in the `clumpy' Poisson distribution (compare Figures \ref{fig:PoiDisk} and \ref{fig:RMTDisk}). This observation led to the work in \cite{LeCHo1990}, where the authors used the eigenvalues of a complex Ginibre matrix to generate a \textit{Voronoi tessellation}, which is more uniform than that generated from Poisson points. A Voronoi tessellation in this context is a tiling made of \textit{Voronoi cells} or polygons, each of which surrounds one eigenvalue and contains all the points closer to that eigenvalue than to any other. The authors compared this tiling to that corresponding to the cellular structure of cucumbers, where they analysed the statistics of the cells including area, perimeter, side length and number of sides. It seems likely that the same analysis (excluding the vegetables) can be performed on a Voronoi tessellation of the sphere, adapting the work in \cite{SafKui1997}. The spherical geometry introduces a richer structure since it is known (by considering the Euler characteristic) that a sphere cannot be tiled by hexagons. This fact implies that some number of other polygons are required to complete the tessellation. For example, if the only other polygon is a pentagon, then there must be 12 of those present, while if pentagons are also allowed, then there must be exactly 12 more pentagons than heptagons. From such considerations one expects to obtain quite different statistics to those seen in the analogous problem on the plane. The complex analogue of the real spherical ensemble provides a way of randomly distributing repulsive points on a sphere, and so it can be used to generate one of these more uniform Voronoi tessellations, which is a convenient place to begin the spherical analysis.

An important fact in the considerations of the circular law and its analogues (the spherical law and the anti-spherical law) is that the number of real eigenvalues grows sub-dominantly; see (\ref{eqn:bigNEN}), (\ref{eqn:SENasy}) and (\ref{eqn:ENT}). Without this fact being true, the circular, spherical and anti-spherical laws could not hold. Clearly, the number of total eigenvalues for an $N\times N$ matrix is $N$, and from (\ref{eqn:bigNEN}) we know that the number of reals in the real Ginibre ensemble is proportional to $\sqrt{N}$. Experimental evidence in \cite{eks1994} suggests that this $\sqrt{N}$ law is universal for an ensemble of matrices with iid entries from any zero mean, finite variance distribution. This is as yet unconfirmed. Further, we see from (\ref{eqn:SENasy}) that there is at least one ensemble (the real spherical ensemble) with non-iid entries that exhibits a similar behaviour. This may lead one to speculate that the $\sqrt{N}$ law extends to non-iid ensembles, however the real truncated ensemble shows that this is not true. We see from (\ref{eqn:ENT}) that the $\sqrt{N}$ law only holds in the weakly orthogonal limits, which as we discussed, just reclaims the real Ginibre results. In the strongly orthogonal limit, the number of reals grows as $\log N$. An investigation of real eigenvalues seems to have been neglected amidst the general coalescence of interest around the distribution of complex eigenvalues, yet is clearly warranted.


\newpage

\appendix
\section*{Appendices}
\addcontentsline{toc}{section}{Appendices}
\section{Probability of $k$ real eigenvalues for the real Ginibre ensemble}
\label{app:GinOE_kernel_elts}
\numberwithin{equation}{section}
\numberwithin{table}{section}

\begin{longtable}{|c|c|c|c|}
\hline
& $\mathrm{Exact}\; p_{N,k}$ & $\mathrm{Decimal} \; p_{N,k}$ & Simulation\\
\hline
\hline
\endfirsthead

\hline
& $\mathrm{Exact}\; p_{N,k}$ & $\mathrm{Decimal} \; p_{N,k}$ & Simulation\\
\hline
\hline
\endhead

$\T p_{2,2}$&$\frac{1}{2}\sqrt{2}$&$0.70711$&$0.70762$\\
$\T p_{2,0}$&$1-\frac{1}{2}\sqrt{2}$&$0.29289$&$0.29238$\\
\hline
$\T p_{3,3}$&$\frac{1}{4}\sqrt{2}$&$0.35355$&$0.35349$\\
$\T p_{3,1}$&$1-\frac{1}{4}\sqrt{2}$&$0.64645$&$0.64651$\\
\hline
$\T p_{4,4}$&$\frac{1}{8}$&$0.125$&$0.12440$\\
$\T p_{4,2}$&$\frac{11}{16}\sqrt{2}-\frac{1}{4}$&$0.72227$&$0.72249$\\
$\T p_{4,0}$&$\frac{9}{8}-\frac{11}{16}\sqrt{2}$&$0.15273$&$0.15311$\\
\hline
$\T p_{5,5}$&$\frac{1}{32}$&$0.03125$&$0.03116$\\
$\T p_{5,3}$&$\frac{13}{32}\sqrt{2} - \frac{1}{16}$&$0.51202$&$0.51300$\\
$\T p_{5,1}$&$\frac{33}{32} - \frac{13}{32}\sqrt{2}$&$0.45673$&$0.45584$\\
\hline
$\T p_{6,6}$&$\frac{1}{256}\sqrt{2}$&$0.00552$&$0.00506$\\
$\T p_{6,4}$&$-\frac{3}{256}\sqrt{2} + \frac{271}{1024}$&$0.24808$&$0.25231$\\
$\T p_{6,2}$&$-\frac{271}{512} + \frac{107}{128}\sqrt{2}$&$0.65290$&$0.64888$\\
$\T p_{6,0}$&$\frac{1295}{1024}-\frac{53}{64}\sqrt{2}$&$0.09350$&$0.09375$\\
\hline
$\T p_{7,7}$&$\frac{1}{2048}\sqrt{2}$&$0.00069$&$0.00073$\\
$\T p_{7,5}$&$\frac{355}{4096} - \frac{3}{2048}\sqrt{2}$&$0.08460$&$0.08343$\\
$\T p_{7,3}$&$-\frac{355}{2048}+\frac{1087}{2048}\sqrt{2}$&$0.66394$&$0.57908$\\
$\T p_{7,1}$&$\frac{4451}{4096} - \frac{1085}{2048}\sqrt{2}$&$0.33744$&$0.33676$\\
\hline
$\T p_{8,8}$&$\frac{1}{16384}$&$0.00006$&$0.00006$\\
$\T p_{8,6}$&$-\frac{1}{4096} + \frac{3851}{262144}\sqrt{2}$&$0.02053$&$0.02052$\\
$\T p_{8,4}$&$\frac{53519}{131072}-\frac{11553}{262144}\sqrt{2}$&$0.34599$&$0.34469$\\
$\T p_{8,2}$&$-\frac{53487}{65536} + \frac{257185}{262144}\sqrt{2}$&$0.57131$&$0.57257$\\
$\T p_{8,0}$&$\frac{184551}{131072} - \frac{249483}{262144}\sqrt{2}$&$0.06210$&$0.06216$\\
\hline
$\T p_{9,9}$&$\frac{1}{262144}$&$0.00000$&$0.00000$\\
$\T p_{9,7}$&$-\frac{1}{65536} + \frac{5297}{2097152}\sqrt{2}$&$0.00356$&$0.00359$\\
$\T p_{9,5}$&$\frac{82347}{524288}-\frac{15891}{2097152}\sqrt{2}$&$0.14635$&$0.14479$\\
$\T p_{9,3}$&$-\frac{82339}{262144} + \frac{1345555}{2097152}\sqrt{2}$&$0.59328$&$0.59056$\\
$\T p_{9,1}$&$\frac{606625}{524288} - \frac{1334961}{2097152}\sqrt{2}$&$0.25681$&$0.26106$\\
\hline
$\T p_{10,10}$&$\frac{1}{8388608}\sqrt{2}$&$0.00000$&$0.00000$\\
$\T p_{10,8}$&$\frac{236539}{536870912}-\frac{5}{8388608}\sqrt{2}$&$0.00044$&$0.00041$\\
\hline
$\T p_{10,6}$&$\frac{35098479}{1073741824}\sqrt{2}-\frac{236539}{134217728}$&$0.04447$&$0.04477$\\
$\T p_{10,4}$&$\frac{149206217}{268435456}-\frac{105292877}{1073741824}\sqrt{2}$&$0.41716$&$0.41561$\\
$\T p_{10,2}$&$\frac{1216831949}{1073741824}\sqrt{2}-\frac{148733139}{134217728}$&$0.49453$&$0.49503$\\
$\T p_{10,0}$&$\frac{834100651}{536870912}-\frac{1146292877}{1073741824}\sqrt{2}$&$0.04341$&$0.04418$\\
\hline
$\T p_{11,11}$&$\frac{1}{268435456}\sqrt{2}$&$0.00000$&$0.00000$\\
$\T p_{11,9}$&$\frac{333123}{8589934592}-\frac{5}{268435456}\sqrt{2}$&$0.00004$&$0.00001$\\
$\T p_{11,7}$&$-\frac{333123}{2147183648} + \frac{60262315}{8589934592}\sqrt{2}$&$0.00977$&$0.01006$\\
$\T p_{11,5}$&$\frac{1020788137}{4294967296} - \frac{180786305}{8589934592}\sqrt{2}$&$0.20791$&$0.20768$\\
$\T p_{11,3}$&-$\frac{1020121891}{2147483648} + \frac{6423679969}{8589934592}\sqrt{2}$&$0.58254$&$0.58174$\\
$\T p_{11,1}$&$\frac{10629845251}{8589934592} - \frac{6303155851}{8589934592}\sqrt{2}$&$0.19975$&$0.20051$\\
\hline
$\T p_{12,12}$&$\frac{1}{8589934592}$&$0.00000$&$0.00000$\\
$\T p_{12,10}$&$\frac{3781485}{2199023255552} \sqrt{2}- \frac{3}{4294967296} $&$0.00000$&$0.00000$\\
$\T p_{12,8}$&$\frac{27511372605} {17592186044416}- \frac{18907425} {2199023255552}$&$0.00155$&$0.00152$\\
$\T p_{12,6}$&$\frac{126455775487}{2199023255552}\sqrt{2}- \frac{27511352125} {4398046511104}$&$0.07507$&$0.07482$\\
$\T p_{12,4}$&$\frac{6237846960567}{8796093022208}- \frac{379291696761} {2199023255552}\sqrt{2}$&$0.46524$&$0.46194$\\
$\T p_{12,2}$&$\frac{356179603371}{274877906944}\sqrt{2} -\frac{6182824264509} {4398046511104}$&$0.42669$&$0.42909$\\
$\T p_{12,0}$&$\frac{29930323227453}{17592186044416} - \frac{1298292889877} {1099511627776}  \sqrt{2}$&$0.03145$&$0.03263$\\
\hline
$\T p_{38,38}$& --- &$0.00000$&$0.00000$\\
$\vdots$ &$\vdots$&$\vdots$&$\vdots$\\
$\T p_{38,16}$& --- &$0.00000$&$0.00000$\\
$\T p_{38,14}$& --- &$0.00001$&$0.00001$\\
$\T p_{38,12}$& --- &$0.00064$&$0.00075$\\
$\T p_{38,10}$& --- &$0.01768$&$0.01751$\\
$\T p_{38,8}$& --- &$0.14961$&$0.14862$\\
$\T p_{38,6}$& --- &$0.40845$&$0.4093$\\
$\T p_{38,4}$& --- &$0.34706$&$0.34735$\\
$\T p_{38,2}$& --- &$0.07474$&$0.07475$\\
$\T p_{38,0}$& --- &$0.00182$&$0.00171$\\
\hline
\caption[Comparison of analytic and simulated probabilities $p_{N,k}$ for the real Ginibre ensemble.]{Comparison of the simulated probabilities from Table \ref{tab:GinOE_pnk} with the analytical results calculated from (\ref{eqn:GinOE_probsGF_pf}) and (\ref{eqn:GinOE_probs_odd}) using the results in Chapter \ref{sec:pnk}. Note that the analytical results for $p_{38,k}$ have been omitted to conserve space.}
\label{tab:pnkxact_sim}
\end{longtable}

\section{Real Ginibre correlation kernel elements}
\subsection{$N$ even}
\label{app:GinOE_kernel_elts_even}

As an aid to the reader, we list here the explicit forms of the correlation kernel elements for $N$ even in the real Ginibre ensemble. The following section, Appendix \ref{app:GinOE_kernel_elts_odd}, contains the kernel elements for $N$ odd. In Chapter \ref{app:Ginsummed} we list the kernels in summed up form and with the various limits discussed in Chapter \ref{sec:Ginkernelts}. The convention used in this appendix, as in the bulk of this work, is $x,y\in\mathbb{R}$ and $w,z\in\mathbb{R}_2^+$.

Define
\begin{align}
\label{def:Phi} \Phi_{j}(x):=\int_{-\infty}^{\infty}\sgn(x-z)e^{-z^2/2}p_j(z)dz,
\end{align}
then the explicit forms of the correlation kernel elements in Definition \ref{def:GinOE_kernel} are
{\footnotesize
\begin{align}
\nonumber S_{r,r}(x,y)&=e^{-x^2/2}\sum_{k=0}^{\frac{N}{2}-1} \frac{1}{r_k}\Bigl[p_{2k+1}(x)\Phi_{2k}(y)-p_{2k}(x) \Phi_{2k+1}(y)\Bigr],\\
\nonumber S_{r,c}(x,w)&=2ie^{-(x^2+\bar{w}^2)/2}\sqrt{\mathrm{erfc}(\sqrt{2}|\mathrm{Im}(w)|)} \sum_{k=0}^{\frac{N}{2}-1} \frac{1}{r_k}\Bigl[p_{2k}(x)p_{2k+1}(\bar{w})-p_{2k+1}(x)p_{2k}(\bar{w})\Bigr],\\
\nonumber S_{c,r}(w,x)&=e^{-w^2/2}\sqrt{\mathrm{erfc}(\sqrt{2}|\mathrm{Im}(w)|)} \sum_{k=0}^{\frac{N}{2}-1} \frac{1}{r_k}\Bigl[p_{2k+1}(w) \Phi_{2k}(x)-p_{2k}(w) \Phi_{2k+1}(x) \Bigr],\\
\nonumber S_{c,c}(w,z)&=2ie^{-(w^2+\bar{z}^2)/2}\sqrt{\mathrm{erfc}(\sqrt{2}|\mathrm{Im}(w)|)} \sqrt{\mathrm{erfc}(\sqrt{2}| \mathrm{Im}(z)|)}\\
\nonumber &\times \sum_{k=0}^{\frac{N}{2}-1} \frac{1}{r_k}\Bigl[p_{2k}(w)p_{2k+1}(\bar{z})-p_{2k+1} (w)p_{2k}(\bar{z}) \Bigr],\\
\nonumber D_{r,r}(x,y)&=2e^{-(x^2+y^2)/2} \sum_{k=0}^{\frac{N}{2}-1} \frac{1}{r_k}\Bigl[p_{2k}(x) p_{2k+1}(y)- p_{2k+1}(x)p_{2k}(y)\Bigr],\\
\nonumber D_{r,c}(x,w)&=2e^{-(x^2+w^2)/2} \sqrt{\mathrm{erfc}(\sqrt{2}|\mathrm{Im}(w)|)} \sum_{k=0}^{\frac{N}{2}-1} \frac{1}{r_k}\Bigl[p_{2k}(x) p_{2k+1}(w)-p_{2k+1}(x) p_{2k}(w)\Bigr],\\
\nonumber D_{c,r}(x,w)&=2e^{-(w^2+x^2)/2}\sqrt{\mathrm{erfc}(\sqrt{2}| \mathrm{Im}(w)|)} \sum_{k=0}^{\frac{N}{2}-1} \frac{1}{r_k}\Bigl[p_{2k}(w)p_{2k+1}(x)-p_{2k+1}(w)p_{2k}(x)\Bigr],\\
\nonumber D_{c,c}(w,z)&=2e^{-(w^2+z^2)/2} \sqrt{\mathrm{erfc}(\sqrt{2}| \mathrm{Im}(w)|)} \sqrt{\mathrm{erfc} (\sqrt{2}|\mathrm{Im}(z)|)}\\
\nonumber &\times\sum_{k=0}^{\frac{N}{2}-1}\frac{1}{r_k} \Bigl[p_{2k}(w)p_{2k+1}(z)-p_{2k+1}(w) p_{2k}(z) \Bigr],\\
\nonumber \tilde{I}_{r,r}(x,y)&= \frac{1}{2}\sum_{k=0}^{\frac{N}{2}-1}\frac{1}{r_k} \Bigl[\Phi_{2k}(x)\Phi_{2k+1}(y)- \Phi_{2k+1}(x) \Phi_{2k}(y)\Bigr]+\frac{1}{2} \mathrm{sgn}(x-y),\\
\nonumber \tilde{I}_{r,c}(x,w)&= ie^{-\bar{w}^2/2} \sqrt{\mathrm{erfc}(\sqrt{2}| \mathrm{Im}(w)|)} \sum_{k=0}^{\frac{N}{2}-1} \frac{1}{r_k}\Bigl[\Phi_{2k+1}(x)p_{2k}(\bar{w}) -\Phi_{2k}(x)p_{2k+1}(\bar{w})\Bigr],\\
\nonumber \tilde{I}_{c,r}(w,x)&=ie^{-\bar{w}^2/2}\sqrt{\mathrm{erfc} (\sqrt{2}|\mathrm{Im}(w)|)} \sum_{k=0}^{\frac{N}{2}-1}\frac{1}{r_k} \Bigl[p_{2k+1}(\bar{w}) \Phi_{2k}(x)-p_{2k}(\bar{w}) \Phi_{2k+1}(x) \Bigr],\\
\nonumber \tilde{I}_{c,c}(w,z)&=2e^{-(\bar{w}^2+\bar{z}^2)/2}\sqrt{\mathrm{erfc}(\sqrt{2}|\mathrm{Im}(w)|)} \sqrt{\mathrm{erfc}(\sqrt{2}|\mathrm{Im}(z)|)}\\
\nonumber &\times \sum_{k=0}^{\frac{N}{2}-1} \frac{1}{r_k}\Bigl[p_{2k+1}(\bar{w}) p_{2k}(\bar{z})-p_{2k}(\bar{w}) p_{2k+1}(\bar{z})\Bigr].
\end{align}
} 

\subsection{$N$ odd}
\label{app:GinOE_kernel_elts_odd}

The explicit forms of the correlation kernel elements for $N$ odd, of Definition \ref{def:GinOE_kernel_odd}, are
{\small
\begin{align}
\nonumber S_{r,r}^{\odd}(x,y)&=e^{-x^2/2} \sum_{k=0}^{\frac{N-1}{2}-1}\frac{1}{r_k} \Bigl[p_{2k+1}(x) \Phi_{2k}(y)-p_{2k}(x) \Phi_{2k+1}(y)\\
\nonumber &-\frac{\bar{\nu}_{2k+1}} {\bar{\nu}_N}\Bigl(p_{2k+1}(x) \Phi_{N-1}(y)- p_{N-1}(x) \Phi_{2k+1}(y) \Bigr)\\
\nonumber &+\frac{\bar{\nu}_{2k+2}}{\bar{\nu}_N} \Bigl(p_{2k}(x) \Phi_{N-1}(y)-p_{N-1}(x) \Phi_{2k}(y)\Bigr)\Bigr] +\frac{e^{-x^2/2}}{\bar{\nu}_N} p_{N-1}(x),\\
\nonumber S_{r,c}^{\odd}(x,w)&=2ie^{-(x^2+\bar{w}^2)/2} \sqrt{\mathrm{erfc}(\sqrt{2}|\mathrm{Im} (w)|)}\\
\nonumber &\times\sum_{k=0}^{\frac{N-1}{2}-1}\frac{1}{r_k} \Bigl[p_{2k}(x) p_{2k+1}(\bar{w})-p_{2k+1}(x) p_{2k} (\bar{w})\\
\nonumber &-\frac{\bar{\nu}_{2k+1}}{\bar{\nu}_N} \Bigl(p_{N-1}(x) p_{2k+1}(\bar{w})- p_{2k+1}(x) p_{N-1}(\bar{w}) \Bigr)\\
\nonumber &+ \frac{\bar{\nu}_{2k+2}}{\bar{\nu}_N} \Bigl(p_{N-1}(x)p_{2k} (\bar{w})-p_{2k}(x) p_{N-1}(\bar{w}) \Bigr)\Bigr],\\
\nonumber S_{c,r}^{\odd}(w,x)&= e^{-w^2/2}\sqrt{\mathrm{erfc} (\sqrt{2}|\mathrm{Im}(w)|)}\\
\nonumber &\times \sum_{k=0}^{\frac{N-1}{2}-1} \frac{1}{r_k} \Bigl[p_{2k+1}(w)\Phi_{2k}(x)- p_{2k}(w) \Phi_{2k+1}(x)\\
\nonumber &-\frac{\bar{\nu}_{2k+1}}{\bar{\nu}_N} \Bigl(p_{2k+1}(w) \Phi_{N-1}(x)- p_{N-1}(w) \Phi_{2k+1}(x)\Bigr)\\
\nonumber &+\frac{\bar{\nu}_{2k+2}} {\bar{\nu}_N} \Bigl(p_{2k}(w)\Phi_{N-1}(x)- p_{N-1}(w)\Phi_{2k}(x) \Bigr)\Bigr]\\
\nonumber &+\frac{e^{-w^2/2}\sqrt{\mathrm{erfc} (\sqrt{2}|\mathrm{Im}(w)|)}}{\bar{\nu}_N} p_{N-1}(w),\\
\nonumber S_{c,c}^{\odd}(w,z)&=2ie^{-(w^2+ \bar{z}^2)/2} \sqrt{\mathrm{erfc} (\sqrt{2}|\mathrm{Im}(w)|)} \sqrt{\mathrm{erfc} (\sqrt{2}| \mathrm{Im}(z)|)}\\
\nonumber &\times\sum_{k=0}^{\frac{N-1}{2}-1} \frac{1}{r_k} \Bigl[p_{2k}(w) p_{2k+1}(\bar{z})-p_{2k+1}(w) p_{2k}(\bar{z})\\
\nonumber & -\frac{\bar{\nu}_{2k+1}} {\bar{\nu}_N}\Bigl(p_{N-1}(w) p_{2k+1}(\bar{z})- p_{2k+1}(w) p_{N-1}(\bar{z})\Bigr)\\
\nonumber &+\frac{\bar{\nu}_{2k+2}} {\bar{\nu}_N}\Bigl(p_{N-1}(w) p_{2k}(\bar{z})- p_{2k}(w) p_{N-1}(\bar{z}) \Bigr)\Bigr],\\
\nonumber D_{r,r}^{\odd}(x,y)&=2e^{-(x^2+y^2)/2} \sum_{k=0}^{\frac{N-1}{2}-1} \frac{1}{r_k} \Bigl[p_{2k}(x) p_{2k+1}(y)- p_{2k+1}(x) p_{2k}(y)\\
\nonumber &-\frac{\bar{\nu}_{2k+1}}{\bar{\nu}_N} \Bigl(p_{N-1}(x)p_{2k+1}(y)- p_{2k+1}(x)p_{N-1}(y)\Bigr)\\
\nonumber &+\frac{\bar{\nu}_{2k+2}} {\bar{\nu}_N} \Bigl(p_{N-1}(x)p_{2k}(y) -p_{2k}(x)p_{N-1}(y)\Bigr) \Bigr],\\
\nonumber D_{r,c}^{\odd}(x,w)&=2e^{-(x^2+w^2)/2} \sqrt{\mathrm{erfc}(\sqrt{2}| \mathrm{Im}(w)|)}\\
\nonumber &\times\sum_{k=0}^{\frac{N-1}{2}-1}\frac{1}{r_k} \Bigl[p_{2k}(x) p_{2k+1}(w)- p_{2k+1}(x) p_{2k}(w)\\
\nonumber &-\frac{\bar{\nu}_{2k+1}}{\bar{\nu}_N} \Bigl(p_{N-1}(x)p_{2k+1}(w)- p_{2k+1}(x) p_{N-1}(w)\Bigr)\\
\nonumber &+\frac{\bar{\nu}_{2k+2}}{\bar{\nu}_N} \Bigl(p_{N-1}(x)p_{2k}(w)- p_{2k}(x)p_{N-1}(w)\Bigr) \Bigr],\\
\nonumber D_{c,r}^{\odd}(w,x)&=2e^{-(w^2+x^2)/2} \sqrt{\mathrm{erfc}(\sqrt{2}| \mathrm{Im}(w)|)}\\
\nonumber &\times\sum_{k=0}^{ \frac{N-1}{2}-1}\frac{1}{r_k} \Bigl[p_{2k}(w)p_{2k+1}(x) -p_{2k+1}(w) p_{2k}(x)\\
\nonumber &-\frac{\bar{\nu}_{2k+1}}{\bar{\nu}_N} \Bigl(p_{N-1}(w) p_{2k+1}(x)-p_{2k+1}(w)p_{N-1}(x) \Bigr)\\
\nonumber &+\frac{\bar{\nu}_{2k+2}}{\bar{\nu}_N}\Bigl(p_{N-1}(w) p_{2k}(x)-p_{2k}(w)p_{N-1}(x) \Bigr)\Bigr],\\
\nonumber D_{c,c}^{\odd}(w,z)&=2e^{-(w^2+z^2)/2}\sqrt{ \mathrm{erfc}(\sqrt{2}|\mathrm{Im}(w)|)} \sqrt{\mathrm{erfc}(\sqrt{2}|\mathrm{Im}(z)|)}\\
\nonumber &\times\sum_{k=0}^{\frac{N-1}{2}-1}\frac{1}{r_k} \Bigl[p_{2k}(w) p_{2k+1}(z)-p_{2k+1}(w) p_{2k}(z)\\
\nonumber &-\frac{\bar{\nu}_{2k+1}}{\bar{\nu}_N} \Bigl(p_{N-1}(w)p_{2k+1}(z)- p_{2k+1}(w)p_{N-1}(z) \Bigr)\\
\nonumber &+\frac{\bar{\nu}_{2k+2}}{\bar{\nu}_N} \Bigl(p_{N-1}(w)p_{2k}(z)- p_{2k}(w)p_{N-1}(z) \Bigr)\Bigr],\\
\nonumber \tilde{I}_{r,r}^{\odd}(x,y)&=\frac{1}{2}\sum_{k=0}^{\frac{N-1}{2}-1} \frac{1}{r_k} \Bigl[\Phi_{2k}(x) \Phi_{2k+1}(y)- \Phi_{2k+1}(x)\Phi_{2k}(y)\\
\nonumber &-\frac{\bar{\nu}_{2k+1}}{\bar{\nu}_N} \Bigl(\Phi_{N-1}(x)\Phi_{2k+1}(y)- \Phi_{2k+1}(x) \Phi_{N-1}(y) \Bigr)\\
\nonumber &+\frac{\bar{\nu}_{2k+2}} {\bar{\nu}_N}\Bigl(\Phi_{N-1}(x) \Phi_{2k}(y)- \Phi_{2k}(x) \Phi_{N-1}(y) \Bigr) \Bigr]\\
\nonumber &+\frac{1}{2}\mathrm{sgn}(x-y)+\frac{\Phi_{N-1}(x)- \Phi_{N-1}(y)} {2\bar{\nu}_N},\\
\nonumber \tilde{I}_{r,c}^{\odd}(x,w)&= ie^{-\bar{w}^2/2}\sqrt{ \mathrm{erfc}(\sqrt{2}|\mathrm{Im}(w)|)}\\
\nonumber &\times\sum_{k=0}^{\frac{N-1}{2}-1} \frac{1}{r_k}\Bigl[\Phi_{2k+1}(x) p_{2k}(\bar{w}) -\Phi_{2k}(x)p_{2k+1} (\bar{w})\\
\nonumber &-\frac{\bar{\nu}_{2k+1}}{\bar{\nu}_N} \Bigl(\Phi_{2k+1}(x)p_{N-1}(\bar{w}) -\Phi_{N-1}(x) p_{2k+1}(\bar{w}) \Bigr)\\
\nonumber &+\frac{\bar{\nu}_{2k+2}}{\bar{\nu}_N} \Bigl(\Phi_{2k}(x)p_{N-1}(\bar{w})- \Phi_{N-1}(x) p_{2k}(\bar{w}) \Bigr)\Bigr]\\
\nonumber &-i\:\frac{e^{-\bar{w}^2/2} \sqrt{\mathrm{erfc}(\sqrt{2}| \mathrm{Im}(w)|)}}{\bar{\nu}_N} p_{N-1}(\bar{w}),\\
\nonumber \tilde{I}_{c,r}^{\odd}(w,x)&= ie^{-\bar{w}^2/2} \sqrt{\mathrm{erfc}(\sqrt{2}| \mathrm{Im} (w)|)}\\
\nonumber &\times\sum_{k=0}^{\frac{N-1}{2}-1}\frac{1}{r_k} \Bigl[p_{2k+1}(\bar{w}) \Phi_{2k}(x)-p_{2k}(\bar{w})\Phi_{2k+1}(x)\\
\nonumber &-\frac{\bar{\nu}_{2k+1}} {\bar{\nu}_N} \Bigl(p_{2k+1}(\bar{w}) \Phi_{N-1}(x)- p_{N-1}(\bar{w}) \Phi_{2k+1}(x)\Bigr)\\
\nonumber &+\frac{\bar{\nu}_{2k+2}}{\bar{\nu}_N} \Bigl(p_{2k}(\bar{w})\Phi_{N-1}(x) -p_{N-1}(\bar{w}) \Phi_{2k}(x)\Bigr)\Bigr]\\
\nonumber &+i\hspace{2pt}\frac{e^{-\bar{w}^2/2} \sqrt{\mathrm{erfc}(\sqrt{2}| \mathrm{Im}(w)|)}}{\bar{\nu}_N} p_{N-1}(\bar{w}),\\
\nonumber \tilde{I}_{c,c}^{\odd}(w,z)&=2e^{-(\bar{w}^2+\bar{z}^2)/2} \sqrt{\mathrm{erfc} (\sqrt{2}|\mathrm{Im}(w)|)} \sqrt{\mathrm{erfc}(\sqrt{2}|\mathrm{Im}(z)|)}\\
\nonumber &\times\sum_{k=0}^{\frac{N-1}{2}-1}\frac{1}{r_k} \Bigl[p_{2k+1}(\bar{w}) _{2k}(\bar{z})-p_{2k}(\bar{w}) p_{2k+1}(\bar{z})\\
\nonumber& -\frac{\bar{\nu}_{2k+1}}{\bar{\nu}_N}\Bigl(p_{2k+1}(\bar{w}) p_{N-1}(\bar{z})- p_{N-1}(\bar{w}) p_{2k+1}(\bar{z}) \Bigr)\\
\nonumber &+\frac{\bar{\nu}_{2k+2}}{\bar{\nu}_N} \Bigl(p_{2k}(\bar{w}) p_{N-1}(\bar{z})- p_{N-1}(\bar{w}) p_{2k}(\bar{z}) \Bigr)\Bigr],
\end{align}
}with $\bar{\nu}_j$ as in Definition \ref{def:GinOE_kernel_odd} and $\Phi_j(x)$ as in (\ref{def:Phi}).

\subsection{Summed and limiting forms of the real Ginibre correlation kernel elements}
\label{app:Ginsummed}

In Chapter \ref{sec:Ginkernelts} we discussed the summed and limiting forms of the correlation kernel elements $S(\mu,\eta)$; here we write out the summed forms of every kernel element, which are contained in \cite[Theorems 8, 10 \& 12]{b&s2009}. Note that they restrict their results to $N$ even, while we allow $N$ to be either even or odd (except for $\tilde{I}_{r,r}(x,y)$, where we give the even and odd forms explicitly).

Starting with the case of finite $N$, we have the summed forms of the kernel elements from Definitions \ref{def:GinOE_kernel} and \ref{def:GinOE_kernel_odd}
{\small
\begin{align}
\nonumber S_{r,r}(x,y)&=\frac{1}{\sqrt{2\pi}}\left[e^{-(x-y)^2/2}\frac{\Gamma(N-1,xy)}{\Gamma(N-1)}+ 2^{(N-3)/2}e^{-x^2/2} x^{N-1}\sgn(y) \frac{\gamma(\frac{N-1}{2},y^2/2)}{\Gamma(N-1)} \right],\\
\nonumber S_{r,c}(x,w)&=\frac{ie^{-(x-\bar{w})^2/2}}{\sqrt{2\pi}} (\bar{w}-x)\frac{\Gamma(N-1,x\bar{w})} {\Gamma(N-1)} \sqrt{\mathrm{erfc}(\sqrt{2}| \mathrm{Im}(w)|)},\\
\nonumber S_{c,r}(w,x)&=\frac{1}{\sqrt{2\pi}}\left[e^{-(w-x)^2/2} \frac{\Gamma(N-1,wx)}{\Gamma(N-1)}+2^{(N-3)/2} e^{-w^2/2}w^{N-1} \sgn(x)\frac{\gamma( \frac{N-1}{2},x^2/2)}{\Gamma(N-1)} \right]\\
\nonumber &\times \sqrt{\mathrm{erfc}(\sqrt{2}| \mathrm{Im}(w)|)},\\
\nonumber S_{c,c}(w,z)&=\frac{ie^{-(w-\bar{z})^2/2}} {\sqrt{2\pi}} (\bar{z}-w) \frac{\Gamma(N-1, w\bar{z})} {\Gamma(N-1)}\sqrt{\mathrm{erfc} (\sqrt{2}|\mathrm{Im}(w)|)} \sqrt{\mathrm{erfc}(\sqrt{2}| \mathrm{Im}(z)|)},\\
\nonumber D_{r,r}(x,y)&=\frac{e^{-(x-y)^2/2}} {\sqrt{2\pi}}(y-x) \frac{\Gamma(N-1,xy)}{\Gamma(N -1)},\\
\nonumber D_{r,c}(x,w)&=\frac{e^{-(x-w)^2/2}}{\sqrt{2\pi}} \sqrt{\mathrm{erfc}(\sqrt{2}|\mathrm{Im} (w)|)} \;(w-x) \frac{\Gamma(N-1,xw)} {\Gamma(N-1)},\\
\nonumber D_{c,r}(w,x)&=\frac{e^{-(w-x)^2/2}}{\sqrt{2\pi}}\sqrt{\mathrm{erfc} (\sqrt{2}|\mathrm{Im} (w)|)} \;(x-w) \frac{\Gamma(N-1,wx)} {\Gamma(N-1)},\\
\nonumber D_{c,c}(w,z)&=\frac{e^{-(w-z)^2/2}} {\sqrt{2\pi}} \sqrt{\mathrm{erfc} (\sqrt{2}|\mathrm{Im}(w)|)} \sqrt{\mathrm{erfc}(\sqrt{2}|\mathrm{Im}(z)|)} \;(z-w)\frac{\Gamma(N-1, wz)}{\Gamma(N -1)},\\
\nonumber \tilde{I}_{r,r}(x,y)&=\frac{e^{-x^2/2}}{\sqrt{2\pi}}\sum_{k=0}^{N/2-1} \frac{x^{2k}}{(2k)!} \; 2^{(2k-1)/2} \gamma \left(\frac{2k+1} {2}, \frac{y^2} {2} \right)\\
\nonumber &-\frac{e^{-y^2/2}}{\sqrt{2\pi}} \sum_{k=0}^{N/2-1} \frac{y^{2k}}{(2k)!} \; 2^{(2k-1)/2} \gamma \left(\frac{2k+1} {2}, \frac{x^2} {2} \right) + \frac{1}{2}\sgn(x-y),\\
\nonumber \tilde{I}_{r,r}^{\odd} (x,y)&=\frac{e^{-x^2/2}}{\sqrt{2\pi}}\sum_{k=0}^{(N-1)/2-1} \frac{x^{2k}}{(2k)!} \left( 2^{(2k-1)/2} \gamma \left(\frac{2k+1} {2}, \frac{y^2} {2} \right) - \frac{\bar{\nu}_{2k+1}} {\bar{\nu}_N} \; 2^{N/2-1} \gamma \left(N/2, y^2/2 \right) \right)\\
\nonumber &-\frac{e^{-y^2/2}}{\sqrt{2\pi}} \sum_{k=0}^{(N-1)/2-1} \frac{y^{2k}}{(2k)!} \left( 2^{(2k-1)/2} \gamma \left(\frac{2k+1} {2}, \frac{x^2} {2} \right) - \frac{\bar{\nu}_{2k+1}} {\bar{\nu}_N} \; 2^{N/2-1} \gamma \left(N/2, x^2/2 \right) \right)\\
\nonumber &+ \frac{2^{N/2-1}}{\bar{\nu}_N}\Big( \gamma \left(N/2, x^2/2 \right) -\gamma \left(N/2, y^2/2 \right) \Big) + \frac{1}{2}\sgn(x-y),\\
\nonumber \tilde{I}_{r,c}(x,w)&=\frac{-i}{\sqrt{2\pi}} \left[e^{-(x-\bar{w})^2/2} \frac{\Gamma(N-1, x\bar{w})} {\Gamma(N-1)}+ 2^{(N-3)/2} e^{-\bar{w}^2/2} \bar{w}^{N-1} \sgn(x)\frac{\gamma( \frac{N-1} {2}, x^2/2)}{ \Gamma(N-1)} \right]\\
\nonumber &\times \sqrt{\mathrm{erfc}(\sqrt{2}| \mathrm{Im}(w)|)},\\
\nonumber \tilde{I}_{c,r}(w,x)&=\frac{i}{\sqrt{2\pi}}\left[e^{-(\bar{w}-x)^2/2} \frac{\Gamma(N-1,\bar{w}x)} {\Gamma(N-1)}+ 2^{(N-3)/2} e^{-\bar{w}^2/2} \bar{w}^{N-1} \sgn(x) \frac{ \gamma(\frac{N-1} {2}, x^2/2)}{\Gamma(N-1)}\right],\\
\nonumber &\times\sqrt{\mathrm{erfc}(\sqrt{2}| \mathrm{Im}(w)|)},\\
\nonumber \tilde{I}_{c,c}(w,z)&=\frac{ie^{-(\bar{w}-\bar{z})^2/2}} {\sqrt{2\pi}}(\bar{w}-\bar{z}) \frac{\Gamma(N-1,\bar{w} \bar{z})}{\Gamma(N-1)} \sqrt{\mathrm{erfc} (\sqrt{2}|\mathrm{Im}(w)|)}\sqrt{ \mathrm{erfc}(\sqrt{2}| \mathrm{Im}(z)|)},
\end{align}
}which includes the equations from (\ref{eqn:Ginsummed}) for completeness.

The limiting kernels at the origin (or anywhere in the bulk near the real line) are
\begin{align}
\nonumber \bK_{rr}^{\bulk}(x,y)&=\left[\begin{array}{cc}
\frac{1}{\sqrt{2\pi}}e^{-(x-y)^2/2} & \frac{1}{\sqrt{2\pi}}(x-y) e^{-(x-y)^2/2}\\
\frac{1}{2}\sgn(x-y) \erfc(|x-y|/ \sqrt{2}) & \frac{1}{\sqrt{2\pi}}e^{-(x-y)^2/2}
\end{array}\right],\\
\nonumber \bK_{rc}^{\bulk}(x,w)&=\sqrt{\frac{\mathrm{erfc}(\sqrt{2}| \mathrm{Im}(z)|)}{2\pi}} \left[\begin{array}{cc}
i(\bar{w}-x) e^{-(x-\bar{w})^2/2} & (x-w)e^{-(x-w)^2/2}\\
-ie^{-(x-\bar{w})^2/2} & e^{-(x-z)^2/2}
\end{array}\right],\\
\label{eqn:Ginbulkall} \bK_{cc}^{\bulk}(w,z)&=\sqrt{\frac{\mathrm{erfc} (\sqrt{2}|\mathrm{Im}(w)|) \mathrm{erfc} (\sqrt{2}|\mathrm{Im}(z)|)} {2\pi}}\\
\nonumber &\times\left[\begin{array}{cc}
i(\bar{z}-w)e^{-(w-\bar{z})^2/2} & (w-z) e^{-(w-z)^2/2}\\
(\bar{w}-\bar{z})e^{-(\bar{w}-\bar{z})^2/2} & i(\bar{w}-z) e^{-(\bar{w}-z)^2/2}
\end{array}\right],
\end{align}
recalling that the block structure of the quaternion determinant correlation kernel is given by (\ref{def:GinOE_correlnK}).
\begin{remark}
We have omitted the complex-real kernel since this is obtained from the real-complex kernel by the necessary anti-symmetry of the Pfaffian. Also note that in \cite{b&s2009} the kernels are written as Pfaffian kernels but here they are kernels of quaternion determinants; the conversion is in (\ref{def:GinPfK}).
\end{remark}

For the real edge here we quote the result of \cite[Theorem 12]{b&s2009} (although we allow $N$ to be odd), which gives slightly more general forms than those of (\ref{eqn:Ginedge}). The extra generality comes from the inclusion of the parameter $u=\pm 1$, which essentially lets you shift to the positive or negative real edge. If we redefine $X_j:=u\sqrt{N}+r_j$ and likewise for $Y,W,Z$ then
\begin{align}
\nonumber S^{\mathrm{edge}}_{r,r}(X,Y)&=\frac{1}{\sqrt{2\pi}}\left[ \frac{e^{-(X-Y)^2/2}}{2} \erfc\Big(u \frac{X+Y}{\sqrt{2}}\Big)+ \frac{e^{-X^2}}{2\sqrt{2}}(1+\erf \:uY)\right],\\
\nonumber S^{\mathrm{edge}}_{r,c}(X,W)&=\frac{i}{2\sqrt{2\pi}}\sqrt{\mathrm{erfc} (\sqrt{2}|\mathrm{Im}(W)|)} \;(\overline{W}-X)e^{-(X-\overline{W})^2/2} \erfc \Big( u\frac{X+ \overline{W}}{\sqrt{2}}\Big),\\
\nonumber S^{\mathrm{edge}}_{c,r}(W,X)&= \frac{1}{\sqrt{2\pi}}\Big[ \frac{e^{-(W-X)^2/2}}{2} \sqrt{\mathrm{erfc} (\sqrt{2}|\mathrm{Im}(W)|)}\;\erfc \Big( u\frac{X+W}{\sqrt{2}}\Big)\\
\nonumber &+\frac{e^{-W^2}}{2\sqrt{2}} (1+\erf\:u X)\Big],\\
\nonumber S^{\mathrm{edge}}_{c,c}(W,Z)&= \frac{i}{2\sqrt{2\pi}}\sqrt{\mathrm{erfc} (\sqrt{2}|\mathrm{Im}(W)|)} \sqrt{\mathrm{erfc} (\sqrt{2}|\mathrm{Im}(Z)|)}\\
\nonumber &\times(\overline{Z}-W) e^{-(\overline{Z}-W)^2/2} \erfc \Big( u\frac{W+ \overline{Z}}{\sqrt{2}} \Big),\\
\nonumber D_{r,r}^{\edge}(X,Y)&=\frac{e^{-(X-Y)^2/2}}{2\sqrt{2\pi}} (Y-X)\erfc\Big(u\frac{X+Y}{\sqrt{2}}\Big),\\
\nonumber D_{r,c}^{\edge}(X,W)&=\frac{e^{-(X-W)^2/2}}{2\sqrt{2\pi}}\sqrt{\mathrm{erfc}(\sqrt{2}|\mathrm{Im}(W)|)}\;(W-X)\erfc\Big(u\frac{X+W}{\sqrt{2}}\Big),\\
\nonumber D_{c,r}^{\edge}(W,X)&=\frac{e^{-(W-X)^2/2}}{2\sqrt{2\pi}}\sqrt{\mathrm{erfc}(\sqrt{2}|\mathrm{Im}(W)|)}\;(X-W)\erfc\Big(u\frac{W+X}{\sqrt{2}}\Big),\\
\nonumber D_{c,c}^{\edge}(W,Z)&=\frac{e^{-(W-Z)^2/2}}{2\sqrt{2\pi}}\sqrt{\mathrm{erfc}(\sqrt{2}|\mathrm{Im}(W)|)}\sqrt{\mathrm{erfc}(\sqrt{2}|\mathrm{Im}(Z)|)}\\
\nonumber &\times (Z-W)\;\erfc\Big(u\frac{W+Z}{\sqrt{2}}\Big),\\
\nonumber \tilde{I}_{r,r}^{\edge}(X,Y)&=\frac{\sgn(X-Y)}{2}\erfc\Bigg(\frac{|X-Y|}{\sqrt{2}}\Bigg),\\
\nonumber \tilde{I}_{r,c}^{\edge}(X,W)&=\frac{-ie^{-(X-\overline{W})^2/2}}{2\sqrt{2\pi}}\sqrt{\mathrm{erfc}(\sqrt{2}|\mathrm{Im}(W)|)}\;\erfc\Bigg(u\frac{X+W}{\sqrt{2}}\Bigg)\\
\nonumber &-\frac{ie^{\overline{W}^2}}{4\sqrt{\pi}}(1+\erf\;uX),\\
\nonumber \tilde{I}_{c,r}^{\edge}(W,X)&=\frac{ie^{-(\overline{W}-X)^2/2}}{2\sqrt{2\pi}}\sqrt{\mathrm{erfc}(\sqrt{2}|\mathrm{Im}(W)|)}\;\erfc\Bigg(u\frac{W+X}{\sqrt{2}}\Bigg)\\
\nonumber &+\frac{ie^{\overline{W}^2}}{4\sqrt{\pi}}(1+\erf\;u X),\\
\nonumber \tilde{I}^{\mathrm{edge}}_{c,c}(W,Z)&=\frac{1}{2\sqrt{2\pi}}\sqrt{\mathrm{erfc}(\sqrt{2}|\mathrm{Im}(W)|)}\sqrt{\mathrm{erfc}(\sqrt{2}|\mathrm{Im}(Z)|)}\\
\label{eqn:Ginedgeall} &\times(\overline{W}-\overline{Z}) e^{-(\overline{W} -\overline{Z})^2/2} \erfc \Big( u\frac{\overline{W}+\overline{Z}}{\sqrt{2}}\Big).
\end{align}

\section{Probability of $k$ real eigenvalues for the partially symmetric real Ginibre ensemble}
\label{app:tGinsimpnk}

\begin{longtable}{|c|c|c|c|}
\hline
$\tau=1/2$ & $\mathrm{Exact}\; p_{N,k,-1/2}$ & $\mathrm{Decimal} \; p_{N,k,-1/2}$ & Simulation\\
\hline
\hline
\endfirsthead

\hline
$\tau=1/2$ & $\mathrm{Exact}\; p_{N,k,-1/2}$ & $\mathrm{Decimal} \; p_{N,k,-1/2}$ & Simulation\\
\hline
\hline
\endhead

$\T p_{2,2}$&$\frac{\sqrt{3}}{2}$&$0.86603$&$0.8669$\\
$\T p_{2,0}$&$1-\frac{\sqrt{3}}{2}$&$0.13397$&$0.1331$\\
\hline
$\T p_{3,3}$&$\frac{3\sqrt{3}}{8}$&$0.649519$&$0.6462$\\
$\T p_{3,1}$&$1-\frac{3\sqrt{3}}{8}$&$0.350481$&$0.3538$\\
\hline
$\T p_{4,4}$&$\frac{27}{64}$&$0.42188$&$0.4216$\\
$\T p_{4,2}$&$\frac{51\sqrt{3}}{64}-\frac{27}{32}$&$0.53648$&$0.5367$\\
$\T p_{4,0}$&$\frac{91}{64}-\frac{51\sqrt{3}}{64}$&$0.04165$&$0.0417$\\
\hline
$\T p_{5,5}$&$\frac{243}{1024}$&$0.237305$&$0.239$\\
$\T p_{5,3}$&$\frac{159\sqrt{3}}{256}-\frac{243}{512}$&$0.601157$&$0.5991$\\
$\T p_{5,1}$&$\frac{1267}{1024}-\frac{159\sqrt{3}}{256}$&$0.161539$&$0.1619$\\
\hline
$\T p_{6,6}$&$\frac{2187\sqrt{3}}{32768}$&$0.1156$&$0.1124$\\
$\T p_{6,4}$&$\frac{115317}{131072}-\frac{6561\sqrt{3}}{32768}$&$0.532$&$0.5377$\\
$\T p_{6,2}$&$\frac{39609}{32768}-\frac{115317}{65536}$&$0.33406$&$0.332$\\
$\T p_{6,0}$&$\frac{246389}{131072}-\frac{35235\sqrt{3}}{32768}$&$0.01735$&$0.0179$\\
\hline
\caption[Comparison of analytic and simulated probabilities $p_{N,k,1/2}$ for the partially symmetric real Ginibre ensemble.]{Exact values of $p_{N,k,1/2}$ for the partially symmetric real Ginibre ensemble with $\tau=1/2$. These exact values are compared with the simulated results of $10000$ independent matrices.}
\label{tab:pnkxact_simp}
\end{longtable}

\begin{longtable}{|c|c|c|c|}
\hline
$\tau=-1/2$ & $\mathrm{Exact}\; p_{N,k,-1/2}$ & $\mathrm{Decimal} \; p_{N,k,-1/2}$ & Simulation\\
\hline
\hline
\endfirsthead

\hline
$\tau=-1/2$ & $\mathrm{Exact}\; p_{N,k,-1/2}$ & $\mathrm{Decimal} \; p_{N,k,-1/2}$ & Simulation\\
\hline
\hline
\endhead

$\T p_{2,2}$&$\frac{1}{2}$&$0.5$&$0.4994$\\
$\T p_{2,0}$&$\frac{1}{2}$&$0.5$&$0.5006$\\
\hline
$\T p_{3,3}$&$\frac{1}{8}$&$0.125$&$0.1248$\\
$\T p_{3,1}$&$\frac{7}{8}$&$0.875$&$0.8752$\\
\hline
$\T p_{4,4}$&$\frac{1}{64}$&$0.01563$&$0.0136$\\
$\T p_{4,2}$&$\frac{41}{64}$&$0.64063$&$0.6455$\\
$\T p_{4,0}$&$\frac{11}{32}$&$0.34375$&$0.3409$\\
\hline
$\T p_{5,5}$&$\frac{1}{1024}$&$0.00098$&$0.006$\\
$\T p_{5,3}$&$\frac{121}{512}$&$0.23633$&$0.2368$\\
$\T p_{5,1}$&$\frac{781}{1024}$&$0.7627$&$0.7626$\\
\hline
$\T p_{6,6}$&$\frac{1}{32768}$&$0.00003$&$0.0001$\\
$\T p_{6,4}$&$\frac{5907}{131072}$&$0.04507$&$0.0462$\\
$\T p_{6,2}$&$\frac{45591}{65536}$&$0.69566$&$0.6955$\\
$\T p_{6,0}$&$\frac{33979}{131072}$&$0.25924$&$0.2582$\\
\hline
\caption[Comparison of analytic and simulated probabilities $p_{N,k,-1/2}$ for the partially symmetric real Ginibre ensemble.]{Exact values of $p_{N,k,-1/2}$ for the partially symmetric real Ginibre ensemble with $\tau=-1/2$. These exact values are compared with the simulated results of $10000$ independent matrices.}
\label{tab:pnkxact_simm}
\end{longtable}

\section{Probability of $k$ real eigenvalues for the real spherical ensemble}
\label{app:Sprobs}

\begin{longtable}{|c|c|c|c|}
\hline
&$\mathrm{Exact}\; p_{N,k}$&$\mathrm{Decimal} \;p_{N,k}$& Simulation\\
\hline
\hline
\endfirsthead

\hline
&$\mathrm{Exact}\; p_{N,k}$&$\mathrm{Decimal} \;p_{N,k}$& Simulation\\
\hline
\hline
\endhead

$p_{2,2}$ \T &$\frac{1}{4}\pi$&$0.785398$&$0.78691$\\
$p_{2,0}$ \T &$1-\frac{1}{4}\pi$&$0.214602$&$0.21309$\\
\hline
$p_{3,3}$ &$\frac{1}{2}$ \T &$0.5$&$0.50051$\\
$p_{3,1}$ &$\frac{1}{2}$ \T &$0.5$&$0.49949$\\
\hline
$p_{4,4}$&$\frac{27}{1024}\pi^2$ \T &$0.260234$&$0.25705$\\
$p_{4,2}$&$\frac{3}{8}\pi-\frac{27}{512}\pi^2$ \T &$0.65763$&$0.66053$\\
$p_{4,0}$&$1-\frac{3}{8}\pi+\frac{27}{1024}\pi^2$ \T &$0.0821365$&$0.08242$\\
\hline
$p_{5,5}$&$\frac{1}{9}$ \T &$0.111111$&$0.11167$\\
$p_{5,3}$&$\frac{11}{18}$ \T &$0.611111$&$0.60969$\\
$p_{5,1}$&$\frac{5}{18}$ \T &$0.277778$&$0.27864$\\
\hline
$p_{6,6}$&$\frac{84375}{67108864}\pi^3$ \T &$0.0389837$&$0.03898$\\
$p_{6,4}$&$\frac{14625}{262144}\pi^2-\frac{253125}{67108864}\pi^3$ \T &$0.433673$&$0.43216$\\
$p_{6,2}$&$\frac{15}{32}\pi-\frac{14625}{131072}\pi^2+\frac{253125}{67108864}\pi^3$ \T &$0.488323$&$0.48873$\\
$p_{6,0}$&$1-\frac{15}{32}\pi+\frac{14625}{262144}\pi^2-\frac{84375}{67108864}\pi^3$ \T &$0.0390194$&$0.04013$\\
\hline
$p_{7,7}$&$\frac{9}{800}$ \T &$0.01125$&$0.01178$\\
$p_{7,5}$&$\frac{39}{160}$ \T &$0.24375$&$0.24244$\\
$p_{7,3}$&$\frac{463}{800}$ \T &$0.57875$&$0.57933$\\
$p_{7,1}$&$\frac{133}{800}$ \T &$0.16625$&$0.16645$\\
\hline
\caption[Comparison of analytic and simulated probabilities $p_{N,k}$ for the real spherical ensemble.]{Calculations of $p_{N,k}$, the probability of finding $k$ real eigenvalues from an $N\times N$ matrix $\bY= \bA^{-1}\bB$. The left column is the analytic calculation, the middle column is the analytic calculation in decimal. These are compared to the right column, which contains the results of a numerical simulation of 100,000 matrices.}\label{table:Spnk}
\end{longtable}

\section{Probability of $k$ real eigenvalues for the real truncated ensemble}
\label{app:TOEpMks}

\begin{longtable}{|c|c|c|c|}
\hline
$L=1$ & $\mathrm{Exact}\; p_{M,k}$ & $\mathrm{Decimal}\; p_{M,k}$ & Simulation\\
\hline
\hline
\endfirsthead

\hline
$L=1$ & $\mathrm{Exact}\; p_{M,k}$ & $\mathrm{Decimal}\; p_{M,k}$ & Simulation\\
\hline
\hline
\endhead

$\T p_{2,2}$&$\frac{2}{\pi}$&$0.63662$&$0.63436$\\
$\T p_{2,0}$&$1-\frac{2}{\pi}$&$0.36338$&$0.36561$\\
\hline
$\T p_{3,3}$&$\frac{2}{3\pi}$&$0.21221$&$0.2127$\\
$\T p_{3,1}$&$1-\frac{2}{3\pi}$&$0.78780$&$0.7873$\\
\hline
$\T p_{4,4}$&$\frac{16}{45 \pi^2}$&$0.03603$&$0.03572$\\
\hline
$\T p_{4,2}$&$\frac{12}{5\pi}- \frac{32}{45\pi^2}$&$0.69190$&$0.69057$\\
$\T p_{4,0}$&$1+\frac{16}{45\pi^2}- \frac{12}{5\pi}$&$0.27209$&$0.27371$\\
\hline
$\T p_{5,5}$&$\frac{16}{525\pi^2}$&$0.00309$&$0.00298$\\
$\T p_{5,3}$&$\frac{20}{21\pi}-\frac{32}{525\pi^2}$&$0.29700$&$0.295$\\
$\T p_{5,1}$&$1+\frac{16}{525\pi^2}- \frac{20}{21\pi}$&$0.69994$&$0.70202$\\
\hline
$\T p_{6,6}$&$\frac{2048}{496125\pi^3}$&$0.00013$&$0.00015$\\
$\T p_{6,4}$&$\frac{7136}{11025\pi^2}- \frac{2048}{165375\pi^3}$&$0.06518$&$0.0656$\\
$\T p_{6,2}$&$\frac{118}{45\pi}-\frac{14272}{11025\pi^2}+ \frac{2048}{165375\pi^3}$&$0.70392$&$0.70276$\\
$\T p_{6,0}$&$1- \frac{2048}{496125\pi^3} +\frac{7136}{11025\pi^2}- \frac{118}{45\pi}$&$0.23077$&$0.23149$\\
\hline
\caption[Comparison of analytic and simulated probabilities $p_{M,k}$ for the real truncated ensemble with $L=1$.]{Exact values of $p_{M,k}$ for the real truncated ensemble with $L=1$. These exact values are compared with the simulated results of $10000$ independent matrices.}
\end{longtable}

\begin{longtable}{|c|c|c|c|}
\hline
$L=2$ & $\mathrm{Exact}\; p_{M,k}$ & $\mathrm{Decimal}\; p_{M,k}$ & Simulation\\
\hline
\hline
\endfirsthead

\hline
$L=2$ & $\mathrm{Exact}\; p_{M,k}$ & $\mathrm{Decimal}\; p_{M,k}$ & Simulation\\
\hline
\hline
\endhead

$\T p_{2,2}$&$\frac{2}{3}$&$0.66667$&$0.66696$\\
$\T p_{2,0}$&$\frac{1}{3}$&$0.33333$&$0.33304$\\
\hline
$\T p_{3,3}$&$\frac{4} {15}$&$0.26667$&$0.26841$\\
$\T p_{3,1}$&$\frac{11} {15}$&$0.73333$&$0.73159$\\
\hline
$\T p_{4,4}$&$\frac{32}{525}$&$0.06095$&$0.06146$\\
$\T p_{4,2}$&$\frac{376}{525}$&$0.71619$&$0.71647$\\
$\T p_{4,0}$&$\frac{39}{175}$&$0.22286$&$0.22207$\\
\hline
$\T p_{5,5}$&$\frac{256} {33075}$&$0.00774$&$0.00755$\\
$\T p_{5,3}$&$\frac{12508} {33075}$&$0.37817$&$0.3749$\\
$\T p_{5,1}$&$\frac{20311} {33075}$&$0.61409$&$0.61755$\\
\hline
$\T p_{6,6}$&$\frac{4069}{7640325}$&$0.00054$&$0.00051$\\
$\T p_{6,4}$&$\frac{95552}{848925}$&$0.11256$&$0.11259$\\
$\T p_{6,2}$&$\frac{1814282}{2546775}$&$0.71239$&$0.71006$\\
$\T p_{6,0}$&$\frac{266683}{1528065}$&$0.17452$&$0.17684$\\
\hline
\caption[Comparison of analytic and simulated probabilities $p_{M,k}$ for the real truncated ensemble with $L=2$.]{Exact values of $p_{M,k}$ for the real truncated ensemble with $L=2$. These exact values are compared with the simulated results of $10000$ independent matrices.}
\end{longtable}

\begin{longtable}{|c|c|c|c|}
\hline
$L=3$ & $\mathrm{Exact}\; p_{M,k}$ & $\mathrm{Decimal}\; p_{M,k}$ & Simulation\\
\hline
\hline
\endfirsthead

\hline
$L=3$ & $\mathrm{Exact}\; p_{M,k}$ & $\mathrm{Decimal}\; p_{M,k}$ & Simulation\\
\hline
\hline
\endhead

$\T p_{2,2}$&$\frac{32}{15\pi}$&$0.67906$&$0.6785$\\
$\T p_{2,0}$&$1- \frac{32}{15\pi}$&$0.32094$&$0.3215$\\
\hline
$\T p_{3,3}$&$\frac{32} {35\pi}$&$0.29103$&$0.29265$\\
\hline
$\T p_{3,1}$&$1- \frac{32} {35\pi}$&$0.70898$&$0.70735$\\
\hline
$\T p_{4,4}$&$\frac{8192}{11025 \pi^2}$&$0.07529$&$0.07474$\\
$\T p_{4,2}$&$\frac{96}{35\pi} -\frac{16384}{11025 \pi^2}$&$0.72251$&$0.7235$\\
$\T p_{4,0}$&$1+\frac{8192}{11025 \pi^2} - \frac{96}{35 \pi}$&$0.20221$&$0.20176$\\
\hline
$\T p_{5,5}$&$\frac{8192} {72765 \pi^2}$&$0.01141$&$0.01137$\\
$\T p_{5,3}$&$\frac{4768} {3465 \pi}- \frac{16384} {72765 \pi^2}$&$0.41520$&$0.41723$\\
$\T p_{5,1}$&$1- \frac{4768}{3465 \pi} +\frac{8192} {72765 \pi^2} $&$0.57340$&$0.5714$\\
\hline
$\T p_{6,6}$&$\frac{33554432}{1092566475 \pi^3}$&$0.00099$&$0.0009$\\
$\T p_{6,4}$&$\frac{73842688}{52026975 \pi^2} -\frac{33554432}{364188825 \pi^3}$&$0.14084$& $0.13993$\\
$\T p_{6,2}$&$\frac{46784}{15015\pi}- \frac{147685376}{52026975 \pi^2}+ \frac{33554432}{364188825\pi^3}$&$0.70715$& $0.70648$\\
$\T p_{6,0}$&$1-\frac{46784}{15015\pi}+\frac{73842688}{52026975 \pi^2}- \frac{33554432} {1092566475 \pi^3}$&$0.15102$&$0.15269$\\
\hline
\caption[Comparison of analytic and simulated probabilities $p_{M,k}$ for the real truncated ensemble with $L=3$.]{Exact values of $p_{M,k}$ for the real truncated ensemble with $L=3$. These exact values are compared with the simulated results of $10000$ independent matrices.}
\end{longtable}

\begin{longtable}{|c|c|c|c|}
\hline
$L=8$ & $\mathrm{Exact}\; p_{M,k}$ & $\mathrm{Decimal}\; p_{M,k}$ & Simulation\\
\hline
\hline
\endfirsthead

\hline
$L=8$ & $\mathrm{Exact}\; p_{M,k}$ & $\mathrm{Decimal}\; p_{M,k}$ & Simulation\\
\hline
\hline
\endhead

$\T p_{2,2}$&$\frac{896}{1287}$&$0.69620$&$0.7003$\\
$\T p_{2,0}$&$\frac{391}{1287}$&$0.30381$&$0.2997$\\
\hline
$\T p_{3,3}$&$\frac{7168} {21879}$&$0.32762$&$0.32883$\\
$\T p_{3,1}$&$\frac{14711} {21879}$&$0.67238$&$0.67117$\\
\hline
$\T p_{4,4}$&$\frac{102760448}{1010569131}$&$0.10167$&$0.10117$\\
$\T p_{4,2}$&$\frac{244424320}{336856377}$&$0.72560$&$0.72606$\\
$\T p_{4,0}$&$\frac{174535723}{1010569131}$&$0.17271$&$0.17277$\\
\hline
$\T p_{5,5}$&$\frac{1174405120} {57602440467}$&$0.02039$&$0.02044$\\
$\T p_{5,3}$&$\frac{27164852224} {57602440467}$&$0.47159$&$0.47229$\\
$\T p_{5,1}$&$\frac{29263183123} {57602440467}$&$0.50802$&$0.50727$\\
\hline
$\T p_{6,6}$&$\frac{4810363371520}{1854356964327153}$&$0.00259$&$0.00266$\\
$\T p_{6,4}$&$\frac{1088085255258112}{5563070892981459}$&$0.19560$&$0.1949$\\
$\T p_{6,2}$&$\frac{3808688634609280} {5563070892981459}$&$0.68464$&$0.68399$\\
$\T p_{6,0}$&$\frac{651865912999507} {5563070892981459}$&$0.11718$&$0.11845$\\
\hline
\caption[Comparison of analytic and simulated probabilities $p_{M,k}$ for the real truncated ensemble with $L=8$.]{Exact values of $p_{M,k}$ for the real truncated ensemble with $L=8$. These exact values are compared with the simulated results of $10000$ independent matrices.}
\label{tab:TOEpnkL8}
\end{longtable}

\subsection{Strongly orthogonal limit of the correlation kernel for the real truncated ensemble}
\label{app:TOElims}

From (\ref{def:Rdecomp}) we see that the limit $M\to\infty,L=1$ corresponds to $\bD$ begin strongly orthogonal. In this limit, the quaternion determinant matrix from Proposition \ref{prop:TOEcorrelne} becomes \cite{KSZ2010, Forrester2010a}
\begin{align}
\nonumber \left[ \begin{array}{cc}
\kappa_{rr}^{L=1}(x_i,x_j) & \kappa_{rc}^{L=1}(x_i,z_m)\\
\kappa_{cr}^{L=1}(z_l,x_j) & \kappa_{rr}^{L=1}(z_l,z_m)
\end{array}
\right]_{{i,j=1,...,n_1\atop l,m=1,...,n_2}},
\end{align}
where
\begin{align}
\nonumber &\kappa_{rr}^{L=1}(x,y):=\bK_{rr}(x,y)_T\Big|_{L=1}\\
\nonumber &=\frac{1}{\pi}\left[\begin{array}{cc}
\frac{\sqrt{1-y^2}}{\sqrt{1-x^2}(1-xy)} & \frac{(x-y)}{\sqrt{(1-x^2)(1-y^2)}(1-xy)^2}\\
\sgn(y-x)\;\mathrm{arcsin}\frac{\sqrt{(1-x^2)(1-y^2)}}{1-xy} & \frac{\sqrt{1-x^2}}{\sqrt{1-y^2}(1-xy)}
\end{array} \right],\\
\nonumber &\kappa_{rc}^{L=1}(x,z):=\bK_{rc}(x,z)_T\Big|_{L=1}\\
\nonumber &=\frac{1}{\pi}\left[\begin{array}{cc}
\frac{-i(x-\bar{z})}{\sqrt{(1-x^2)|1-z^2|}(1-x\bar{z})^2} & \frac{(x-z)}{\sqrt{(1-x^2)|1-z^2|}(1-xz)^2}\\
\frac{-i\sqrt{1-x^2}}{\sqrt{|1-\bar{z}^2|}(1-x\bar{z})} & \frac{\sqrt{1-x^2}}{\sqrt{|1-z^2|}(1-xz)}
\end{array} \right],\\
\nonumber &\kappa_{cr}^{L=1}(z,x):=\bK_{cr}(z,x)_T\Big|_{L=1}\\
\nonumber &=\frac{1}{\pi}\left[\begin{array}{cc}
\frac{\sqrt{(1-x^2)}}{\sqrt{|1-z^2|}(1-zx)^2} & \frac{(z-x)}{\sqrt{(1-x^2)|1-z^2|}(1-zx)^2}\\
\frac{i\sqrt{1-x^2}}{\sqrt{|1-\bar{z}^2|}(1-\bar{z}x)} & \frac{i(\bar{z}-x)}{\sqrt{(1-x^2)|1-z^2|}(1-\bar{z}x)^2}
\end{array} \right],\\
\nonumber &\kappa_{cc}^{L=1}(z_1,z_2):=\bK_{cc}(z_1,z_2)_T\Big|_{L=1}\\
\label{eqn:TlimL1} &=\frac{1}{\pi}\left[\begin{array}{cc}
\frac{i(\bar{z}_2-z_1)}{\sqrt{|1-z_1^2||1-z_2^2|}(1-z_1\bar{z}_2)^2} & \frac{(z_1-z_2)}{\sqrt{|1-z_1^2||1-z_2^2|}(1-z_1 z_2)^2}\\
\frac{(\bar{z}_1-\bar{z}_2)}{\sqrt{|1-z_1^2||1-z_2^2|}(1-\bar{z}_1 \bar{z}_2)^2} & \frac{i(\bar{z}_1-z_2)}{\sqrt{|1-z_1^2||1-z_2^2|}(1-\bar{z}_1 z_2)^2}
\end{array} \right].
\end{align}

\end{document}